\documentclass[useAMS,usenatbib]{mn2e}
\usepackage{float} 
\usepackage{txfonts}
\usepackage{graphicx}
\usepackage{textcomp}
\usepackage{natbib}
\usepackage{lscape}
\usepackage{deluxetable}
\usepackage{mathrsfs} 

\newcommand{\um}{\textmu m }
\newcommand{\uu}{\textmu m}

\bibpunct[]{(}{)}{;}{a}{}{,}

\title[NIR to MIR spectrum of QSOs]{The near-to-mid infrared spectrum of quasars}

\author[A. Hern\'an-Caballero et al.]{
Antonio Hern\'an-Caballero,$^{1}$\thanks{E-mail: ahc@denebola.org} 
Evanthia Hatziminaoglou,$^2$
Almudena Alonso-Herrero,$^3$\newauthor
and Silvia Mateos$^4$\\
$^{1}$Departamento de Astrof\'isica y CC. de la Atm\'osfera, Facultad de CC. F\'isicas, Universidad Complutense de Madrid, E-28040 Madrid, Spain\\
$^{2}$European Southern Observatory, Karl-Schwarzschild-Str. 2, 85748 Garching bei M\"unchen, Germany\\
$^{3}$Centro de Astrobiolog\'ia, (CAB, CSIC-INTA), ESAC Campus, E-28692 Villanueva de la Ca\~nada, Madrid, Spain\\
$^{4}$Instituto de F\'isica de Cantabria, CSIC-UC, Avenida de los Castros s/n, 39005, Santander, Spain}

\begin{document}
\date{Accepted ........ Received ........;}

\pagerange{\pageref{firstpage}--\pageref{lastpage}} \pubyear{2016}

\maketitle

\label{firstpage}

\begin{abstract}
We analyse a sample of 85 luminous ($\log$($\nu L_\nu$(3\uu)/erg s$^{-1}$)$>$45.5) quasars with restframe $\sim$2--11 \um spectroscopy from AKARI and \textit{Spitzer}. Their high luminosity allows a direct determination of the near-infrared quasar spectrum free from host galaxy emission. 
A semi-empirical model consisting of a single template for the accretion disk and two blackbodies for the dust emission successfully reproduces the 0.1--10 \um spectral energy distributions (SEDs).
Excess emission at 1--2 \um over the best-fitting model suggests that hotter dust is necessary in addition to the $\sim$1200 K blackbody and the disk to reproduce the entire near-infrared spectrum. Variation in the extinction affecting the disk and in the relative strength of the disk and dust components accounts for the diversity of individual SEDs. 
Quasars with higher dust-to-disk luminosity ratios show slightly redder infrared continua and less prominent silicate emission.
We find no luminosity dependence in the shape of the average infrared quasar spectrum.  
We generate a new quasar template that covers the restframe range 0.1--11 \uu, and separate templates for the disk and dust components. Comparison with other infrared quasar composites suggests that previous ones are less reliable in the 2--4 \um range. Our template is the first one to provide a detailed view of the infrared emission on both sides of the 4 \um bump.
\end{abstract}

\begin{keywords}
galaxies:active -- infrared:galaxies -- quasars: general -- quasars: emission lines
\end{keywords} 

\section{Introduction} 

The current paradigm of an optically thick accretion disk around a super-massive black hole, described for the first time in \citet{Shakura73}, predicts the characteristic blue continuum of quasar spectra to extend all the way to the near-infrared (NIR). A large fraction of this emission is thought to be absorbed and then re-radiated in the NIR and mid-infrared (MIR) by dust surrounding the nucleus \citep[e.g.][]{Rees69,Neugebauer79,Barvainis87} that masks the underlying continuum and makes thus a direct identification of the exact shape of the intrinsic NIR continuum nearly impossible. However, both theoretical considerations \citep[e.g.][]{Hubeny01} and measurements of polarised light \citep{Kishimoto08} favour a power-law extension of the optical spectrum to the NIR, though the actual spectral shape remains unclear. At the same time, and while the geometry and composition of the dust surrounding the nucleus are still a matter of debate \citep[e.g.][]{Antonucci93,Netzer15}, several properties of the innermost regions of active galactic nuclei (AGN) can be constrained by NIR observations, such as the size of the region where the dust is confined \citep{Suganuma06}, the dust covering factor \citep{Mor09,Mateos16}, and its temperature \citep{Glikman06,Kim15}.

An extra complication in defining the shape of the NIR emission of AGN comes from the contamination by stellar emission in the host galaxy. 
The stellar emission peaks at $\sim$1.6 \um while the AGN output reaches a minimum at $\sim$1 \uu. This implies that even if the AGN dominates over the host galaxy in bolometric luminosity, stellar emission may dominate in the NIR \citep[see ][ and references therein]{Merloni15}.
Removing contamination from the host requires either resolving the emission on sub-kpc scales (challenging even for low redshift galaxies) or performing some kind of spectral decomposition, the outcome of which depends on assumptions about the shape of the spectrum of both the host galaxy and the AGN \citep{Hernan-Caballero15}.
Spectral indices such as the equivalent width of the 2.3 \um CO absorption band can constraint the luminosity of the stellar component \citep{Burtscher15}, but the observations required are currently viable only for local AGN.

The contrast between AGN and their host galaxies increases with AGN luminosity. In a sample of hard X-ray selected AGN with restframe 2--10 keV intrinsic luminosities between 10$^{42}$ and 10$^{46}$ erg s$^{-1}$ \citet{Mateos15} found that the median contribution from the host galaxy to the observed 3.4 \um flux in type 1 AGN is $\sim$40\% at 10$^{43}$ erg s$^{-1}$ but only $\sim$10\% at 10$^{45}$ erg s$^{-1}$.
Therefore luminous quasars are the ideal targets to study the NIR emission of the AGN clean of contamination from the host galaxy.

Broadband observations with the InfraRed Array Camera (IRAC) onboard \textit{Spitzer} revealed the prevalence in quasar spectral energy distributions (SEDs) of a broad bump centered at $\sim$3--4 \um \citep{Hatziminaoglou05,Richards06}.
This bump is also evident in the \textit{Spitzer} Infrared Spectrograph (IRS) spectra of high redshift quasars \citep[e.g.][]{Hernan-Caballero09,Mor09,Deo11,Hernan-Caballero11} as well as the AKARI 2.5--5 \um spectra of some low redshift ones \citep{Kim15}.

The first composite NIR (0.58--3.5 \uu) quasar spectrum was obtained from the individual spectra of 27 low-redshift quasars observed with the NASA Infrared Telescope Facility \citep{Glikman06}. In this work, the authors showed the curvature in the composite spectrum to be well reproduced by the combination of a single power-law that dominates the optical emission, and a hot ($\sim$1250 K) black body component, that accounts for the 3 \um bump. On the other hand, \citet{Kim15} found that two blackbodies at $\sim$1100 K and 220 K are sufficient to reproduce the IR SEDs and AKARI 2.5--5 \um spectra of 83 nearby quasars and type 1 Seyferts after subtraction of the disk emission, with no need for a hotter dust component.
Such reported differences may be in part caused by variations in the spectral coverage and/or luminosity of the samples. \citet{Gallagher07} found an increase in the 1.8--8 \um slope with luminosity in a sample of radio-quiet SDSS quasars, but in a sample of 180 SDSS quasars with \textit{Spitzer}/IRS spectroscopy \citet{Hill14} concluded that the intrinsic MIR quasar continuum is luminosity-independent, and that differences in the observed spectrum are consistent with higher contamination from the host galaxy at low luminosities.

In this work we overcome the main drawbacks of previous studies of the NIR emission of quasars by using a large sample of luminous quasars with spectroscopic observations in the restframe NIR and MIR ranges as well as uniform broadband photometry covering from the UV to the MIR. The high luminosity of the sample ensures negligible contamination from the host galaxy, while a careful subtraction of the accretion disk emission allows for a detailed analysis of the dust contribution.
In \S2 we describe the sample selection, followed by details on the stitching of the AKARI and IRS spectra (\S3) and the compilation of the broadband photometry (\S4). \S5 discusses the relation between the optical and NIR spectral indices and the NIR to optical luminosity ratio. \S6 describes the fitting of the observed SEDs with a disk+dust model of the quasar emission. In \S7 we present the composite NIR to MIR spectrum of the whole sample and discuss trends with the quasar luminosity and NIR-to-optical luminosity ratio, while in \S8 we analyse the strength of the NIR hydrogen recombination lines and PAH features. \S9 presents our new quasar template as well as templates for the disk and dust components. Finally, \S10 summarises our main results.

\section{Sample selection}

Our parent sample is the list of extragalactic sources with \textit{Spitzer}/IRS spectra in the version 7 of the Cornell Atlas of \textit{Spitzer}/IRS Sources \citep[CASSIS;][]{Lebouteiller11,Lebouteiller15}.
A careful identification of the optical counterpart was performed as part of the preparation work for the Infrared Database of Extragalactic Observables with \textit{Spitzer} (IDEOS; Spoon et al. in prep.). 
Redshifts were compiled from the NASA Extragalactic Database (NED), and validated with the \textit{Spitzer}/IRS spectrum using cross-correlation with templates \citep[see ][ for details]{Hernan-Caballero12,Hernan-Caballero16}.

We select from CASSIS all the sources meeting the following criteria:
i) optical spectroscopic redshift; ii) classified as quasar or Seyfert 1 in NED and/or the SDSS Data Release 12 (DR12)\footnote{Four sources with unclear or disputed classifications were excluded from the sample. These are: 3C 273, 3C 318, IRAS F10214+4724, and [HB89] 1556+335.}; iii) NIR luminosity $\log$ $\nu$L$_\nu$(3\uu)/erg s$^{-1}$ $>$ 45.5; iv) the \textit{Spitzer}/IRS observations cover at least the restframe 2.5--5 \um range.
Condition iv) is met by sources with observations at least for the SL2 (5.15--7.5 \uu) and SL1 (7.5--14 \uu) modules of IRS for $z$$>$1, as well as those with SL1 and LL2 (14-21 \uu) observations for $z$$>$2. 

There are 76 CASSIS sources meeting all four criteria. In addition, there are nine low redshift quasars meeting all but iv) which also were observed in the 2.5--5 \um range with AKARI \citep{Kim15}. Since their combined AKARI+IRS spectra cover the entire 2.5--5 \um range in the restframe, we also include these nine sources in the sample.

\begin{figure} 
\includegraphics[width=8.4cm]{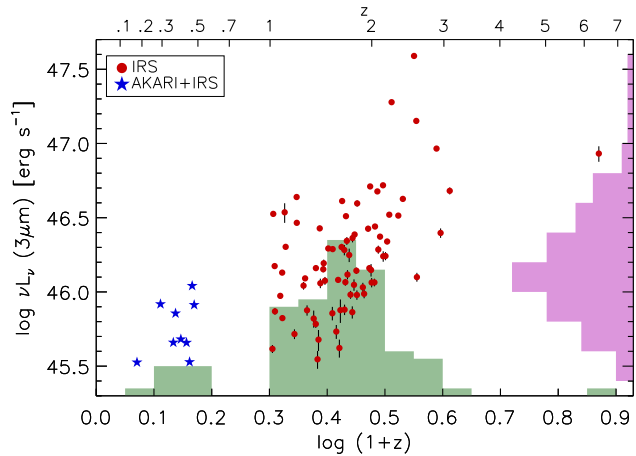}
\caption[]{Distribution in the restframe 3 \um luminosity vs redshift plane for the sources in our sample. The solid histograms represent the number counts in same-sized bins of $\log$(1+$z$) and $\log$ $\nu L_\nu$(3\uu).\label{fig:Lz}}
\end{figure}

The final sample containing 85 sources at 0.17$<$$z$$<$6.42 (average $z$: 1.66) is listed in Table \ref{sampletable}.
Figure \ref{fig:Lz} shows the redshift and near-IR luminosity distributions of the sample.

\section{Stitching}\label{sec_stitching}

We stitch the different orders in the \textit{Spitzer}/IRS spectra following the procedure described in \citet{Hernan-Caballero16}. Very briefly, we take the LL2 spectrum as reference and scale the three SL modules to match the SL1 and LL2 spectra in the overlap region. Further adjustments are made to the scaling factors of SL3 and SL2, if needed. Because the IR emission from the galaxy is outshone by the AGN, which is unresolved in \textit{Spitzer} observations, differences in aperture size between the SL and LL modules are unimportant and the scaling factors applied are usually close to unity. However, inaccurate slit placement for one of the modules may cause a significant loss in flux, which would translate into a large scaling factor. We find no such cases in our sample.

We also stitch together the AKARI and \textit{Spitzer} spectra for the 9 low redshift quasars. The AKARI spectra were obtained using a squared slit 1' wide and extracted in a 1' $\times$ 7.5'' aperture, which includes most of the extended emission from the host galaxy \citep[see][ for details]{Kim15}. However, the 3.3 \um PAH feature is undetected in the 9 sources, indicating that the AGN overwhelms the emission from the host. A gap of $\sim$0.2 \um between the long wavelength end of the AKARI spectrum and the start of the \textit{Spitzer} spectrum complicates the stitching. We overcome this by fitting with a straight line the 5.15--7 \um range of the \textit{Spitzer} spectrum and stitching the AKARI spectrum to its extrapolation at 4.9 \uu. 
The mean (median) value of the scaling factors for the AKARI spectra is 0.997 (1.000), which suggests that the contamination from the host galaxy is negligible. The standard deviation is 0.135, fully consistent with uncertainties of $\sim$10\% in the absolute flux calibration of both \textit{Spitzer} and AKARI spectra.

\section{Broadband photometry}

In addition to the \textit{Spitzer} and AKARI spectra, we compiled optical and NIR broadband photometry for the sources in the sample.

We retrieved PSF photometry in the $u$, $g$, $r$, $i$, and $z$ bands from the SDSS DR12 SkyServer\footnote{http://skyserver.sdss.org/dr12}.
We found matches for 71 of our sources using a 6'' search radius. 
For the remaining 14 sources outside of the sky area covered by SDSS we searched for any available optical photometry in NED. We found measurements at least in the $R$ or $V$ bands for all but two sources.
We corrected for Galactic extinction following \citet{Schlegel98}. The median extinction corrections for the $u$, $g$, $r$, $i$, and $z$ bands are 0.051, 0.038, 0.027, 0.020, and 0.015 magnitudes, respectively.

We queried the NASA/IPAC Infrared Science Archive (IRSA) for $JHK$ photometry from the Two Micron All Sky Survey \citep[2MASS;][]{Skrutskie06} Point Source Catalog (PSC). Using a search radius of 6'' we obtained results for 47 sources (all of them within 2'' of the \textit{Spitzer}/IRS position).
We took $JHK$ photometry for 15 additional sources within the SDSS area from the SDSS Data Release 7 (DR7) quasar catalogue \citep{Schneider10} and the SDSS Data Release 9 (DR9) quasar catalogue \citep{Paris12}. These catalogues contain forced aperture photometry on the 2MASS images at the SDSS coordinates, and include all $>$2$\sigma$ detections (compared to $>$5$\sigma$ for the 2MASS PSC).
For 14 sources in the Southern hemisphere we retrieved deeper $JHKs$ photometry from the VISTA Hemisphere Survey \citep[VHS;][]{McMahon13} Data Release 3.
In addition, we obtained $YJHK$ photometry for 24 sources in the areas covered by the UKIRT Infrared Deep Sky Survey \citep[UKIDSS;][]{Lawrence07}. Since the VHS and UKIDSS data are deeper than 2MASS, we use them in the cases where both are available.
There are 9 sources with no NIR photometry from either VHS, UKIDSS or 2MASS. We searched NED for any NIR measurements from pointed observations and found results for three of them (including one K-band upper limit).

In summary, we obtained measurements in five optical bands for $\sim$85\% of the sample, and in 3 to 4 NIR bands for $\sim$90\% of the sources. This ensures a robust estimate of the luminosity and optical spectral index of the accretion disk component. For the three sources with only one or two data points in the restframe optical range, we obtain a rough estimate of the disk luminosity by fixing the spectral index (see \S\ref{sec:disk}).

\begin{figure*} 
\includegraphics[width=5.8cm]{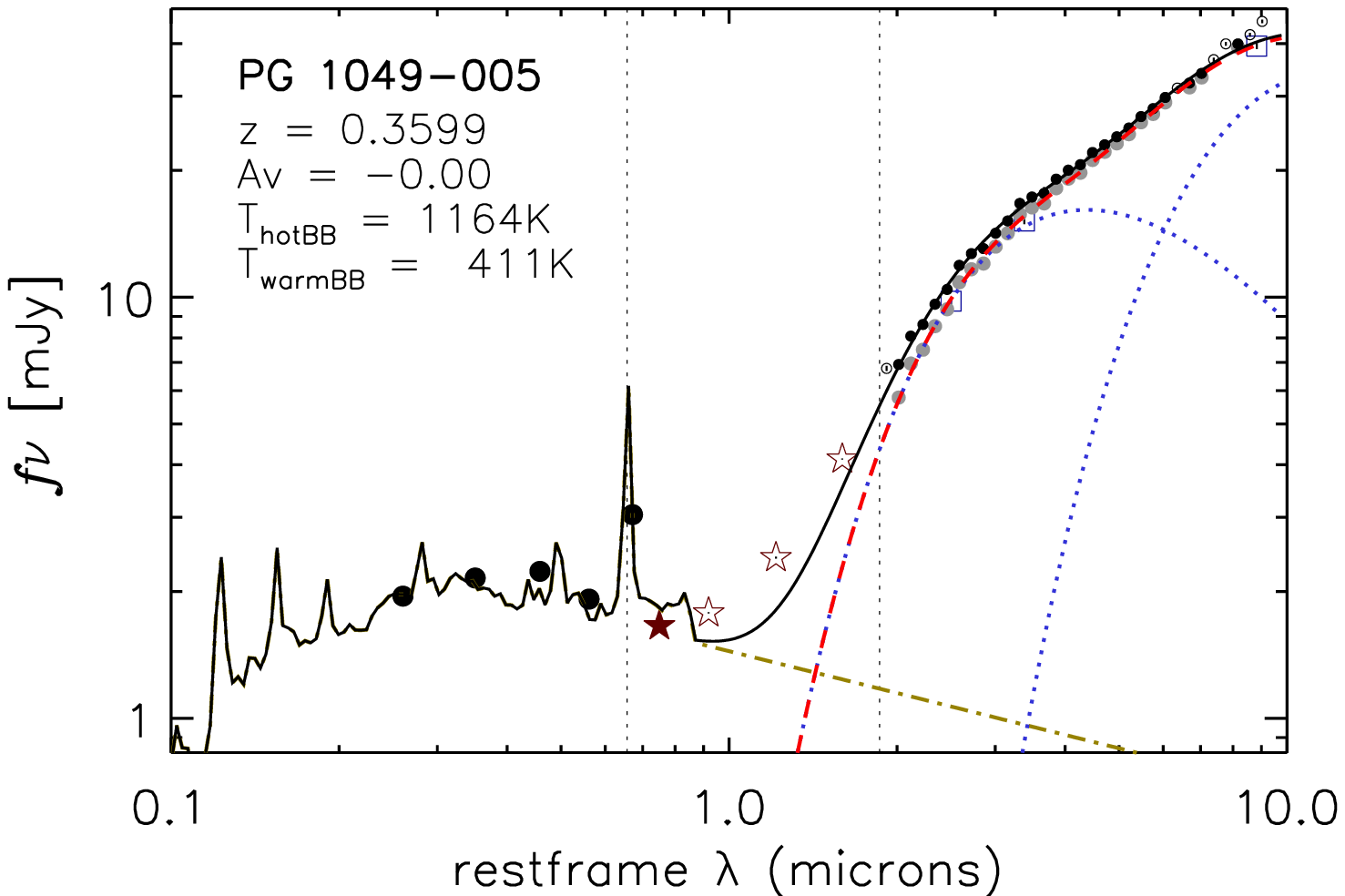}
\includegraphics[width=5.8cm]{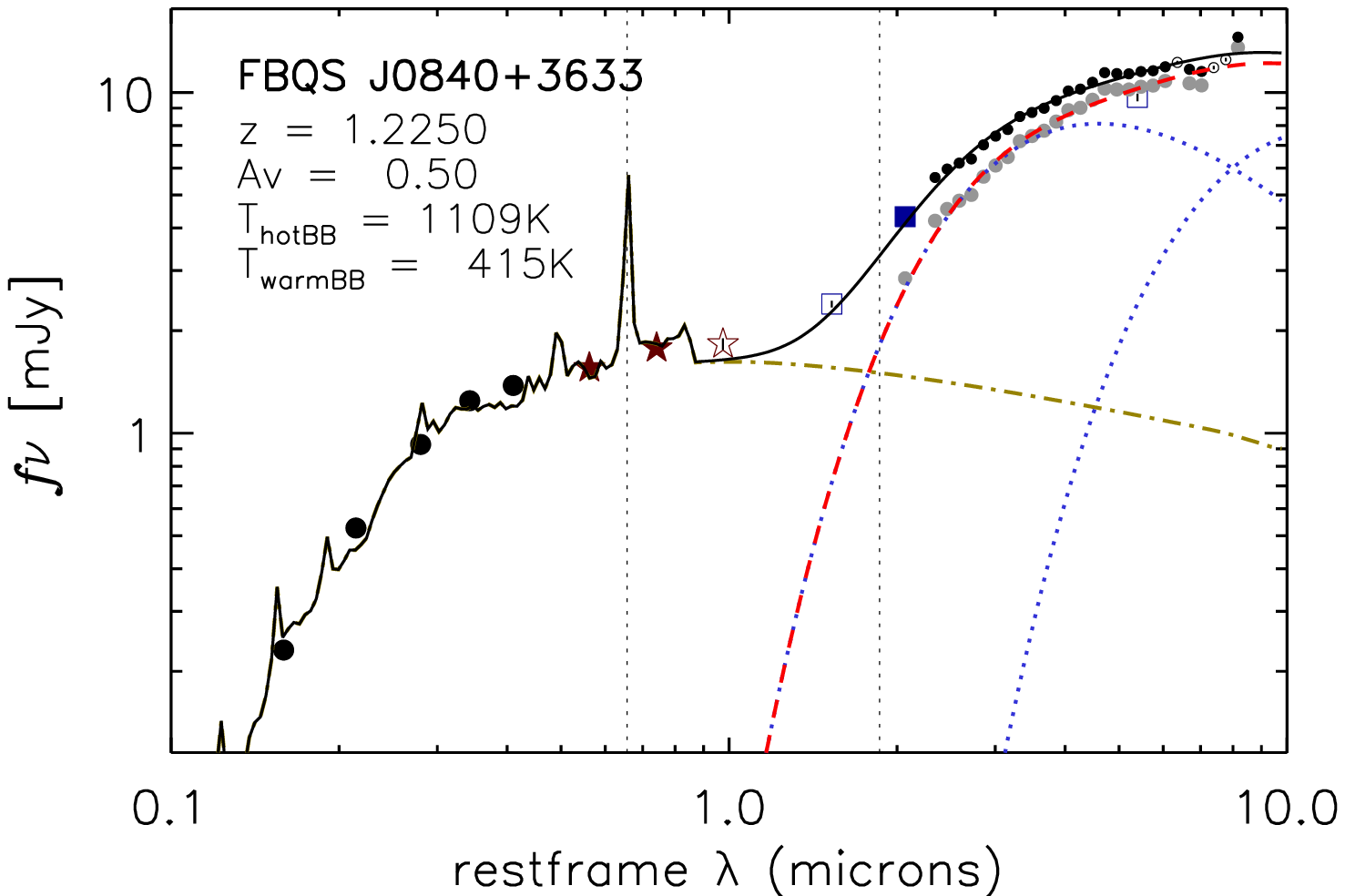}
\includegraphics[width=5.8cm]{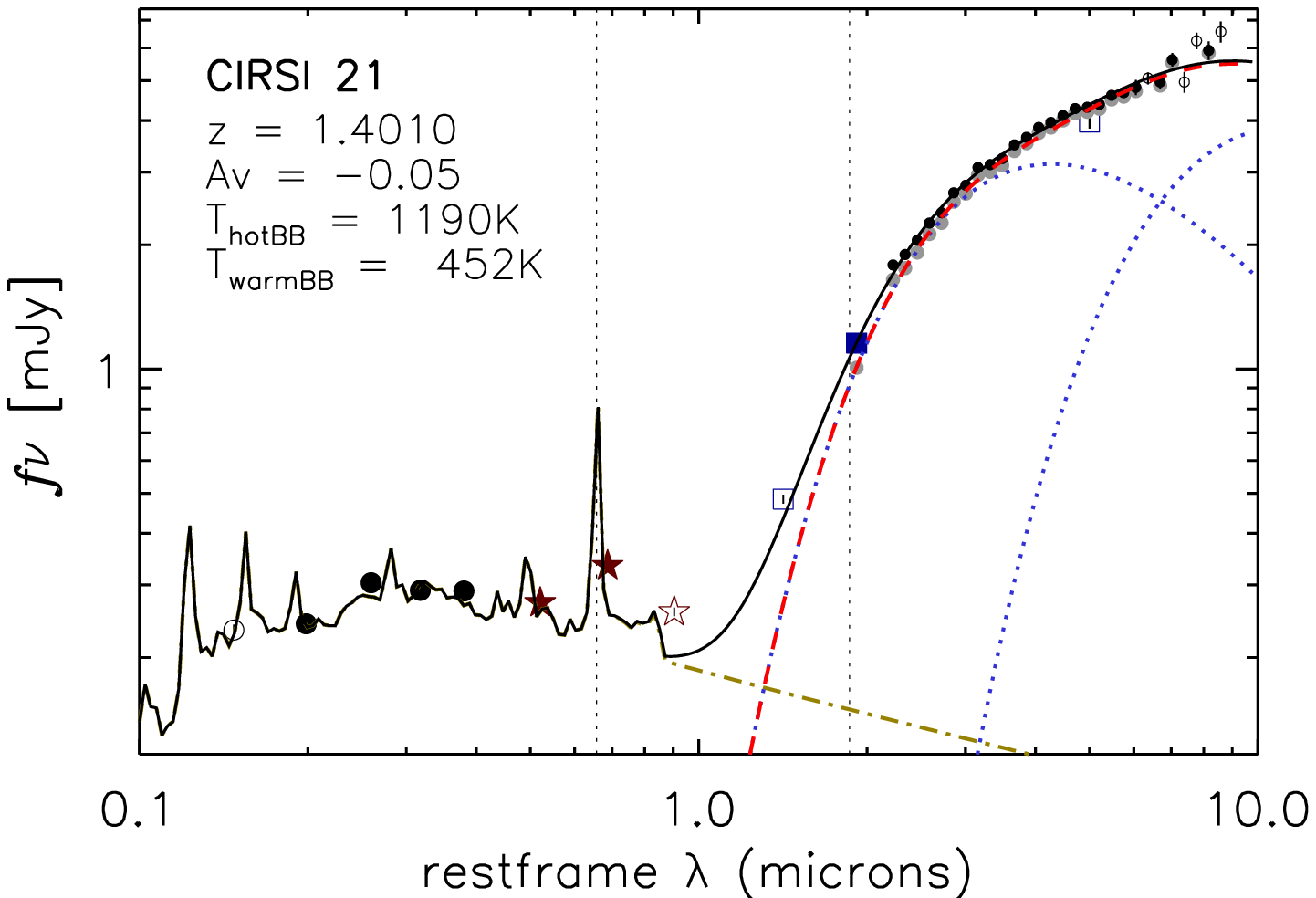}
\includegraphics[width=5.8cm]{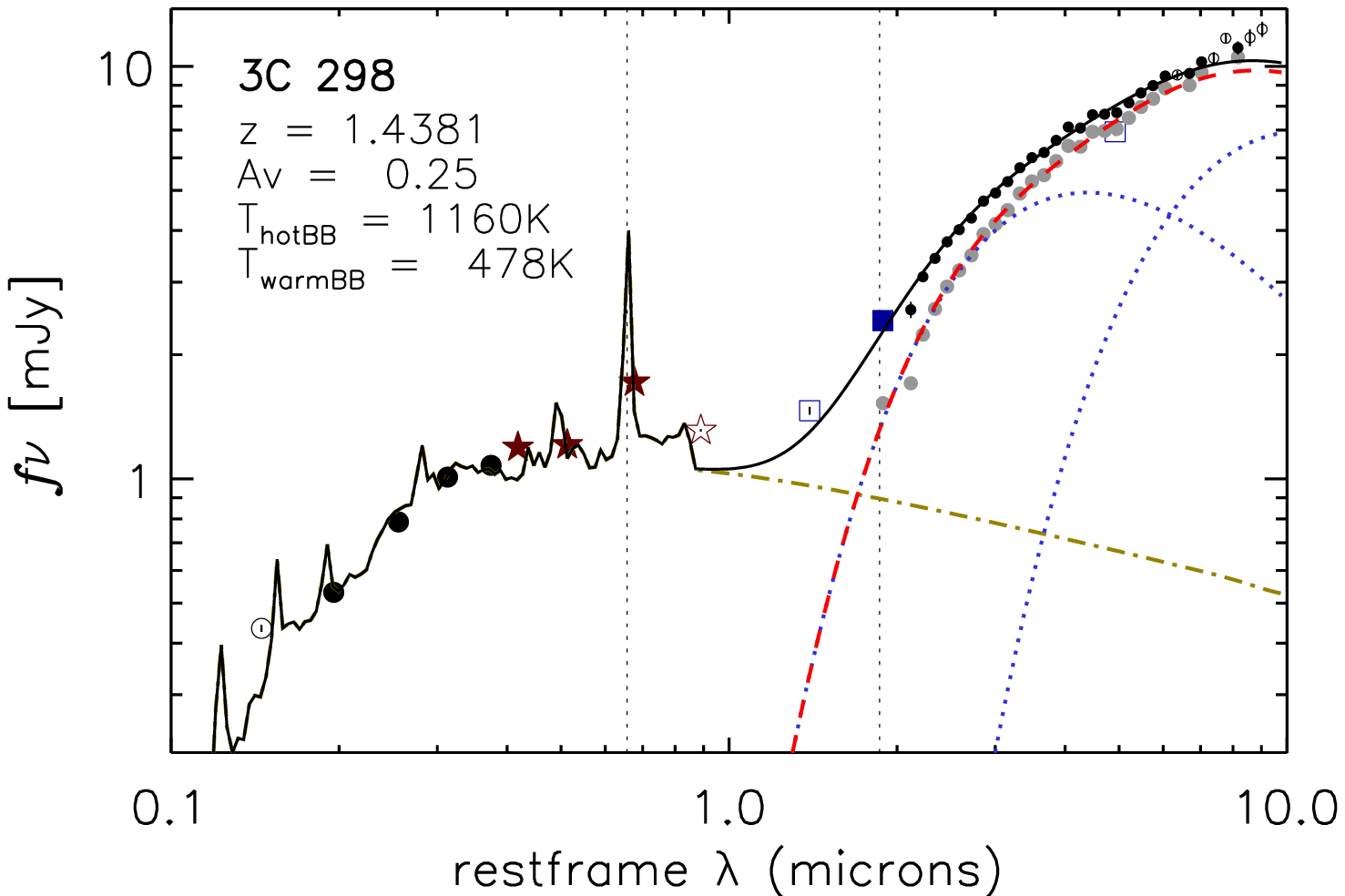}
\includegraphics[width=5.8cm]{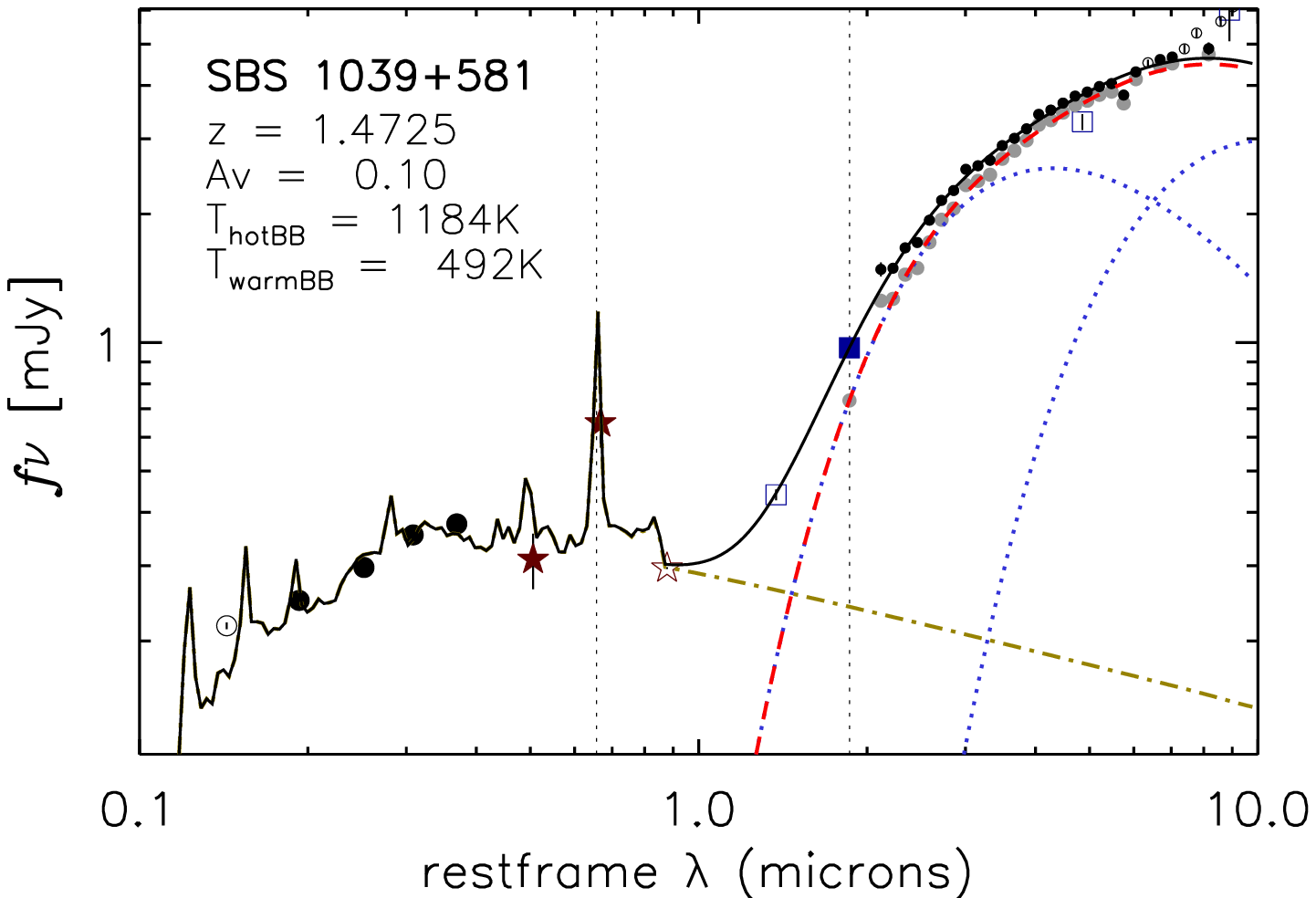}
\includegraphics[width=5.8cm]{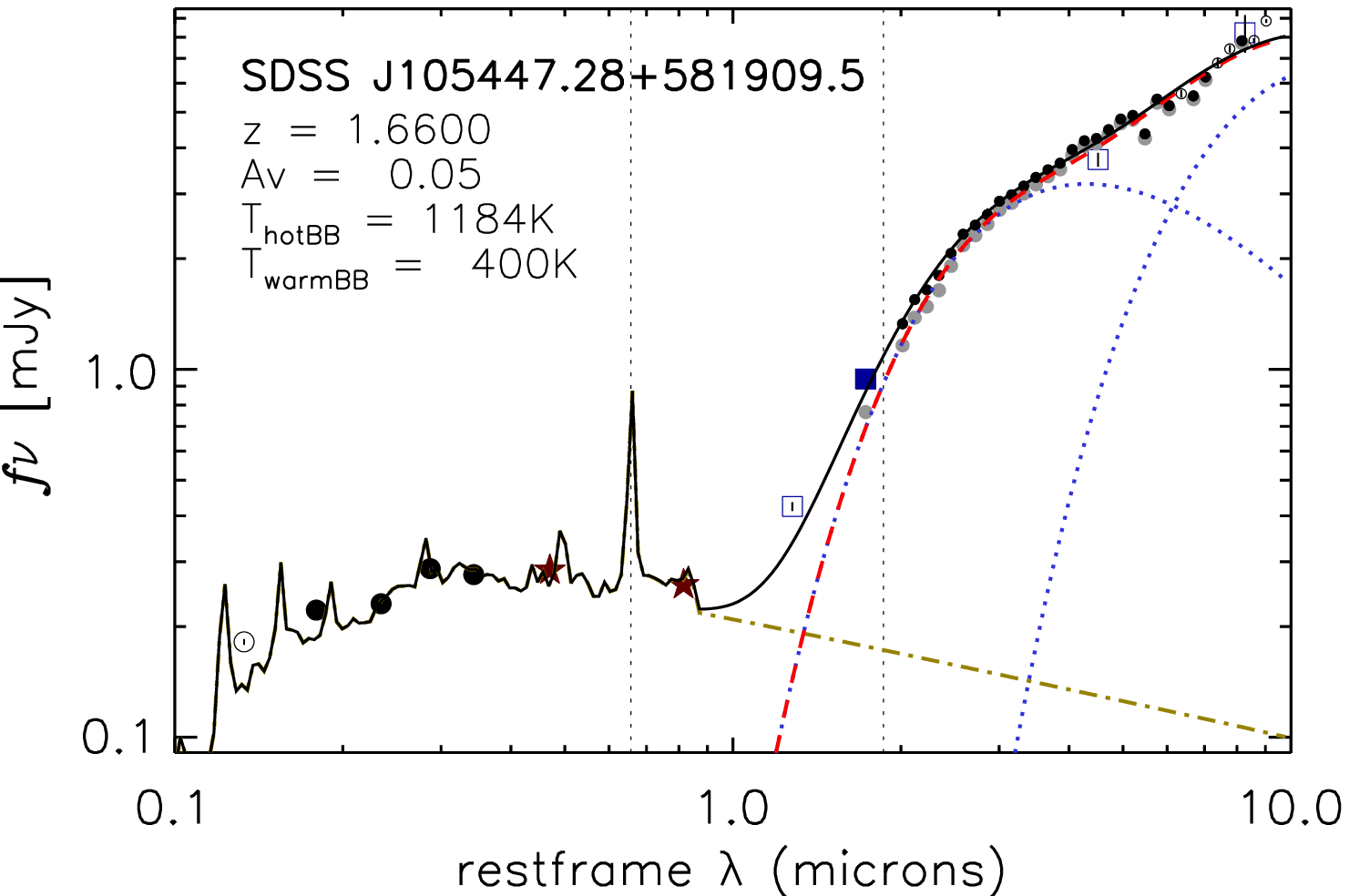}
\includegraphics[width=5.8cm]{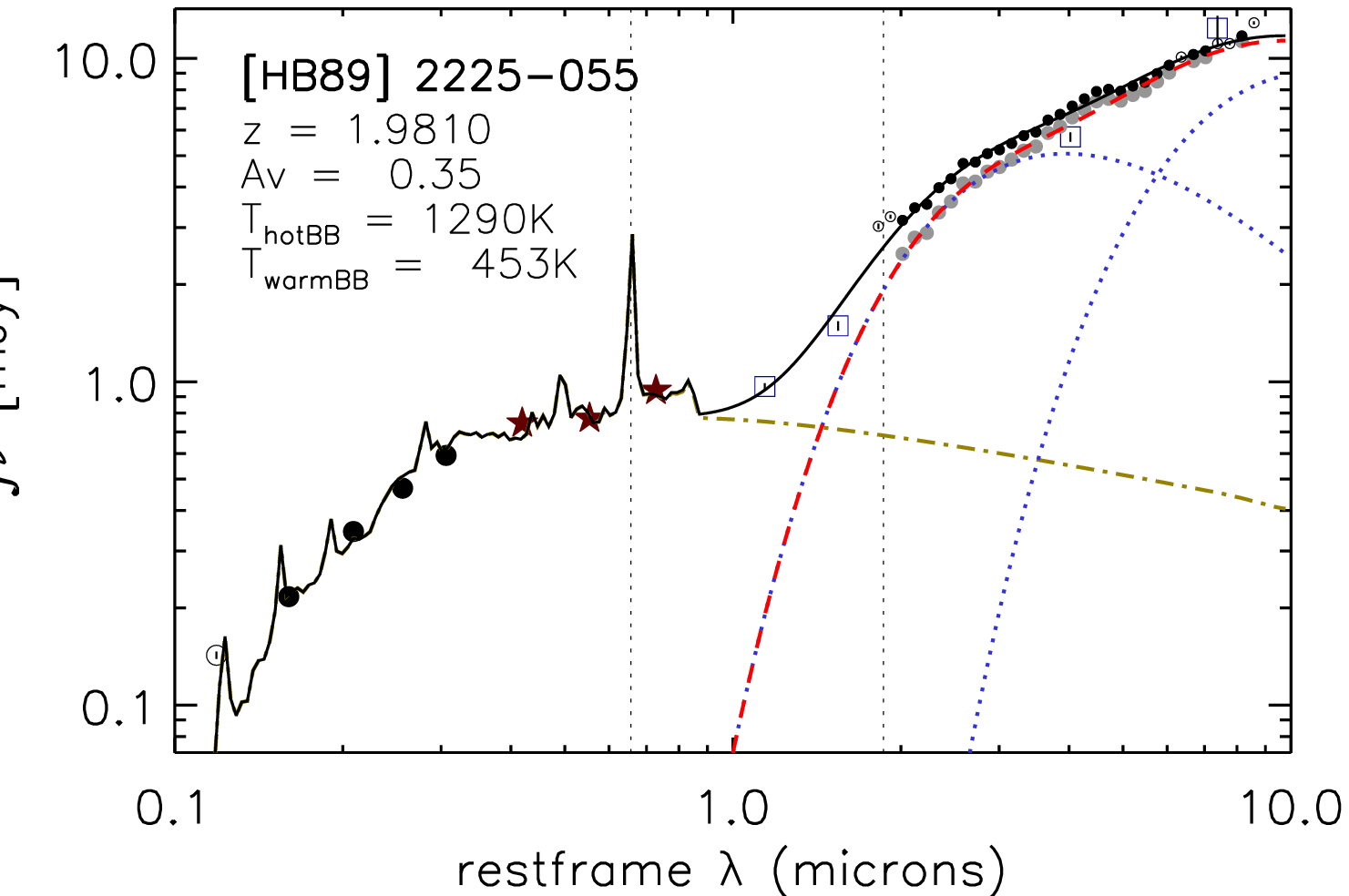}
\includegraphics[width=5.8cm]{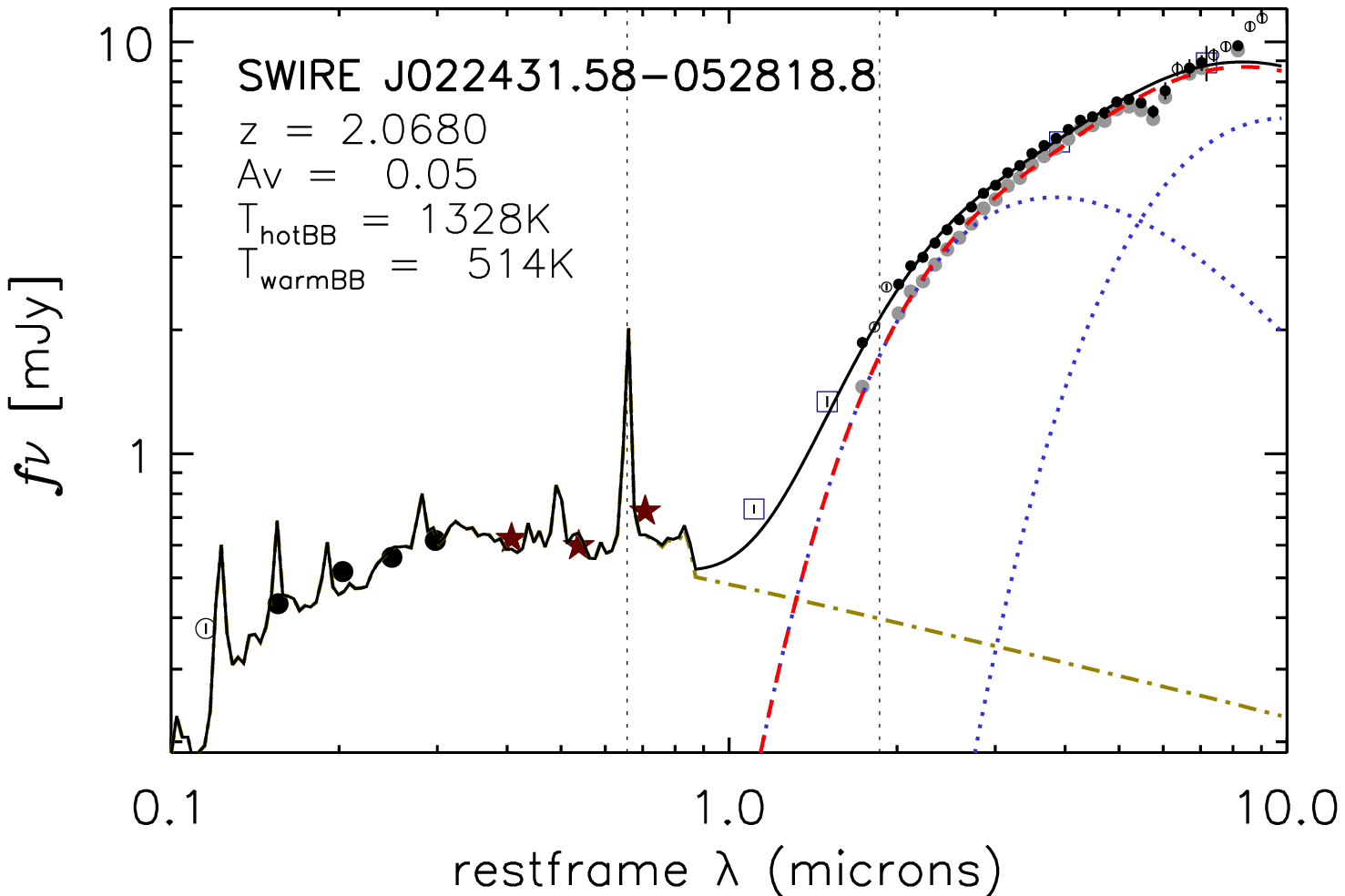}
\includegraphics[width=5.8cm]{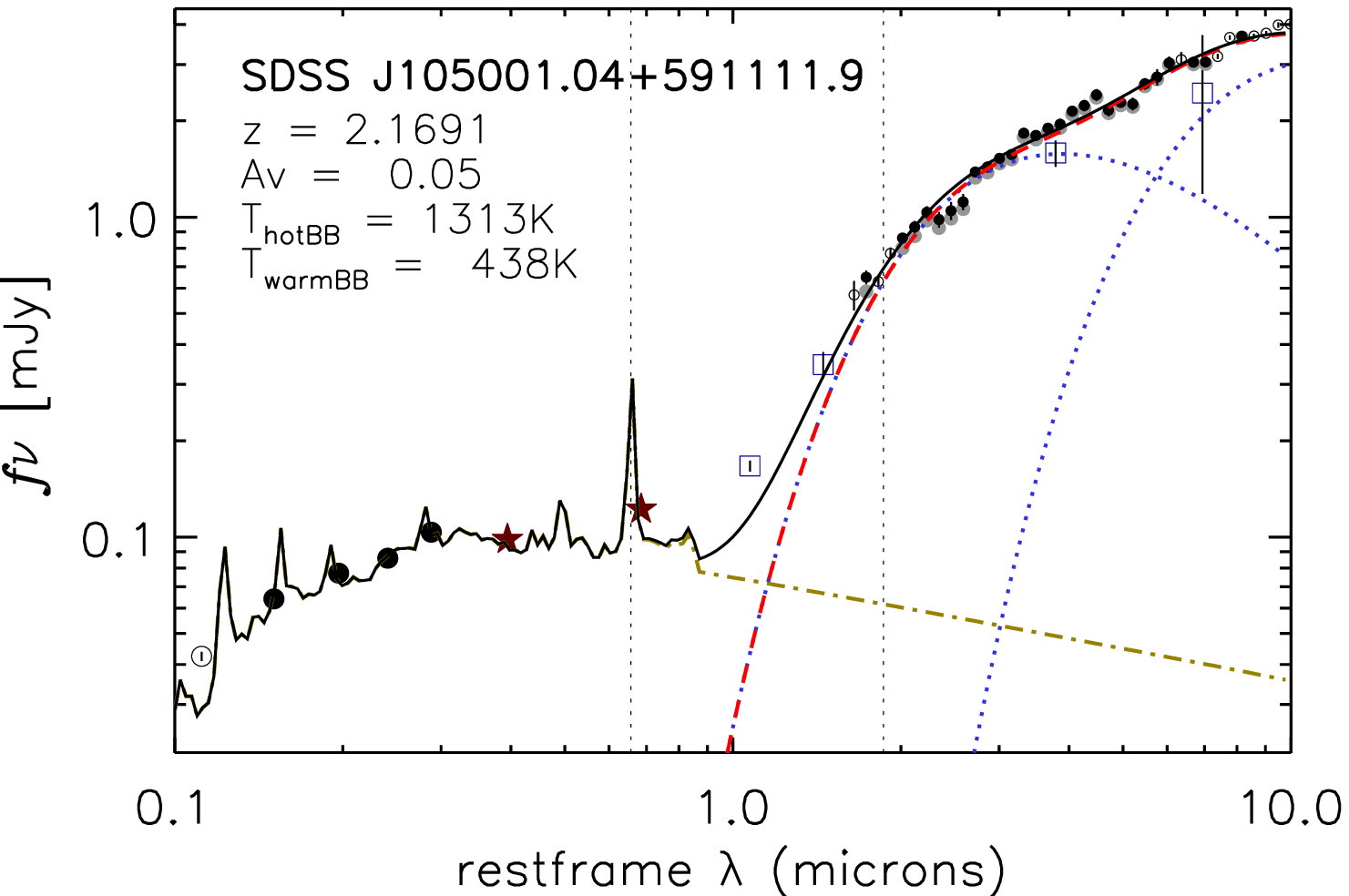}
\includegraphics[width=5.8cm]{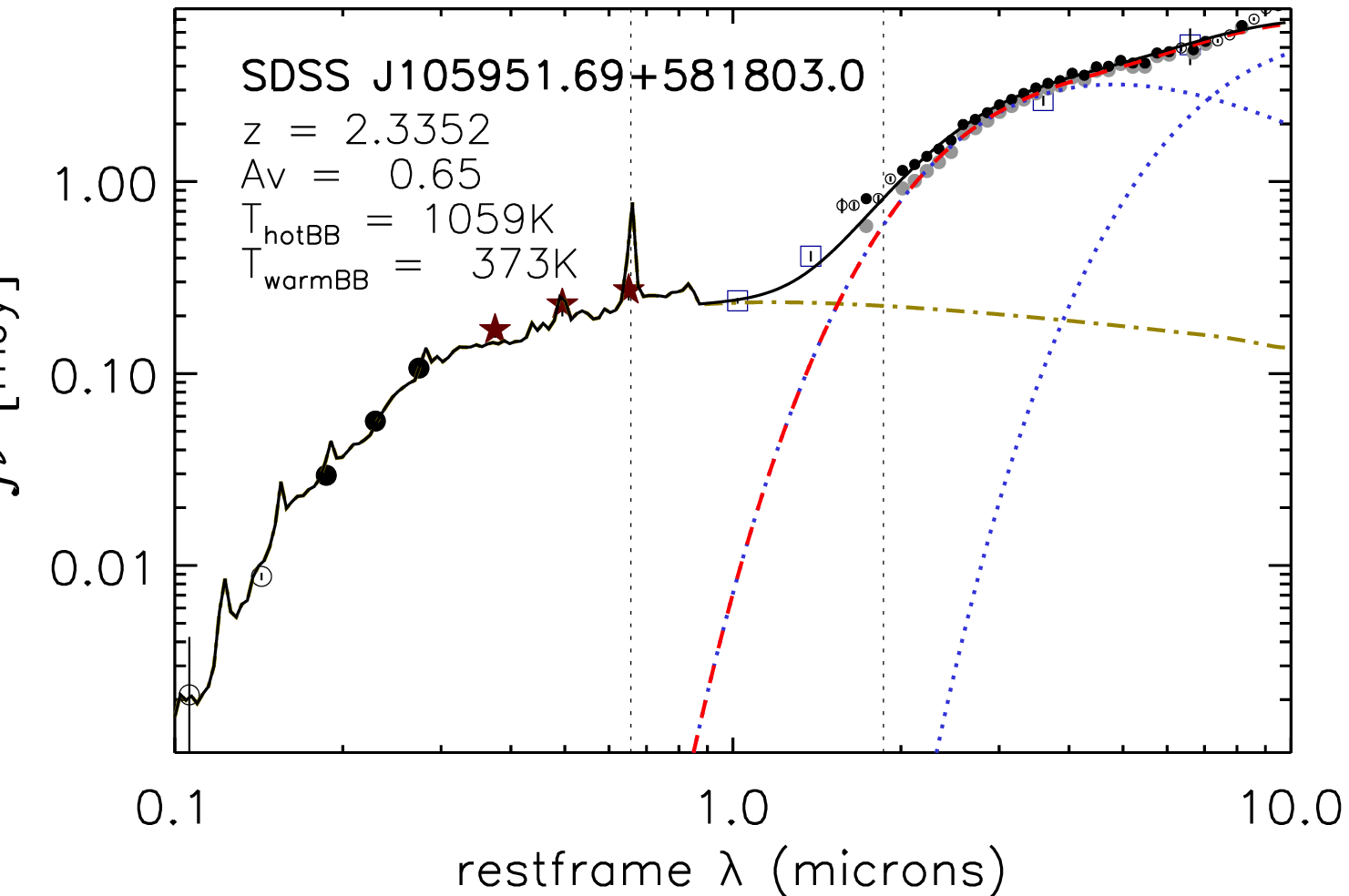}
\includegraphics[width=5.8cm]{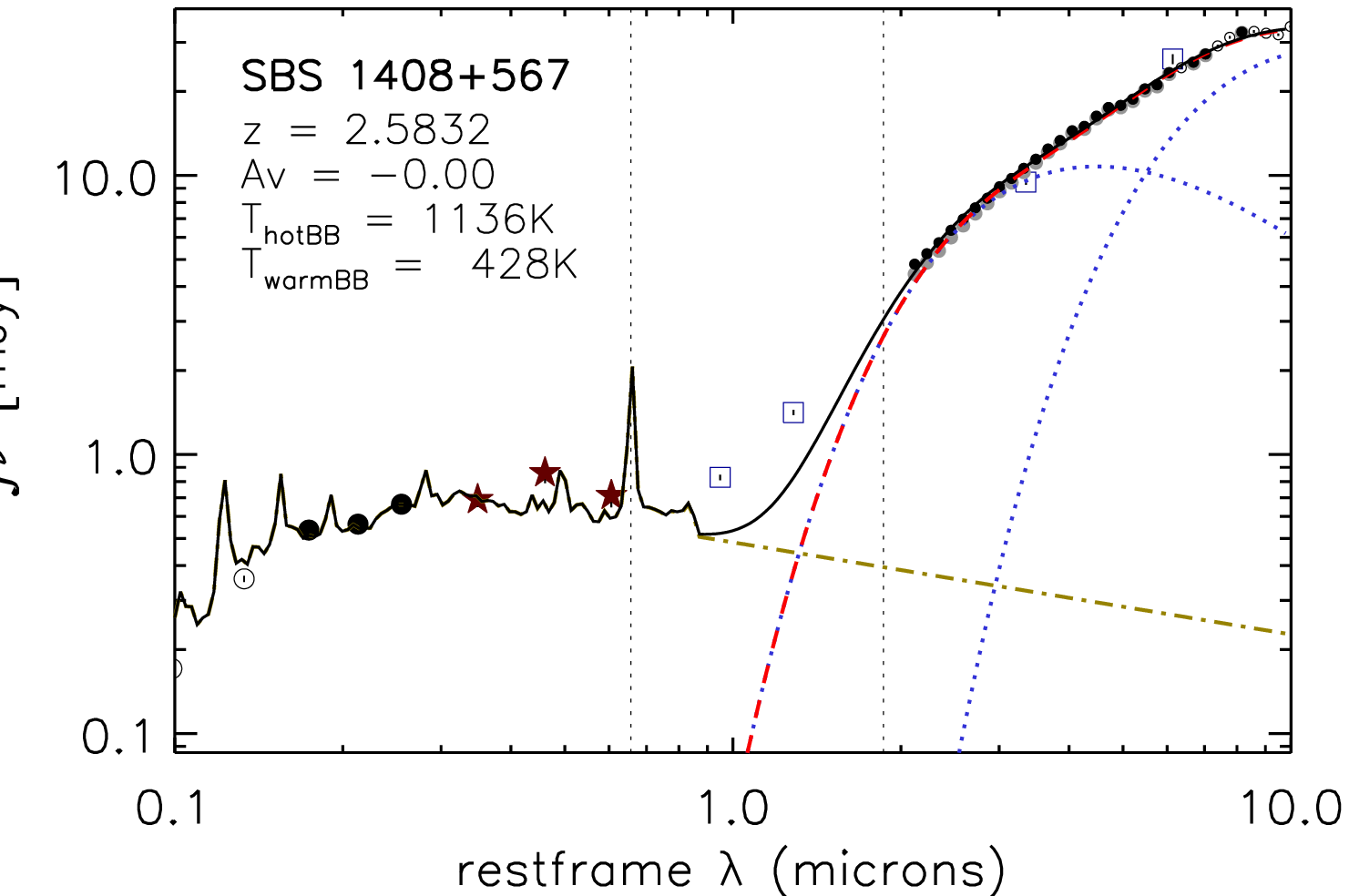}
\includegraphics[width=5.8cm]{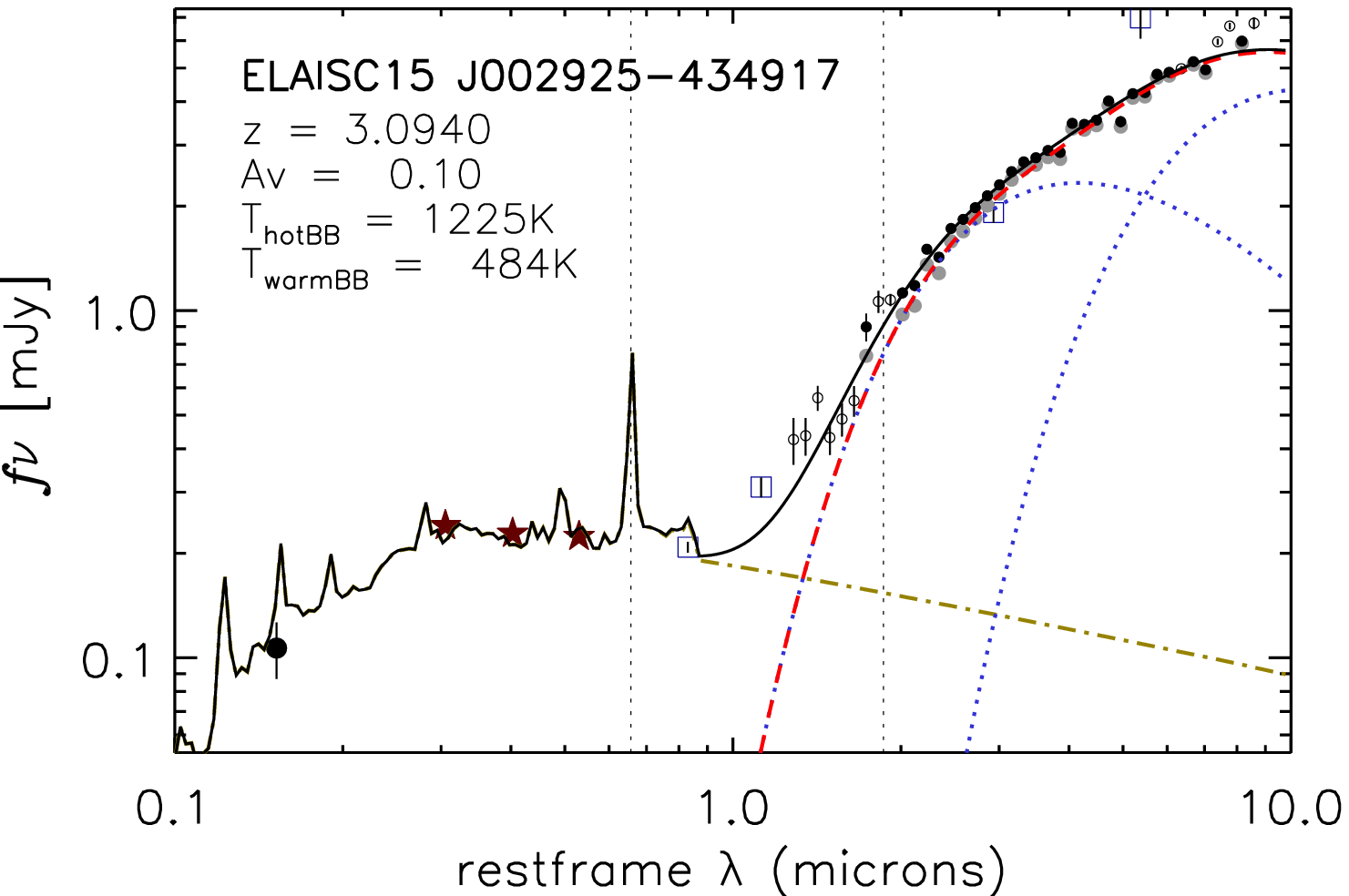}
\caption[]{Examples of SEDs and their best fitting disk+dust model for sources in the sample. Circles, stars, and squares represent broadband photometry in the observed-frame optical (from SDSS), NIR (2MASS/UKIDSS/VHS), and MIR (WISE), respectively. Filled symbols indicate bands used to fit the disk component (yellow dot-dashed line). The (AKARI+)IRS spectra resampled at $\lambda$/$\Delta\lambda$ = 20 are shown with small black dots. Grey dots below the (AKARI+)IRS spectra represent the dust spectrum obtained after subtraction of the disk component. The model for the dust spectrum (red dashed line) is the linear combination of two black-bodies at adjustable temperatures (blue dotted lines). The combined disk+dust model is represented by the black solid line. The vertical dotted lines mark the restframe wavelength of the H$\alpha$ and Pa$\alpha$ recombination lines. The entire sample is shown in Appendix \ref{appendix:fits}.\label{fig:fit-examples}}
\end{figure*}

In addition to optical and NIR fluxes, we retrieved from IRSA photometry in the 3.4, 4.6, 12, and 22 \um bands of the Wide-field Infrared Survey Explorer \citep[WISE;][]{Wright10}. 
We obtained measurements for all but one sources. While the longer wavelength WISE data points overlap with the IRS spectra, the 3.4 and 4.6 \um bands are useful to constrain the shape of the restframe 1--2.5 \um SED in the 76 IRS-only $z$$>$1 sources.
AKARI and WISE calibrations are tied to Spitzer by having absolute calibrators in common \citep{Cutri12}. However, the WISE 3.4 \um fluxes are known to be systematically lower by 10 per cent compared to AKARI fluxes \citep{Kim15}, though the reasons are not well understood. 
SEDs for all the sources in the sample, including the broadband photometry and the (AKARI+)IRS spectra are shown in Figure \ref{fig:fit-examples} and Appendix \ref{appendix:fits}.

\section{Optical and NIR spectral indices}\label{sec:spindices}

We first characterise the optical and NIR segments of the quasar spectra by measuring spectral indices. We define the optical spectral index as: 

\begin{equation}
\alpha_{opt} = \frac{\log(f_{0.3}/f_{1})}{\log(1\mu m/0.3\mu m)}
\end{equation}

\noindent where $f_{0.3}$ and $f_{1}$ are the flux densities at the restframe wavelengths 0.3 and 1 \uu. Similarly, we define the NIR spectral index as:

\begin{equation}
\alpha_{NIR} = \frac{\log(f_{1}/f_{3})}{\log(3\mu m/1\mu m)}
\end{equation}
\noindent with $f_{3}$ being the flux density at restframe 3 \uu. 

These indices correspond to the exponent $\alpha$ of a power-law of the form: $f_\nu$ $\propto$ $\nu^\alpha$.
Since the observed bands sample different restframe wavelengths depending on the redshift of the source, we use power-law interpolation to obtain $f_{0.3}$ and $f_{1}$ from the broadband photometry. 
Because the actual SEDs of the quasars are more complex than a simple power-law, 
we discard interpolations from sparsely sampled SEDs and only consider $f_{0.3}$ and $f_{1}$ measurements in sources with at least one photometric point within the restframe 0.2--0.4 \um and 0.75--1.25 \um ranges, respectively. On the other hand, $f_{3}$ is measured directly on the AKARI or IRS spectrum.

\begin{figure} 
\includegraphics[width=8.4cm]{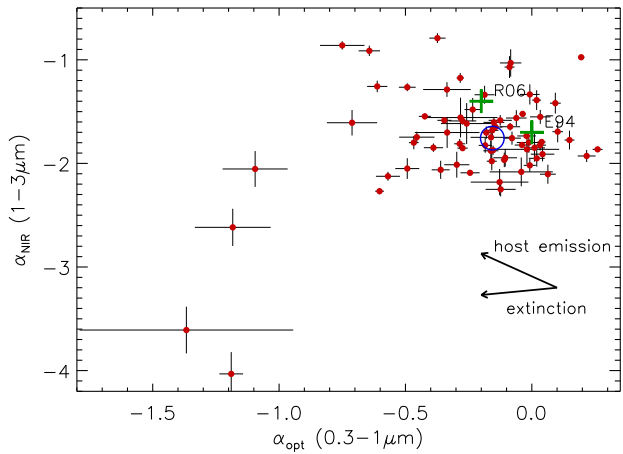}
\caption[]{Restframe NIR (1--3 \uu) versus optical (0.3--1 \uu) spectral indices for the individual sources with valid $f_{0.3}$ and $f_{1}$ measurements. The large open circle marks the median value for the sample, while the large plus signs indicate the values corresponding to the mean quasars SEDs from \citet{Richards06} and \citet{Elvis94}.\label{fig:alpha-opt-NIR}}
\end{figure}

Figure \ref{fig:alpha-opt-NIR} compares $\alpha_{opt}$ and $\alpha_{NIR}$ for the 68 sources with valid measurements. We exclude [HB89] 1700+518, which has an extremely low $\alpha_{opt}$=-3 based on just two photometric measurements in the $V$ and $J$ bands. 
For the remaining sources, the values of $\alpha_{opt}$ concentrate around the median values $\tilde{\alpha}_{opt}$ = -0.16 and $\tilde{\alpha}_{NIR}$ = -1.76, with standard deviations of 0.25 and 0.36, respectively. This is roughly consistent with the mean SED of 259 SDSS quasars with \textit{Spitzer}/IRAC detections from \citet{Richards06} ($\alpha_{opt}$=-0.2, $\alpha_{NIR}$=-1.4) and the radio-quiet AGN template from \citet{Elvis94} ($\alpha_{opt}$=0.0, $\alpha_{NIR}$=-1.7).

The dispersion in $\alpha_{opt}$ values, $\sigma$($\alpha_{opt}$) = 0.25, is consistent with that found by \citet{Richards01} in a much larger sample of SDSS quasars. 
Several factors may contribute to this dispersion: variations in the intrinsic spectral index of the disk emission, extinction by material in the nuclear region, in the host galaxy, or in other galaxies along the line of sight, and contamination by emission from the host galaxy. 
Studies using samples of less luminous quasars have found significantly higher dispersion \citep[e.g.][]{Hao10,Bongiorno12}, that is interpreted as caused by a combination of extinction and contamination by stellar emission in the host galaxy. 
However, the latter is unlikely to contribute significantly in our sample:  
assuming that the host galaxy has a typical stellar mass of 10$^{11}$ M$_\odot$ and a \citet{Salpeter55} Initial Mass Function, its restframe 3 \um luminosity is $\sim$10$^{43.5}$ erg s$^{-1}$, or 1\% of our luminosity threshold. Even at 1 \uu, where the fractional contribution from the host should be highest, the contamination from stellar emission would be only $\lesssim$10\% of the observed flux for the less luminous quasar of the sample. 
Furthermore, while extinction increases both $\alpha_{opt}$ and $\alpha_{NIR}$ (the latter by a much smaller amount), stellar emission modifies them by quantities that are comparable in magnitude but opposite in direction (see arrows in Figure \ref{fig:alpha-opt-NIR}). As a consequence, an anticorrelation between both indices is observed in samples with significant levels of contamination by stellar emission \citep{Hao10,Bongiorno12}. The lack of correlation in our sample confirms our expectation that there is no significant contamination from the host galaxy at these AGN luminosities.
Therefore, the dispersion in $\alpha_{opt}$ in our sample must arise from differences in the intrinsic slope and/or the extinction level affecting the quasar continuum. However, it is difficult to tell them apart, because both of them modify optical spectra in similar ways.
Recently, \citet{Krawczyk15} used a bayesian marginalisation method to break the degeneracy and estimated the intrinsic dispersion in the optical spectral index of SDSS quasars at $\sim$0.2, suggesting that this would be the dominant factor.
 
Since the impact of extinction is much lower at NIR wavelengths and we rule out a significant contamination by the host galaxy, the dispersion in $\alpha_{NIR}$ must be caused by either differences in the SED of the dust emission or differences in the relative luminosity of the disk and dust components. The former would be caused by variation in the distribution of temperatures of the hot dust grains at small radii, while the later depends mostly on the geometrical covering factor and the viewing angle.
To quantify the source-to-source variation in the SED of the dust, we need to remove first the contribution from the disk at NIR wavelengths. This is the subject of the next section.

We define the NIR-to-optical luminosity ratio as $r_{NO}$ = $\nu L_\nu$(3\uu)/$\nu L_\nu$(0.5 \uu), which can be considered as a proxy for the dust to disk luminosity ratio, a measure of the `apparent' covering factor of the dust \citep{Mor09,Alonso-Herrero11,Stalevski16}. Because $\lambda$ = 0.5 \um and $\lambda$ = 3 \um are both safely distant from the $\sim$1.6 \um peak of the stellar emission, this ratio is insensitive to a small contamination from the host galaxy. 

\begin{figure} 
\includegraphics[width=8.4cm]{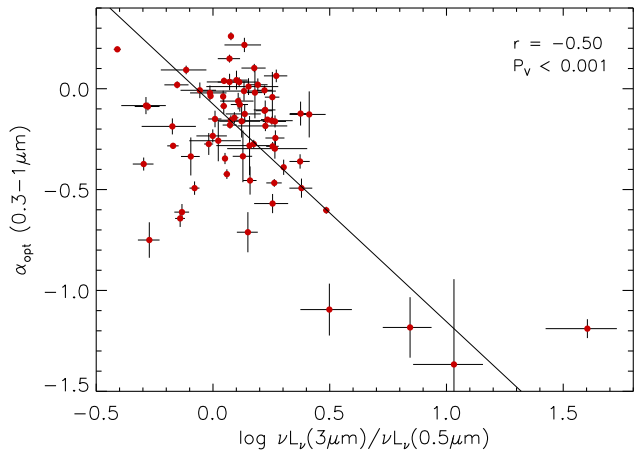}
\caption[]{Restframe optical (0.3--1 \uu) spectral index versus NIR-to-optical luminosity ratio ($r_{NO}$). The solid line marks the bisector of the X vs Y and Y vs X fits. $r$ and $P_v$ represent, respectively, the weighted correlation coefficient and the probability of finding such a high (absolute) value of $r$ if the two variables are uncorrelated.\label{fig:alpha-opt-rNO}}
\end{figure}

Figure \ref{fig:alpha-opt-rNO} shows a loose but strongly significant ($>$3$\sigma)$ anti-correlation between $\alpha_{opt}$ and $r_{NO}$. There are two factors that likely contribute to redder $\alpha_{opt}$ in sources with higher $r_{NO}$. The first one is increased contribution from the dust to the 1 \um emission due to higher relative luminosity of the dust component (see Figure \ref{fig:rdisk-distrib}). The second is increased reddening of the disk component due to higher dust column density in the line of sight towards the disk. The latter is supported by a clear correlation between the MIR-to-UV flux ratio and the slope of the UV spectrum \citep{Ma13}, and the increase in the average geometrical covering factor along the sequence of Seyfert types 1, 1.8--1.9, and 2 \citep{Mateos16}.

\section{Modelling of the continuum emission}\label{sec:modelling}

The temperature of silicate and graphite grains, believed to be the main ingredients of dust around AGN, is limited by sublimation to $\lesssim$1500 K. Therefore, the restframe UV and optical emission of the dust is negligible, and nearly all the continuum emission at $\lambda$$<$0.85 \um arises in the accretion disk. The relative contribution of the disk to the total emission is expected to decline quickly at $\lambda$$>$1 \um due to steeply rising emission from the dust. However, emission from the accretion disk could still be significant at 2--3 \uu.

The shape of the NIR SED of the disk is uncertain due to the difficulty in disentangling the emission from the dust, but both theoretical considerations \citep[e.g.][]{Hubeny01} and polarized light observations \citep{Kishimoto08}, favour power-law SEDs with $\alpha$$\sim$1/3 at NIR wavelengths.
A power-law extension of the rest-frame optical disk emission into the NIR is a common assumption for the primary source that is used as input for radiative transfer codes that model the torus emission \citep[e.g.][]{Granato94,Nenkova02,Fritz06,Stalevski12}, but it is not universal. In particular, some torus models assume a steep fall of the disk spectrum longwards of 1 \um \citep[e.g.][]{Hoenig06,Nenkova08,Siebenmorgen15} which makes its NIR emission negligible compared to that of the torus even for face-on views.
Since the NIR range contributes a small fraction of the bolometric emission from the primary source, its impact on the spectrum of the dust-reprocessed radiation must be small. However, the NIR SED of the \textit{total} AGN emission (disk+dust) depends substantially on the spectral shape assumed for the NIR spectrum of the disk \citep{Feltre12}. In particular, modelling with steeply falling disk emission at $\lambda$$>$1 \um could contribute to the NIR excess observed in type 1 AGN relative to the best-fitting disk+torus model \citep[e.g.][]{Mor09,Alonso-Herrero11}.
In a sample of 26 Palomar-Green quasars, \citet{Mor09} found that their \textit{Spitzer}/IRS spectra cannot be reproduced with a combination of the clumpy torus models from \citet{Nenkova08} and emission from the narrow line region, and a third component of very hot dust ($T$$\sim$1400 K) is required. In the analysis of the \textit{Spitzer}/IRS spectra of 25 $z$$\sim$2 quasars, \citet{Deo11} concluded that the additional hot dust component is required for fitting simultaneously the NIR SED and the $\sim$10 \um silicate feature with clumpy torus models.
However, \citet{Mateos15} were able to fit the nuclear SEDs of X-ray selected quasars using a non-truncated disk and the torus models from \citet{Nenkova08}, with no NIR excesses.

\subsection{Emission from the accretion disk}\label{sec:disk}
 
Previous studies modelled the optical emission of the disk with a single power-law with the form $f_\nu \propto \nu^\alpha$ of adjustable \citep[e.g.:][]{Kim15} or fixed \citep[e.g.:][]{Mateos15} spectral index $\alpha$. While this usually provides acceptable fits in the restframe optical range (0.3--0.85 \uu), its extrapolation into the restframe UV cannot reproduce the usually much redder observed SED (this change in slope originates the so-called big blue bump at $\sim$0.3 \uu), and a double power-law is often preferred.
Another difficulty in the case of broadband photometry is introduced by the contamination from strong emission lines like H$_\alpha$. 
Since the number of photometric data points at restframe UV to optical wavelengths in our sources is between 2 and 7, a model with two power-laws plus emission lines would have too many free parameters. 
Accordingly, we choose to use instead a single empirical quasar template. To reproduce the variety of optical spectral indices in our sample, we allow for 
varying amount of extinction to affect the template. Therefore our model has only two free parameters: the normalisation of the template and $A_V$. 

We use the extinction law of the Small Magellanic Cloud Bar \citep{Gordon03}, which is often preferred for quasars because it lacks the 2175\AA{} absorption feature. We choose the composite quasar spectrum from \citet{Shen16} as our universal UV+optical quasar template.
Unlike previous quasar composites \citep[e.g.][]{VandenBerk01}, the Shen composite is derived from observed-frame NIR spectroscopy of luminous ($L_{\rm{bol}}$ = 10$^{46.2-48.2}$ erg s$^{-1}$) high redshift (1.5$<z<$3.5) quasars.
This composite is significantly bluer in the optical range compared to that from Vanden Berk. Its slope at $\lambda$$>$0.3 \um is close to $\alpha$=1/3, and therefore its extension into the NIR with the same slope is consistent with theoretical expectations.
Redwards of the H$_\alpha$ line the composite spectrum is somewhat redder than $\alpha$=1/3. Because this could indicate some residual contamination from the host or the onset of the dust emission, we extrapolate the template into the NIR range by scaling an $\alpha$=1/3 power-law to fit the 0.3--0.6 \um range.

We obtain reddened versions of the quasar template by applying different levels of extinction, from A$_V$ = -0.4 to A$_V$ = 1.0 in steps of 0.05. Negative values of A$_V$ are needed to obtain good fits for some of the quasars with bluer continua, since the template is the average of many quasar spectra affected by different levels of extinction. In this respect, the A$_V$ values should be considered as relative to that of the average template.

We find the best fitting template for the disk of each quasar by least-squares fitting the photometric points corresponding to bands with effective wavelengths between 0.15 and 0.85 \um in the restframe. This means that we use up to 7 bands from SDSS and 2MASS/UKIDSS/VHS to fit a model that effectively has only two free parameters (A$_V$ and the normalisation constant for the template). For the three sources with only one or two data points in the fitting range we force A$_V$=0. We also force A$_V$=0 for five additional sources with three or four data points but all of them at $\lambda_{rest}<$4000\AA.
The values of the `relative' extinction obtained (considering only the 77 sources with free A$_V$) range from A$_V$ = -0.1 to 0.9 mag (median value: 0.05), but 90\% of them are within the [-0.1,0.4] interval.
Given the simplicity of the model, the quality of the fits is remarkable (see Figure \ref{fig:fit-examples}), with residuals apparently dominated by the varying strengths of the broad emission lines. 

\begin{figure} 
\begin{center}
\includegraphics[width=8.4cm]{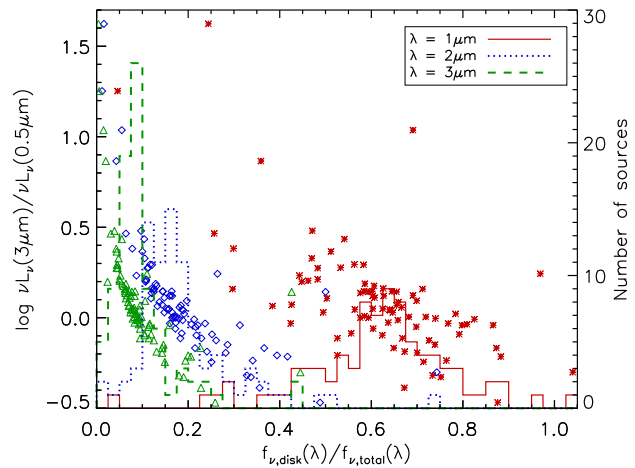}
\end{center}
\caption[]{Relation between the NIR-to-optical luminosity ratio, $r_{NO}$ = $\nu L_\nu$(3\uu)/$\nu L_\nu$(0.5\uu), and the estimated contribution from the disk to the total emission, $r_{disk}(\lambda)$, at restframe 1\um (red stars), 2 \um (blue diamonds), and 3\um (green triangles). The histograms show the distributions of $r_{disk}(\lambda)$ for each of the three restframe NIR wavelengths.\label{fig:rdisk-distrib}}
\end{figure}

The extension into the restframe NIR of the disk model allows to estimate the contribution from the accretion disk to the observed restframe NIR emission, $r_{disk}(\lambda)$ = $f_{\nu,disk}(\lambda)$/$f_{\nu,total}(\lambda)$. Values vary widely from source to source due to differences in the relative luminosity of the disk and dust components, but they show a consistent and rapid decline in the relative contribution of the disk to the total emission with increasing wavelength (Figure \ref{fig:rdisk-distrib}). The median values of $r_{disk}(\lambda)$ at the restframe wavelengths 1, 2, and 3 \um are 63\%, 17\%, and 8\%, respectively.

\subsection{Emission from the dust}\label{sec:continuum}

Assuming that the NIR and MIR emission from the host galaxy is negligible, all the flux remaining after subtraction of the disk component corresponds to reprocessed radiation from dust heated by the AGN. 
This dust may be located in a toroidal structure that surrounds the central engine (the so-called dusty torus) or in the narrow-line region, or both. 
Since our aim is not to test physical models of the AGN dust emission, but to characterise the observed NIR spectrum in a geometry-agnostic way, we resort to fitting a simple analytical function defined as the sum of two blackbody profiles:

\begin{equation}
f_{\nu}(\lambda) = a B_{\nu}(\lambda,T_{hot}) + b B_{\nu}(\lambda,T_{warm})
\end{equation}

\noindent where the temperatures $T_{hot}$ and $T_{warm}$ are free to vary within the intervals 850--2000 K and 150--900 K, respectively, and the normalisation factors $a$ and $b$ must be non-negative.

In the fit, we use photometric and spectroscopic data points within the 1.7--8.4 \um range in the restframe. This is to avoid complications due to possible residual contamination from the host galaxy or the disk and the onset of the 10 \um silicate feature at shorter and longer wavelengths, respectively. We also remove from the fit data points in the 6.10--6.45 \um and 7.3--8.1 \um intervals to prevent any contamination from the PAH features. Prior to the fit, we remove the disk component derived in the previous section and resample the (AKARI+)IRS spectrum with a uniform resolution $R$ = $\lambda$ /$\Delta\lambda$ = 20 (a factor 2--4 lower than the original data) in order to improve the signal to noise ratio (SNR).

The best fitting values of the four free parameters ($a$, $b$, $T_{hot}$, $T_{warm}$) are calculated using the IDL routine MPFIT \citep{Markwardt09}. We repeat the fit 100 times for each source with different initial values for $T_{hot}$ and $T_{warm}$ to ensure that the fit converges to a global minimum. We obtain remarkably good fits for all sources, with reduced $\chi^2$ values around unity. 

\begin{figure} 
\includegraphics[width=8.4cm]{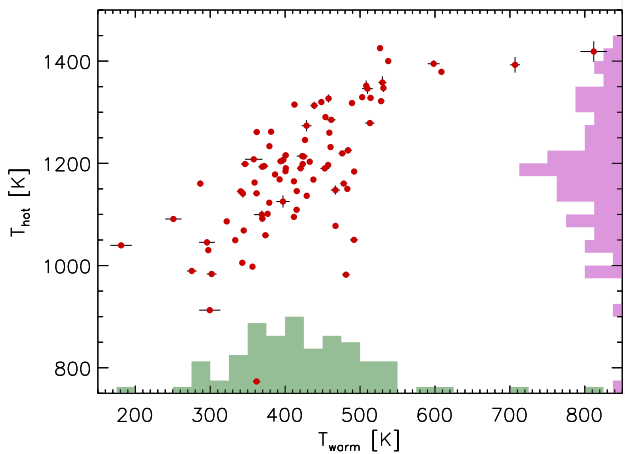}
\caption[]{Relation between the temperatures for the hot and warm dust components in individual sources. Error bars indicate the nominal 1-$\sigma$ uncertainties derived by MPFIT. The solid histograms represent the distribution of $T_{hot}$ and $T_{warm}$ in the sample.\label{fig:Thot-Twarm}}
\end{figure} 

Figure \ref{fig:Thot-Twarm} shows the temperatures of the hot and warm dust components of the best fitting model for each individual source. Values of $T_{hot}$ range between 900 and 1425 K (mean: 1188 K, median: 1190 K). The only outlier is SDSS J100552.63+493447.9 at $T_{hot}$ = 773 K, which has extremely red optical and NIR spectra.
The distribution for $T_{warm}$ ranges from 180 to 815 K (mean: 424 K, median: 415 K), but only 5 sources are outside the 250--550 range. 

The ratio between the bolometric luminosity of the hot and warm components ($r_{HW}$) ranges from 0.3 to 4.4 (mean: 1.66, median: 1.54). There is a correlation between the temperatures of the hot and warm dust components, which is probably a consequence of modelling a continuous distribution of temperatures with only two discrete values. However, neither $T_{hot}$ or $T_{warm}$ correlate with $r_{HW}$, $\alpha_{NIR}$, $r_{NO}$ or $A_V$.

The best fitting two-blackbody model for the disk-subtracted continuum and the combined disk+dust models are shown in Figure \ref{fig:fit-examples}. The disk+dust model reproduces correctly the optical SED and the IR spectrum within the restframe 1.7--8.4 \um interval used to adjust the parameters of the blackbody components. However, the measurements within the $\sim$0.9--1.7 \um range (which were excluded from both the disk and dust modelling) are systematically underestimated by the disk+dust model. 

To understand why the broadband restframe NIR photometry departs from our otherwise succesful model, we compute the relative excess at 1.2 \um as:
$e_{NIR}$ = ($f_{obs}$ - $f_{model}$)/$f_{model}$, where $f_{obs}$ is the flux density at restframe 1.2 \um interpolated in the observed SED and $f_{model}$ is the prediction from the best-fitting disk+dust model. Because the wavelength coverage in the restframe 1.0--1.5 \um range is sparse in many sources, we avoid biasing our analysis by considering only sources with at least one photometric point in the 1.1--1.3 \um interval or, alternatively, one in each of the intervals 1.0--1.2 and 1.2--1.4 \uu. There are 46 such sources in the sample. Their relative excesses range between 3\% and 198\% (median: 36\%). 
 
\begin{figure*} 
\includegraphics[width=17cm]{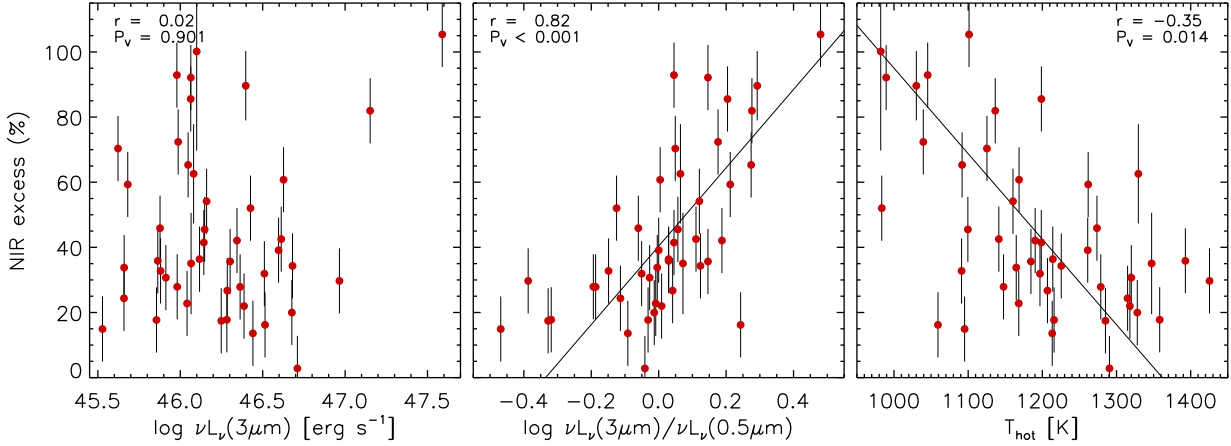}
\caption[]{Correlations (or lack thereof) of the relative excess at 1.2 \um after subtraction of the best-fitting disk+dust model with the restframe 3 \um monochromatic luminosity (left), the NIR-to-optical luminosity ratio (centre),
and the temperature of the hot blackbody component (right). The error-weighted Pearson correlation coefficient and p-value (probability of obtaining a value of $r$ greater than observed if the null hypothesis -no correlation- is true) are given in the top left corner of each panel. The plots exclude the outlier {[}HB89{]} 1700+518.\label{fig:eNIR-correlations}}
\end{figure*}

We consider three possible scenarios for the origin of this excess: i) contamination by stellar emission in the host galaxy, ii) underestimation of the NIR emission of the accretion disk, iii) underestimation of the NIR emission of the dust. To identify the more likely one, we look for correlations with the 3 \um luminosity and the NIR-to-optical luminosity ratio (Figure \ref{fig:eNIR-correlations}). If the host galaxy were responsible for the excess we should find an anti-correlation between $e_{NIR}$ and $\nu L_{\nu}$(3\uu), but we find none. The correlation between $e_{NIR}$ and $r_{NO}$ should be negative (anti-correlation) if the excess originated in the disk, and positive if it belonged to the dust. We find a clear positive correlation (Figure \ref{fig:eNIR-correlations}, center panel), and therefore scenario iii) is preferred. 

Since the emission at $\lambda$$>$1.7 \um is well reproduced by the model, the excess at shorter wavelengths must originate in very hot dust at or near the sublimation temperature. The fitting procedure is largely blind to such dust because data points at longer wavelengths dominate the fit. This arguably favours lower values for $T_{hot}$ than would be obtained using the full NIR SED. 
Such interpretation is backed by the loose but strongly significant anti-correlation between $e_{NIR}$ and $T_{hot}$ (Figure \ref{fig:eNIR-correlations}, right panel), which shows that strong NIR excesses are more likely in sources with colder $T_{hot}$ values.

To check whether a two blackbody model can also reproduce the $\lambda$$<$1.7 \um spectrum of the dust, we repeat the fits including also data points in the restframe 1.2--1.7 \um range. We find that the resulting higher $T_{hot}$ values decrease the 1.2 \um excess by $\sim$50\% on average, but only at the expense of significantly worse fits at longer wavelengths. In some cases $T_{hot}$ gets extreme high values up to 1750 K. This, together with the large dispersion of the excesses for any given value of $T_{hot}$ shown in Figure \ref{fig:eNIR-correlations} suggest that a single temperature hot blackbody component cannot reproduce accurately the entire NIR spectrum of the dust. 

We note that while the optical, NIR, and MIR data were obtained at different epochs, sometimes several years apart, our results are probably not affected by variability in the quasar emission. It is well established that the amplitude of the optical variability is strongly anti-correlated to the quasar luminosity \citep[e.g.:][]{Hook94,Meusinger13,Kozlowski16} and for the luminosity of our sample the amplitude is expected to be only $\sim$0.1 mag on timescales of years, comparable to the photometric uncertainty of our data. 
In addition, the component sampled by the MIR data (black body) is energetically decoupled from the disk component in our fitting method (i.e. there is no energy balance, since this is not a physical model).
We have evaluated the impact of optical variability in the reddening determination by increasing/decreasing the SDSS fluxes by 10\% and repeating the fits with the disk model. All the sources obtain their new best fit with the same A$_V$ as before or $\pm$0.05 magnitudes, indicating that the change in A$_V$ is smaller than our discretization step.

\section{Composite spectra}\label{sec:composites}

Further insight into the origin of variation in the NIR spectrum of quasars can be gained from the comparison of the composite spectra of adequately selected subsamples. As composite spectra are model independent, this is a complementary approach to the model-fitting in the previous section.

\begin{figure} 
\includegraphics[width=8.4cm]{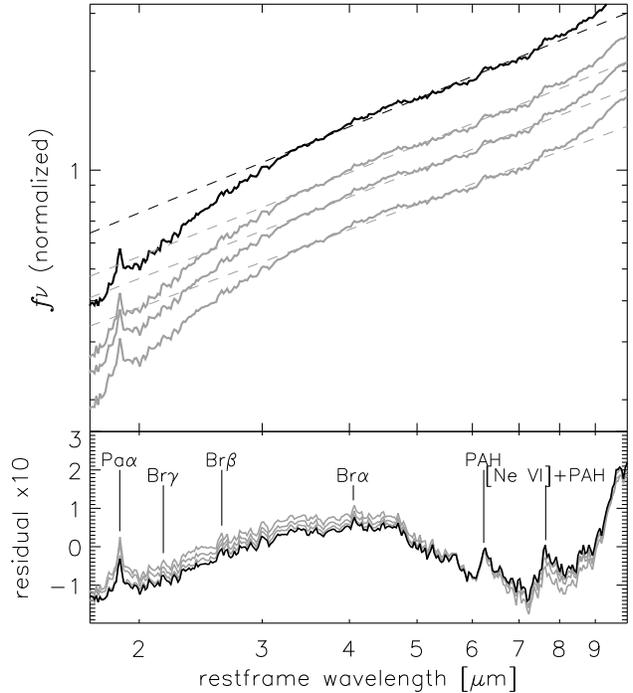}
\caption[]{Top panel: final composite spectra after several iterations of the normalisation procedure for initial normalisation at 3 \um (black solid line), or 4, 5, and 7 \um (grey solid lines). The dashed lines represent the best-fitting power-law for each composite in the the 3--7 \um spectral range. Bottom: residuals obtained by subtracting from the composite spectra a power-law with $\alpha$ = -1. Labels mark some significant spectral features.\label{fig:subtract-powerlaw}}
\end{figure}

We obtain the composite (AKARI+)IRS spectrum of the whole sample with the following procedure:

First, the spectra are shifted to the rest-frame and resampled to a common grid of wavelengths. Each point in the resampled grid represents an equal-sized interval in log($\lambda$).  
We normalise the spectra at a given restframe wavelength.  
For every wavelength in the grid, we obtain the flux of the composite spectrum as the weighted average of the fluxes of the individual sources, excluding the sources with no data at that wavelength. 
The shape of this preliminary composite spectrum depends on the normalisation wavelength, since a normalisation at short wavelengths increases the relative contribution of the redder spectra at long wavelengths, and therefore overestimates the slope of the continuum. Conversely, normalisation at long wavelengths favours the bluer ones and leads to underestimation of the continuum slope.
To compensate for this, we use the composite spectrum as a template to re-normalise the individual spectra by fitting with a least squares procedure the entire spectral range covered by each of them. We combine the re-normalised spectra using the same weighted average procedure to obtain a new composite spectrum, and iterate. We find that changes in the composite spectrum are negligible after the second or third iterations, and the wavelength chosen for the initial normalisation has a negligible impact on the final result (see Figure \ref{fig:subtract-powerlaw}).

The uncertainties in the composite spectrum are computed with a bootstrap resampling: we obtain composite spectra for 100 random samples extracted with replacement from the whole sample, each containing 85 sources. Then the 1-$\sigma$ confidence interval for the composite spectrum at each wavelength is given by the 16$^{th}$ and 84$^{th}$ percentiles of the distribution for the 100 bootstrap iterations.

The composite spectrum shows that the 3--7 \um SED is well reproduced by a power-law with $\alpha$ = -1. This is in agreement with previous estimates of the MIR spectral index of high redshift quasars \citep[e.g.][]{Lutz08,Hernan-Caballero09} as well as low redshift ones \citep[e.g.][]{Netzer07}.
However, at shorter wavelengths the continuum slope is increasingly steep and interrupted only by a prominent Paschen $\alpha$ line. Other hydrogen recombination lines as well as PAH features can also be identified in the spectrum (see \S\ref{sec:lines} for a discussion).

\subsection{Dependence with AGN luminosity}

\begin{figure} 
\includegraphics[width=8.4cm]{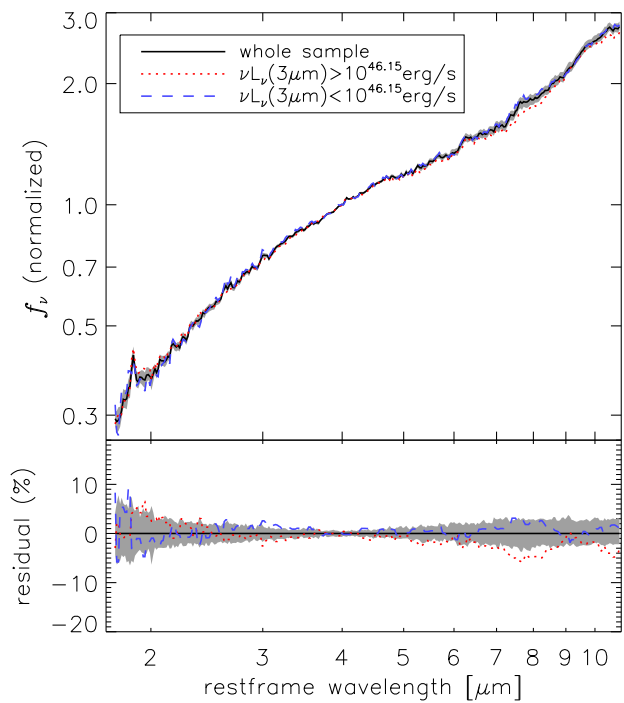}
\caption[]{Comparison between the composite spectra for the low and high luminosity subsamples. The shaded area represents the 1-$\sigma$ uncertainty for the composite template of the whole sample. The bottom panel shows the residuals of subtracting the whole sample composite from the high and low luminosity composites, normalised by the flux of the whole sample composite.\label{fig:composites-Lum}}
\end{figure}

We obtain composite spectra for the subsamples with lower ($\nu L_\nu$(3\uu) $=$ 10$^{45.5-46.15}$ erg s$^{-1}$) and higher ($\nu L_\nu$(3\uu) $>$ 10$^{46.15}$ erg s$^{-1}$) NIR luminosity, that contain the same number of sources. 
If the (AKARI+)IRS spectra were contaminated by stellar emission in the host galaxies of the quasars, the impact should be higher in the lower luminosity composite, that would appear bluer at short wavelengths compared to the higher luminosity one. However, we find no significant differences between the two composites (Figure \ref{fig:composites-Lum}). The lack of divergence at short wavelengths reinforces our previous claim that the stellar emission is negligible for the sources in the sample.

At longer wavelengths we find that the divergence is marginally significant around the $\sim$7.7 \um PAH complex. The residual suggests a weaker contribution from star formation to the spectra of more luminous quasars, in agreement with previous results \citep[e.g.][]{Haas03,Maiolino07,Hernan-Caballero09,Barger15}.
If this small divergence is attributed to star formation in the host galaxy, then the entire 2--10 \um spectrum of the quasar itself is independent on the NIR luminosity, at least within the luminosity range of our sample. This is contrary to previous results based on broadband photometry that suggested a luminosity dependence in the NIR-to-MIR quasar SED, and in particular a more prominent 4 \um bump and bluer MIR continuum for the more luminous quasars \citep{Richards06,Gallagher07}.

\subsection{Dependence with the NIR-to-optical luminosity ratio}

\begin{figure} 
\includegraphics[width=8.4cm]{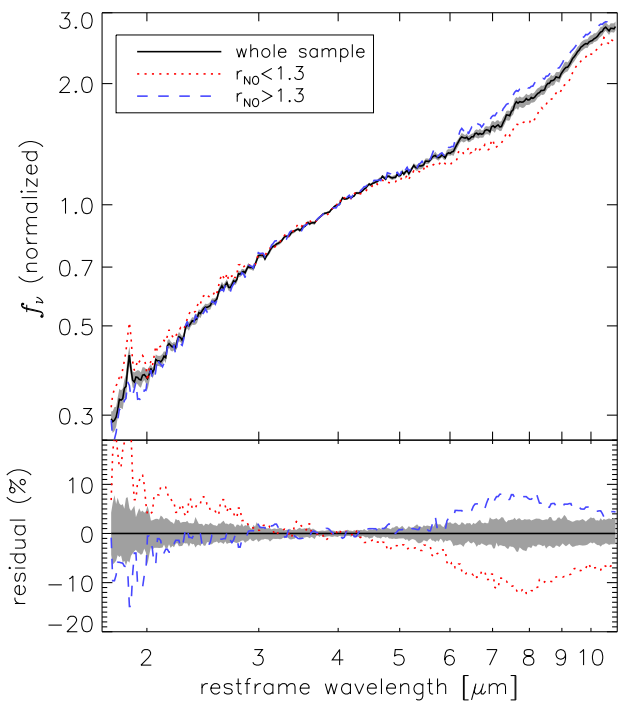}
\caption[]{Comparison between the composite spectra for the subsamples with high and low values of the restframe 3 \um to 0.5 \um luminosity ratio, $r_{NO}$. Symbols as in Figure \ref{fig:composites-Lum}.\label{fig:composites-rNO}}
\end{figure}

The uncertain contribution from the accretion disk to the NIR and MIR emission should manifest itself as differences in the composite spectrum between the high $r_{NO}$ and low $r_{NO}$ subsamples, because the contribution from the disk should be higher in the latter.
We split the sample into two same-sized subsamples with $r_{NO}$$<$1.3 and $r_{NO}$$>$1.3. We compare their composites in Figure \ref{fig:composites-rNO}.
The $r_{NO}$$>$1.3 composite is significantly redder at short wavelengths, consistent with a smaller contribution from the accretion disk to the NIR emission. We also note the lack of a clear Pa$\alpha$ emission line in this composite, which is also expected for a high dust-to-disk luminosity ratio if the luminosity of the hydrogen recombination lines correlates with that of the disk continuum (see \S\ref{sec:lines}).
If the relative luminosities of the accretion disk and dust were the only factor affecting $r_{NO}$, the two composites should converge to the same spectrum at MIR wavelengths, where the disk contribution is expected to be negligible. However, we find highly significant divergence also at $>$5 \uu. The high $r_{NO}$ composite is flatter (more power-law like) and redder than the low $r_{NO}$ one, which presents a blue 5--8 \um continuum and a stronger silicate emission feature. This suggests that the spectrum of the dust component is slightly different in high $r_{NO}$ and low $r_{NO}$ quasars, with a bluer continuum and a more pronounced 4 \um bump in the latter. This is in good qualitative agreement with the results by \citet{Mateos16} in a large X-ray selected AGN sample, who found redder MIR SEDs for the tori in type 1 AGN with higher geometrical covering factors.
 
\subsection{Dependence with the optical spectral index}

\begin{figure} 
\includegraphics[width=8.4cm]{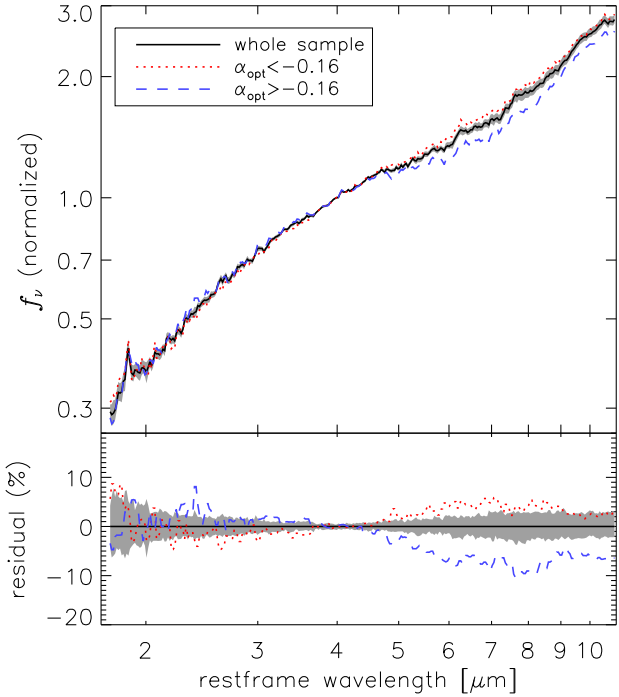}
\caption[]{Comparison between the composite spectra for the subsamples with high and low values of the optical spectral index, $\alpha_{opt}$. Symbols as in Figure \ref{fig:composites-Lum}.\label{fig:composites-aopt}}
\end{figure}

Finally, we checked for a dependence of the composite spectrum with $\alpha_{opt}$, that relates on both the intrinsic slope and the extinction affecting the disk component. The composite spectra for $\alpha_{opt}>$-0.16 and $\alpha_{opt}<$-0.16 (Figure \ref{fig:composites-aopt}) are not significantly different at NIR wavelengths. However, there is a clear divergence at $\lambda>$4 \uu, with the composite for the blue optical AGN having a more pronounced 4 \um bump and a bluer MIR spectrum. 
The shape of the residuals resemble that of the high and low $r_{NO}$ composites (except for the collapse in the divergence at $\lambda$$\lesssim$2.5 \uu), but the departure from the full sample composite is less pronounced. This suggests that the variation with $\alpha_{opt}$ is mostly an indirect consequence of the variation with $r_{NO}$ enabled by the (weak) correlation between $r_{NO}$ and $\alpha_{opt}$. 

\subsection{Comparison with other IR composites}\label{subsec:compare-composites}

\begin{figure} 
\includegraphics[width=8.4cm]{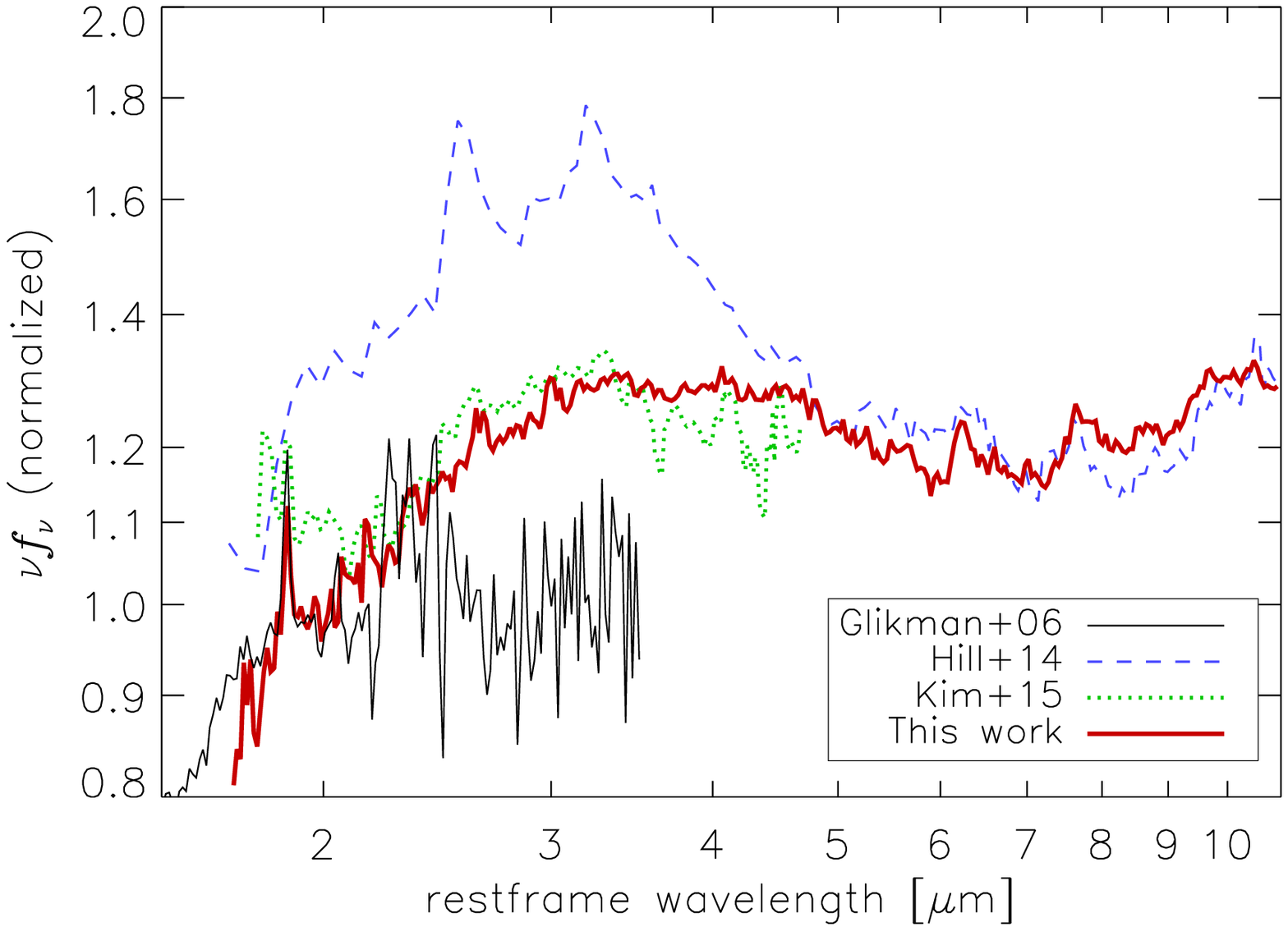}
\caption[]{Comparison of composite infrared spectra of quasars. The thin black line is the composite NIR spectrum of 27 low redshift ($z$$<$0.42) bright (K$_s$$<$14.5) SDSS quasars from \citet{Glikman06}. The blue dashed line is the composite \textit{Spitzer}/IRS spectrum of luminous ($\log$(L$_{5.6\mu m}$/erg s$^{-1}$) = 44.8--46.1) SDSS quasars from \citet{Hill14}. 
The green dotted line is the composite AKARI spectrum of 48 PG QSOs from \citet{Kim15}.  
The thick red line is our own composite quasar spectrum.\label{fig:compare-other}}
\end{figure}

We now compare our IR quasar composite spectrum with other composites from the literature with spectral coverage in the $\sim$2--4 \um range (Figure \ref{fig:compare-other}).

The composite from \citet{Glikman06} shows a prominent Paschen $\alpha$ line that is comparable in strength to that in our template, but the continuum longwards of the line is flat (in $\nu f_\nu$) while our template shows a prominent 3\um bump. The discrepancy is most likely due to the fact that the $\lambda$$>$2 \um part of their composite is obtained from the noisy L-band spectra of a few sources (between 4 and 15 depending on wavelength). On the other hand, at $\lambda$$\lesssim$2\um the Glikman composite is in good agreement with our composite broadband SED (see \S\ref{sec:template}).

The composite from \citet{Hill14}
was obtained from the \textit{Spitzer}/IRS spectra of 61 SDSS quasars with 5.6 \um luminosities between 10$^{44.8}$ and 10$^{46.1}$ erg s$^{-1}$. We have normalised their composite to match ours at 5.6 \uu, where their coverage is highest.
The two spectra have similar shapes in the 4--11 \um range. The main difference is the larger EW of the 6.2 \um and 7.7 \um PAH features in their composite, probably caused by a lower by $\sim$0.5 dex median luminosity compared to our sample ($\sim$10$^{45.5}$ erg s$^{-1}$ vs 10$^{46.15}$ erg s$^{-1}$). Their $\lambda$$<$4 \um spectrum shows a substantially more pronounced bump compared to ours. The differences may be due to the low number of objects contributing to this part of their composite. That might also be the cause of the spikes at $\sim$2.5 and $\sim$3.1 \uu, which do not match any known emission feature.

The composite AKARI spectrum of 48 PG QSOs from \citet{Kim15} shows a better match to the 2--4 \um range of our composite, with departures only at the extremes of the range. Our IR composite is therefore consistent with the others in the ranges where they are the more robust. This highlights the homogeneity of optically selected quasar samples and suggests that previous discrepancies in the average quasar spectrum were mostly due to observational limitations.

\section{Emission lines}\label{sec:lines}

\begin{figure} 
\includegraphics[width=8.4cm]{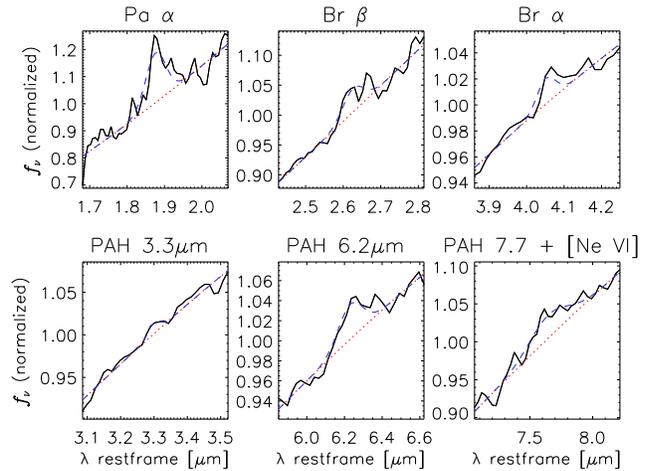}
\caption[]{Emission lines and PAH features observed in the composite quasar spectrum for the full sample. The continuum (dotted line) and feature (dashed line) are modelled with a straight line and a Gaussian profile, respectively.\label{fig:fit-lines}}
\end{figure}

At the spectral resolution of the \textit{Spitzer}/IRS low resolution modules (R$\sim$60--120) and AKARI (R$\sim$60) the emission lines of quasars are usually washed out by the strong continuum emission in individual spectra. Even the broader PAH features are undetected in most quasars because they contribute a tiny fraction of the flux density per resolution element, and therefore get masked by photometric noise. However, by averaging many spectra, we obtain a very high SNR in the composite, which allows to detect several emission lines and PAH bands.

We measure the strength of the emission lines and PAH features in the composite spectrum by fitting simultaneously a Gaussian profile and a linear local continuum. We set free the height and width of the Gaussian but fix the central wavelength at the theoretical value. The width of the fitting interval (between 0.4 and 1.2 \uu) is enough for a good determination of the continuum and not as large as to make curvature effects important. Figure \ref{fig:fit-lines} shows the best fits for the hydrogen recombination lines Paschen $\alpha$, Bracket $\alpha$, and Bracket $\beta$, as well as the PAH features at 3.3, 6.2 and 7.7 \uu.

We estimate uncertainties for the line strengths using the distribution of the measured values in 100 random composite spectra obtained through bootstrap resampling of the full sample.
The median values and 1-$\sigma$ confidence intervals (percentiles 16$^{th}$ and 84$^{th}$) for the equivalent width and the FWHM are given in Table \ref{table:emission-lines}. 
All the features except the 3.3 \um PAH are detected at the $>$3$\sigma$ level. The strongest detections correspond to the 7.7 \um PAH complex ($\sim$4.7$\sigma$) that includes the [Ne {\sc vi}] 7.642 \um line, and Paschen $\alpha$, that is detected at the $\sim$4.5$\sigma$ level in spite of having only 25 sources (5 at $z$$<$1 and 20 at $z$$>$1) with enough spectral coverage to observe it. 

The three hydrogen lines have comparable values for the FWHM ($\sim$0.065-0.075 \uu). They correspond to velocity dispersions between 10000 km/s for Paschen $\alpha$ and 5000 km/s for Bracket $\alpha$. These values are not corrected for the instrumental broadening, which in the case of IRS spectra is $\sim$4000 km/s for Paschen $\alpha$ and $\sim$2000 km/s for Bracket $\alpha$.
Also, the uncertainty in the wavelength calibration and the redshifts of the individual sources contributing to the composites may produce some additional broadening. Therefore we consider these values to be fairly consistent with expectations for the composite spectra of luminous QSOs.

Since the EW is insensitive to the normalisation, the EW of features in the composite spectrum represent the average EW in the individual spectra.
We obtain an EW of 124$\pm$30 \AA{} and 17$\pm$4 \AA{} for the  Paschen $\alpha$ and Bracket $\alpha$ lines, respectively. 
The Paschen $\alpha$ EW is consistent with a mean value of 104$\pm$43 \AA{} found by \citet{Landt08} for a sample of 23 well known type 1 AGN at $z$$<$0.4 with NIR spectra from the NASA Infrared Telescope Facility. It is also in good agreement with the 
EW = 120 \AA{} found by \citet{Soifer04} for the Paschen $\alpha$ line in the \textit{Spitzer}/IRS spectrum of the strongly lensed $z$=3.92 quasar APM 08279+5255, that was not included in our sample.
In contrast, \citet{Glikman06} give a Paschen $\alpha$ EW of just 12.7 \AA{} for the composite spectrum of 27 bright ($K_s$$<$14.5) SDSS quasars at $z$$<$0.42. However, we have re-measured the EW of several lines on their composite spectrum and found that the values listed in their Table 6 are systematically underestimated by a factor 3-4. Our estimate of the Paschen $\alpha$ EW on their template is 55 \AA.
This is still a factor $\sim$2 lower than the value found in our composite, which is striking.
Significant contamination from the stellar emission of the host galaxy is ruled out by \citet{Glikman06}, who estimate a contribution from the host galaxy of just $\sim$6\% at 1 \uu.  
Higher average extinction in the BLR for the \citet{Glikman06} sample is also unlikely because their quasars were selected to be blue and with little to no extinction, and the amount of extinction required to halve the Paschen $\alpha$ flux would cause a factor $\sim$40 decrease in H$\alpha$.
We can also rule out a reduction in the Paschen $\alpha$ EW due to stronger dust emission relative to the disk in their sample, since the NIR-to-optical ratio (as measured by $r_{NO}$) is in fact slightly higher in our sample. Therefore, the difference in EW suggests that the Paschen $\alpha$ line in our sample is a factor $\sim$2 more luminous relative to both the disk and the dust continuum emissions.

The composite broadband SED (connected dots in Figure \ref{fig:full-template}) also shows tentative evidence for strong Paschen $\beta$ 1.282 \um and/or Paschen $\gamma$ 1.094 \um in the form of a small peak at $\sim$1.2 \um (see \S\ref{sec:template}). 
Futher observations in the restframe NIR with the NIRSpec and MIRI instruments onboard JWST will be essential to confirm and understand these results.

\section{Quasar template}\label{sec:template}

\begin{figure} 
\includegraphics[width=8.4cm]{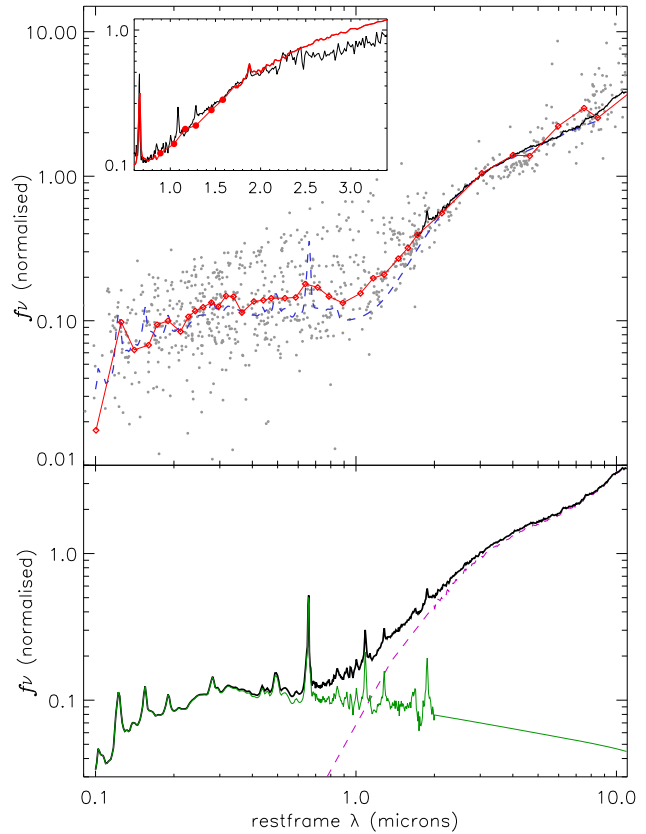}
\caption[]{Generation of a quasar template from our sample. Top panel: the (AKARI+)IRS composite spectrum (solid black line) is extended to shorter wavelengths using the 0.85--1.7 \um range of the broadband SED composite (red connected dots), and the $<$0.85 \um range of the model composite (blue dashed line). The small grey dots represent the broadband photometry of individual sources normalised at restframe 3 \uu. Inset plot: comparison of our template (red thick line and connected dots) with the quasar composite from \citet{Glikman06} (black thin line).
Bottom panel: The thick black line is the final quasar template after replacing the 0.85--1.7\um range from our broadband SED with the Glikman composite. We also obtain templates for the dust emission (pink dashed line) and the disk emission (including emission lines; green thin line). See text for details.\label{fig:full-template}}
\end{figure}

In this section we present a new template for the 0.1--11\um spectrum  of quasars based on our sample.
Since the composite (AKARI+)IRS spectrum constructed in \S\ref{sec:composites} is defined only at $\lambda$$>$1.7 \uu, we also build composites from the broadband SEDs and the best-fitting disk+dust models of the individual sources.
The broadband SED composite is obtained by normalising the individual SEDs at restframe 3.0 \um and obtaining median fluxes in bins containing 20 data points each. The model composite is the median of the individual best-fitting models, after normalisation at 3.0 \uu. 

The top panel in Figure \ref{fig:full-template} shows the three composites. There is a good correspondence between the (AKARI+)IRS (black solid line) and the broadband (connected dots) composites, in spite of the poor sampling of the broadband SED at $\lambda$$>$2 \uu. The broadband SED also reproduces  accurately the composite model in the optical and UV ranges (which is ultimately derived from the empirical quasar template from \citet{Shen16}, as described in \S\ref{sec:disk}), in spite of the large dispersion among individual sources that is a consequence of their varying NIR-to-optical luminosity ratios. 
The main discrepancy is at $\sim$1.2 \uu, where the model underpredicts the flux in the broadband SED (see \S\ref{sec:continuum} for a discussion).

We generate a quasar template by stitching together the $\lambda$=1.7--11 \um composite (AKARI+)IRS spectrum, the 0.85--1.7 \um interval of the composite broadband SED, and the 0.10--0.85 \um section of the composite model. We compare the 0.6--3.4\um range of this template with the composite spectrum from \citet{Glikman06} in the inset in Figure \ref{fig:full-template}. 
While the Glikman composite diverges from ours at $\lambda$$>$2 \uu, probably because it relies on noisy L-band spectra of a few sources (see  \S\ref{subsec:compare-composites}), the two templates are in good agreement for $\lambda$$\lesssim$2\um in spite of the poor sampling of our broadband SED. This confirms the robustness of our broadband NIR data and suggests that both samples have equally low contamination by the host galaxy emission. 
We capitalise on this good agreement by replacing the 0.85--1.7\um section of our template with the Glikman composite, which includes several interesting emission lines in this range. Therefore our final 0.1--11\um template is derived entirely from spectroscopic observations.

We also obtain separate templates for the dust and the disk (plus emission lines) as follows:
We subtract from the full template the composite model spectrum of the disk component in the individual sources. This provides a first approximation to the spectrum of the dust, which is however contaminated by the NIR emission lines since they were not included in our power-law extrapolation of the Shen template.
We remove the emission lines from this preliminary dust template by fitting a broken power-law in the $\lambda$$<$2\um range, excluding the intervals affected by the main emission lines: He {\sc ii} 1.083 \um + Pa$\gamma$ 1.094 \uu, Pa$\beta$ 1.282 \uu, and Pa$\alpha$ 1.876 \uu. The best fit is obtained for a spectral index $\alpha$ = -2.3 between 1.35 and 2.0 \um and a steeper $\alpha$=-3.2 for $\lambda$$<$1.35\uu. 
Finally we subtract this line-free dust template from the full template to obtain a template for the disk emission that now includes the NIR emission lines. The final disk, dust, and full (disk+dust) templates are shown in the bottom panel of Figure \ref{fig:full-template} and in Table \ref{table:templates}.

\section{Summary and conclusions}

We present the analysis of a sample of 85 luminous ($\log$($\nu L_\nu$(3\uu)/erg s$^{-1}$) = 45.5--47.5) quasars with available restframe NIR and MIR spectroscopy from AKARI and \textit{Spitzer}/IRS. We complement the spectroscopic data with broadband photometry from SDSS, 2MASS, UKIDSS, and WISE. 

The optical and NIR spectral indices of individual sources concentrate at $\alpha_{opt}$=-0.2 and $\alpha_{NIR}$=-1.75, consistent with the quasar templates from \citet{Elvis94} and \citet{Richards06}. The small dispersion and lack of correlation between the two indices indicate little to no contamination from the stellar population in the host galaxy. 
We find the restframe UV and optical broadband SEDs of individual sources to be well reproduced by a single quasar template affected by varying levels of extinction.
The analytical extrapolation of this template into the NIR as a power-law with $\alpha$=1/3, together with a simple model for the dust emission consisting on the superposition of two blackbodies, provides good fits for the restframe 1.7--8.4 \um spectrum in all sources, with temperatures for the hot blackbody component ranging between 950 and 1450 K.
However, the extrapolation at shorter wavelengths of the best fitting model systematically underpredicts the observed flux near the 1 \um minimum of the quasar SED. We find a correlation between the relative strength of this NIR excess and the NIR to optical luminosity ratio ($r_{NO}$), which indicates that the excess is not related to the disk or host galaxy emissions but originates in dust near the sublimation radius. 
We combine the (AKARI+)IRS spectra of the individual sources to obtain composite spectra with very high SNR. The signal is high enough to detect several NIR hydrogen recombination lines and PAH features, in spite of their extremely low EW.
Contrary to previous results, we find no luminosity dependence in the NIR and MIR shape of the quasar spectrum within the range of our sample. We find a small but highly significant dependence with $r_{NO}$, which suggests that quasars with higher apparent covering factors have a redder NIR-to-MIR SED and a less prominent silicate emission feature. 
We generate a new spectroscopic quasar template from our sample that covers the restframe range 0.1--11 \uu. We also produce templates for the disk and dust components. Comparison of our template with other spectroscopic quasar composites shows that previous composites are less reliable in the 2--4 \um range, probably due to the low number of sources with sufficient spectral coverage and/or noisy data in this range. Our template is the first one that provides a detailed and consistent view of the IR emission on either sides of the 4 \um bump.
 
\section*{Acknowledgements}

We wish to thank the anonymous referee for his/her useful suggestions that helped to improve this paper.
A.H.-C. acknowledges funding by the Spanish Ministry of Economy and Competitiveness (Mineco) under grants AYA2015-70815-ERC and AYA2012-31277.  
A.A.-H. acknowledges funding by the Mineco grant AYA2015-64346-C2-1-P which is partly funded by the FEDER programme. A.H.-C., A.A.-H., and S.M. acknowledge funding by the Mineco grant AYA2012-31447, which is partly funded by the FEDER programme.
This work is based on observations made with the \textit{Spitzer Space
Telescope}, which is operated by the Jet Propulsion Laboratory, Caltech
under NASA contract 1407.
The Cornell Atlas of \textit{Spitzer}/IRS Sources (CASSIS) is a product of the Infrared Science Center at Cornell University, supported by NASA and JPL.
This publication makes use of data products from the Two Micron All Sky Survey, which is a joint project of the University of Massachusetts and the Infrared Processing and Analysis Center/California Institute of Technology, funded by the National Aeronautics and Space Administration and the National Science Foundation.
This research has made use of the NASA/IPAC Infrared Science Archive, which is operated by the Jet Propulsion Laboratory, California Institute of Technology, under contract with the National Aeronautics and Space Administration.

\appendix

\section{SEDs and model fits for individual sources}\label{appendix:fits}

\onecolumn

\setcounter{table}{0}
\renewcommand{\thetable}{\arabic{table}}

\begin{deluxetable}{rlrrr r@{$\pm$}l ll ll}
\tabletypesize{\scriptsize}
\tablewidth{0pc}
\tablecolumns{9}
\tablecaption{Sources included in the quasar sample.\label{sampletable}}
\tablehead{\colhead{AOR ID} & \colhead{Source name} & \colhead{RA(J2000.0)} & \colhead{Dec(J2000.0)} & \colhead{$z$} &
\multicolumn{2}{c}{$\nu L_\nu$(3\uu)\tablenotemark{a}} & \colhead{opt. phot.\tablenotemark{b}} & \colhead{NIR phot.\tablenotemark{c}} & \colhead{spec. class\tablenotemark{d}} & \colhead{spec. ref.\tablenotemark{e}}}
\startdata
  4734464\_0 &               PG 1116+215 & 169.78622 &  21.32168 &   0.1765 &   3.35 & 0.026 & SDSS & 2MASS &       QSO &                 SDSS DR12 \\
  4675840\_0 &           [HB89] 1700+518 & 255.35359 &  51.82235 &   0.2920 &   8.29 & 0.032 & NED & 2MASS &   BAL QSO &   \citet{Schmidt99} \\
 14192640\_0 &               PG 1049-005 & 162.96443 &  -0.85495 &   0.3599 &   4.57 & 0.043 & SDSS & UKIDSS &       QSO &                 SDSS DR12 \\
 22199552\_0 &              SBS 1704+608 & 256.17242 &  60.74185 &   0.3719 &   7.17 & 0.052 & SDSS & 2MASS &       QSO &                 SDSS DR12 \\
 14200576\_0 &           [HB89] 1543+489 & 236.37604 &  48.76920 &   0.4014 &   4.80 & 0.044 & SDSS & 2MASS &       QSO &                 SDSS DR12 \\
 14203136\_0 &           [HB89] 2308+098 & 347.82401 &  10.13773 &   0.4333 &   4.56 & 0.058 & SDSS & UKIDSS &       QSO &     \citet{Veron06} \\
 14188288\_0 &               PG 0003+158 &   1.49663 &  16.16352 &   0.4509 &   3.37 & 0.044 & SDSS & UKIDSS &       QSO &     \citet{Veron06} \\
 10949376\_0 &           [HB89] 2112+059 & 318.71909 &   6.12855 &   0.4660 &   11.0 & 0.095 & SDSS & 2MASS &       QSO &     \citet{Veron06} \\
 14195456\_0 &              SBS 1259+593 & 195.30411 &  59.03509 &   0.4778 &   8.18 & 0.054 & SDSS & 2MASS &       QSO &                 SDSS DR12 \\
 10494720\_0 &   2MASS J17075650+5630136 & 256.98541 &  56.50387 &   1.0190 &   4.13 &  0.25 & - & 2MASS &       QSO &     \citet{Veron06} \\
  4732160\_0 &               PG 1254+047 & 194.24974 &   4.45948 &   1.0256 &   33.6 &   1.5 & SDSS & UKIDSS &       QSO &                 SDSS DR12 \\
 18983168\_0 &           [HB89] 2230+114 & 338.15161 &  11.73081 &   1.0370 &   14.9 &  0.25 & SDSS & 2MASS &    HP QSO &     \citet{Veron06} \\
 16347392\_0 &  SDSS J143132.13+341417.3 & 217.88399 &  34.23819 &   1.0396 &   7.39 &  0.37 & SDSS & 2MASS &       QSO &                 SDSS DR12 \\
 10493696\_0 &   2MASS J14265292+3323231 & 216.72050 &  33.38977 &   1.0831 &   9.41 &  0.35 & SDSS & 2MASS &       QSO &                 SDSS DR12 \\
 11345664\_0 &    2MASSi J1640100+410522 & 250.04230 &  41.08961 &   1.0990 &   13.5 &  0.35 & SDSS & 2MASS &       QSO &     \citet{Veron06} \\
 24194560\_2 &  SDSS J105404.10+574019.7 & 163.51711 &  57.67220 &   1.1009 &   6.66 &  0.29 & SDSS & UKIDSS &       QSO &                 SDSS DR12 \\
  4738560\_0 &  SDSS J100552.63+493447.9 & 151.46931 &  49.58018 &   1.1200 &   34.4 &   5.2 & SDSS & NED &      Sy 1 &     \citet{Rowan-Robinson00} \\
 21649408\_0 &  SDSS J172704.67+593736.6 & 261.76947 &  59.62681 &   1.1284 &   20.1 &  0.41 & SDSS & 2MASS &       QSO &                 SDSS DR12 \\
 21655552\_0 &  SDSS J143220.15+331512.2 & 218.08406 &  33.25351 &   1.2043 &   5.20 &  0.39 & SDSS & 2MASS &       QSO &                 SDSS DR12 \\
  4671488\_0 &               PG 0946+301 & 147.42142 &  29.92204 &   1.2234 &   43.6 &   1.7 & SDSS & 2MASS &       QSO &                 SDSS DR12 \\
 10950400\_0 &           FBQS J0840+3633 & 130.18513 &  36.55790 &   1.2250 &   29.2 &  0.38 & SDSS & 2MASS &   BAL QSO &       \citet{Trump06} \\
 24169984\_0 &  SDSS J103803.35+572701.6 & 159.51402 &  57.45052 &   1.2855 &   11.0 &  0.66 & SDSS & 2MASS &       QSO &                 SDSS DR12 \\
 24190464\_3 &  SDSS J160835.64+542329.3 & 242.14841 &  54.39159 &   1.3000 &   12.3 &  0.43 & SDSS & UKIDSS &       QSO &                 SDSS DR12 \\
 11341056\_0 &  SDSS J164018.34+405813.0 & 250.07646 &  40.97036 &   1.3177 &   7.52 &  0.60 & SDSS & 2MASS &       QSO &                 SDSS DR12 \\
 11355392\_0 & APMUKS(BJ) B003833.18-442 &  10.23160 & -44.21375 &   1.3800 &   6.62 &  0.87 & NED & VHS &       QSO &     \citet{Veron06} \\
 11354880\_0 &  ESIS J003814.10-433314.9 &   9.55853 & -43.55415 &   1.4000 &   6.08 &  0.36 & NED & VHS &       QSO &     \citet{Veron06} \\
 24194304\_2 &                  CIRSI 21 &  34.62738 &  -4.93969 &   1.4010 &   14.5 &  0.61 & SDSS & VHS &       QSO &                 SDSS DR12 \\
 11344640\_0 &  SDSS J163739.43+414347.9 & 249.41437 &  41.73003 &   1.4168 &   3.52 &  0.48 & SDSS & - &       QSO &                 SDSS DR12 \\
 11344384\_0 &  SDSS J163739.29+405643.6 & 249.41373 &  40.94553 &   1.4260 &   4.77 &  0.77 & SDSS & - &       QSO &     \citet{Veron06} \\
 10828032\_0 &                    3C 298 & 214.78418 &   6.47638 &   1.4381 &   26.8 &   1.1 & SDSS & UKIDSS &       QSO &                 SDSS DR12 \\
 10495488\_0 &   2MASS J20145083-2710429 & 303.71185 & -27.17859 &   1.4440 &   11.4 &  0.84 & - & 2MASS &       QSO &     \citet{Veron06} \\
 24160256\_1 &              SBS 1039+581 & 160.73190 &  57.93058 &   1.4725 &   14.2 &  0.65 & SDSS & UKIDSS &       QSO &                 SDSS DR12 \\
 11351040\_0 &  SDSS J160418.97+541524.3 & 241.07915 &  54.25678 &   1.4770 &   15.6 &  0.91 & SDSS & UKIDSS &       QSO &                 SDSS DR12 \\
 24192768\_0 &  SDSS J171124.22+593121.4 & 257.85095 &  59.52258 &   1.4893 &   11.9 &  0.72 & SDSS & 2MASS &   BAL QSO &       \citet{Trump06} \\
 24173312\_0 &              SBS 1034+583 & 159.35327 &  58.08696 &   1.5225 &   19.6 &  0.73 & SDSS & 2MASS &       QSO &                 SDSS DR12 \\
 11353088\_0 &   ELAISC15 J003014-430332 &   7.56249 & -43.05930 &   1.5640 &   7.18 &  0.75 & NED & VHS &       QSO &     \citet{Veron06} \\
 11355136\_0 &  ESIS J003829.91-434454.3 &   9.62455 & -43.74841 &   1.5670 &   19.4 &  0.71 & NED & VHS &       QSO &     \citet{Veron06} \\
 11345408\_0 &  SDSS J163952.84+410344.8 & 249.97020 &  41.06256 &   1.6046 &   5.41 &  0.58 & SDSS & - &       QSO &                 SDSS DR12 \\
 24163072\_0 &  SDSS J172238.73+585107.0 & 260.66147 &  58.85196 &   1.6237 &   12.0 &  0.44 & SDSS & 2MASS &       QSO &                 SDSS DR12 \\
 11353856\_0 &               IRAC 251351 &   8.14316 & -43.32727 &   1.6370 &   4.19 &  0.58 & NED & VHS &       QSO &     \citet{Veron06} \\
 25390336\_0 &  SDSS J143102.94+323927.8 & 217.76225 &  32.65788 &   1.6474 &   7.54 &   1.4 & SDSS & 2MASS &       QSO &                 SDSS DR12 \\
 25387264\_0 &  SDSS J105447.28+581909.5 & 163.69690 &  58.31935 &   1.6600 &   20.1 &   1.0 & SDSS & UKIDSS &       QSO &                 SDSS DR12 \\
 25976832\_0 &  SDSS J100401.27+423123.0 & 151.00519 &  42.52314 &   1.6657 &   41.0 &   1.5 & SDSS & 2MASS &       QSO &                 SDSS DR12 \\
 25977856\_0 &  SDSS J105951.05+090905.7 & 164.96277 &   9.15153 &   1.6897 &   19.2 &   1.5 & SDSS & UKIDSS &       QSO &                 SDSS DR12 \\
 11347200\_0 &  SDSS J161543.52+544828.7 & 243.93134 &  54.80809 &   1.6920 &   7.60 &  0.61 & SDSS & UKIDSS &       QSO &     \citet{Veron06} \\
 11343104\_0 &  SDSS J163425.11+404152.4 & 248.60474 &  40.69806 &   1.7005 &   11.6 &  0.61 & SDSS & 2MASS &       QSO &                 SDSS DR12 \\
 11353600\_0 & GALEX 2533910445613399575 &   8.05468 & -43.76488 &   1.7070 &   32.4 &  0.73 & NED & VHS &       QSO &     \citet{Veron06} \\
 25390848\_0 &  SDSS J160950.71+532909.5 & 242.46120 &  53.48606 &   1.7165 &   22.1 &   1.3 & SDSS & UKIDSS &   BAL QSO &       \citet{Trump06} \\
 25388800\_0 &  SDSS J104155.15+571603.2 & 160.47987 &  57.26747 &   1.7235 &   13.1 &  0.92 & SDSS & UKIDSS &       QSO &                 SDSS DR12 \\
 25977600\_0 &           [HB89] 0947+482 & 147.69807 &  48.01322 &   1.7420 &   17.7 &   1.8 & SDSS & 2MASS &       QSO &                 SDSS DR12 \\
 11345920\_0 &  SDSS J164016.08+412101.2 & 250.06699 &  41.35041 &   1.7564 &   9.58 &  0.59 & SDSS & 2MASS &       QSO &                 SDSS DR12 \\
 11344896\_0 &  SDSS J163847.42+421141.6 & 249.69757 &  42.19501 &   1.7760 &   7.30 &  0.69 & SDSS & - &       QSO &                 SDSS DR12 \\
 25977344\_0 &  SDSS J132120.48+574259.4 & 200.33549 &  57.71642 &   1.7789 &   23.1 &   1.5 & SDSS & 2MASS &       QSO &                 SDSS DR12 \\
 25389824\_0 &  SDSS J142730.19+324106.4 & 216.87598 &  32.68522 &   1.7951 &   11.2 &  0.99 & SDSS & 2MASS &       QSO &                 SDSS DR12 \\
 14847744\_0 &    2MASSi J0812004+402815 & 123.00184 &  40.47057 &   1.8017 &   24.4 &  0.29 & SDSS & 2MASS &       QSO &                 SDSS DR12 \\
 24143104\_4 &  SDSS J105158.52+590652.0 & 162.99397 &  59.11444 &   1.8218 &   13.9 &  0.69 & SDSS & UKIDSS &   BAL QSO &       \citet{Trump06} \\
 25388544\_0 &  SDSS J103931.14+581709.4 & 159.87962 &  58.28600 &   1.8276 &   9.54 &  0.69 & SDSS & UKIDSS &       QSO &                 SDSS DR12 \\
 25977088\_0 &  SDSS J151307.74+605956.9 & 228.28229 &  60.99923 &   1.8306 &   39.5 &   1.9 & SDSS & 2MASS &       QSO &                 SDSS DR12 \\
 25391104\_0 &  SDSS J172522.05+595250.9 & 261.34195 &  59.88067 &   1.8955 &   10.7 &  0.78 & SDSS & - &       QSO &                 SDSS DR12 \\
 25389056\_0 &  SDSS J104114.48+575023.9 & 160.31013 &  57.84004 &   1.9100 &   9.71 &  0.65 & SDSS & UKIDSS &       QSO &                 SDSS DR12 \\
 25389312\_0 &              SBS 1040+567 & 160.98119 &  56.46585 &   1.9569 &   26.7 &   1.2 & SDSS & 2MASS &       QSO &                 SDSS DR12 \\
 25388032\_0 &  SDSS J105153.76+565005.6 & 162.97382 &  56.83500 &   1.9712 &   14.4 &  0.75 & SDSS & UKIDSS &       QSO &                 SDSS DR12 \\
  4671232\_0 &           [HB89] 2225-055 & 337.12653 &  -5.31518 &   1.9810 &   51.4 &  0.93 & SDSS & VHS &       QSO &     \citet{Veron06} \\
 25390592\_0 &  SDSS J143605.07+334242.6 & 219.02115 &  33.71188 &   1.9917 &   14.1 &   1.4 & SDSS & 2MASS &       QSO &                 SDSS DR12 \\
 25388288\_0 &  SDSS J160004.32+550429.8 & 240.01790 &  55.07499 &   1.9987 &   11.6 &  0.83 & SDSS & UKIDSS &       QSO &                 SDSS DR12 \\
 11346688\_0 &  SDSS J161007.11+535814.0 & 242.52946 &  53.97067 &   2.0321 &   11.6 &  0.68 & SDSS & UKIDSS &       QSO &                 SDSS DR12 \\
 11352832\_0 & SWIRE4 J002959.22-434835. &   7.49671 & -43.80980 &   2.0390 &   27.6 &  0.90 & NED & VHS &       QSO &     \citet{Veron06} \\
 24152064\_1 & SWIRE J022431.58-052818.8 &  36.13160 &  -5.47192 &   2.0680 &   47.5 &   1.1 & SDSS & VHS &       QSO &                 SDSS DR12 \\
 25390080\_0 &  SDSS J142954.70+330134.7 & 217.47786 &  33.02631 &   2.0794 &   19.3 &   1.2 & SDSS & 2MASS &       QSO &                 SDSS DR12 \\
 11353344\_0 &   ELAISC15 J003059-442133 &   7.74852 & -44.35920 &   2.1010 &   23.6 &  0.91 & NED & VHS &       QSO &     \citet{Veron06} \\
  4733184\_0 &    2MASSi J0845387+342043 & 131.41103 &  34.34542 &   2.1378 &   52.3 &   1.9 & SDSS & 2MASS &   BAL QSO &       \citet{Trump06} \\
 25387776\_0 &  SDSS J161238.26+532255.0 & 243.15956 &  53.38190 &   2.1404 &   17.4 &   1.4 & SDSS & UKIDSS &       QSO &                 SDSS DR12 \\
 25389568\_0 &  SDSS J105001.04+591111.9 & 162.50430 &  59.18669 &   2.1691 &   17.4 &   1.1 & SDSS & UKIDSS &   BAL QSO &       \citet{Trump06} \\
 11354624\_0 &   ELAISC15 J003715-423515 &   9.31502 & -42.58731 &   2.1900 &   21.9 &  0.90 & NED & VHS &       QSO &     \citet{Veron06} \\
 10950656\_0 &            LBQS 2212-1759 & 333.88184 & -17.73546 &   2.2170 &   33.1 &  0.78 & NED & 2MASS &   BAL QSO &     \citet{Clavel06} \\
  4733440\_0 &           [HB89] 1246-057 & 192.30774 &  -5.98869 &   2.2470 &   190. &   2.2 & SDSS & VHS &       QSO &     \citet{Veron06} \\
 24192000\_0 &  SDSS J105951.69+581803.0 & 164.96539 &  58.30089 &   2.3352 &   32.7 &  0.70 & SDSS & - &       QSO &                 SDSS DR12 \\
 14847232\_0 &    2MASSi J1612399+471157 & 243.16621 &  47.19920 &   2.3961 &   42.3 &  0.54 & SDSS & 2MASS &       QSO &     \citet{Veron06} \\
  3732992\_0 &           [HB89] 1413+117 & 213.94263 &  11.49558 &   2.5511 &   389. &   6.4 & SDSS & UKIDSS &       QSO &                 SDSS DR12 \\
 14847488\_0 &              SBS 1408+567 & 212.48164 &  56.47412 &   2.5832 &   142. &   1.1 & SDSS & 2MASS &       QSO &                 SDSS DR12 \\
 11343872\_0 &  SDSS J163655.77+405910.0 & 249.23256 &  40.98629 &   2.5918 &   12.6 &  0.86 & SDSS & NED &   red QSO &     \citet{Willott03} \\
  4732416\_0 &              SBS 1524+517 & 231.47449 &  51.61357 &   2.8841 &   92.4 &   3.0 & SDSS & 2MASS &   BAL QSO &       \citet{Trump06} \\
 11346432\_0 &  SDSS J160637.87+535008.4 & 241.65781 &  53.83570 &   2.9476 &   25.0 &   1.8 & SDSS & UKIDSS &   BAL QSO &       \citet{Trump06} \\
 11352576\_0 &   ELAISC15 J002925-434917 &   7.35764 & -43.82210 &   3.0940 &   48.0 &   2.6 & NED & VHS &       QSO &     \citet{Veron06} \\
  6310400\_0 &           SDSS J1148+5251 & 177.06952 &  52.86394 &   6.4189 &   85.5 &   10. & SDSS & NED &       QSO &                 SDSS DR12 \\
\enddata
\tablenotetext{a}{Monochromatic luminosity at restframe 3\um in units of 10$^{45}$ erg s$^{-1}$.}
\tablenotetext{b}{Source of the optical broadband photometry.}
\tablenotetext{c}{Source of the near infrared broadband photometry.}
\tablenotetext{d}{Optical spectroscopic classification.}
\tablenotetext{d}{Reference for the optical spectroscopic classification.}
\end{deluxetable}

\begin{deluxetable}{l r r r r r r r} 
\tabletypesize{\small}
\tablecaption{Emission lines in composite spectrum\label{table:emission-lines}}
\tablewidth{0pt}
\tablehead{\colhead{line} & \colhead{$\lambda_{rest}$} & \multicolumn{3}{c}{EW [\uu]} & \multicolumn{3}{c}{FWHM [\uu]} \\
\colhead{} & \colhead{[\uu]} & \colhead{median} & \colhead{p16} & \colhead{p84} & \colhead{median} & \colhead{p16} & \colhead{p84}}
\startdata
Pa $\alpha$
 &    1.876 &   0.0124 &   0.0101 &   0.0168 &    0.063 &   0.052 &   0.077\\
Br $\beta$
 &    2.626 &   0.0023 &   0.0016 &   0.0030 &    0.076 &   0.060 &   0.090\\
Br $\alpha$
 &    4.052 &   0.0017 &   0.0013 &   0.0022 &    0.067 &   0.057 &   0.086\\
PAH 3.3
 &    3.300 &   0.0025 &   0.0014 &   0.0040 &    0.146 &   0.081 &   0.225\\
PAH 6.2
 &    6.230 &   0.0037 &   0.0029 &   0.0046 &    0.147 &   0.132 &   0.161\\
PAH 7.7+[Ne {\sc vi}]
 &    7.642 &   0.0138 &   0.0117 &   0.0175 &    0.267 &   0.222 &   0.315\\
\enddata
\end{deluxetable}

\begin{deluxetable}{c c c c} 
\tabletypesize{\small}
\tablecaption{Composite quasar spectrum\label{table:templates}}
\tablewidth{0pt}
\tablehead{\colhead{$\lambda_{rest}$ [\uu]} & \colhead{$L_\nu$(total)/$L_{3\mu m}$} & \colhead{$L_\nu$(disk)/$L_{3\mu m}$} & \colhead{$L_\nu$(dust)/$L_{3\mu m}$}}
\startdata
   0.1000 &   0.03356 &   0.03352 &   0.00004\\
   0.1011 &   0.03746 &   0.03742 &   0.00004\\
   0.1023 &   0.04670 &   0.04665 &   0.00005\\
   0.1034 &   0.04504 &   0.04499 &   0.00005\\
   0.1046 &   0.04264 &   0.04259 &   0.00005\\
   0.1058 &   0.04214 &   0.04209 &   0.00005\\
   0.1070 &   0.04185 &   0.04180 &   0.00005\\
   0.1082 &   0.03957 &   0.03951 &   0.00006\\
   0.1094 &   0.03701 &   0.03695 &   0.00006\\
   0.1107 &   0.03719 &   0.03713 &   0.00006\\
   0.1119 &   0.03819 &   0.03813 &   0.00006\\
\enddata
\tablecomments{This is a sample of the full table which is available in the online version of the article.}
\end{deluxetable}

\begin{figure*}
\includegraphics[width=8.4cm]{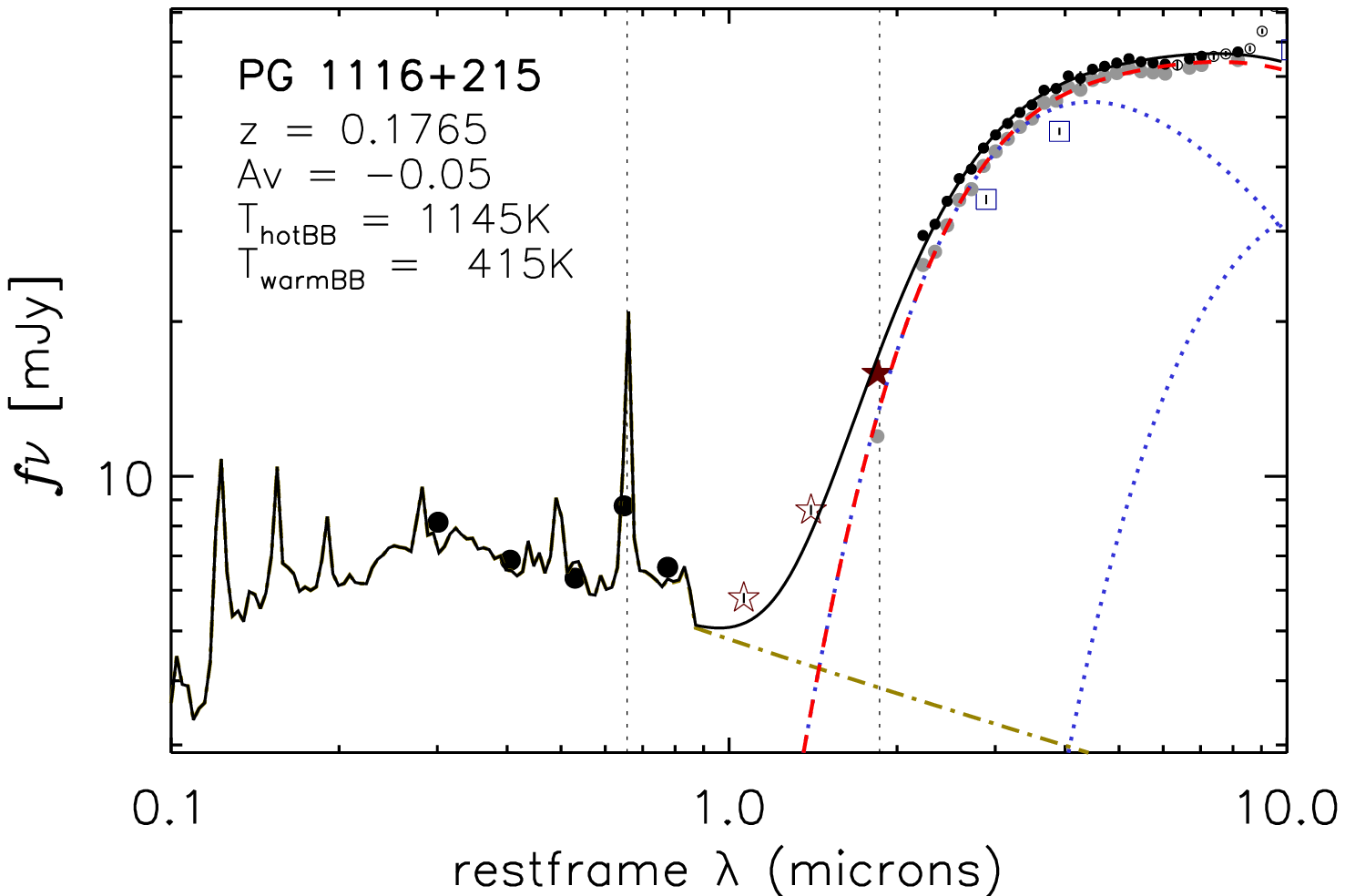}
\includegraphics[width=8.4cm]{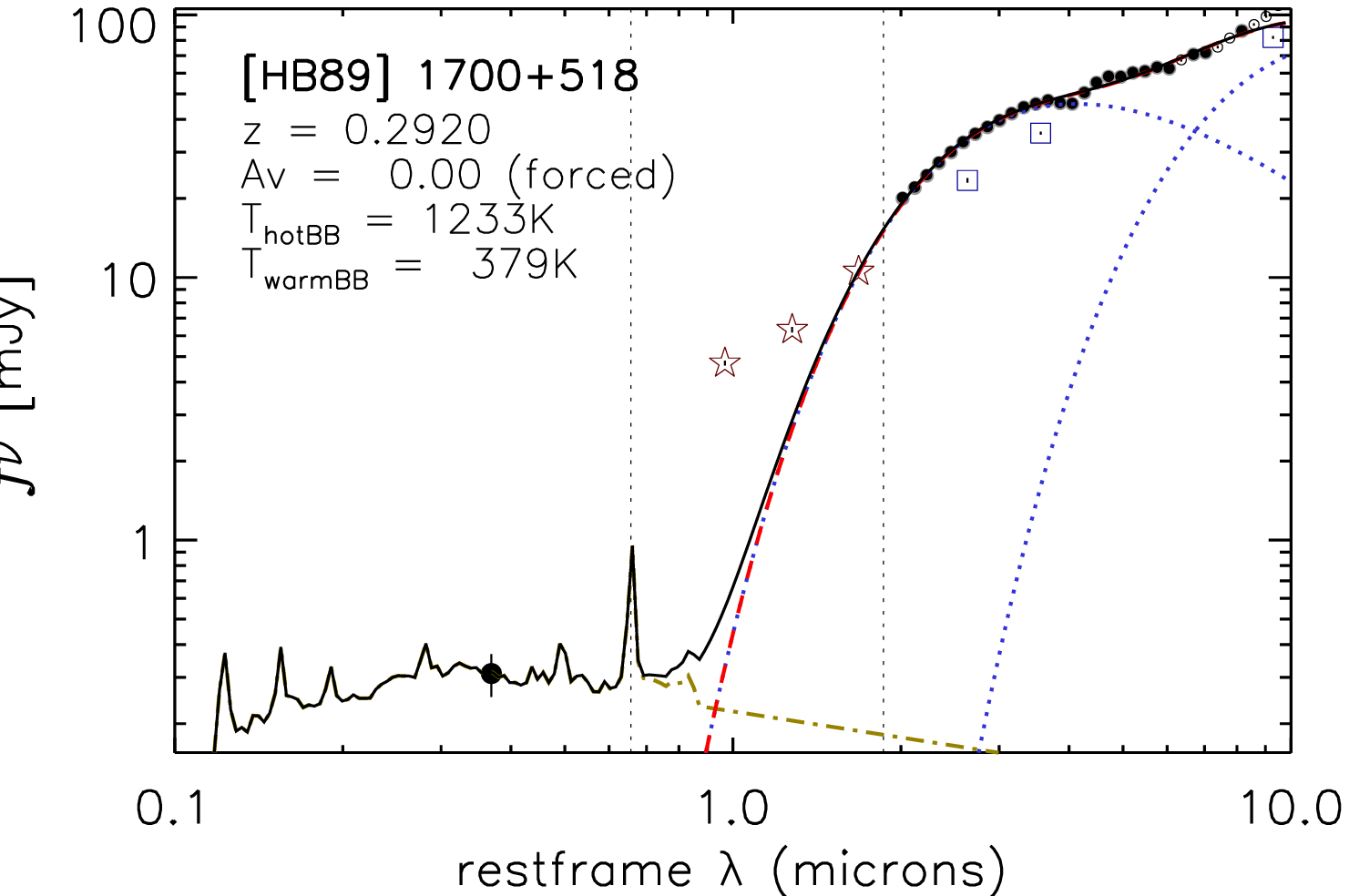}\vspace{0.4cm}
\includegraphics[width=8.4cm]{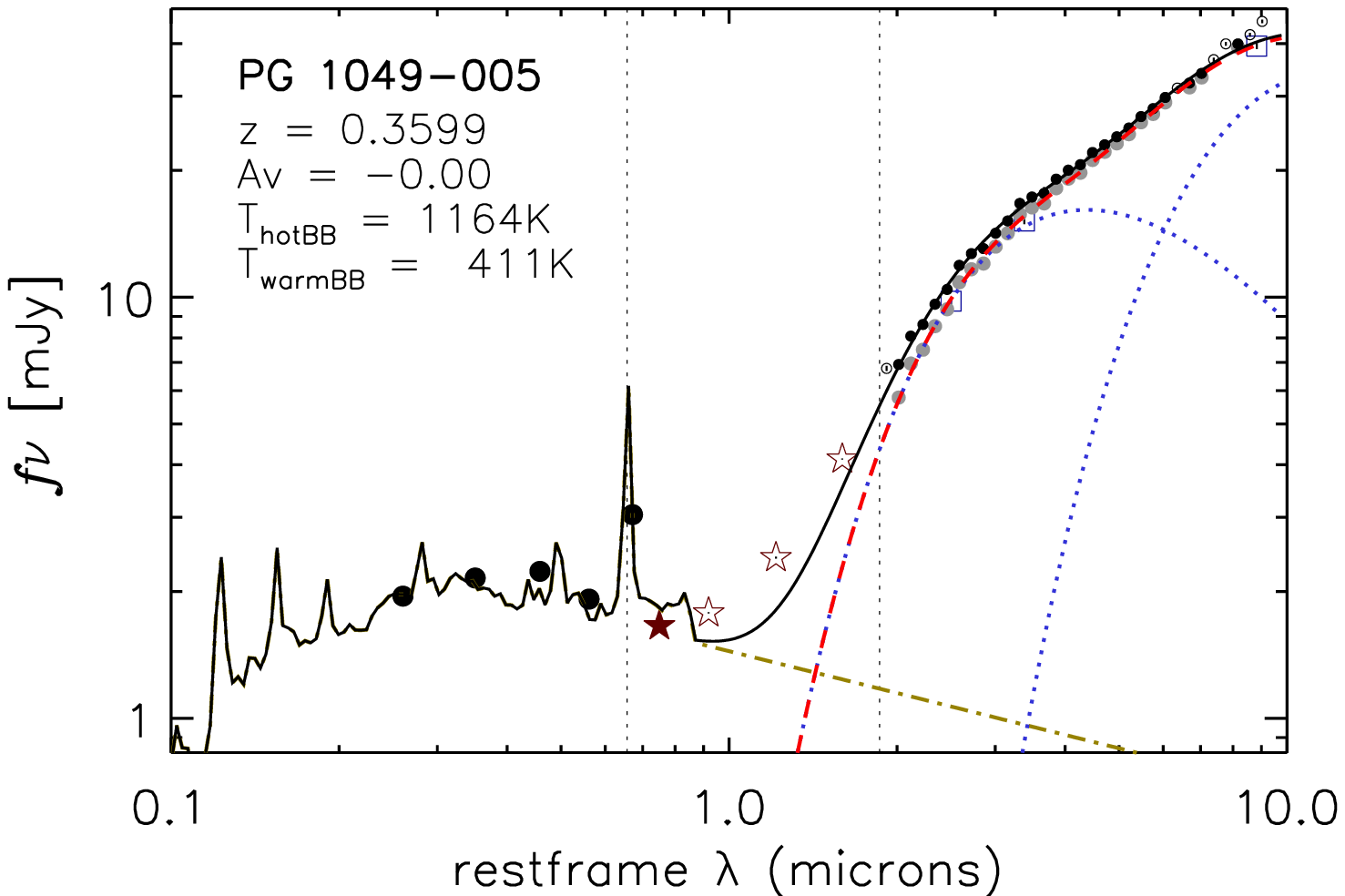}
\includegraphics[width=8.4cm]{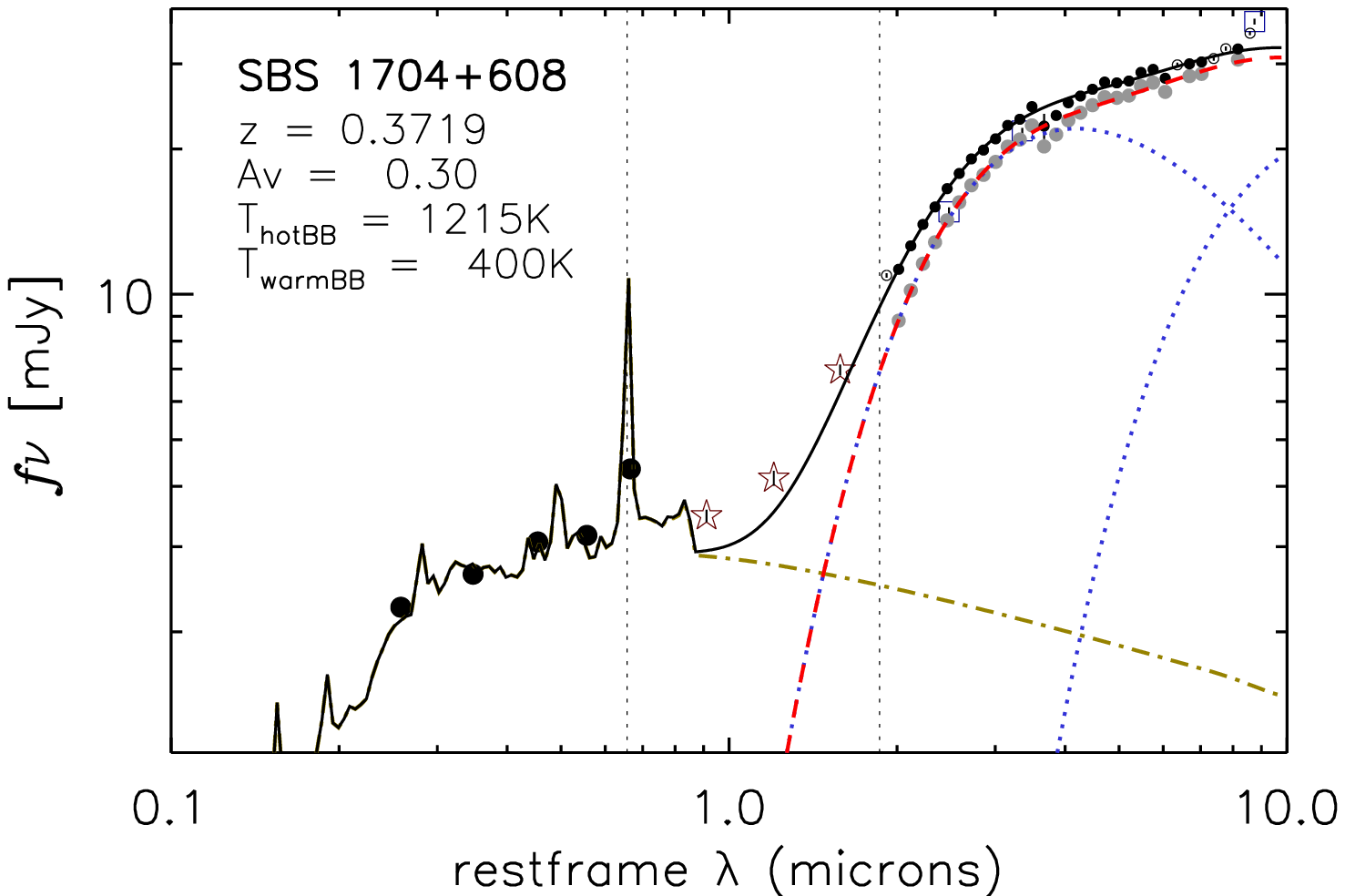}\vspace{0.4cm}
\includegraphics[width=8.4cm]{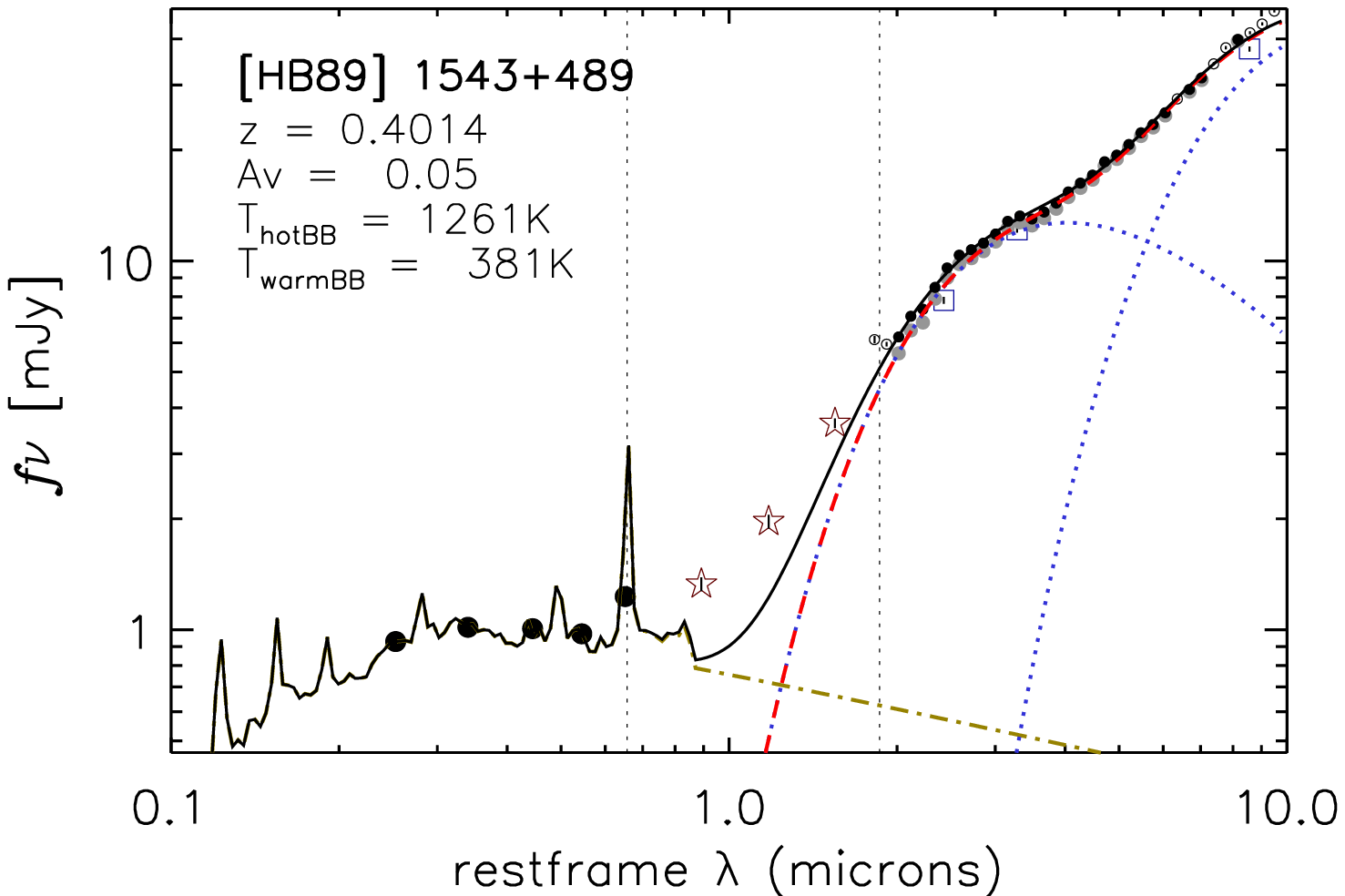}
\includegraphics[width=8.4cm]{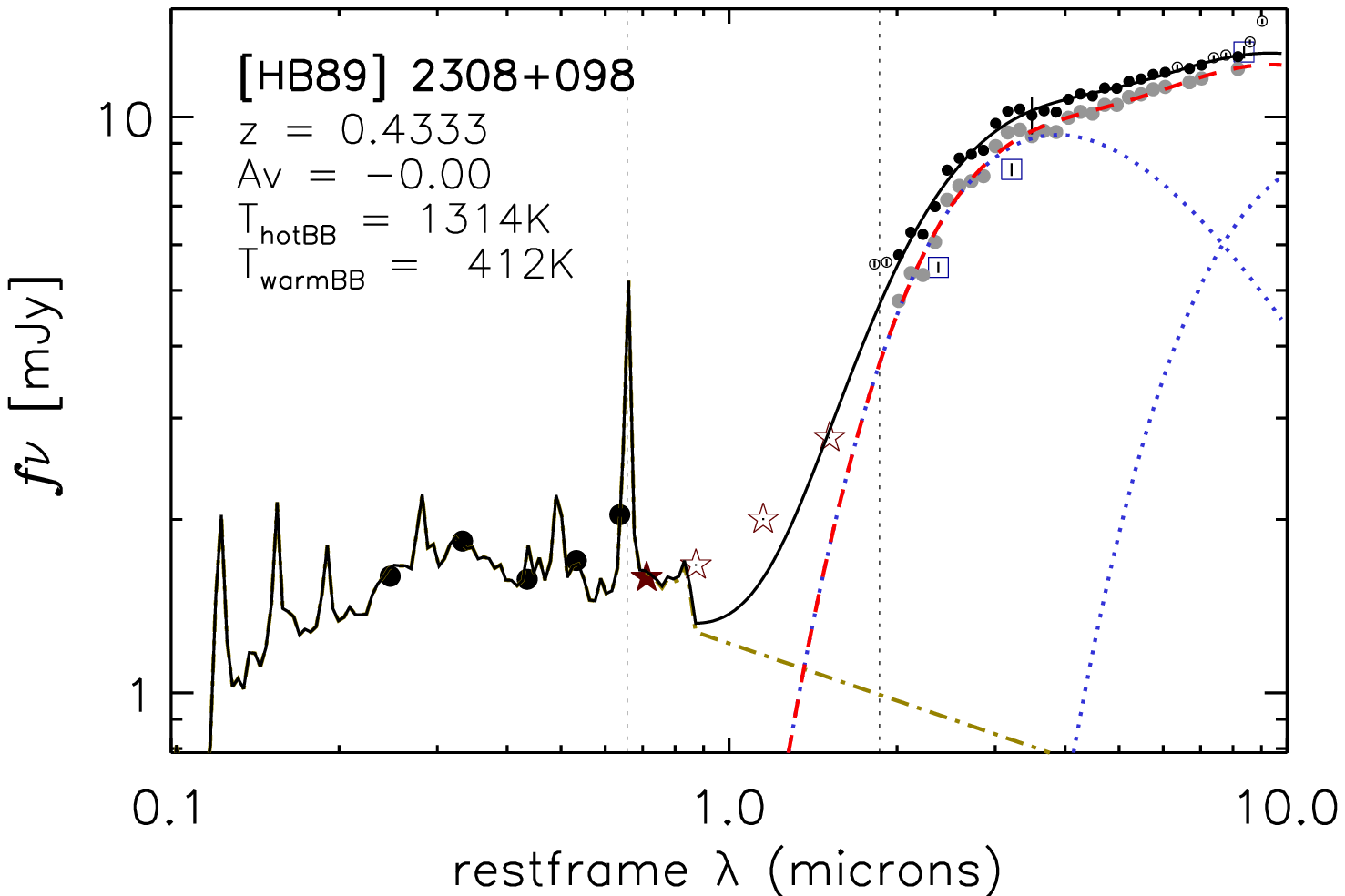}
\caption[]{Spectral energy distributions of the sources in the sample. Circles, stars, and squares represent broadband photometry in the observed-frame optical (from SDSS), NIR (2MASS/UKIDSS/VHS), and MIR (WISE), respectively. Filled symbols indicate bands used to fit the disk component (yellow dot-dashed line). The (AKARI+)IRS spectra resampled at $\lambda$/$\Delta\lambda$ = 20 are shown with small black dots. Grey dots below the (AKARI+)IRS spectra represent the dust spectrum obtained after subtraction of the disk component. The model for the dust spectrum (red dashed line) is the linear combination of two black-bodies at adjustable temperatures (blue dotted lines). The combined disk+dust model is represented by the black solid line. The vertical dotted lines mark the restframe wavelength of the H$\alpha$ and Pa$\alpha$ recombination lines.}
\end{figure*}

\addtocounter{figure}{-1}
\begin{figure*}
\includegraphics[width=8.4cm]{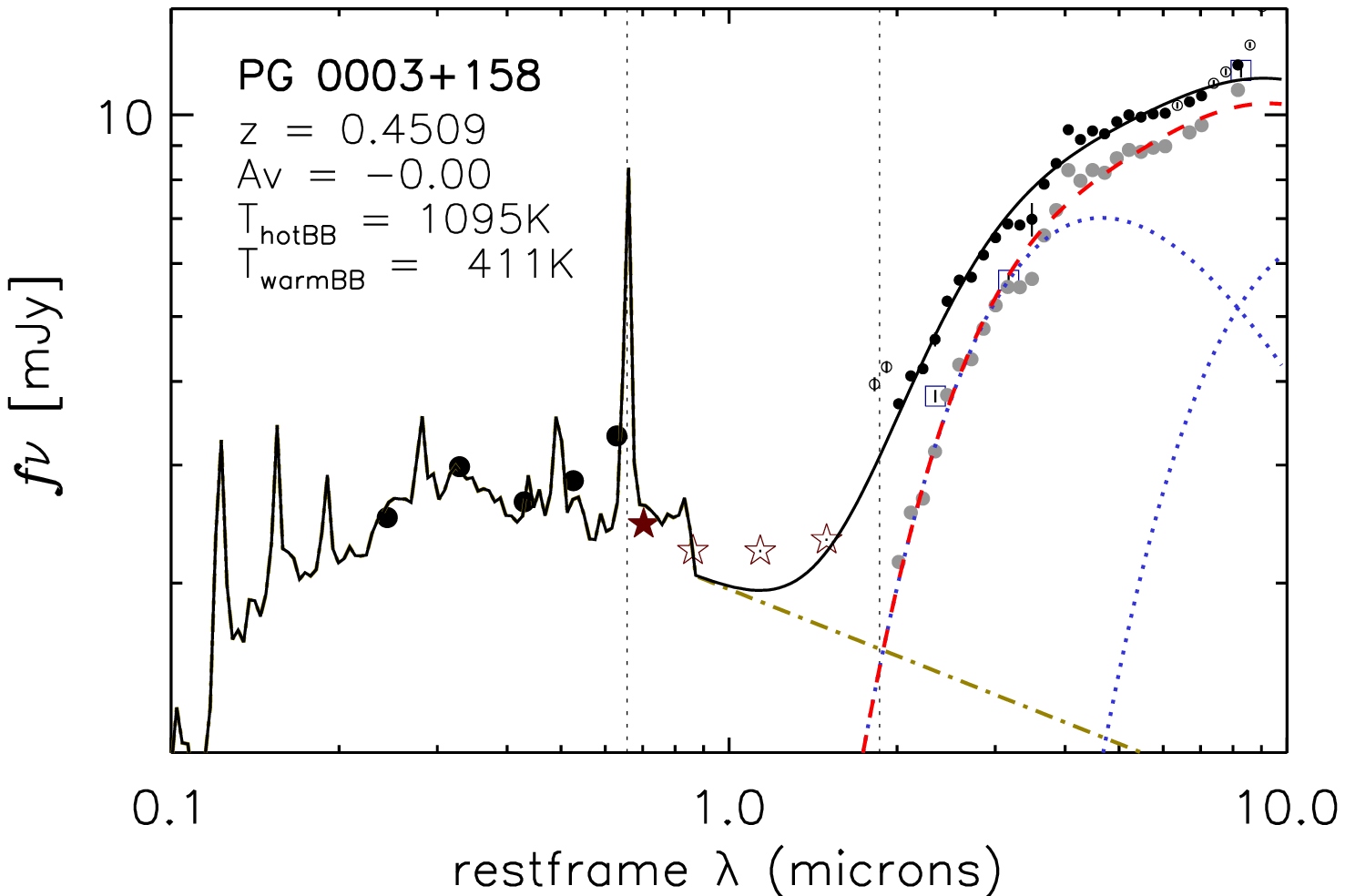}
\includegraphics[width=8.4cm]{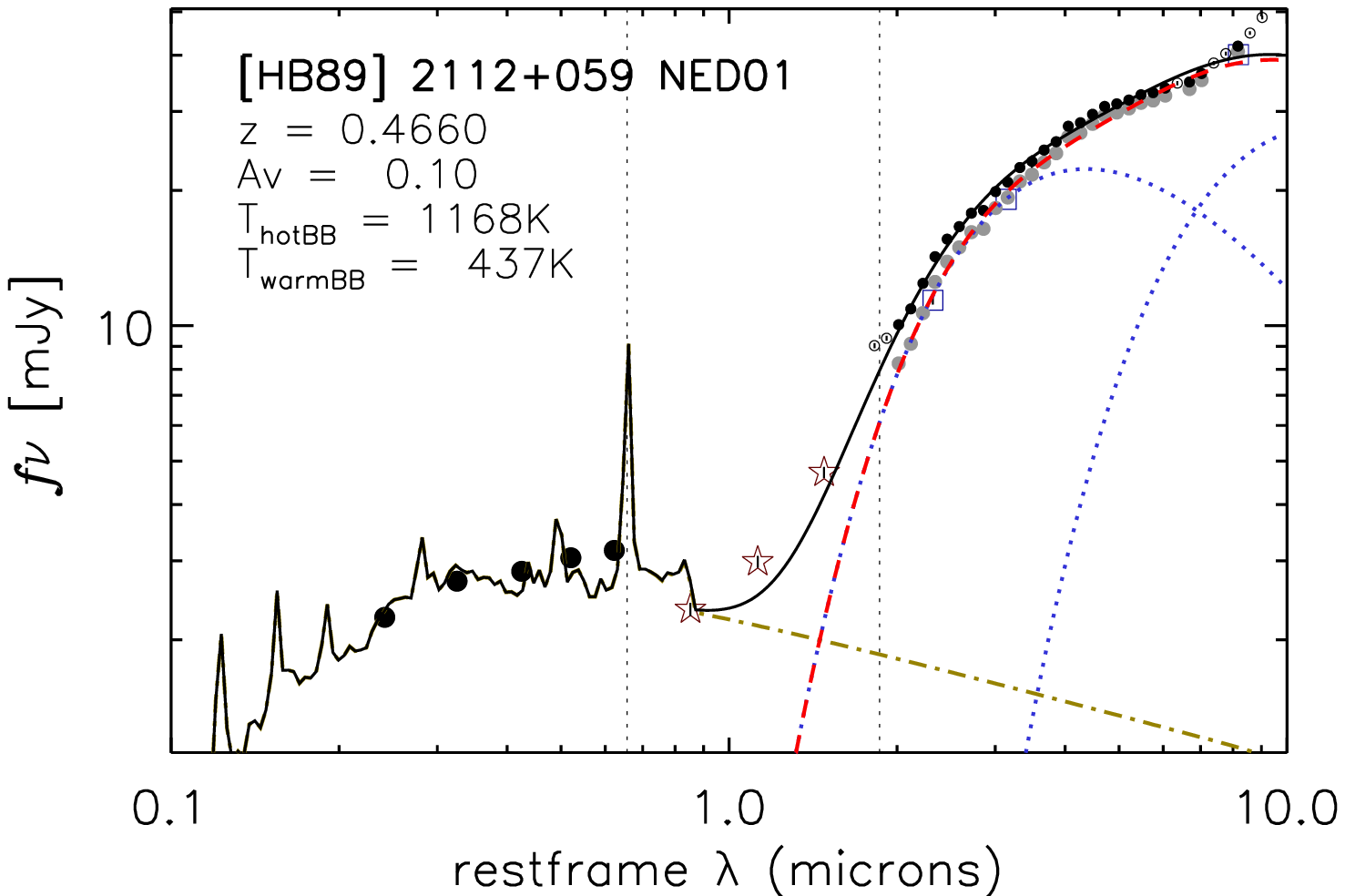}\vspace{0.4cm}
\includegraphics[width=8.4cm]{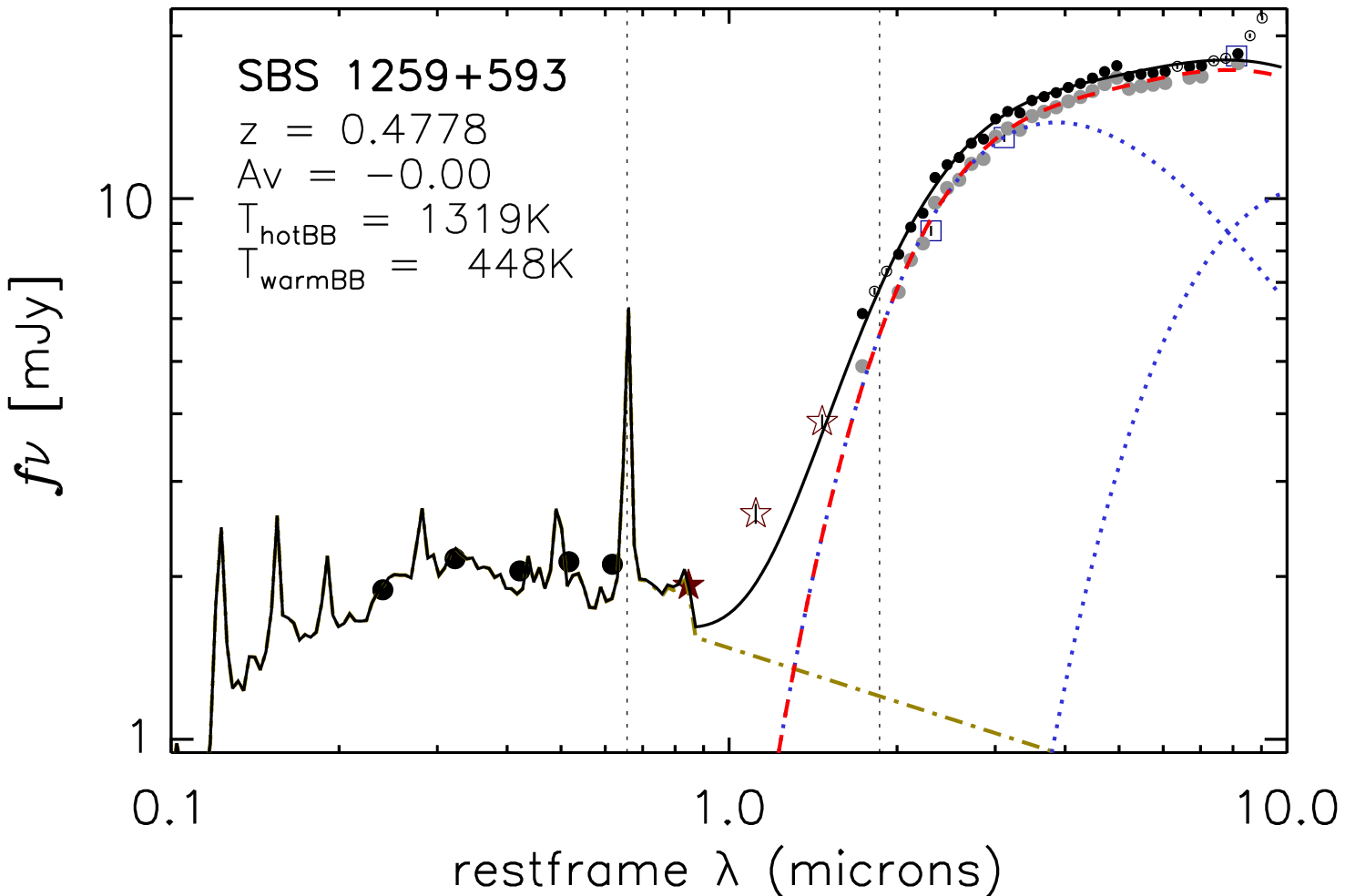}
\includegraphics[width=8.4cm]{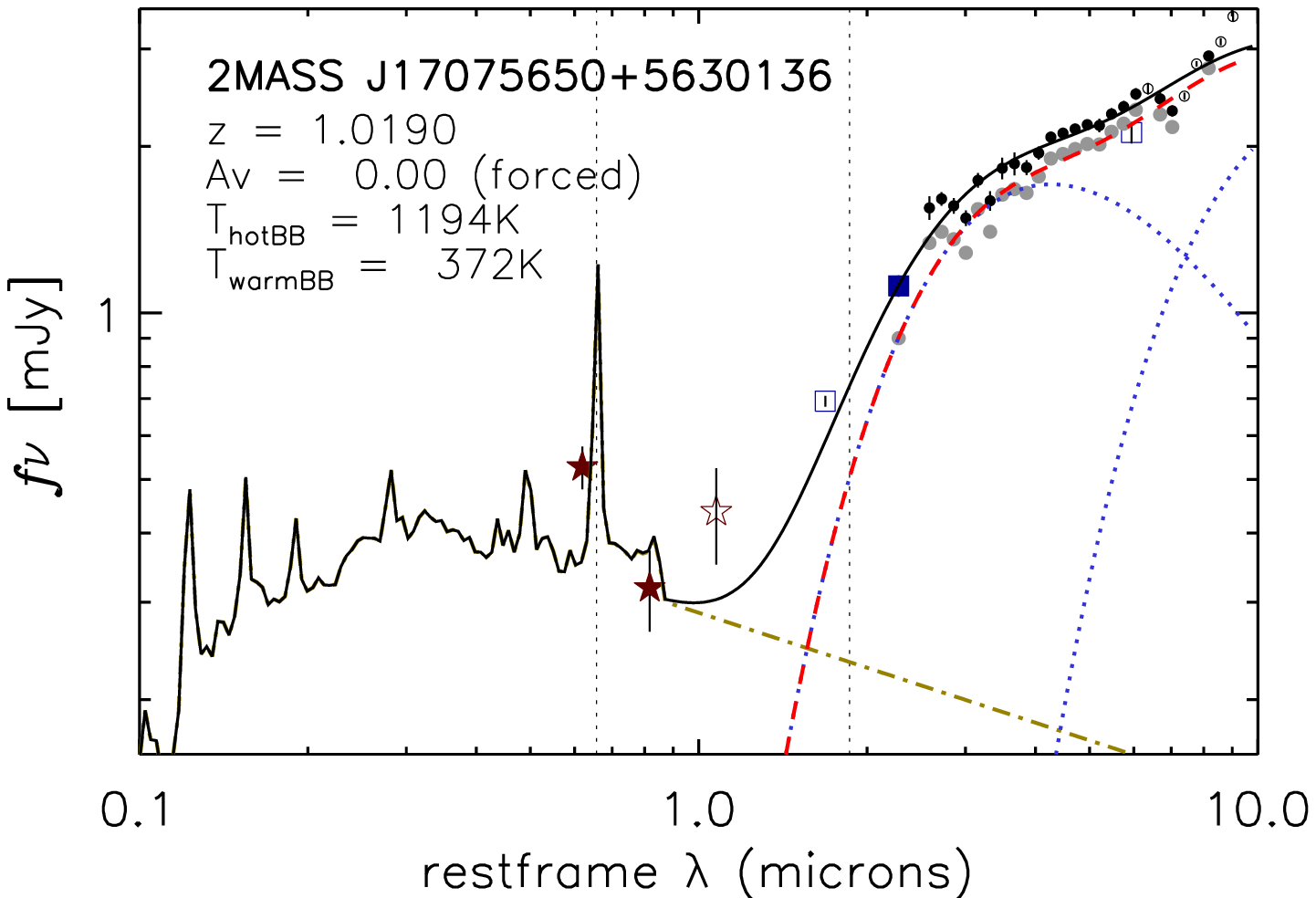}\vspace{0.4cm}
\includegraphics[width=8.4cm]{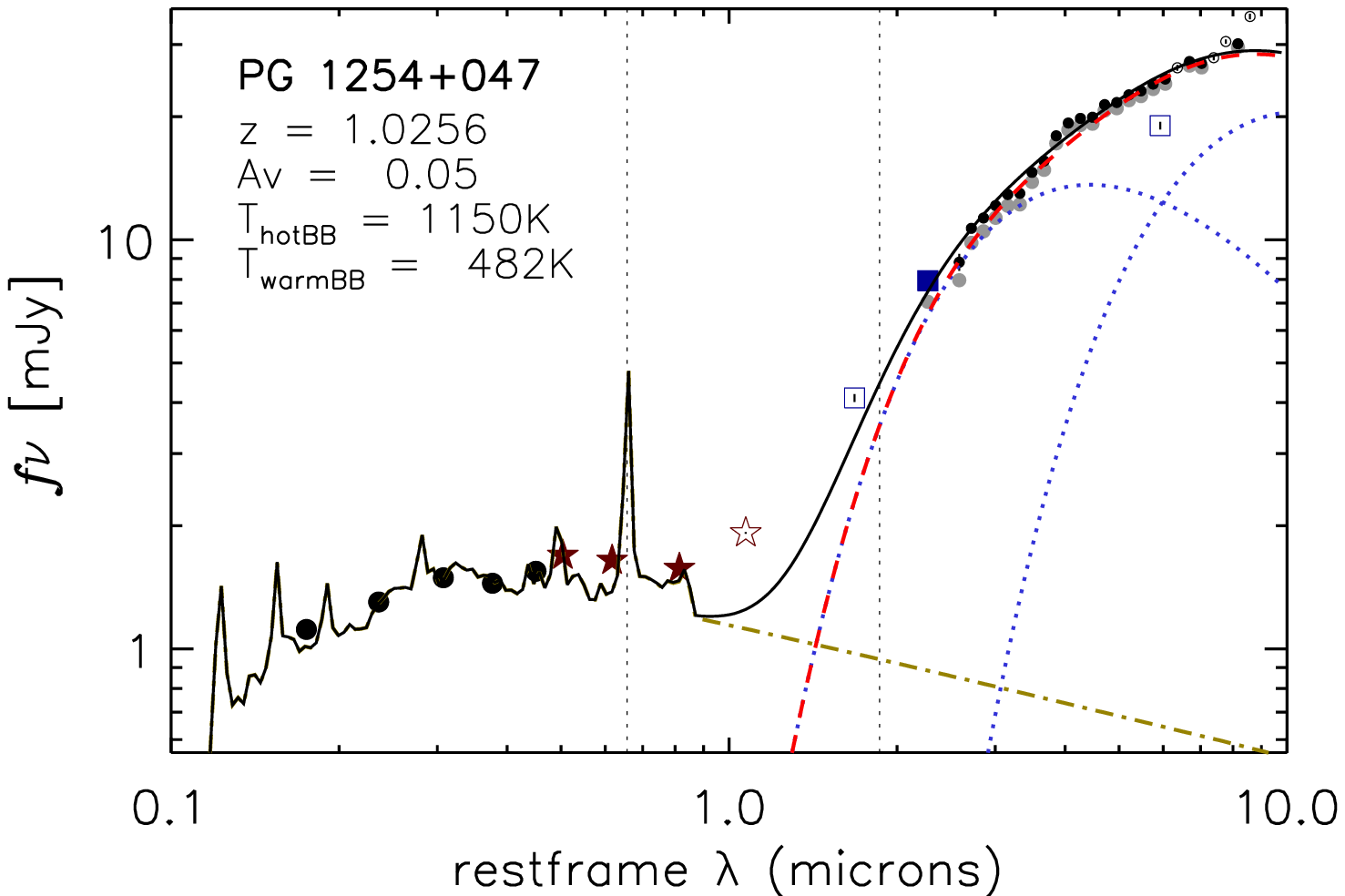}
\includegraphics[width=8.4cm]{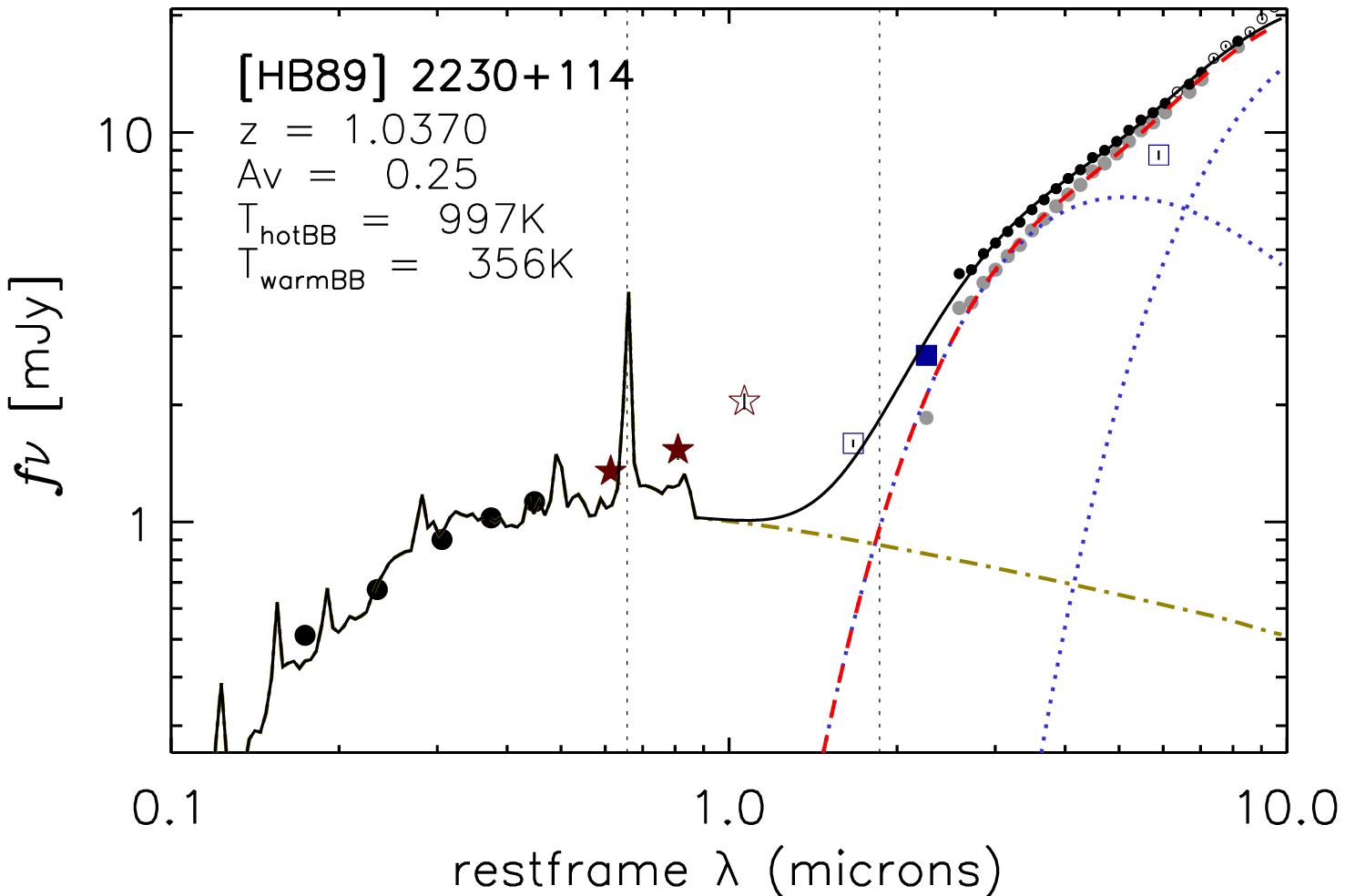}\vspace{0.4cm}
\includegraphics[width=8.4cm]{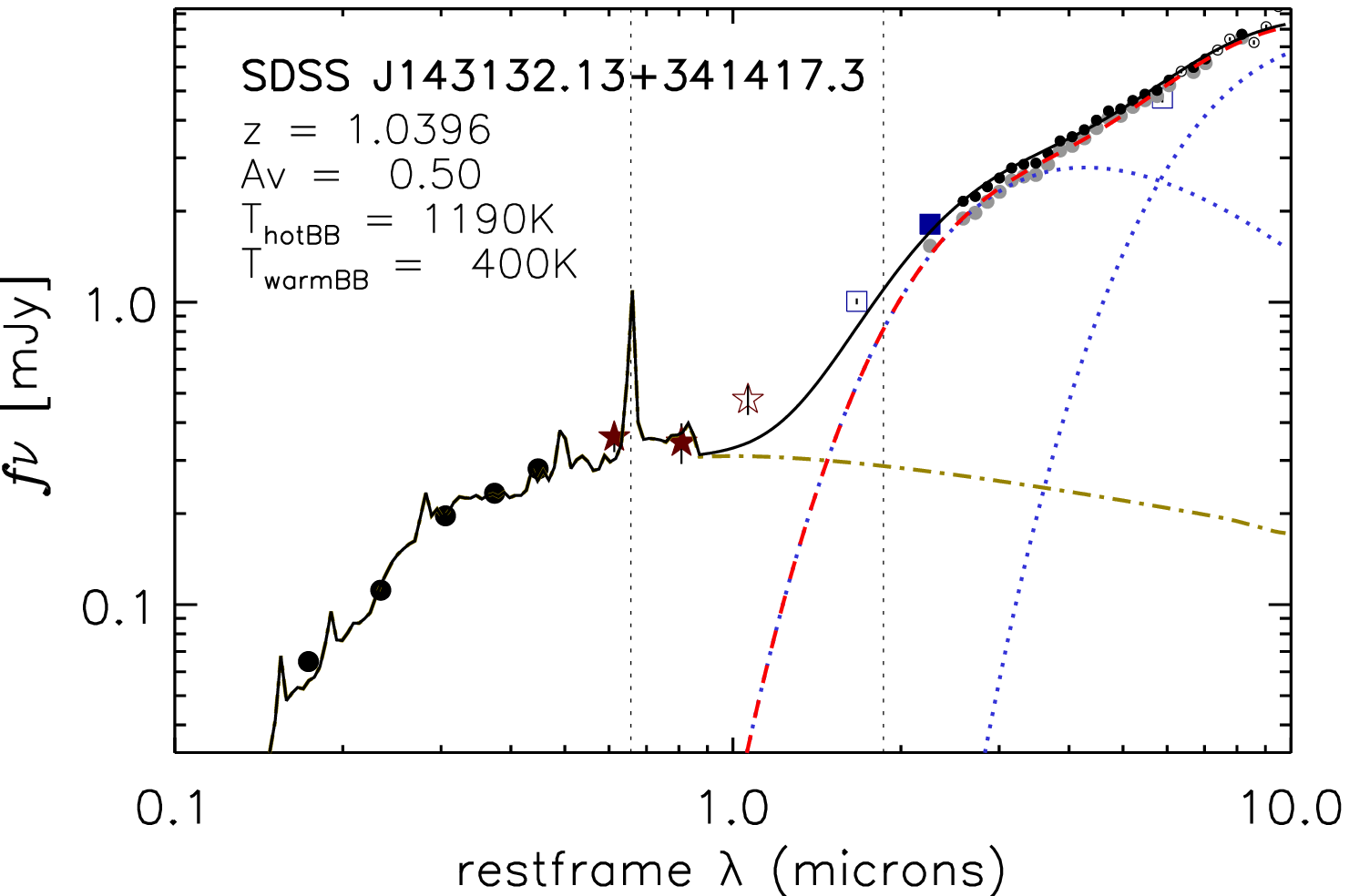}
\includegraphics[width=8.4cm]{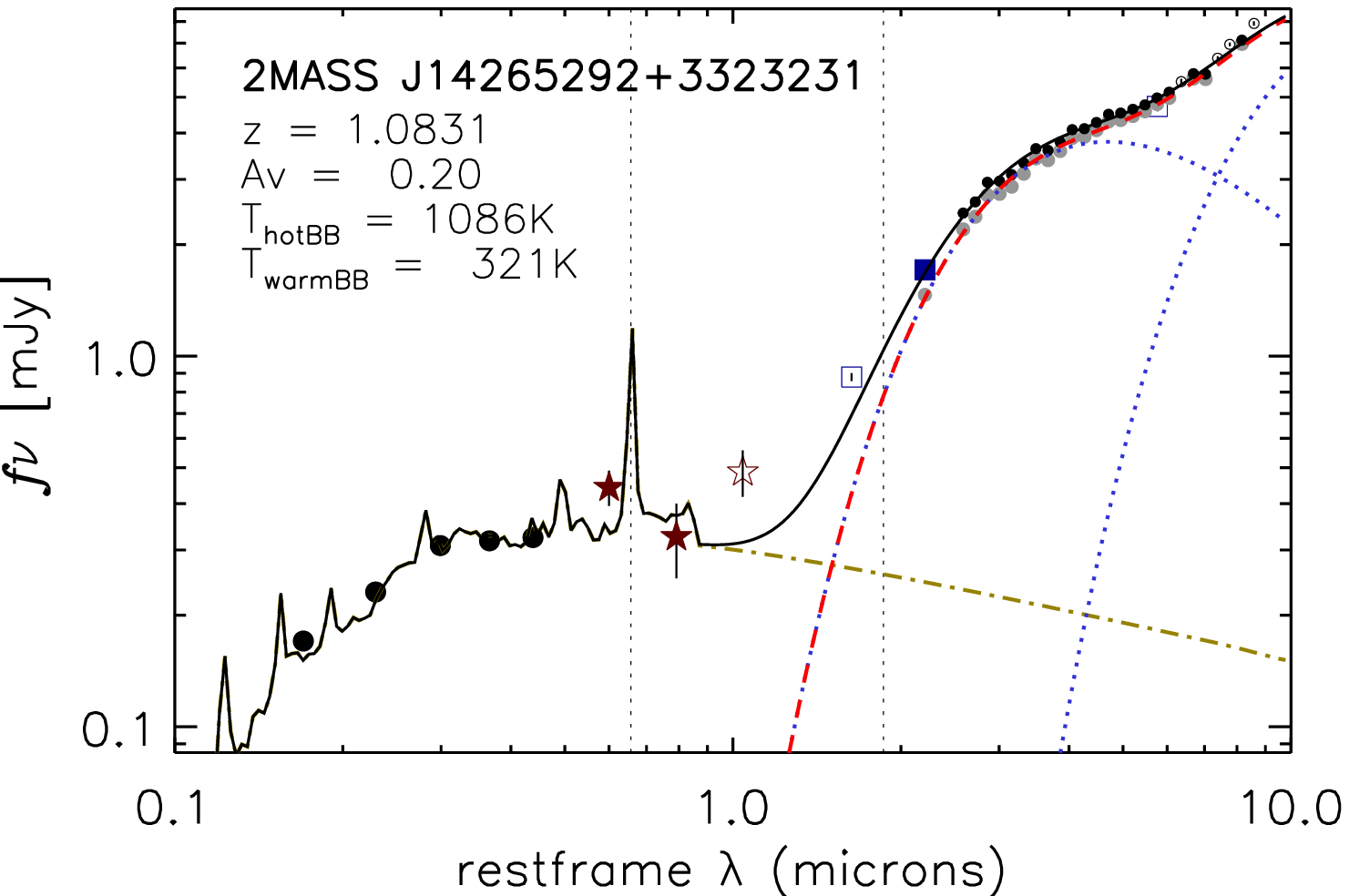}
\caption{continued}
\end{figure*}

\addtocounter{figure}{-1}
\begin{figure*}
\includegraphics[width=8.4cm]{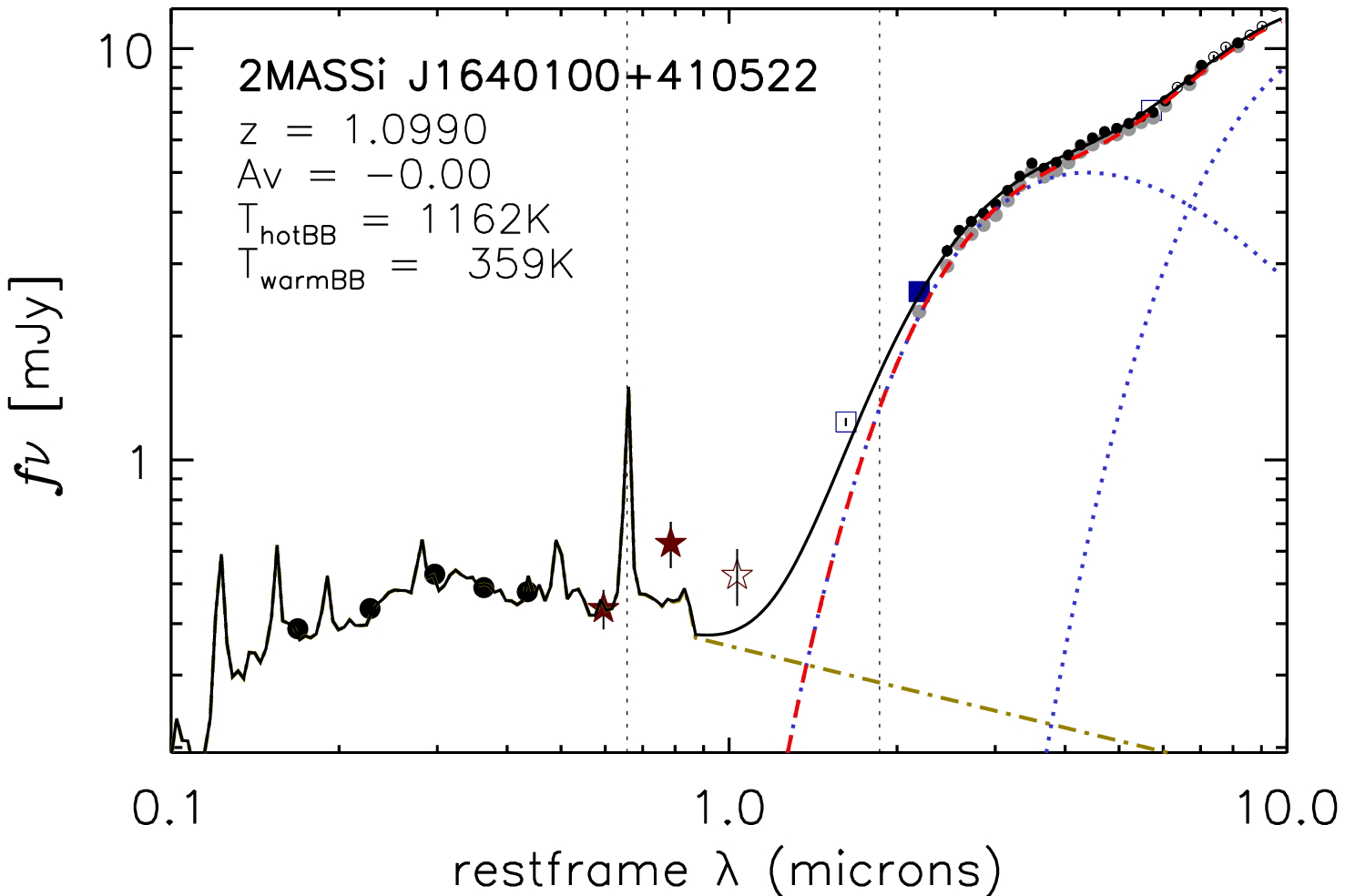}
\includegraphics[width=8.4cm]{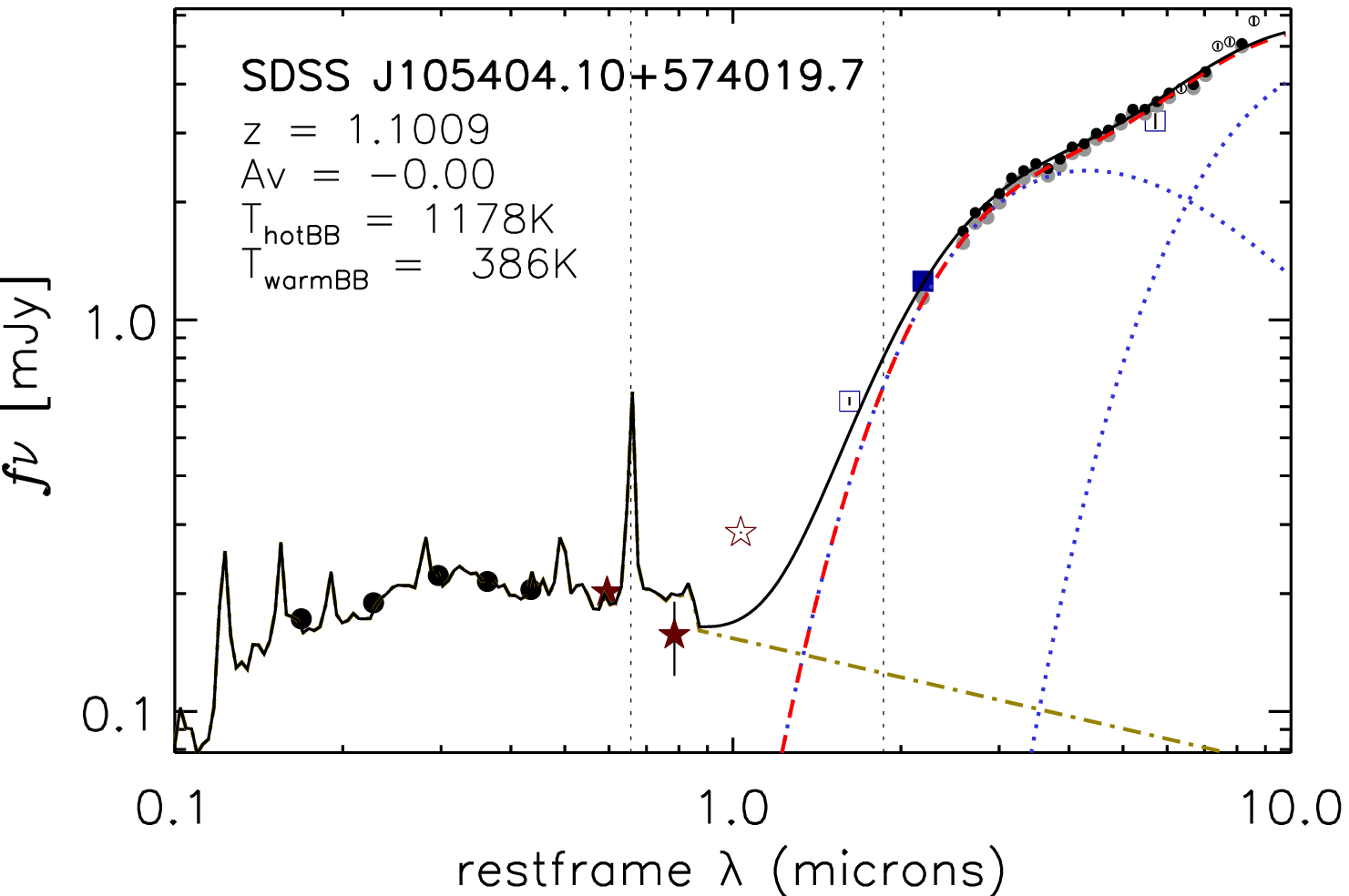}\vspace{0.4cm}
\includegraphics[width=8.4cm]{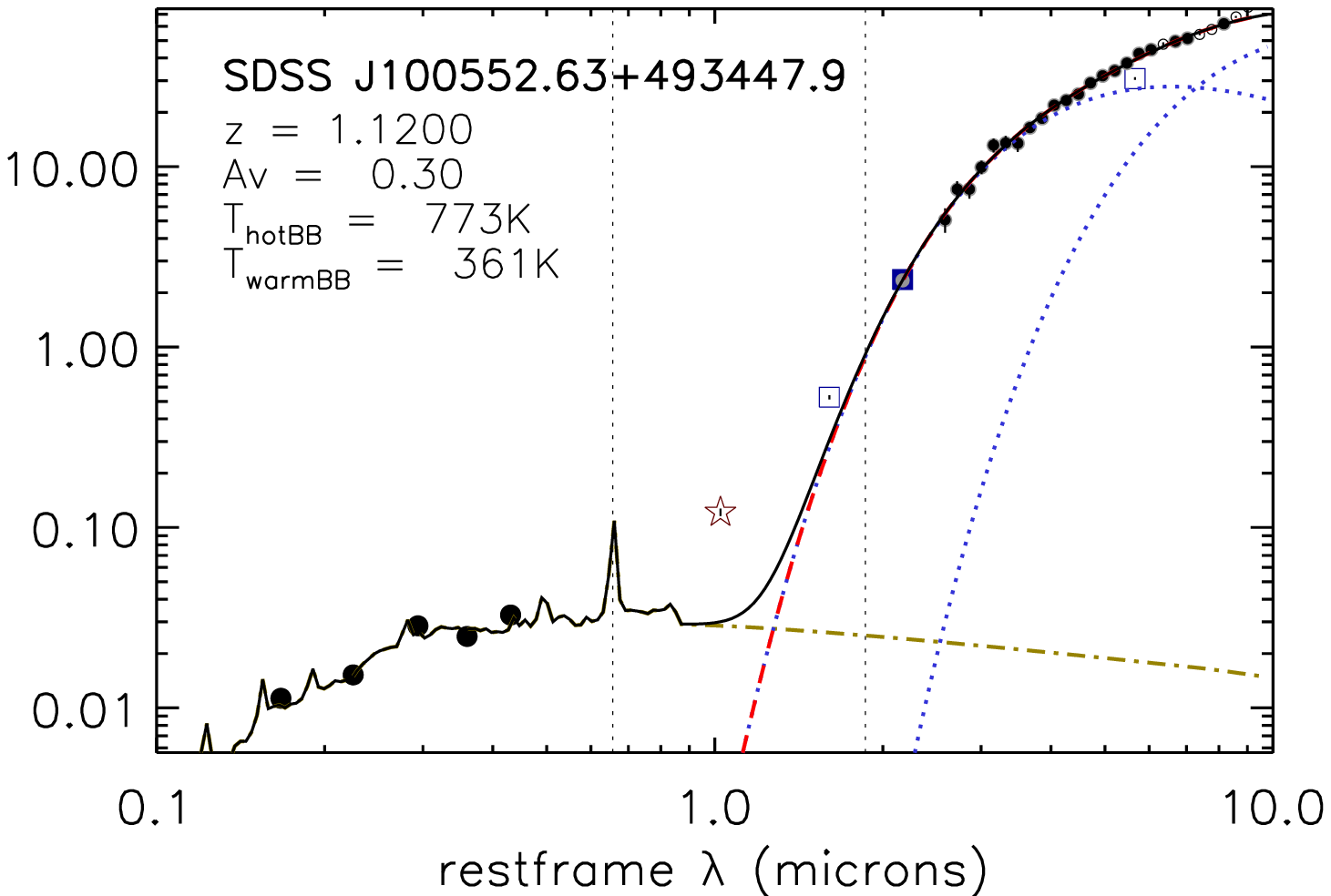}
\includegraphics[width=8.4cm]{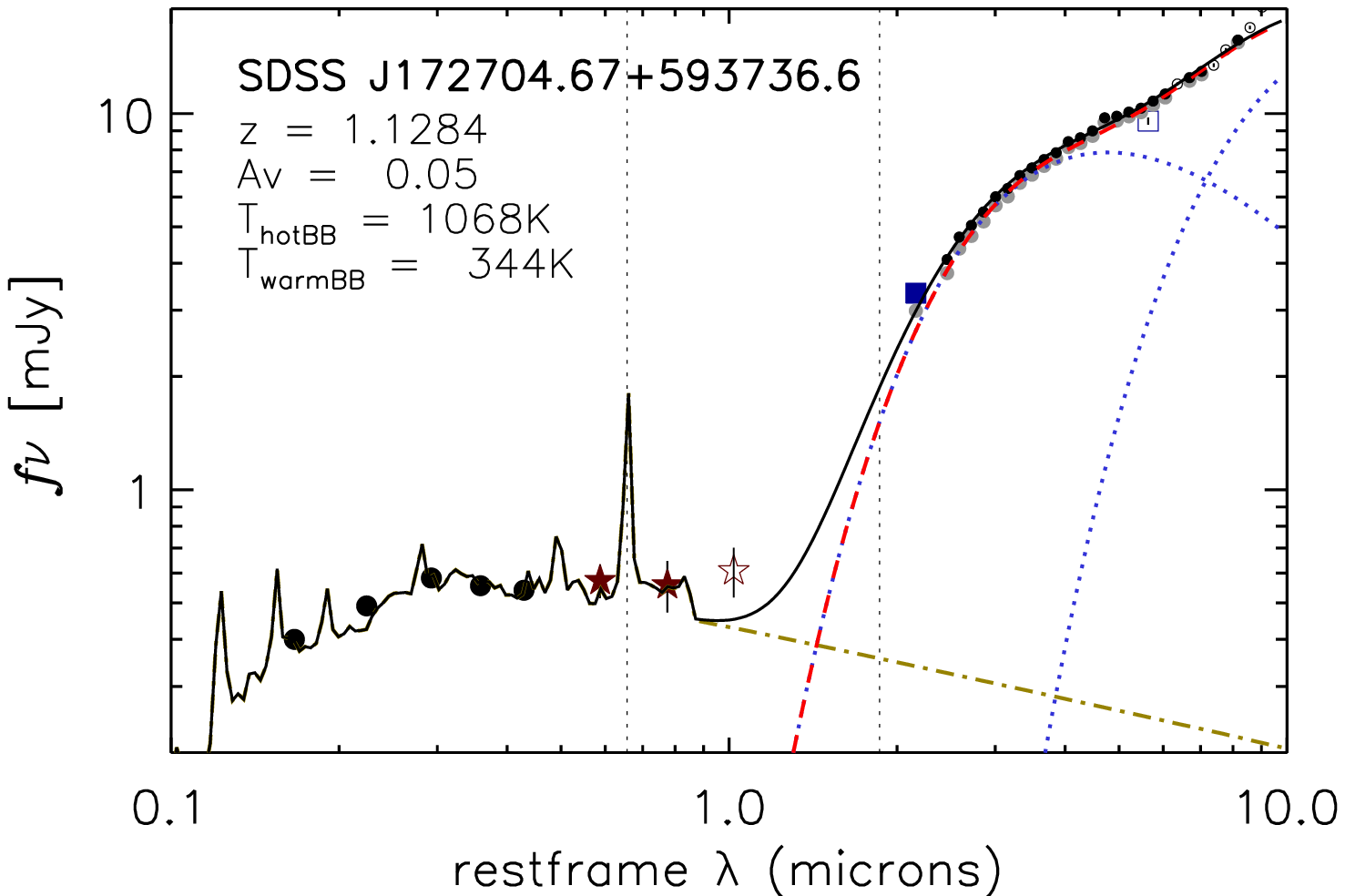}\vspace{0.4cm}
\includegraphics[width=8.4cm]{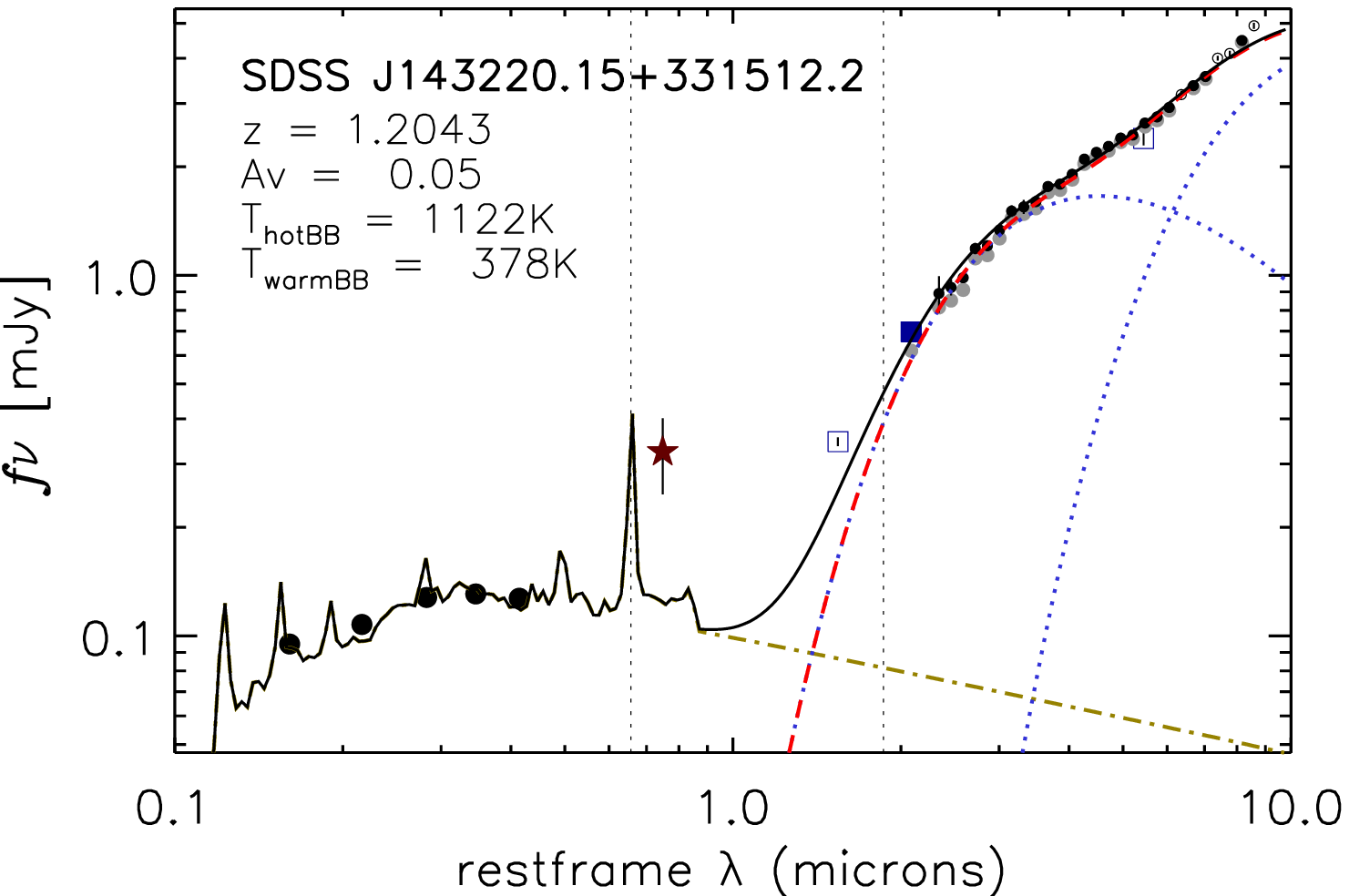}
\includegraphics[width=8.4cm]{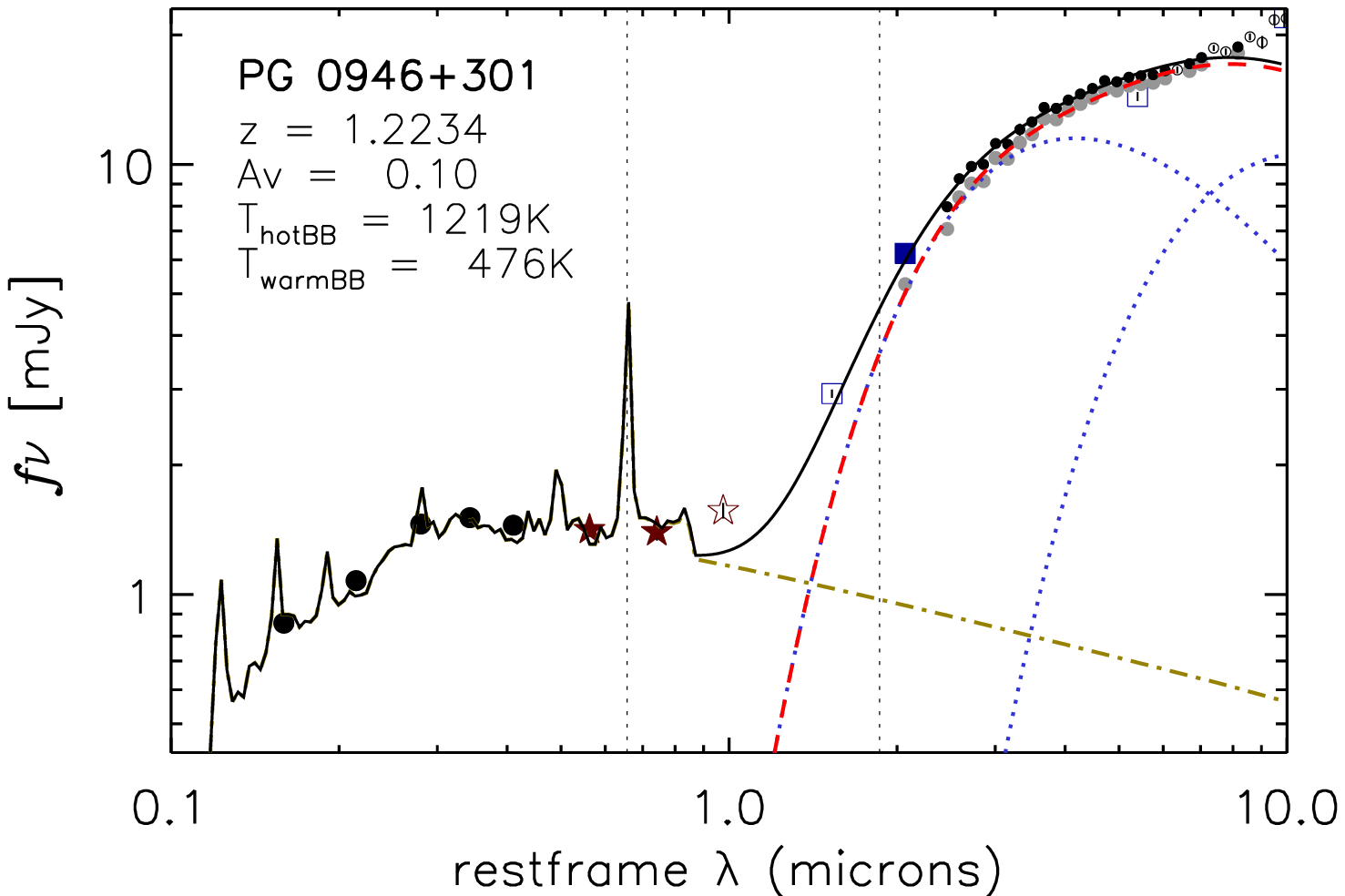}\vspace{0.4cm}
\includegraphics[width=8.4cm]{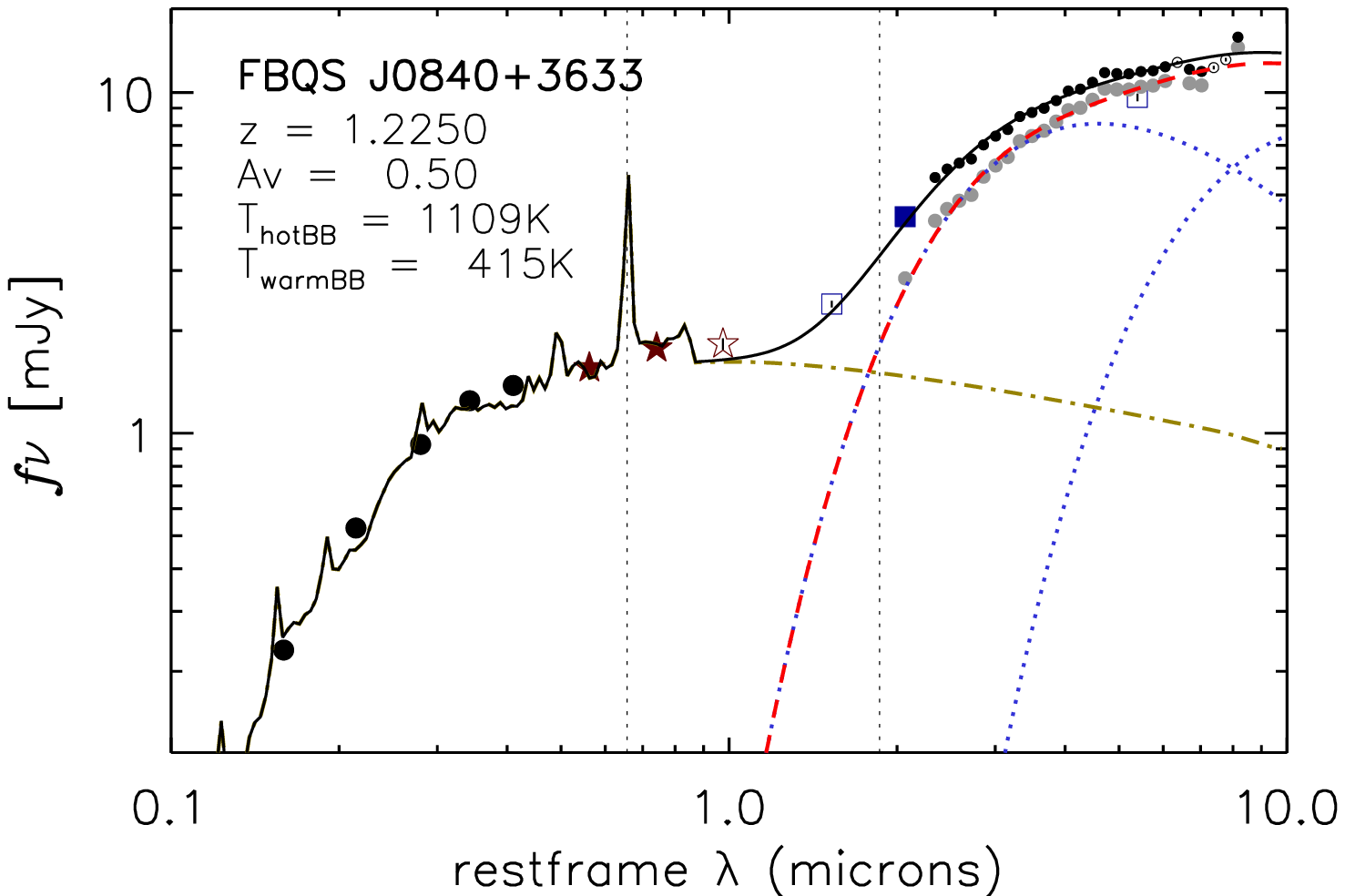}
\includegraphics[width=8.4cm]{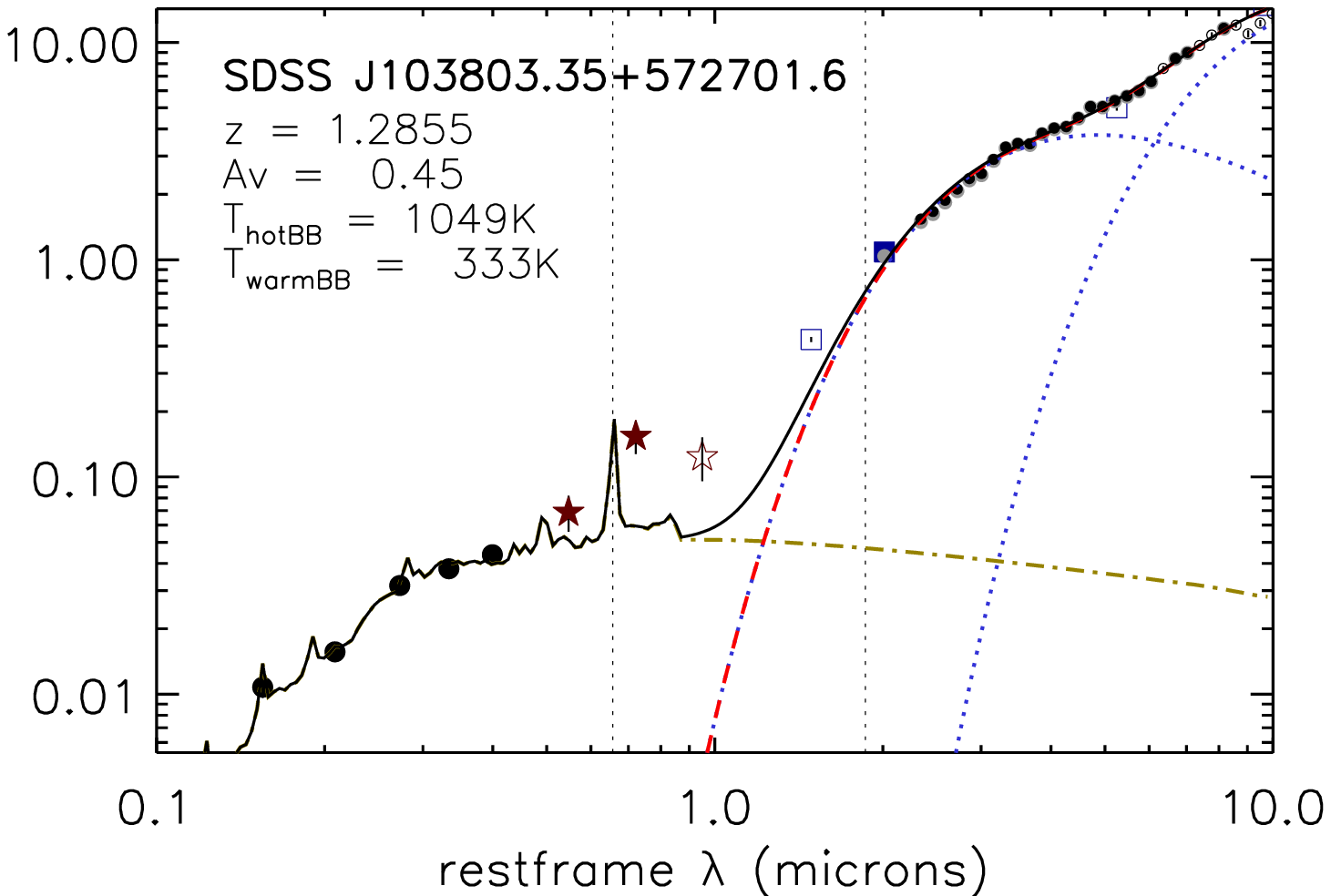}
\caption{continued}
\end{figure*}

\addtocounter{figure}{-1}
\begin{figure*}
\includegraphics[width=8.4cm]{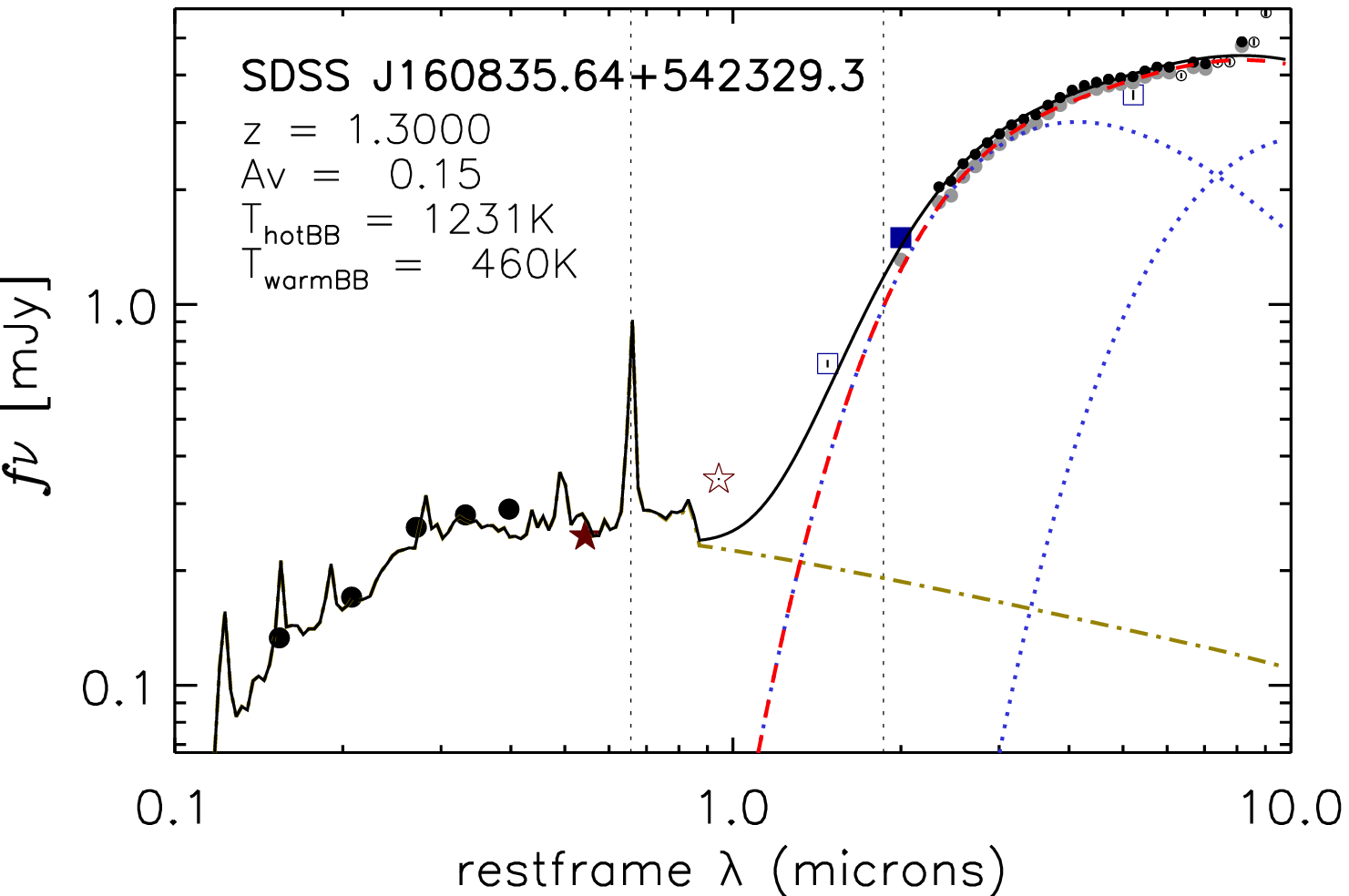}
\includegraphics[width=8.4cm]{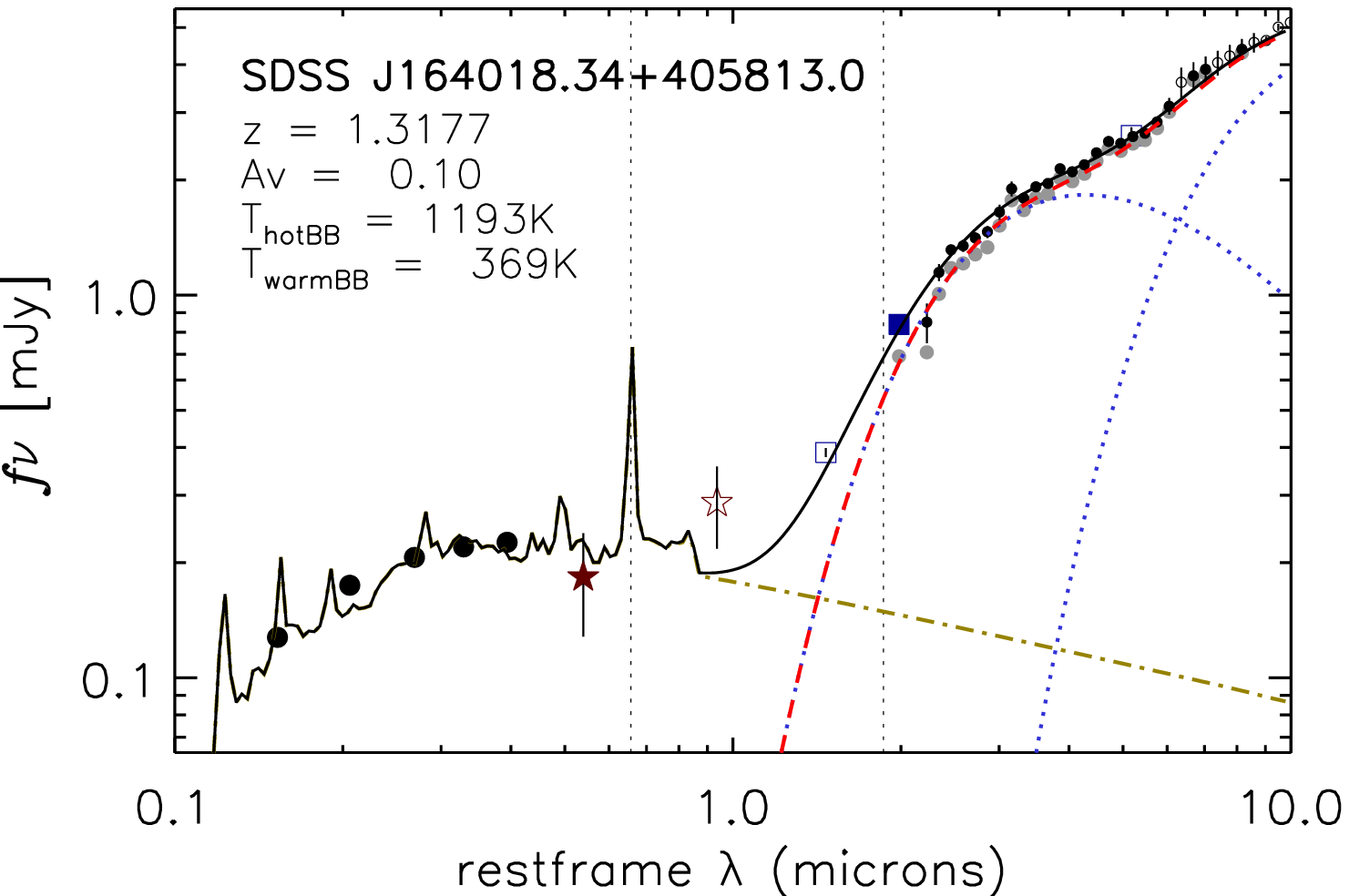}\vspace{0.4cm}
\includegraphics[width=8.4cm]{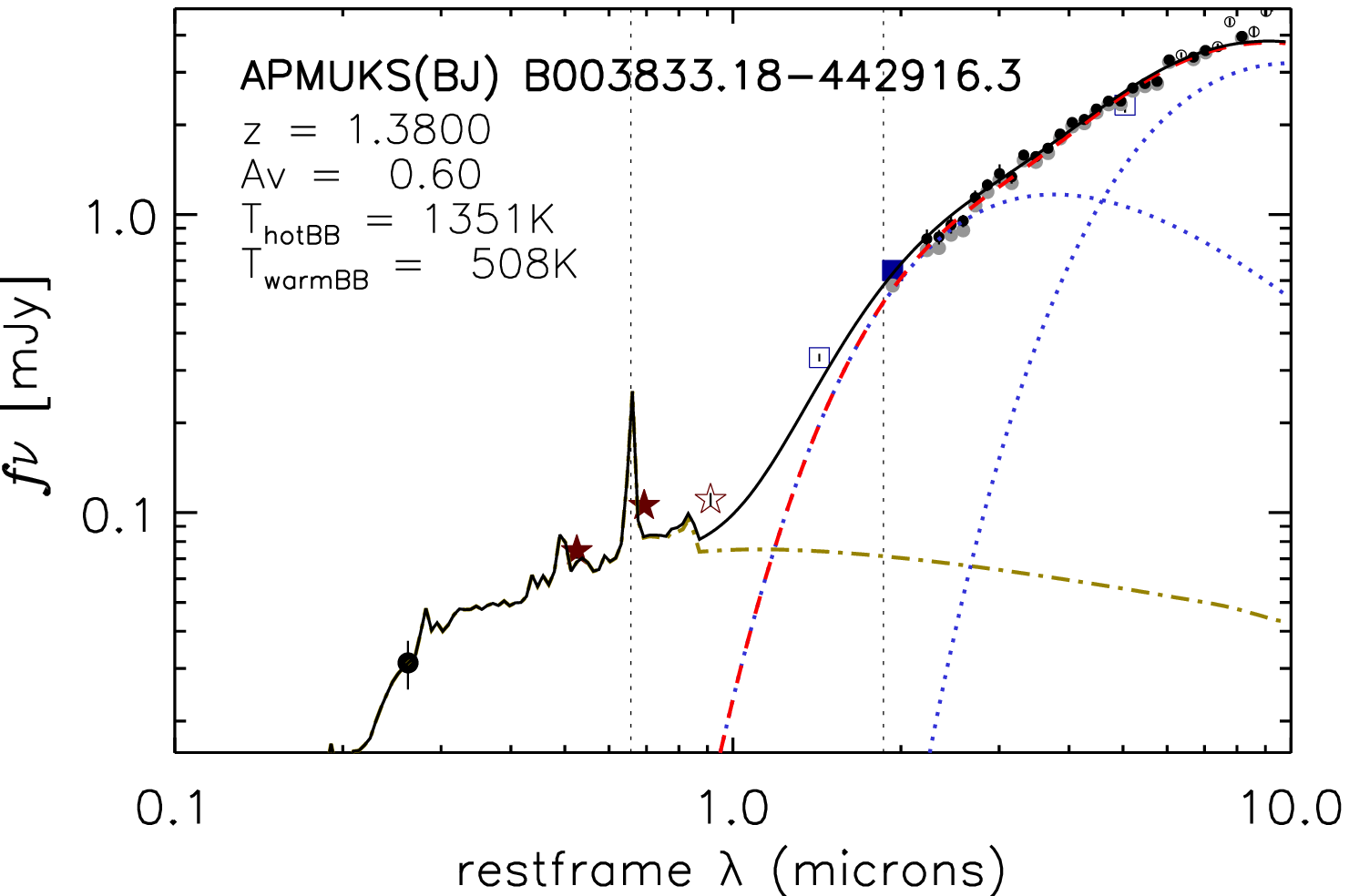}
\includegraphics[width=8.4cm]{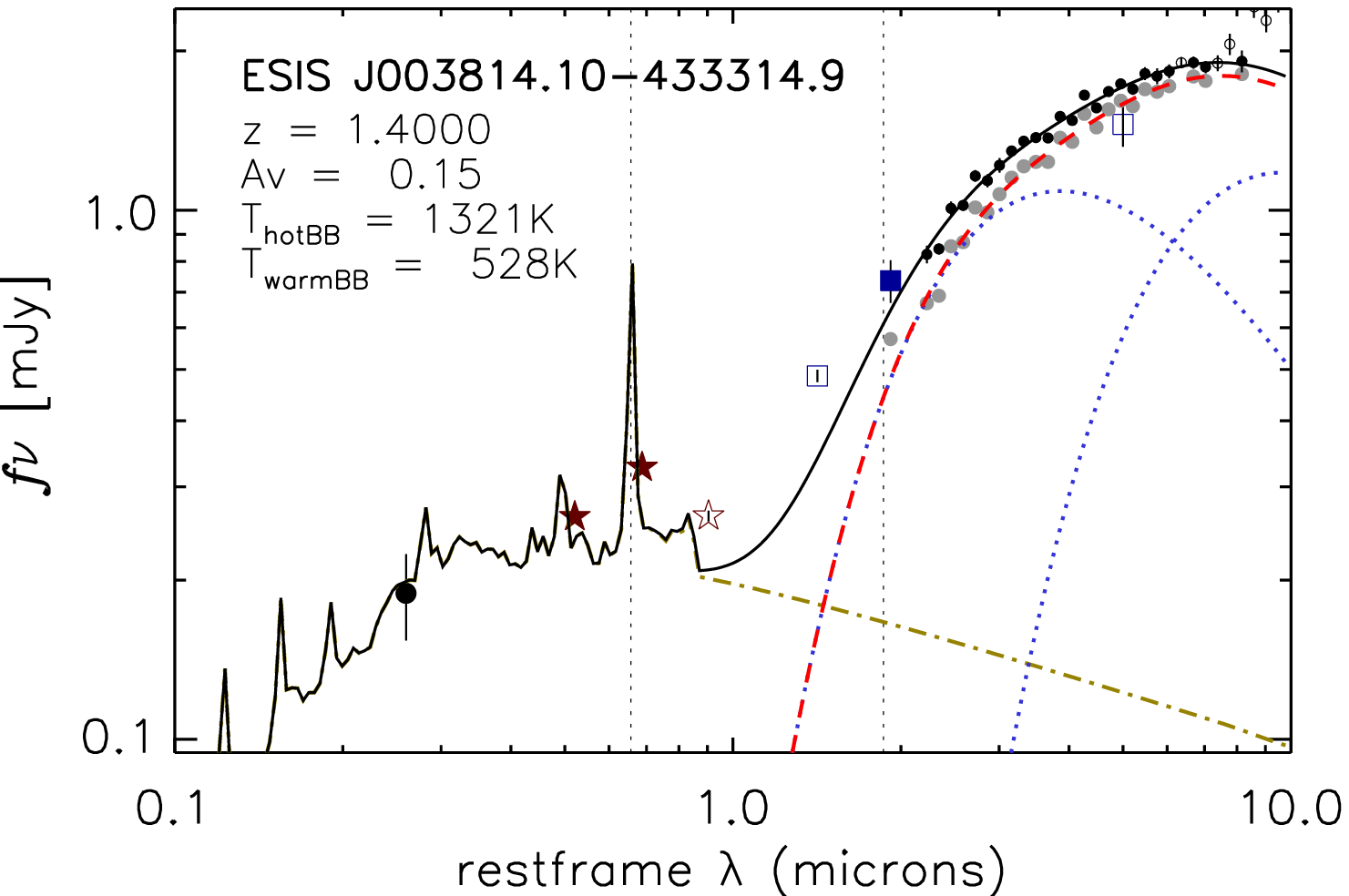}\vspace{0.4cm}
\includegraphics[width=8.4cm]{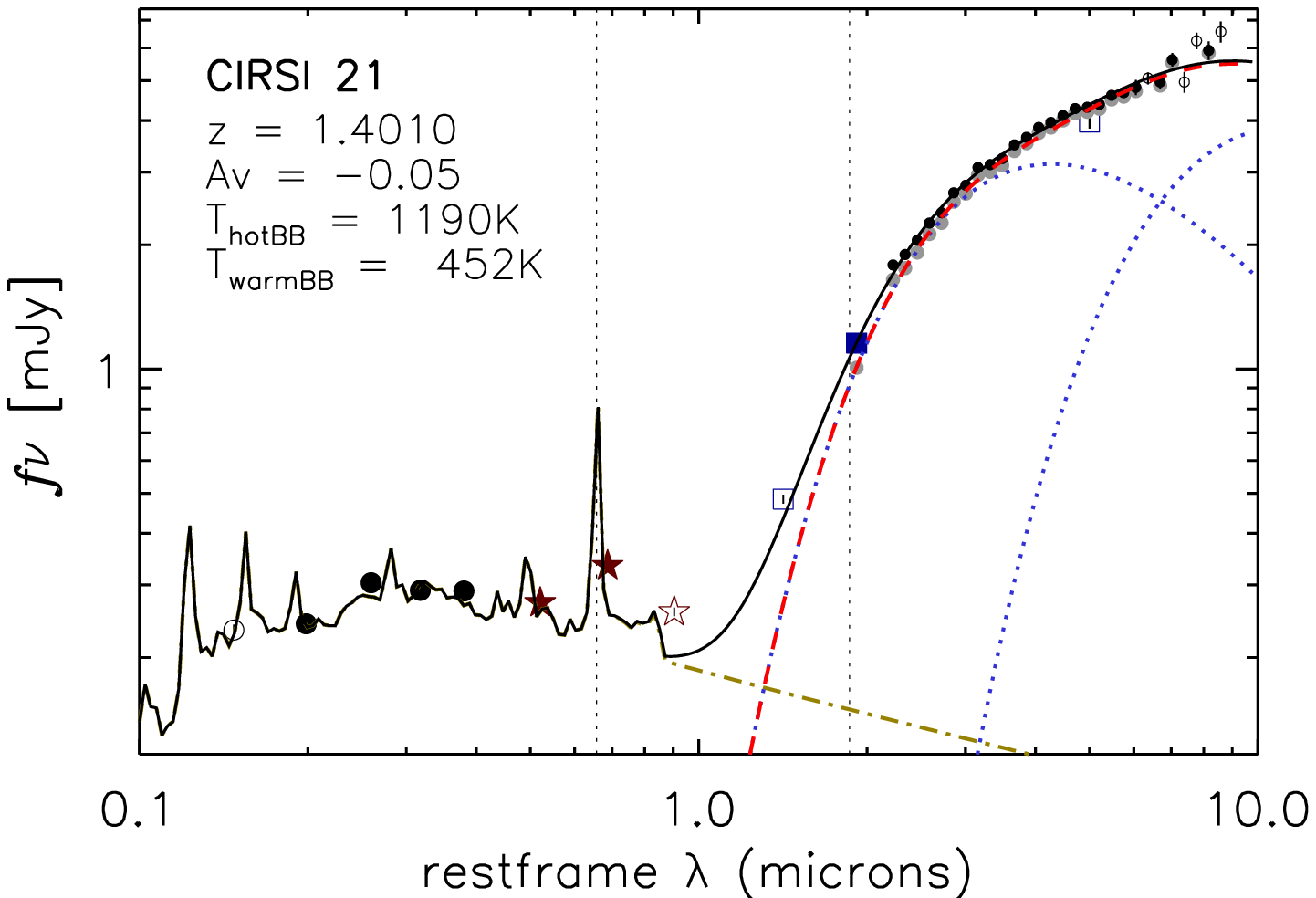}
\includegraphics[width=8.4cm]{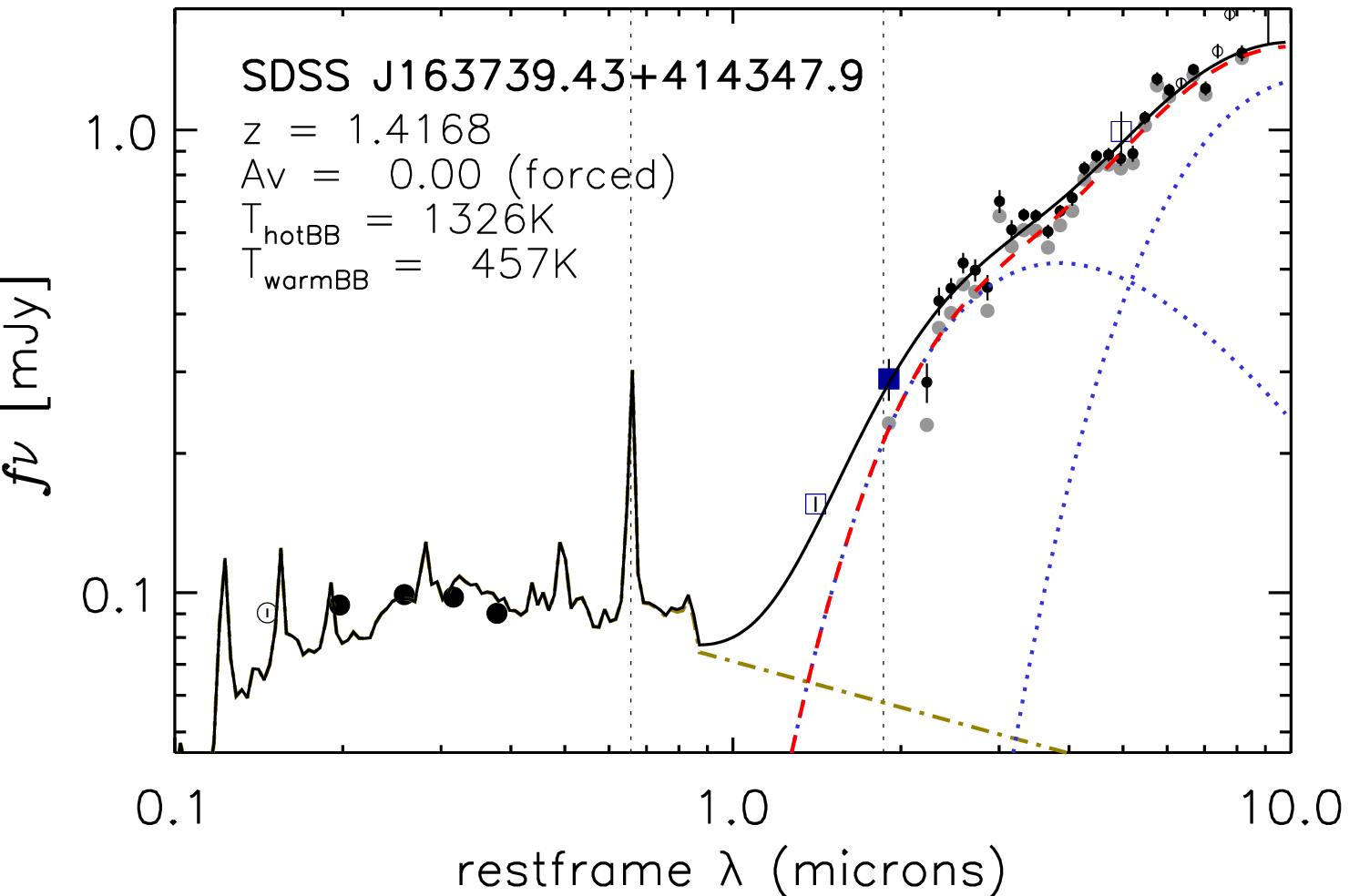}\vspace{0.4cm}
\includegraphics[width=8.4cm]{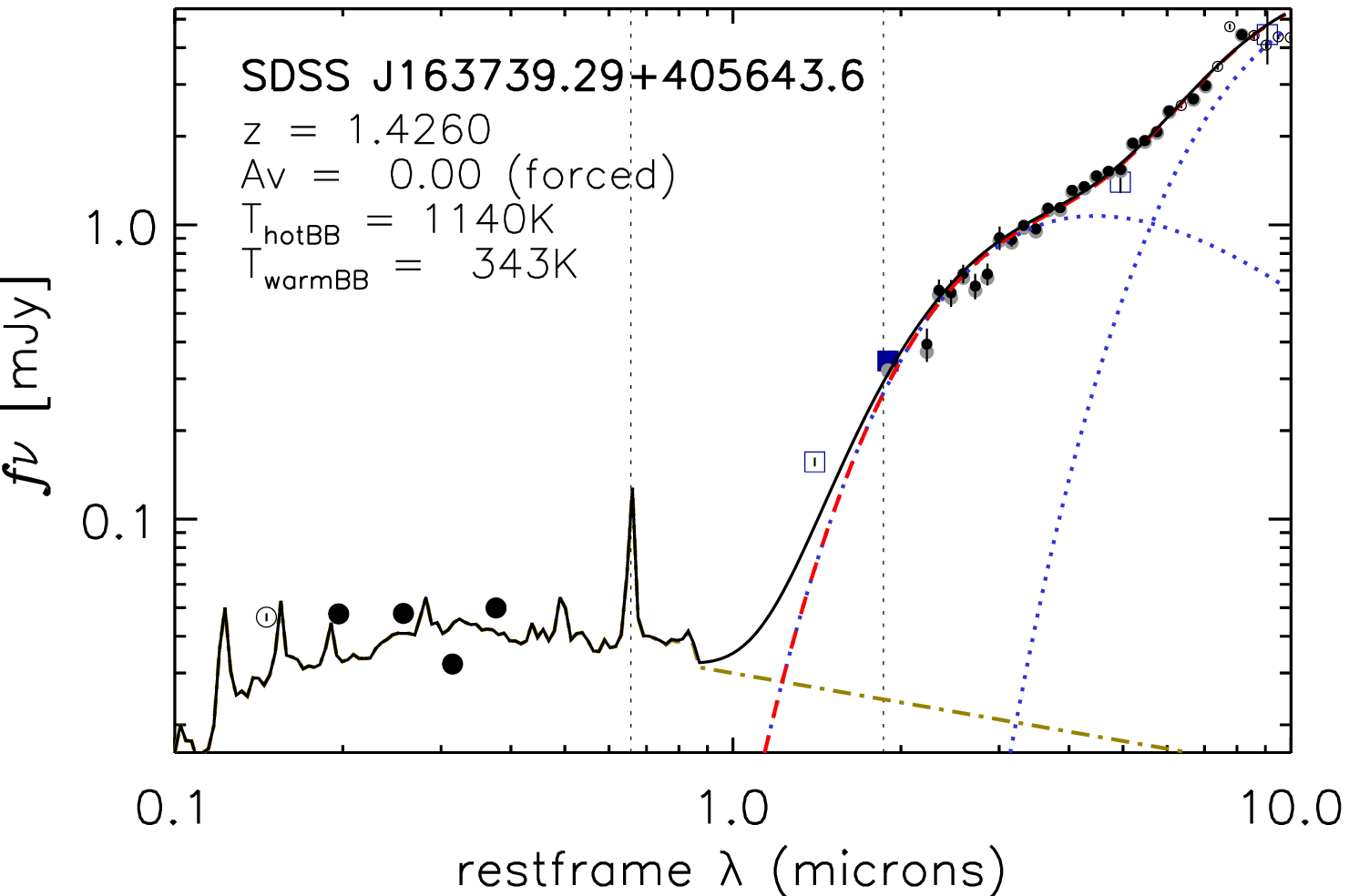}
\includegraphics[width=8.4cm]{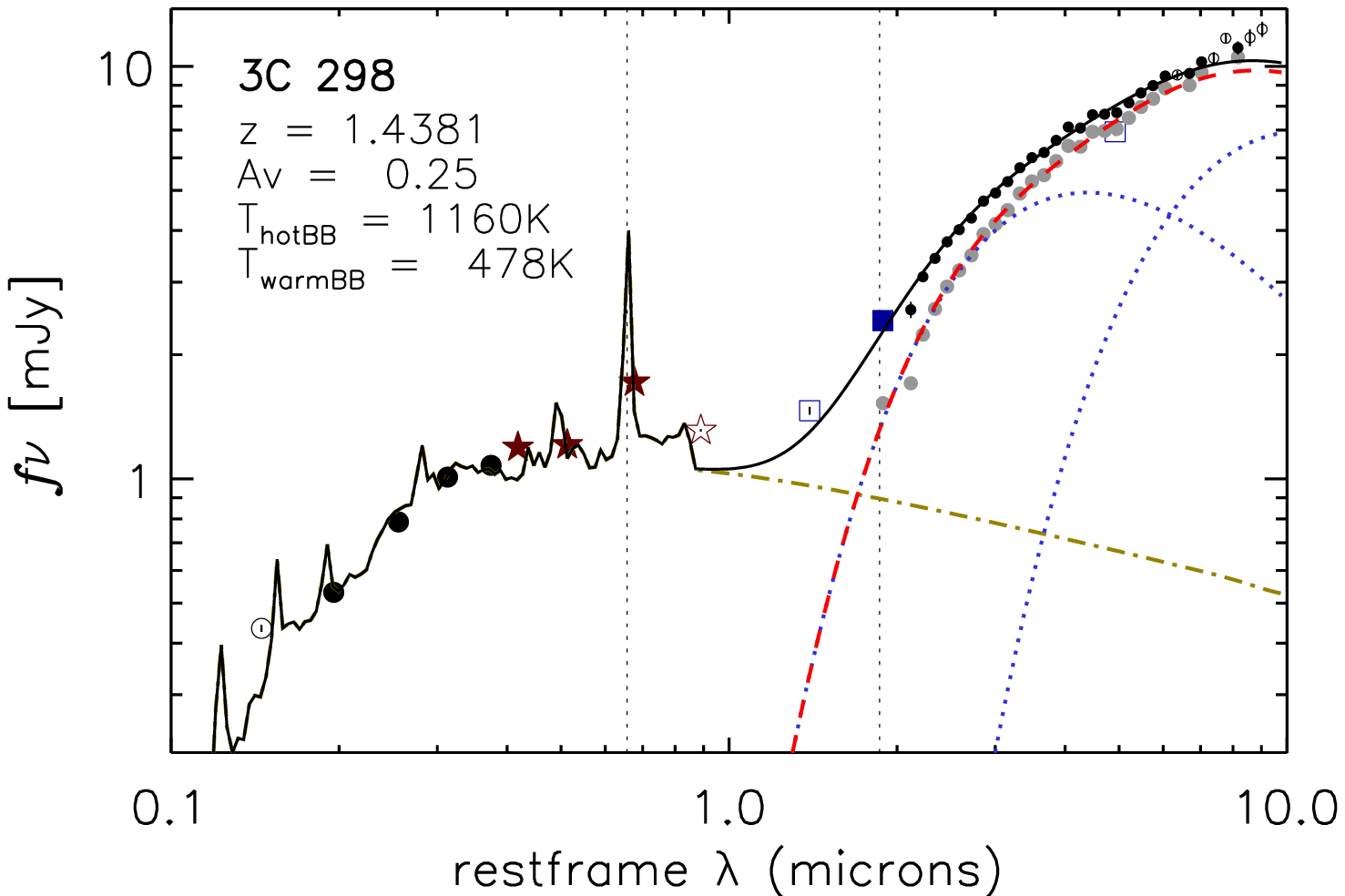}
\caption{continued}
\end{figure*}

\addtocounter{figure}{-1}
\begin{figure*}
\includegraphics[width=8.4cm]{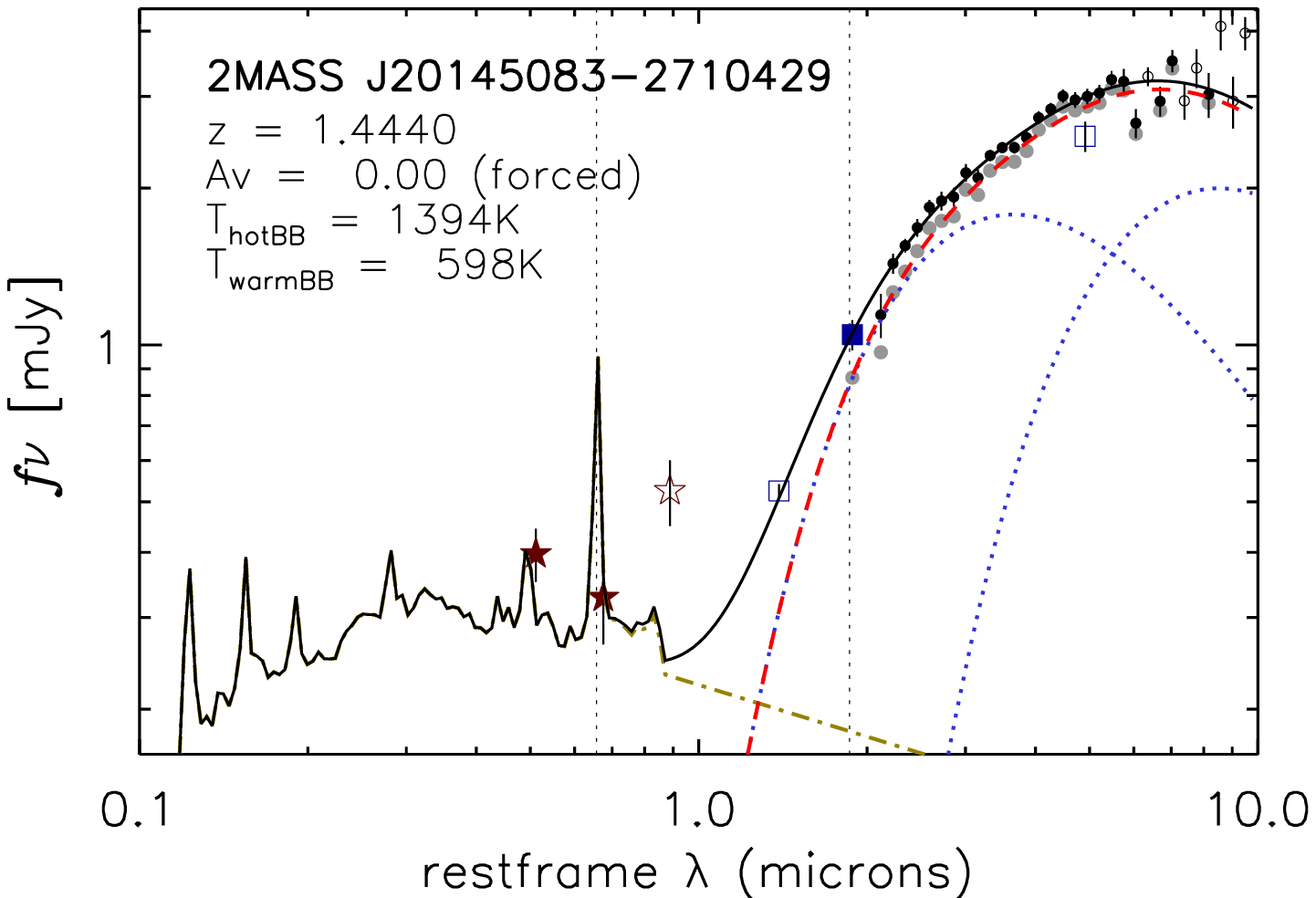}
\includegraphics[width=8.4cm]{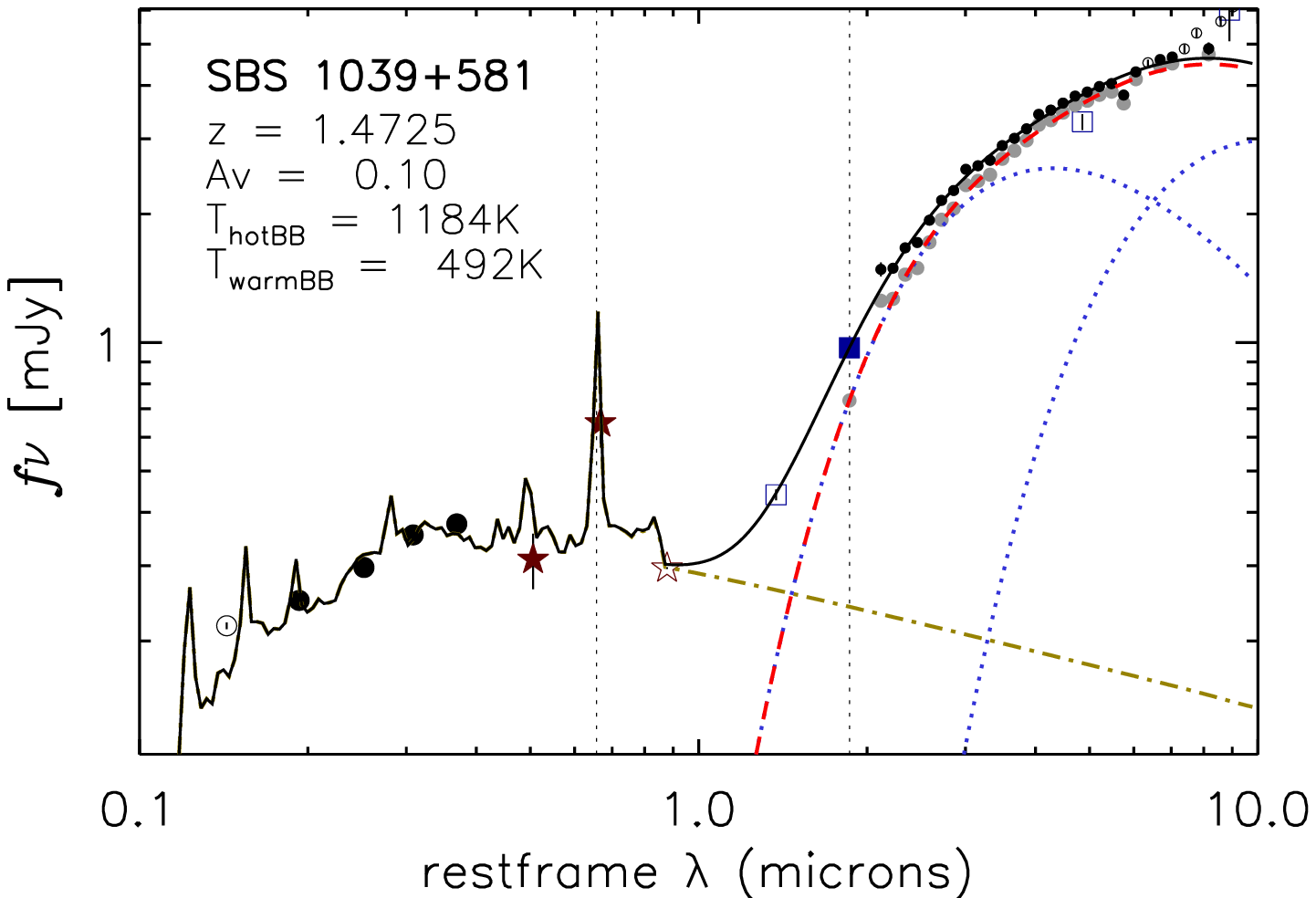}\vspace{0.4cm}
\includegraphics[width=8.4cm]{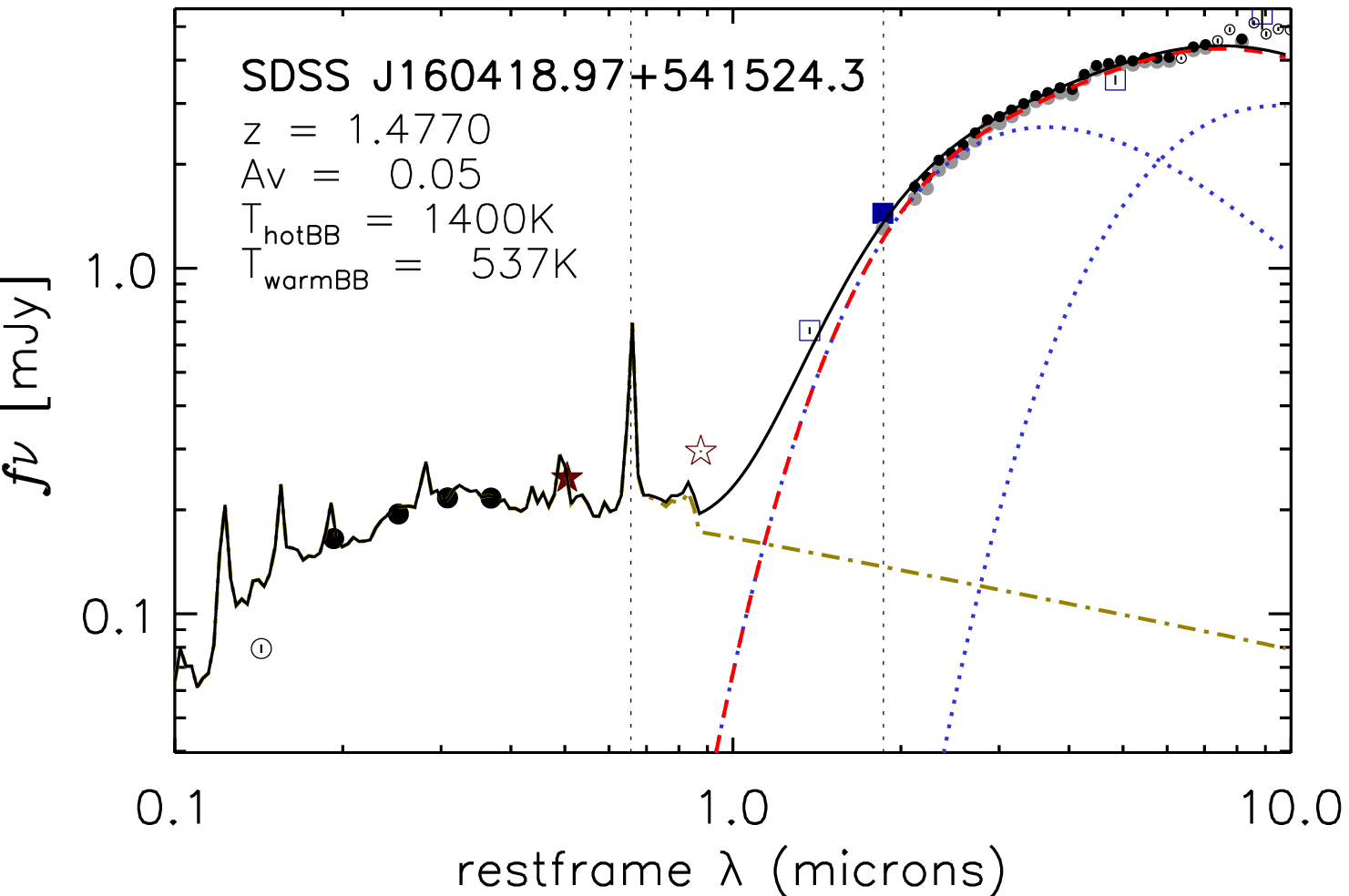}
\includegraphics[width=8.4cm]{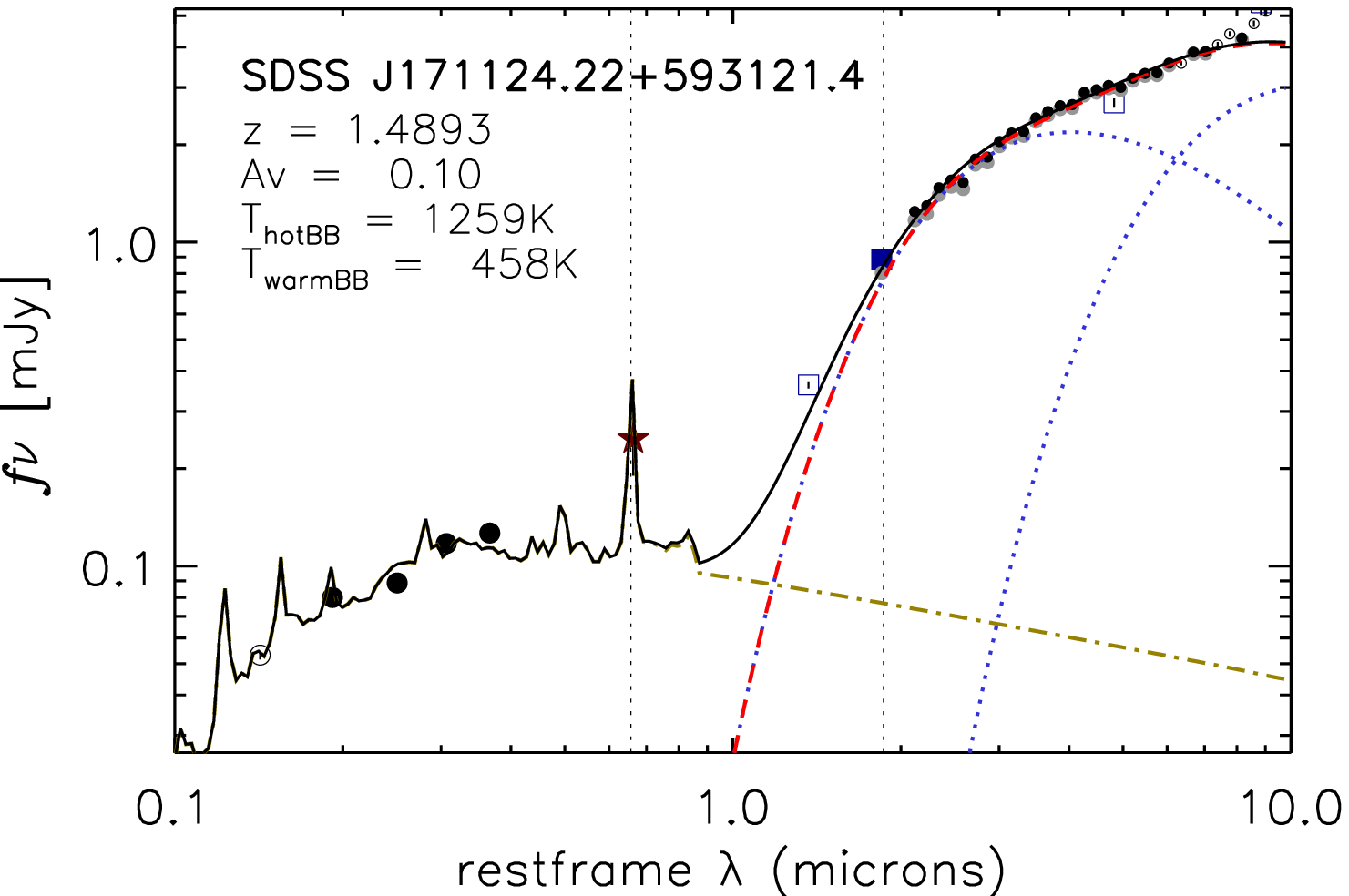}\vspace{0.4cm}
\includegraphics[width=8.4cm]{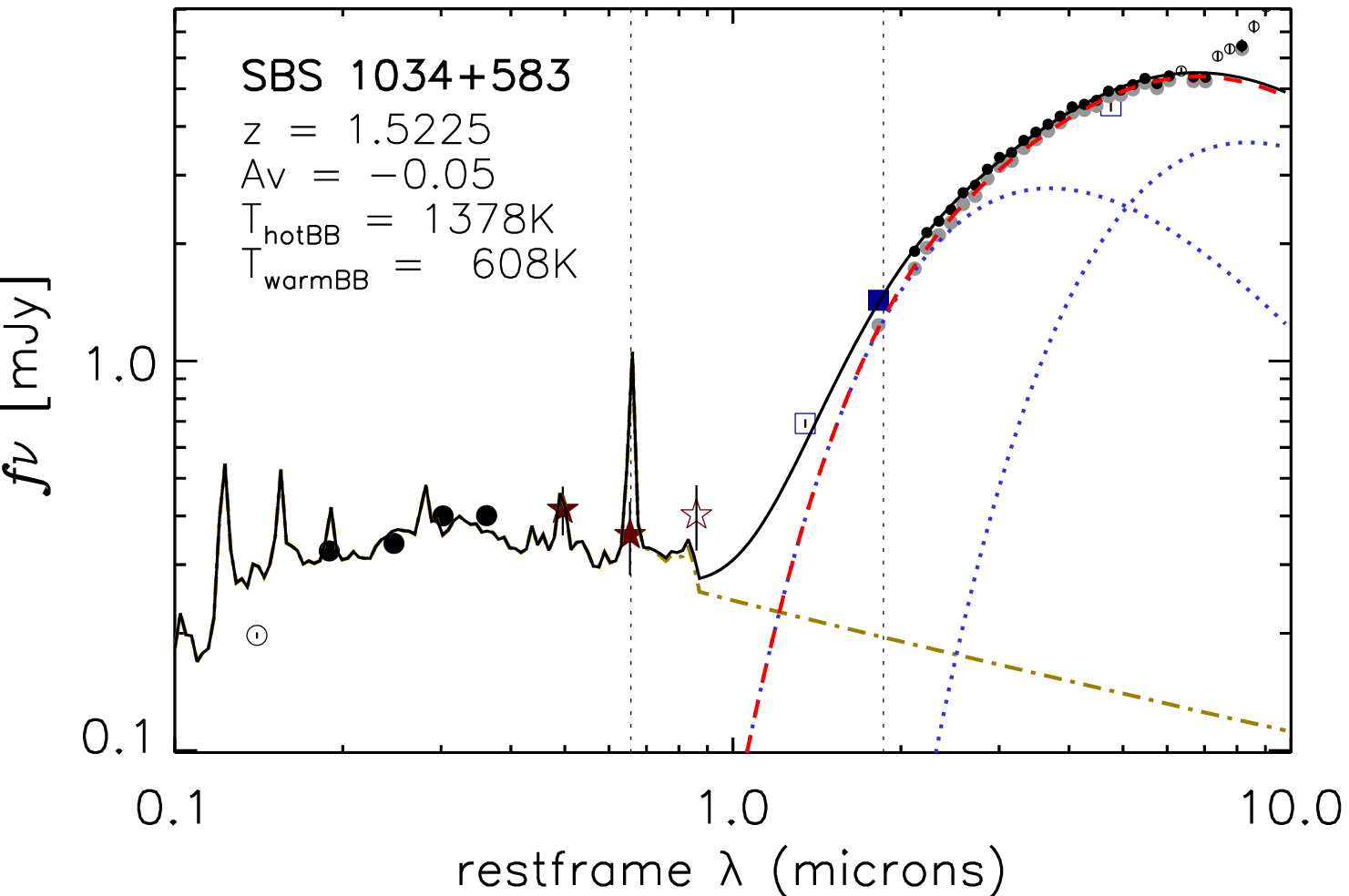}
\includegraphics[width=8.4cm]{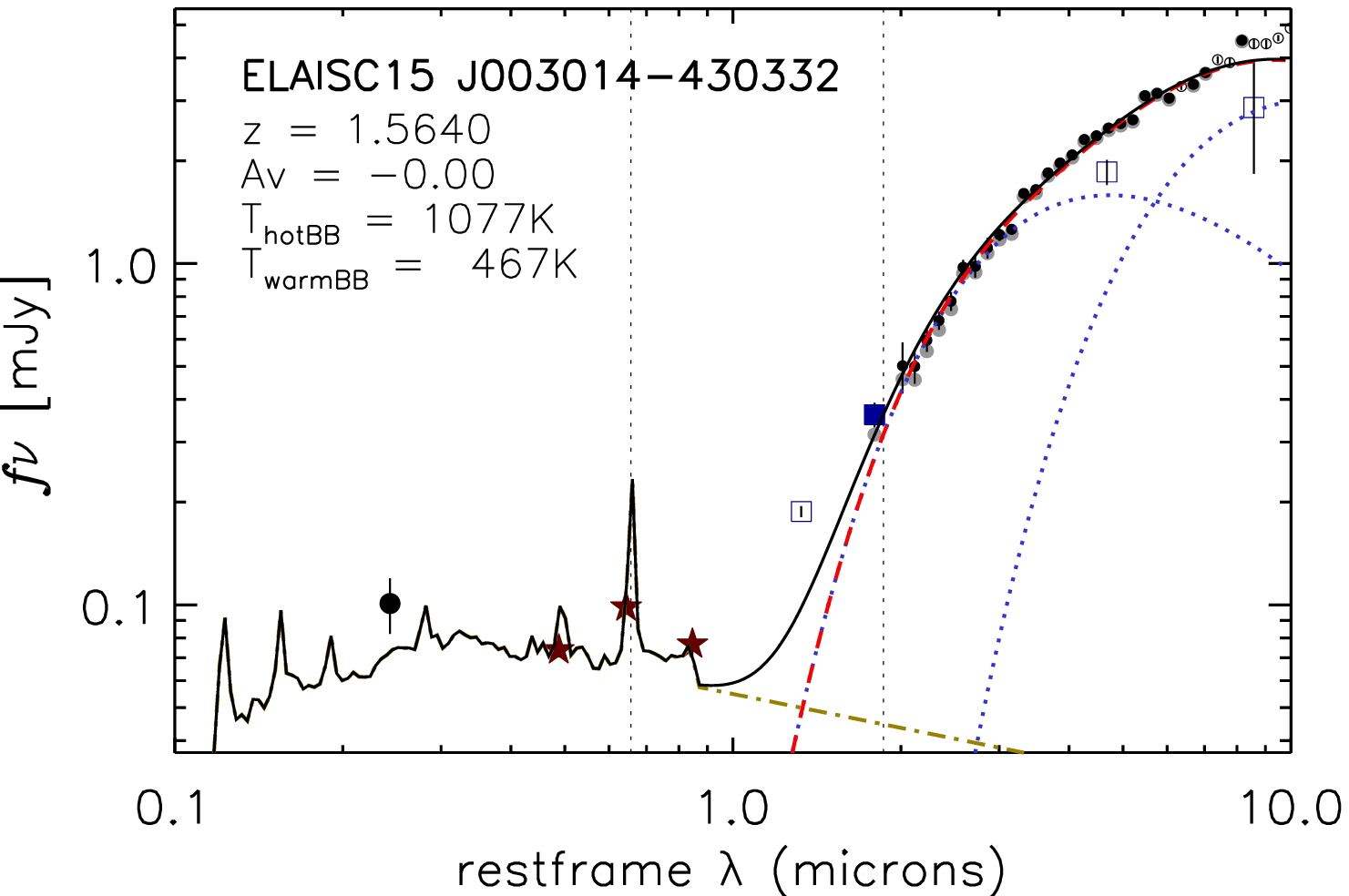}\vspace{0.4cm}
\includegraphics[width=8.4cm]{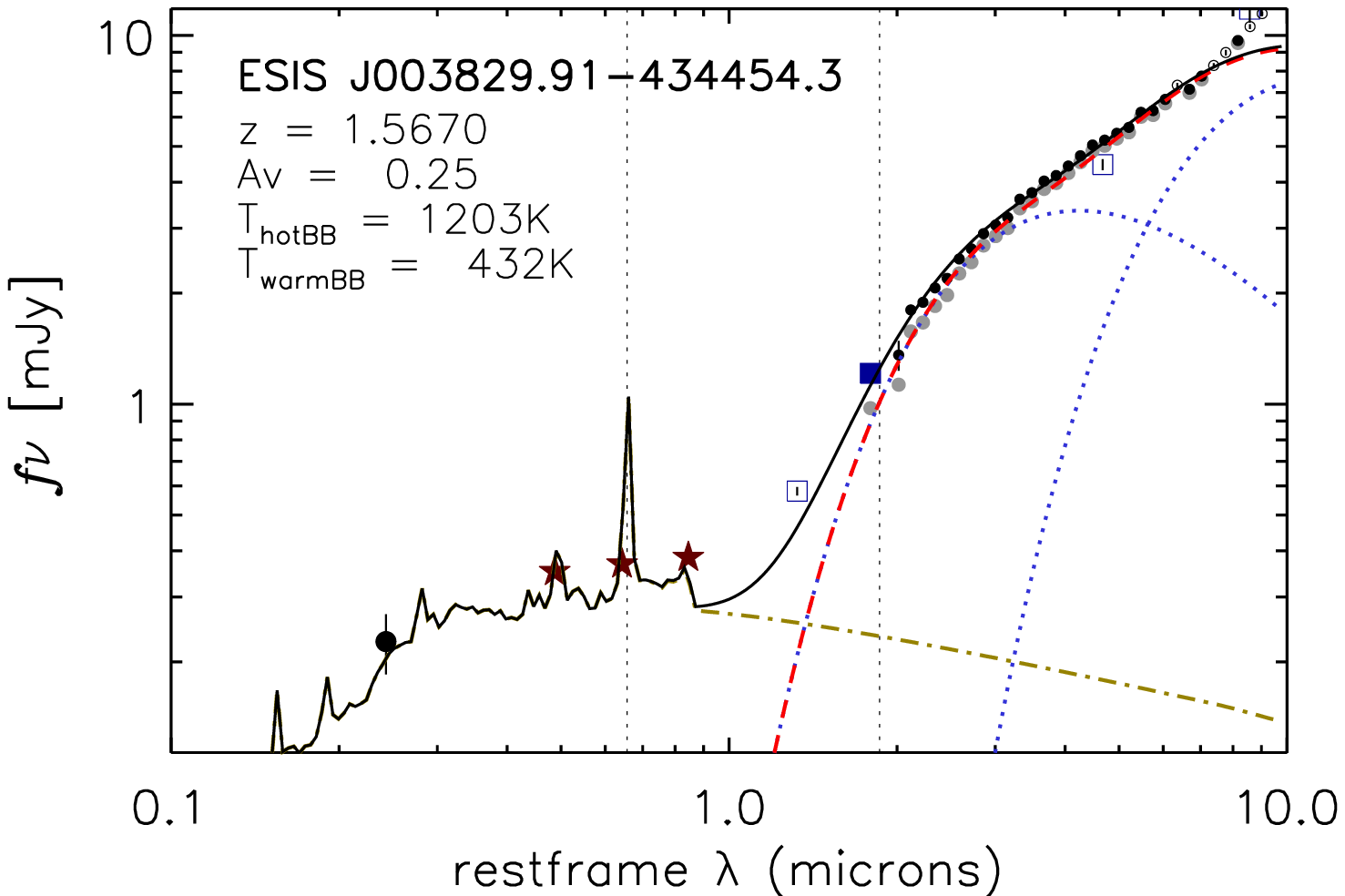}
\includegraphics[width=8.4cm]{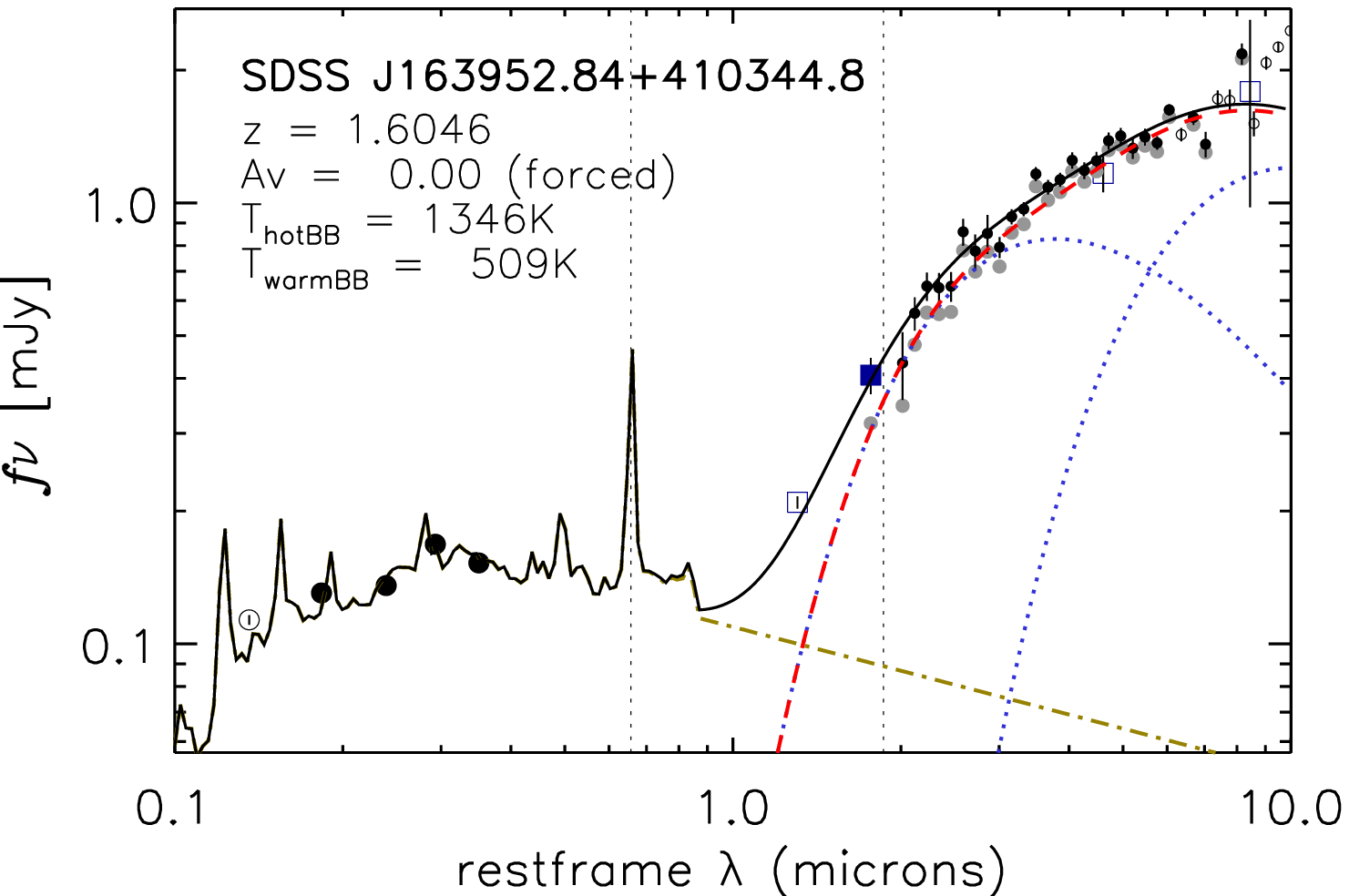}
\caption{continued}
\end{figure*}

\addtocounter{figure}{-1}
\begin{figure*}
\includegraphics[width=8.4cm]{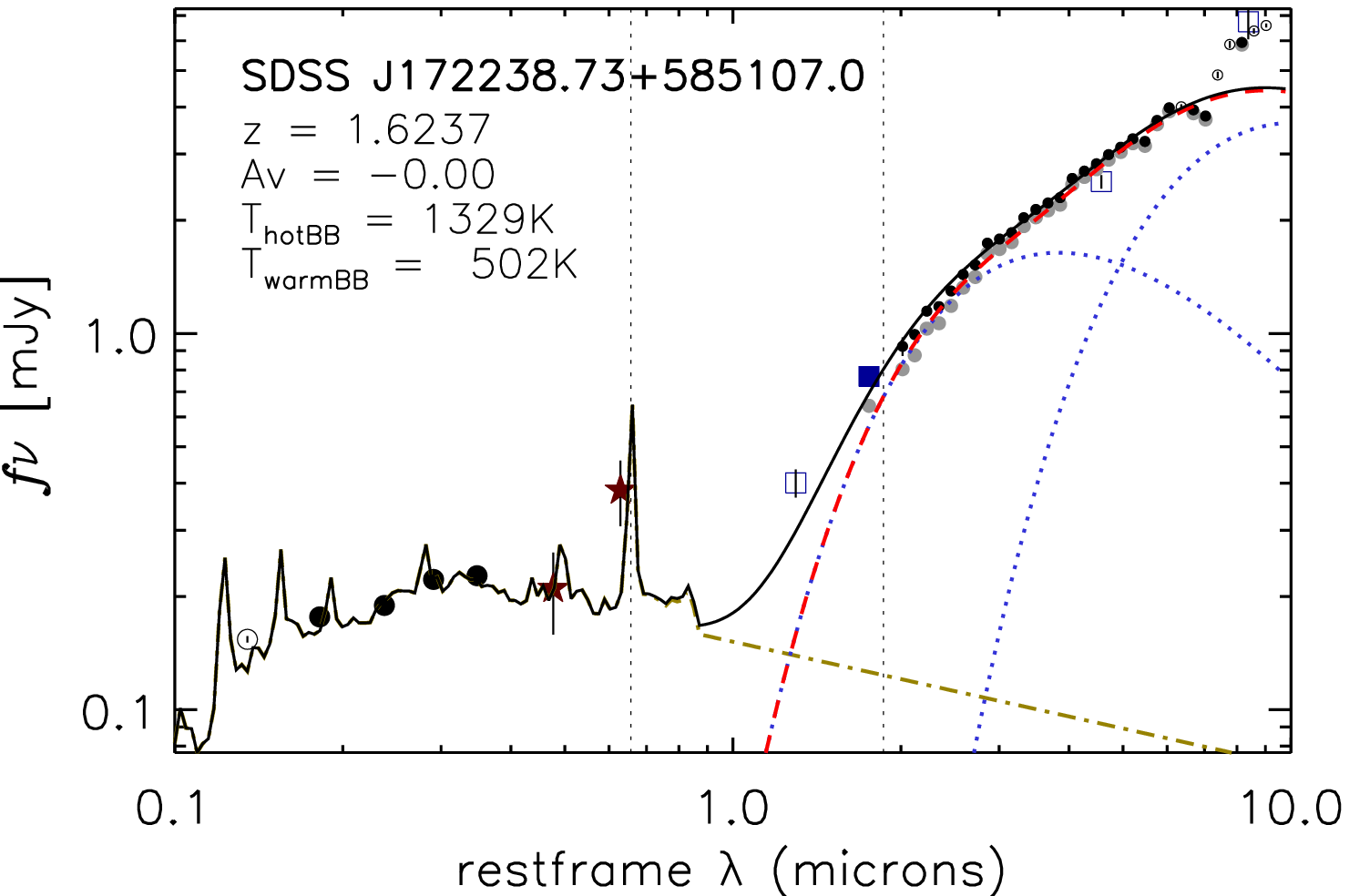}
\includegraphics[width=8.4cm]{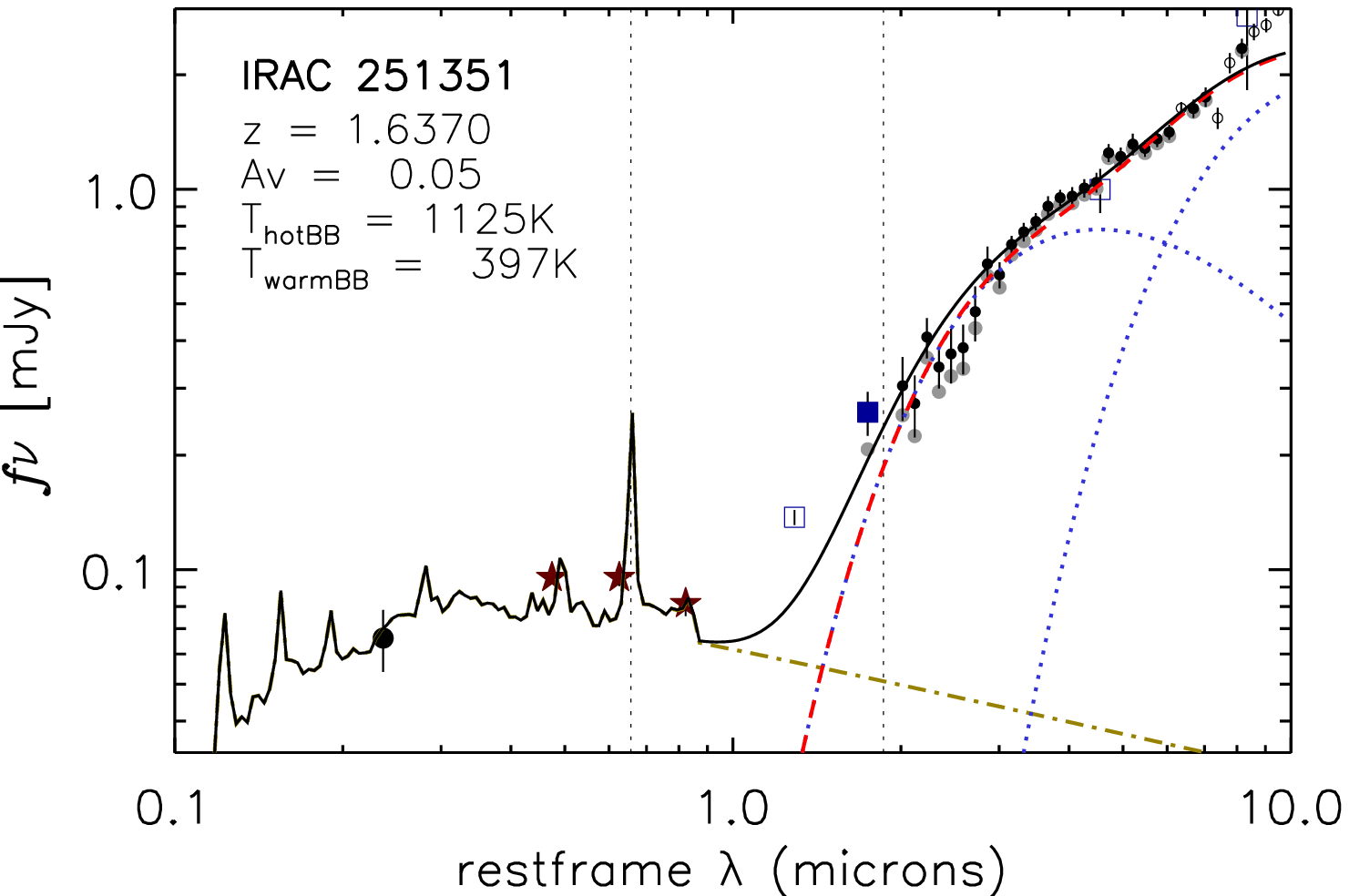}\vspace{0.4cm}
\includegraphics[width=8.4cm]{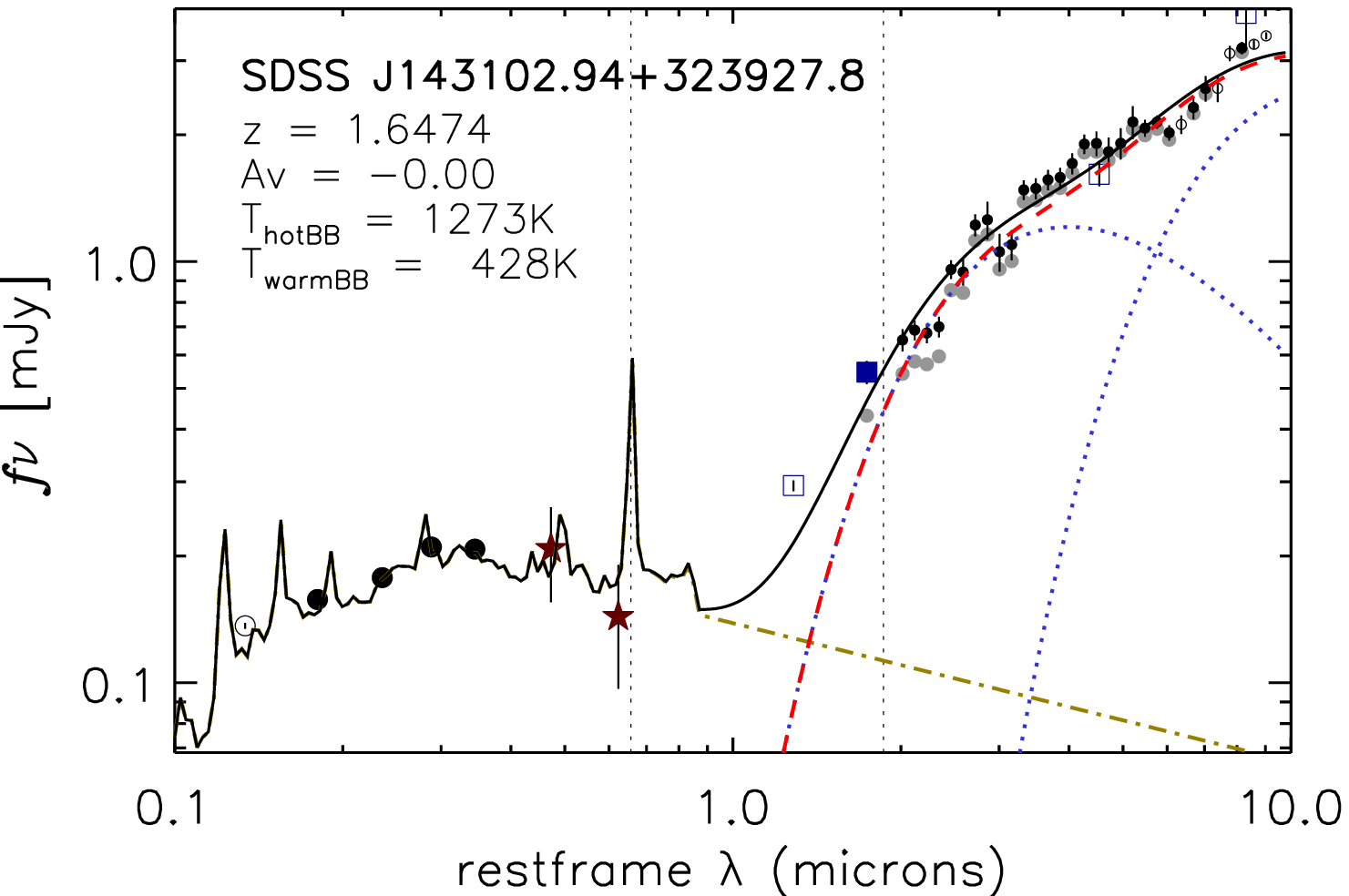}
\includegraphics[width=8.4cm]{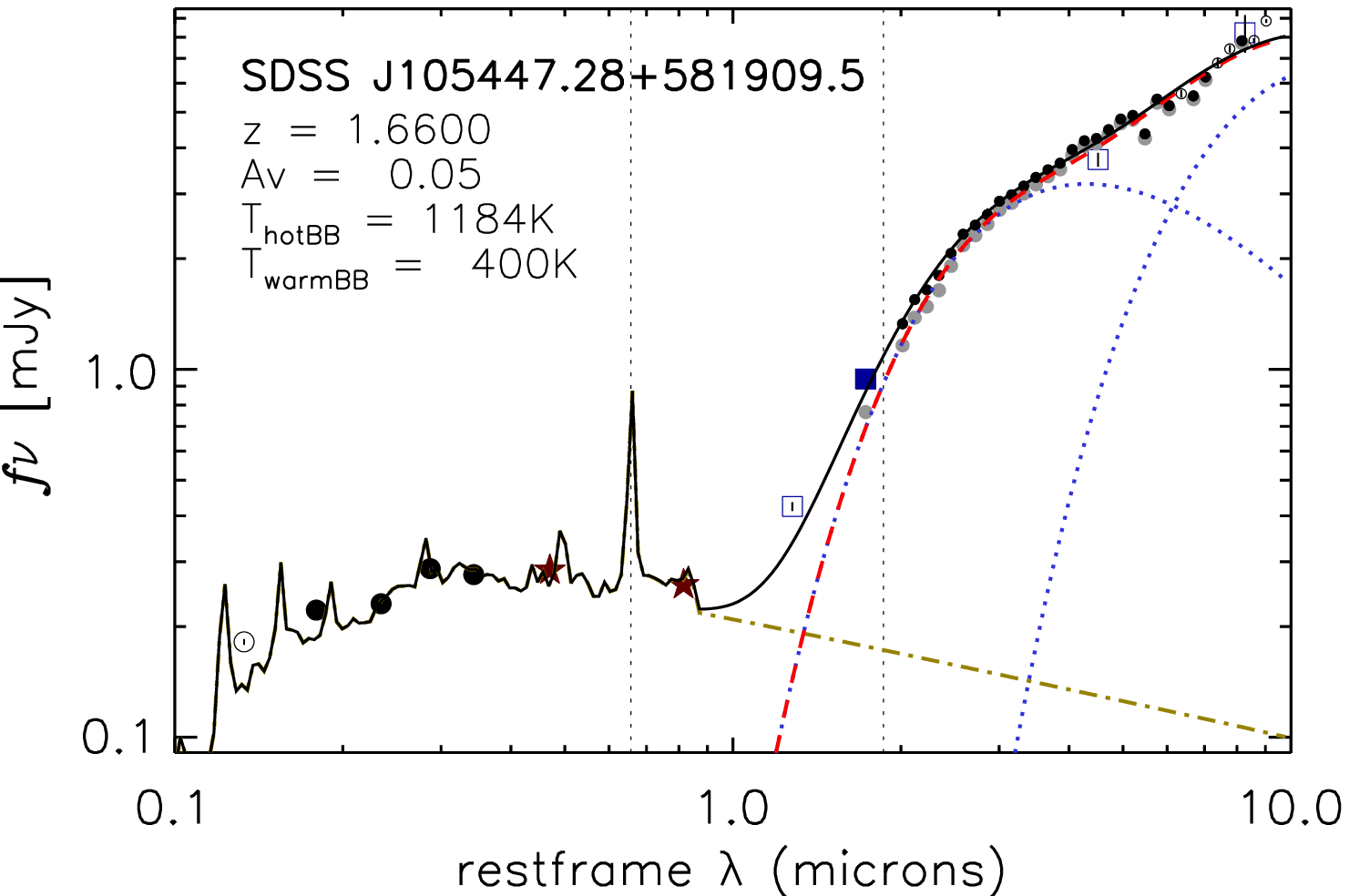}\vspace{0.4cm}
\includegraphics[width=8.4cm]{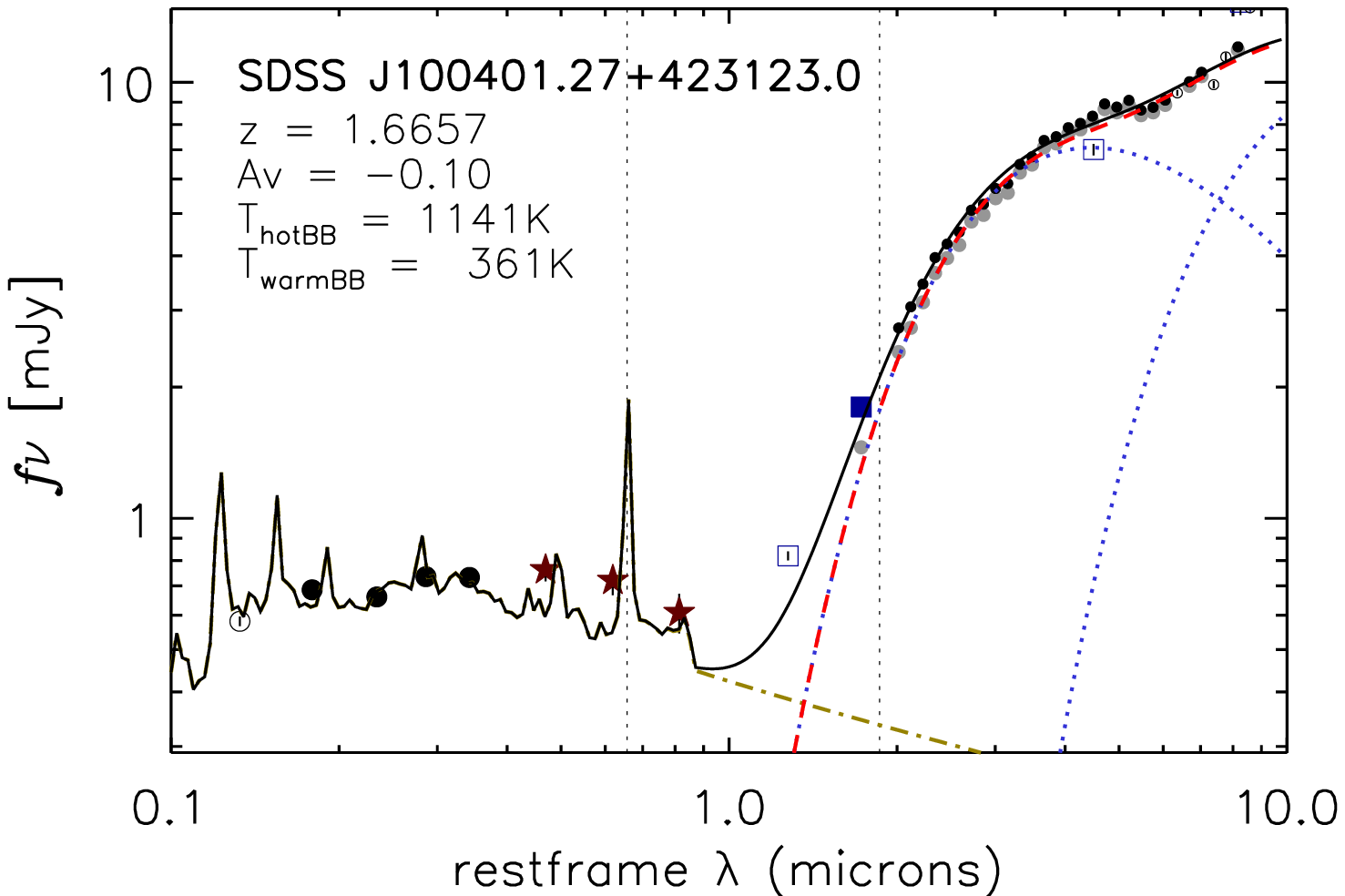}
\includegraphics[width=8.4cm]{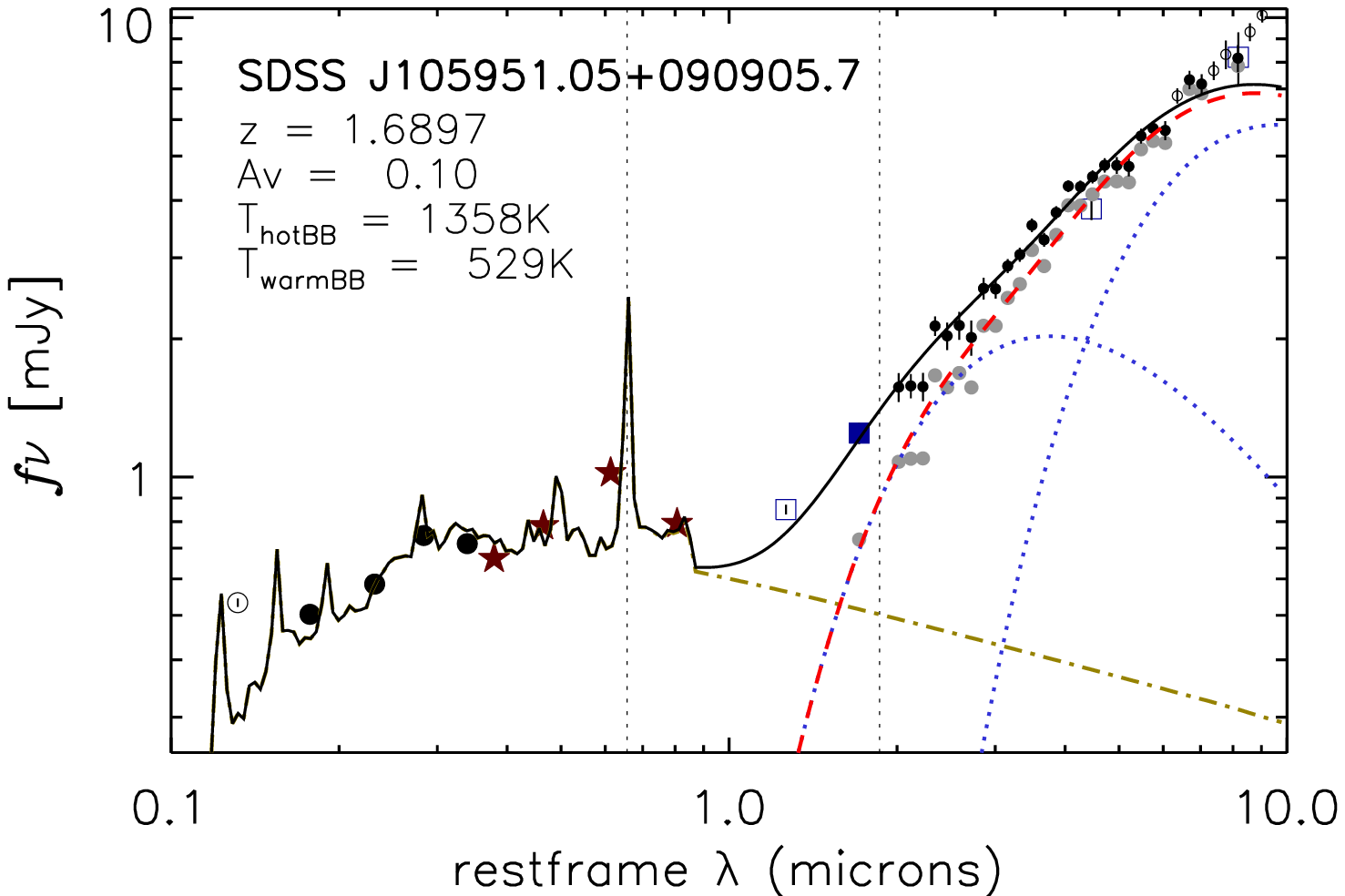}\vspace{0.4cm}
\includegraphics[width=8.4cm]{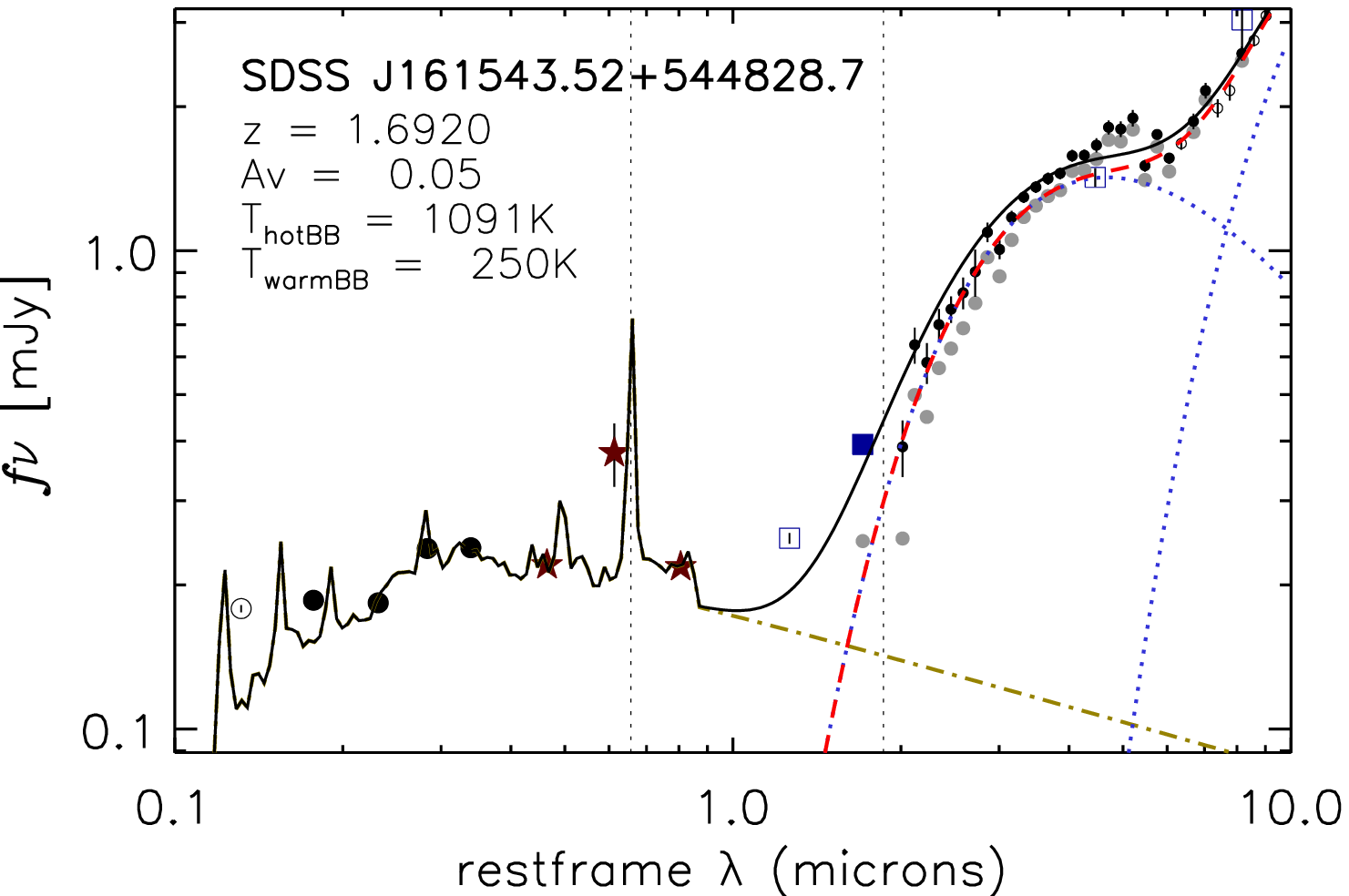}
\includegraphics[width=8.4cm]{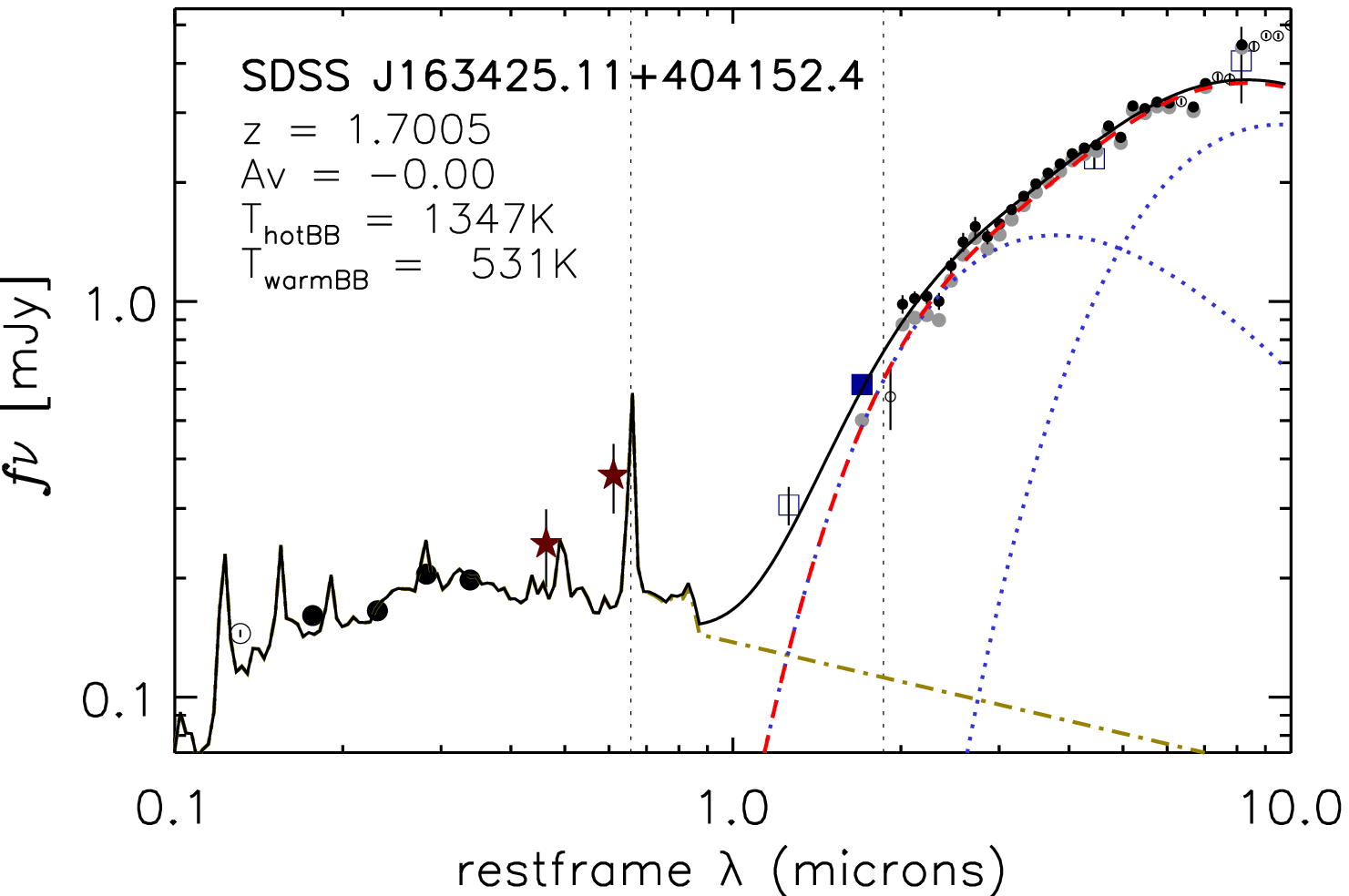}
\caption{continued}
\end{figure*}

\addtocounter{figure}{-1}
\begin{figure*}
\includegraphics[width=8.4cm]{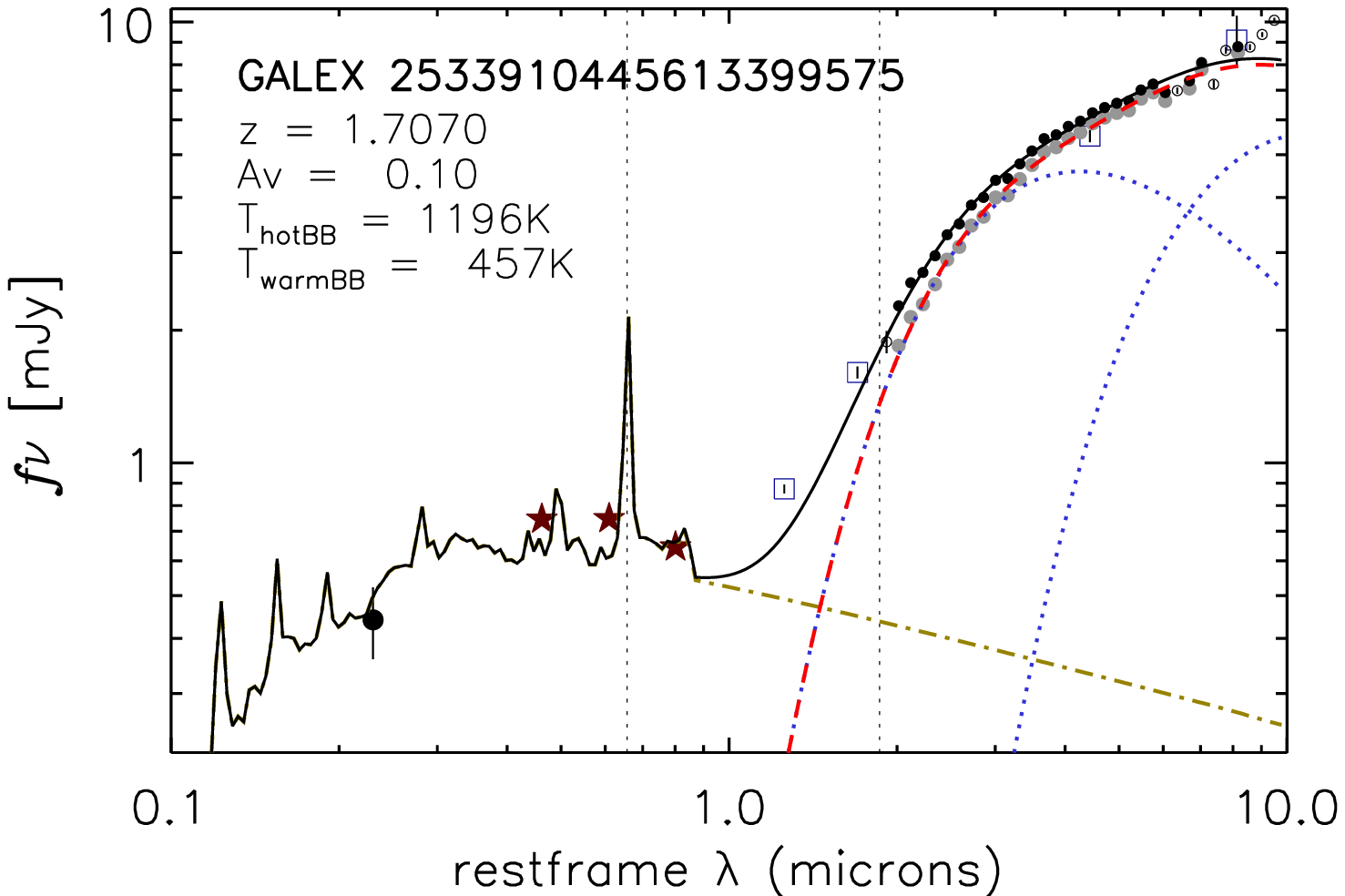}
\includegraphics[width=8.4cm]{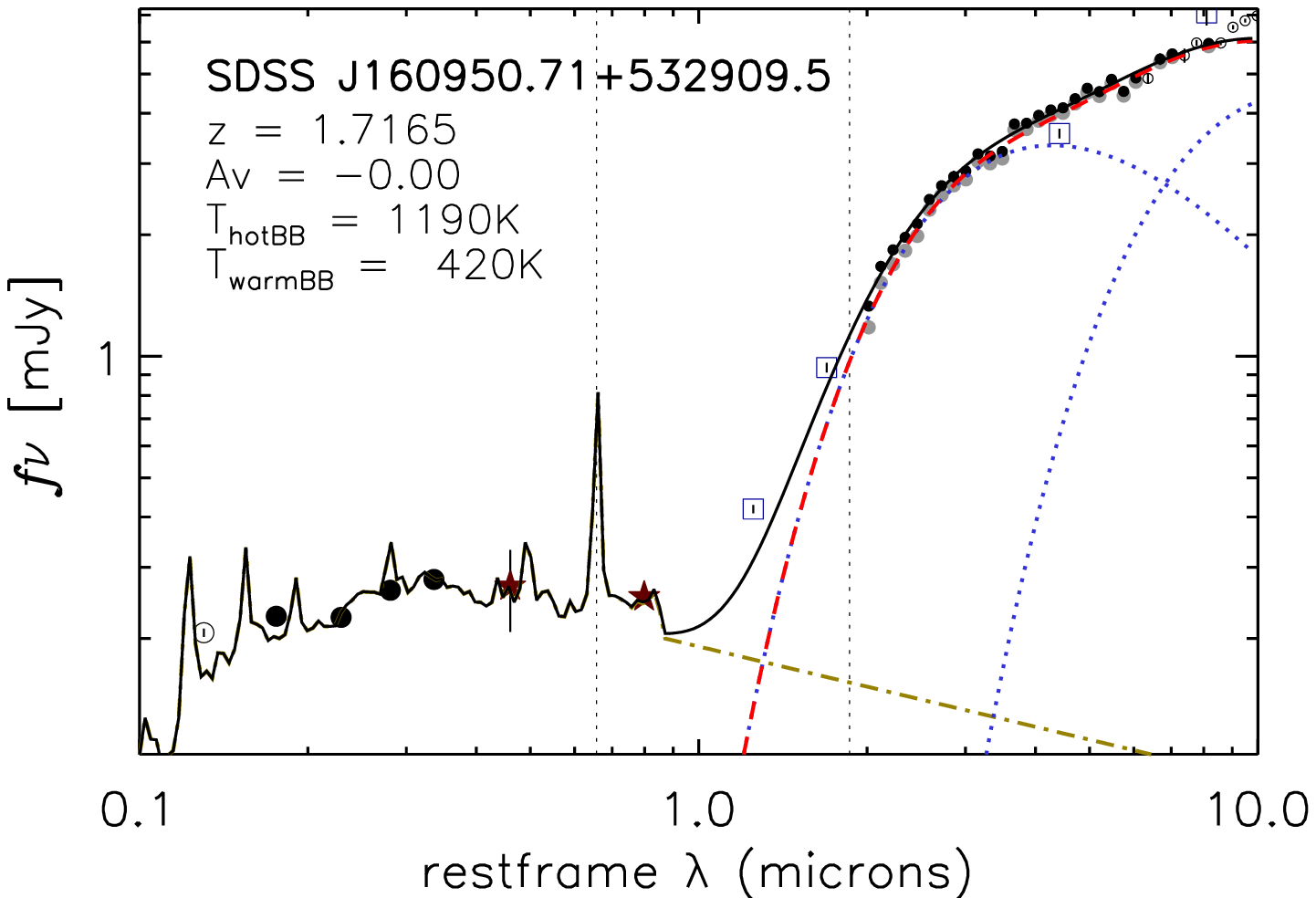}\vspace{0.4cm}
\includegraphics[width=8.4cm]{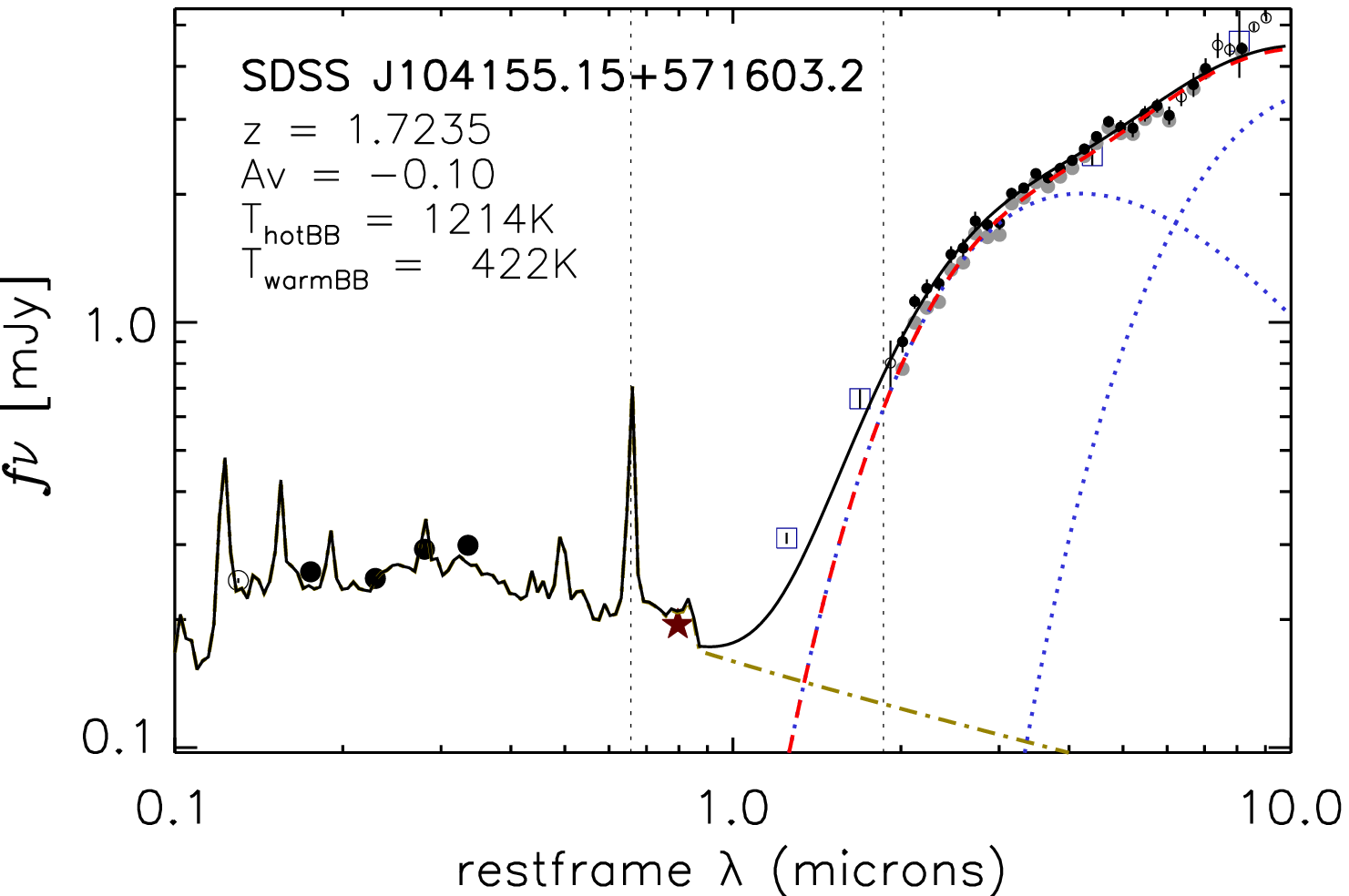}
\includegraphics[width=8.4cm]{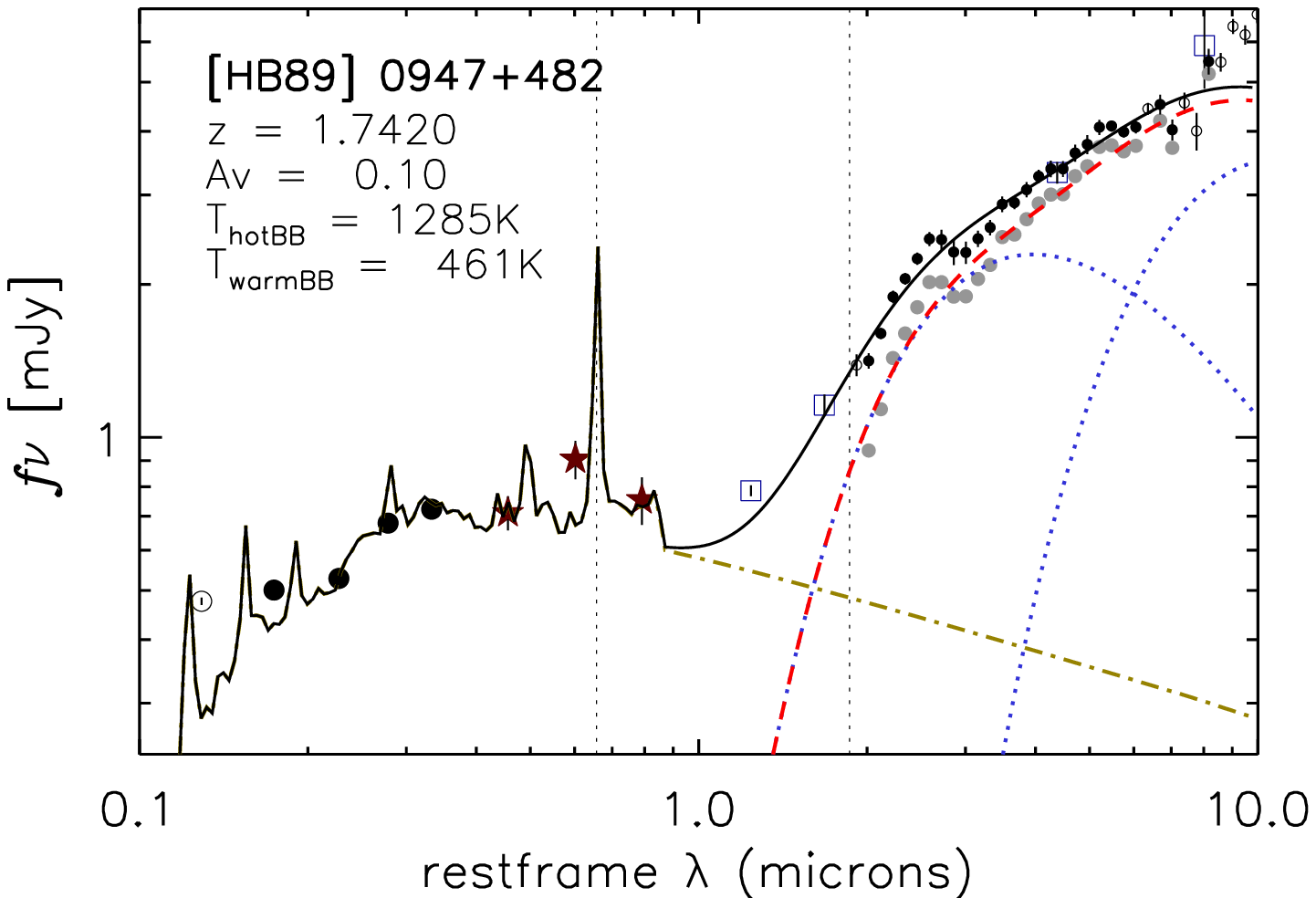}\vspace{0.4cm}
\includegraphics[width=8.4cm]{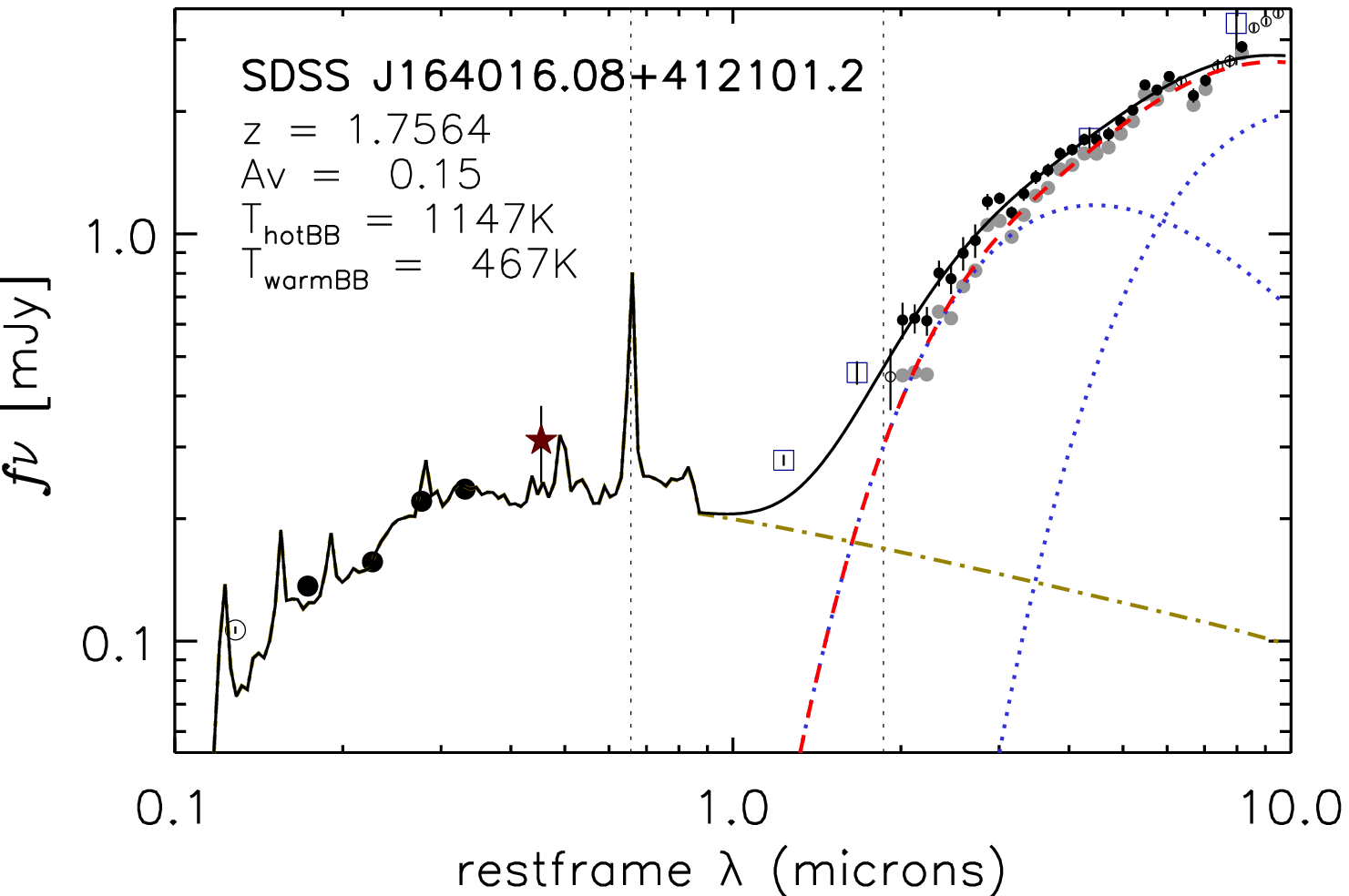}
\includegraphics[width=8.4cm]{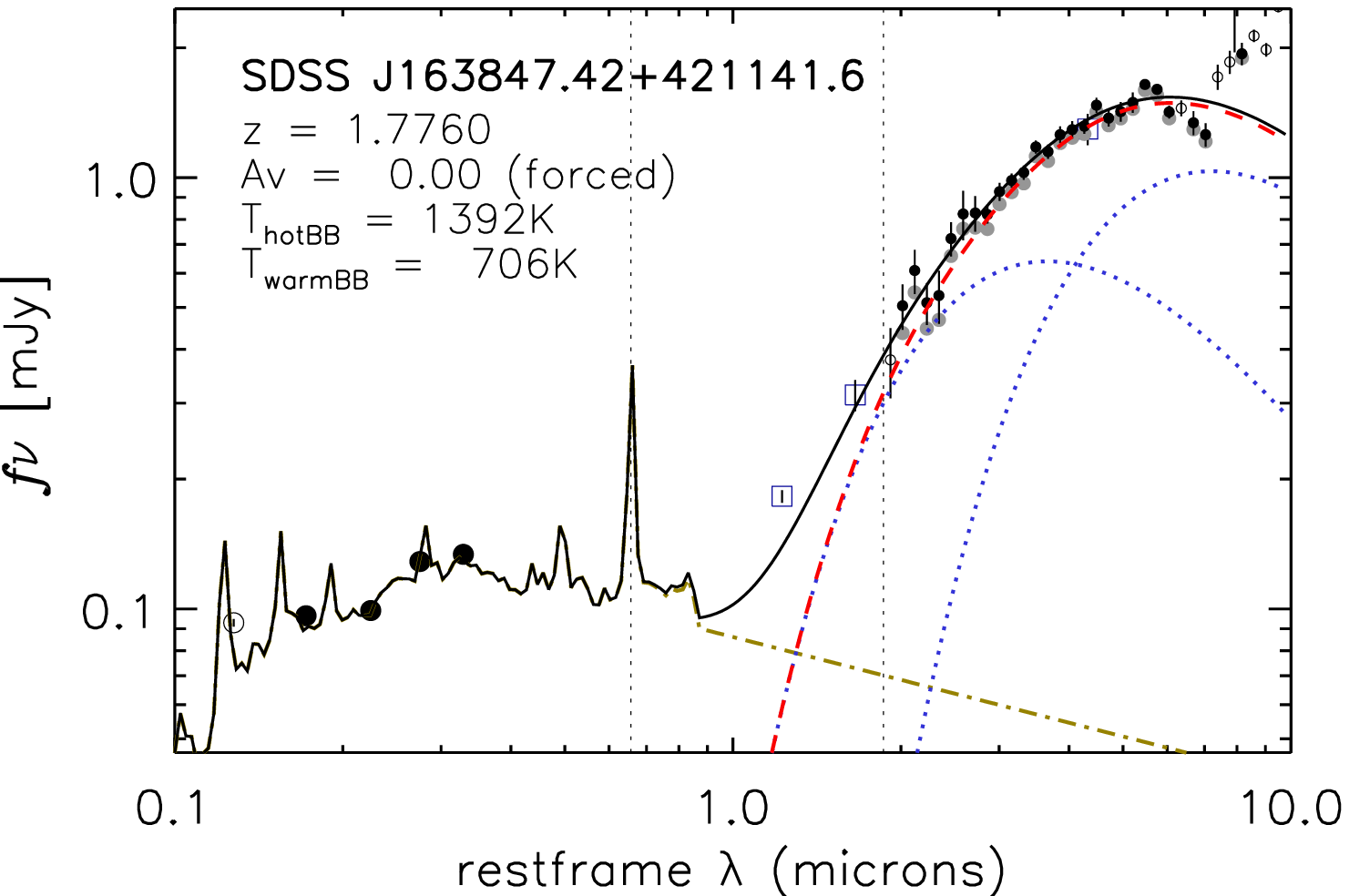}\vspace{0.4cm}
\includegraphics[width=8.4cm]{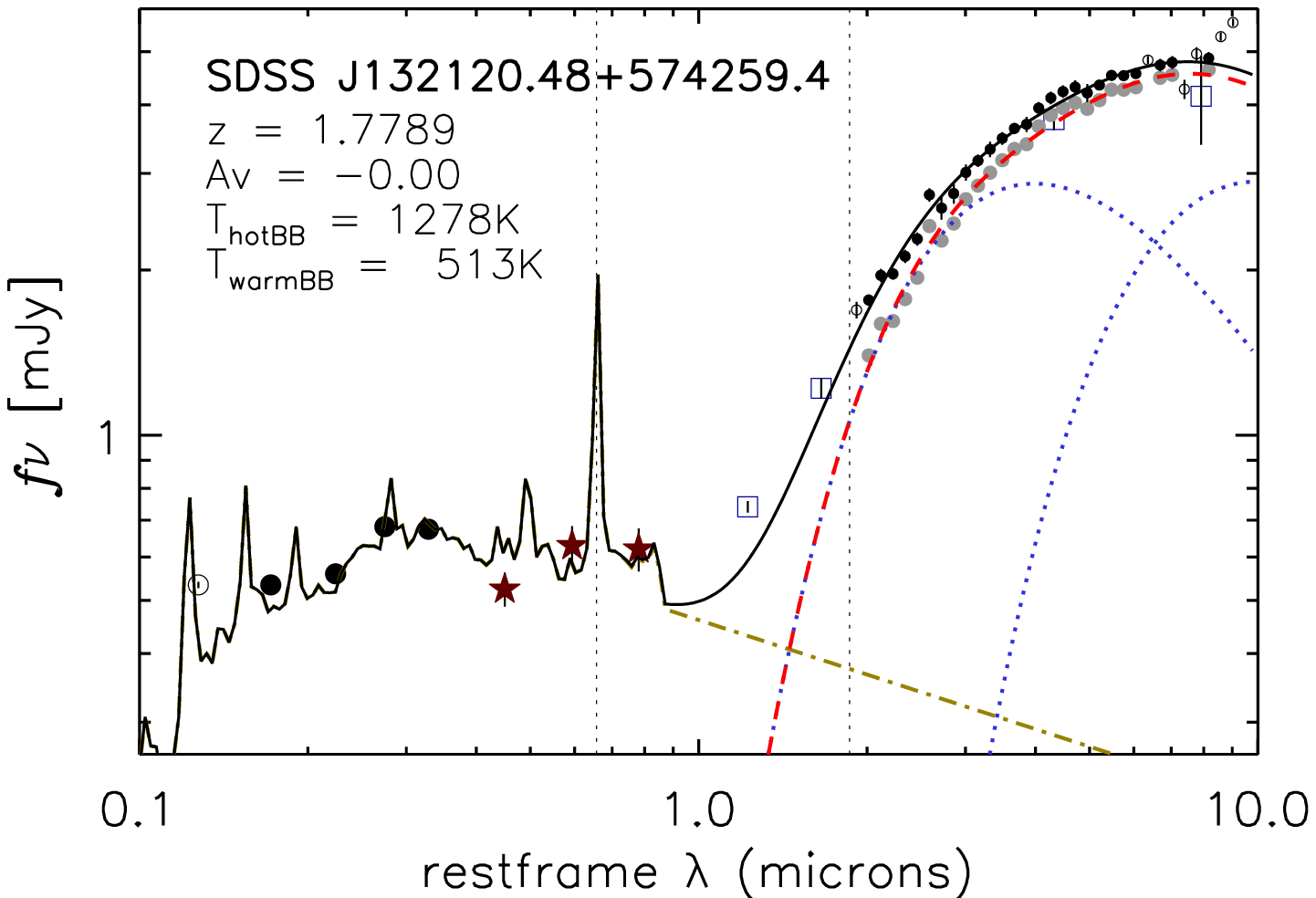}
\includegraphics[width=8.4cm]{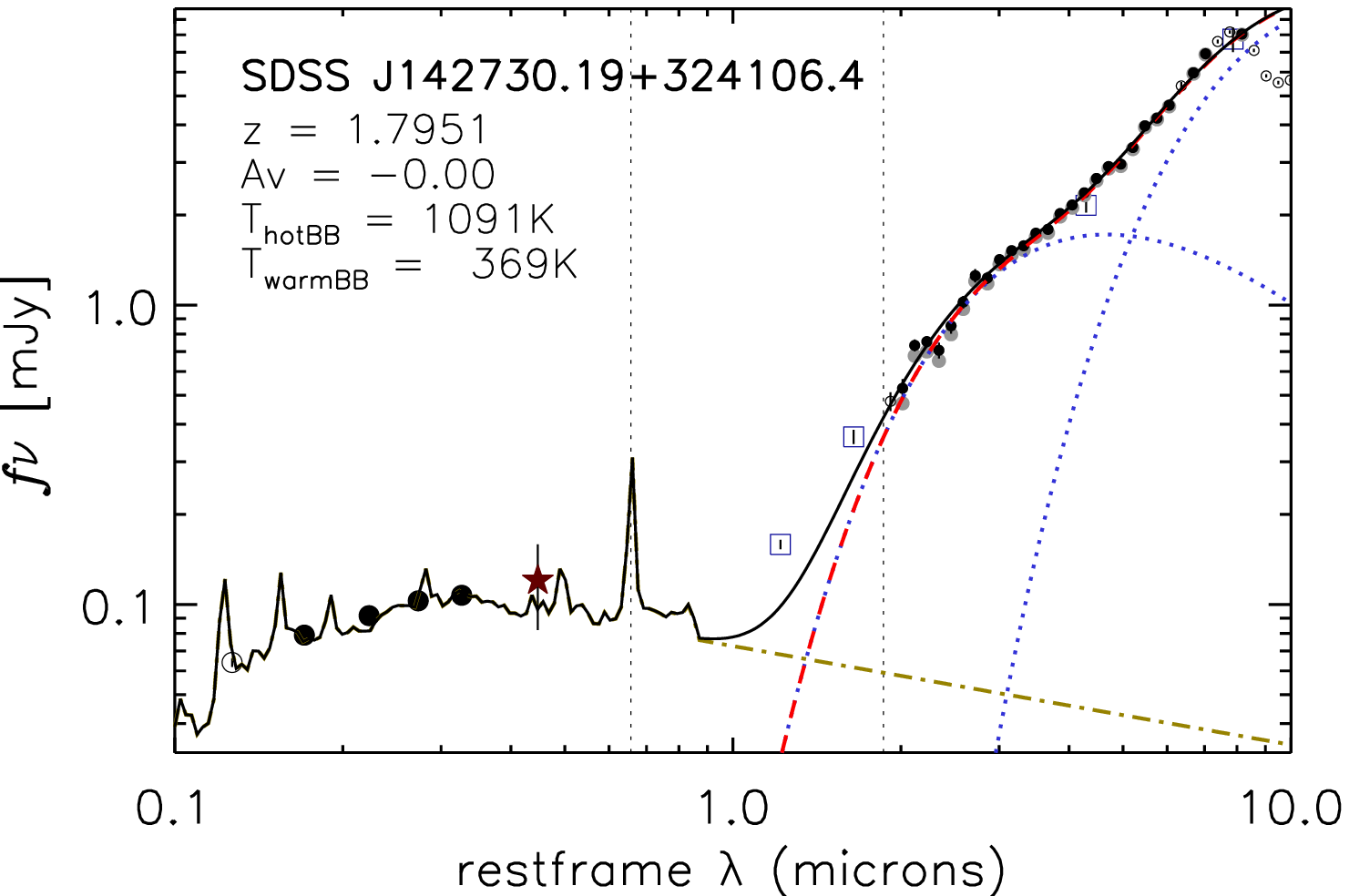}
\caption{continued}
\end{figure*}

\addtocounter{figure}{-1}
\begin{figure*}
\includegraphics[width=8.4cm]{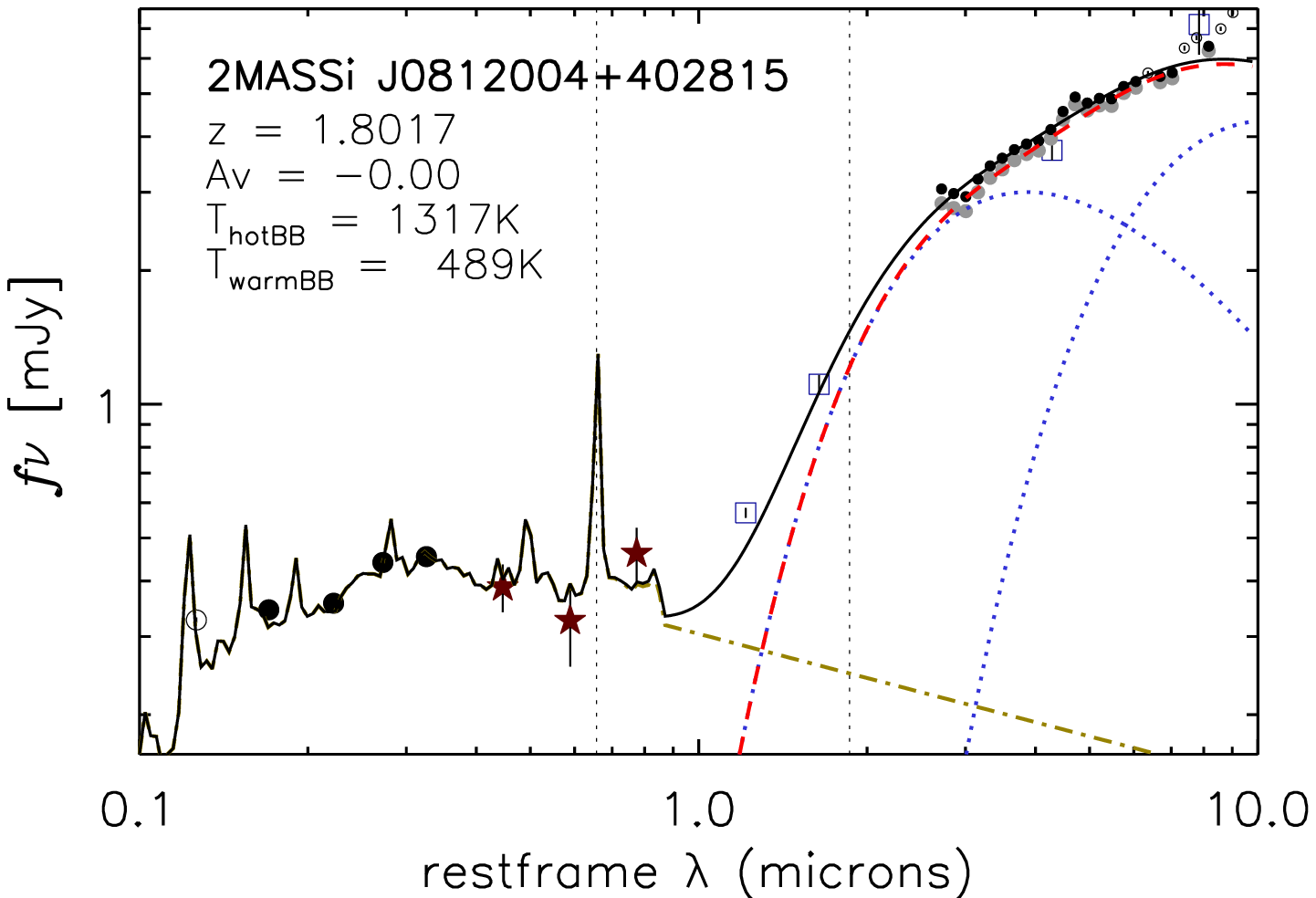}
\includegraphics[width=8.4cm]{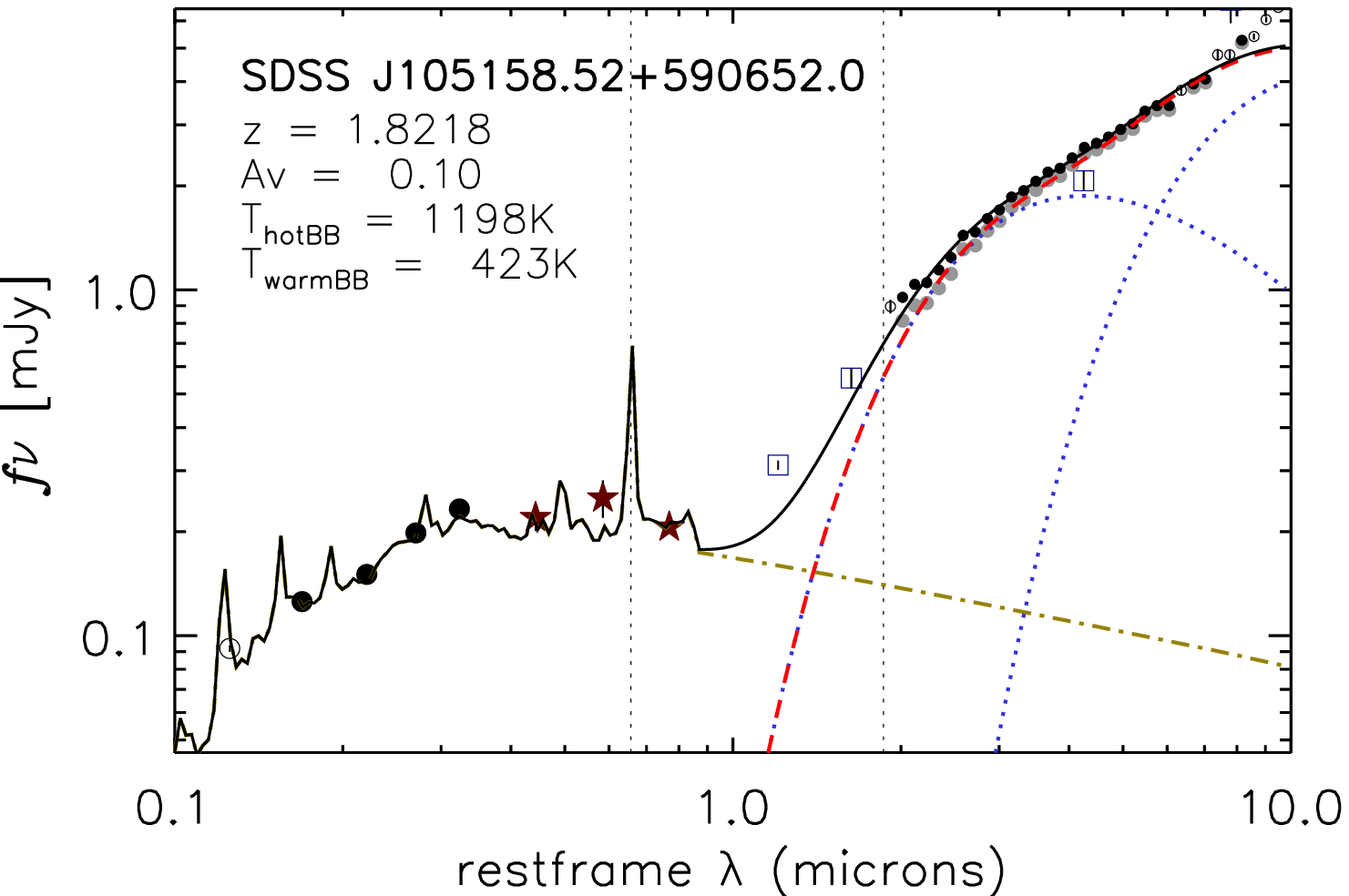}\vspace{0.4cm}
\includegraphics[width=8.4cm]{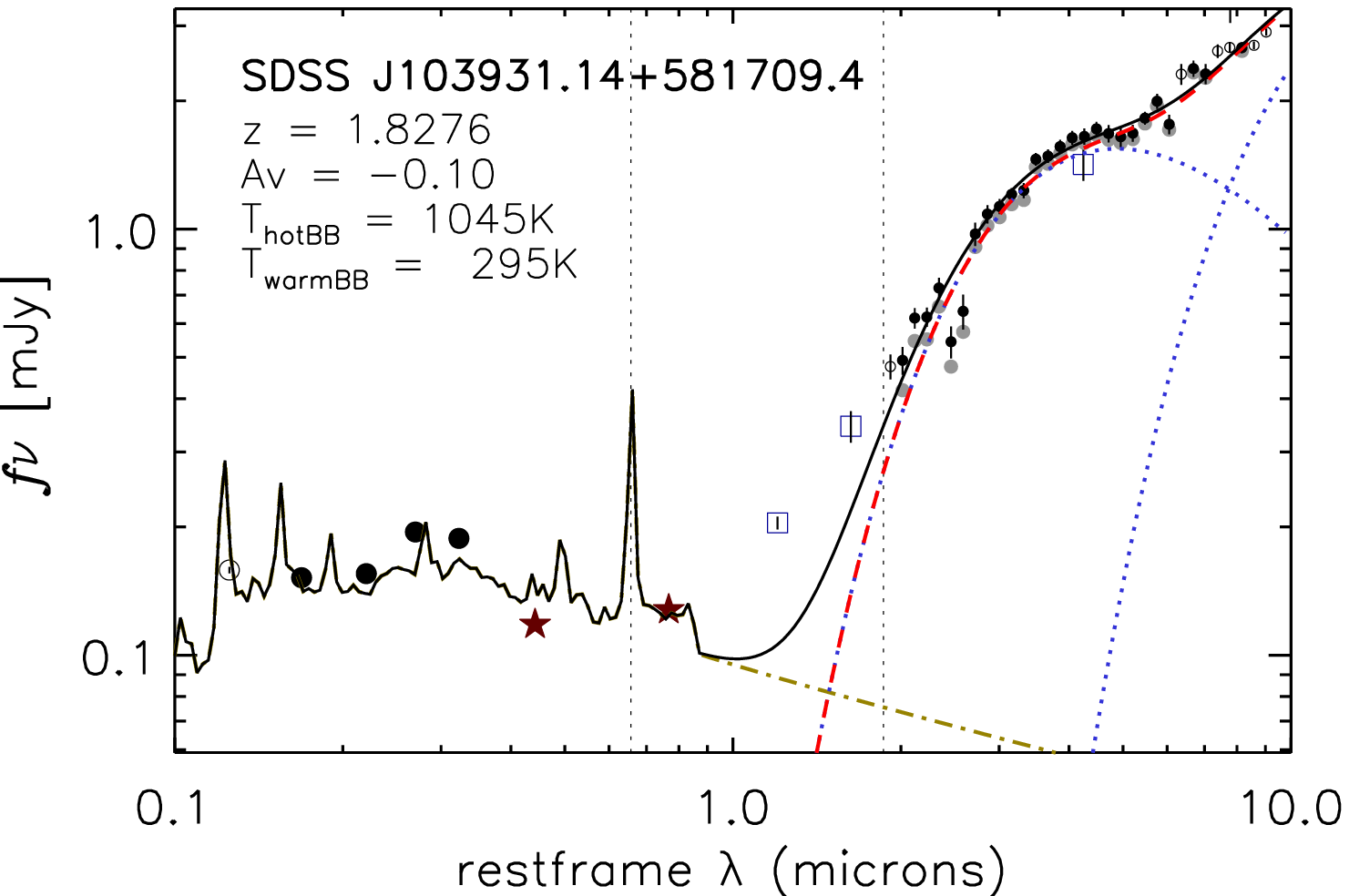}
\includegraphics[width=8.4cm]{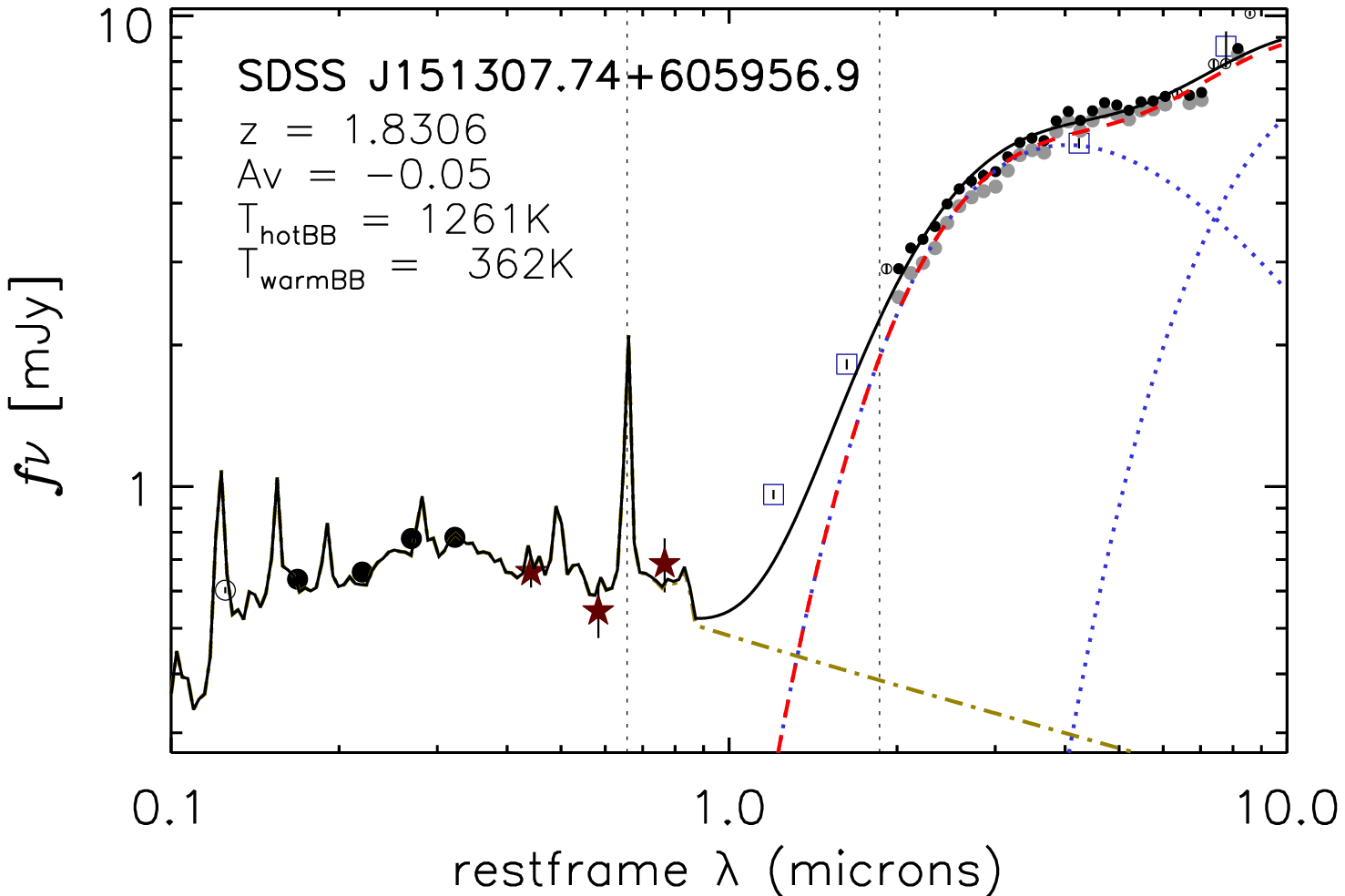}\vspace{0.4cm}
\includegraphics[width=8.4cm]{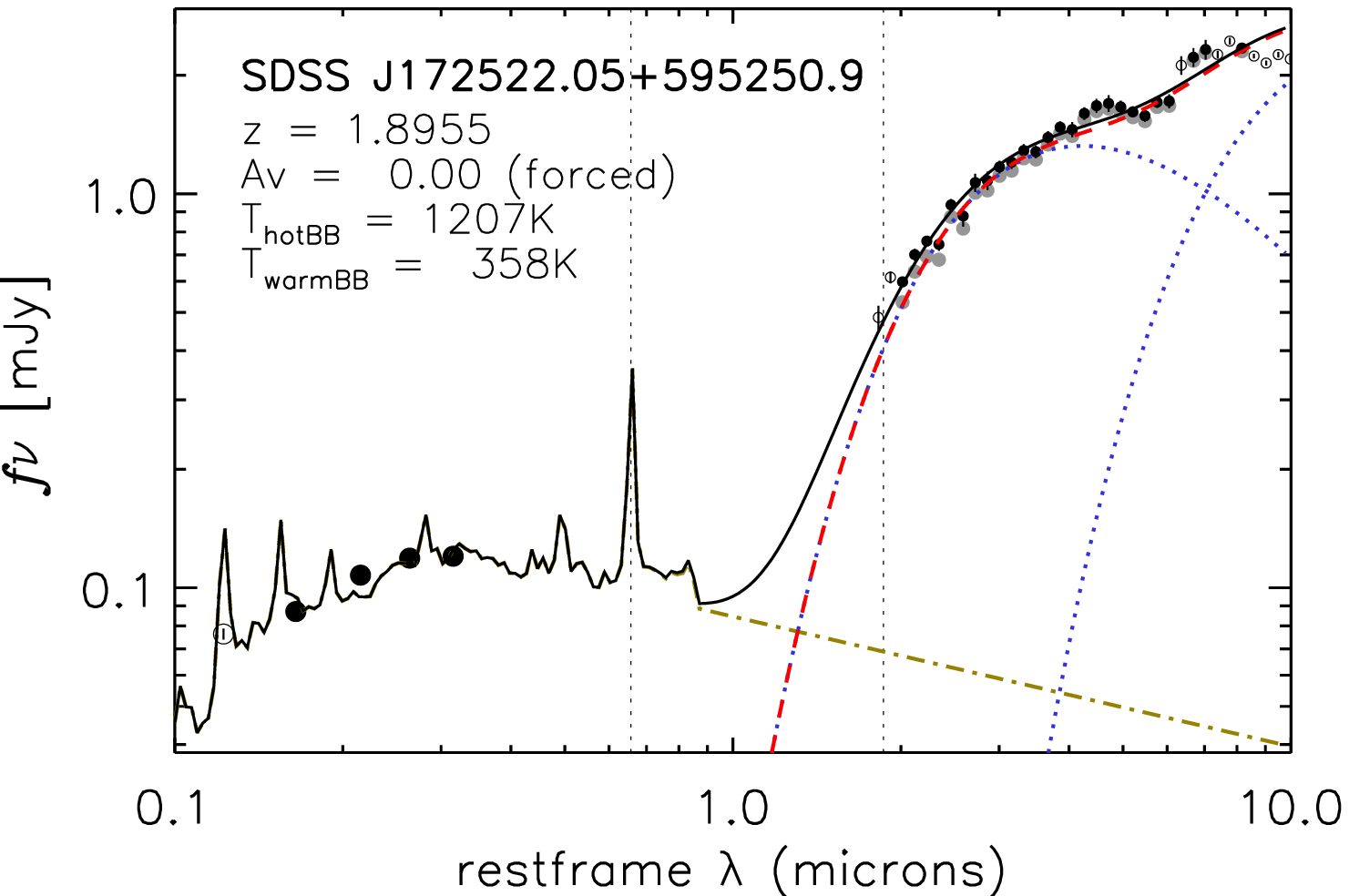}
\includegraphics[width=8.4cm]{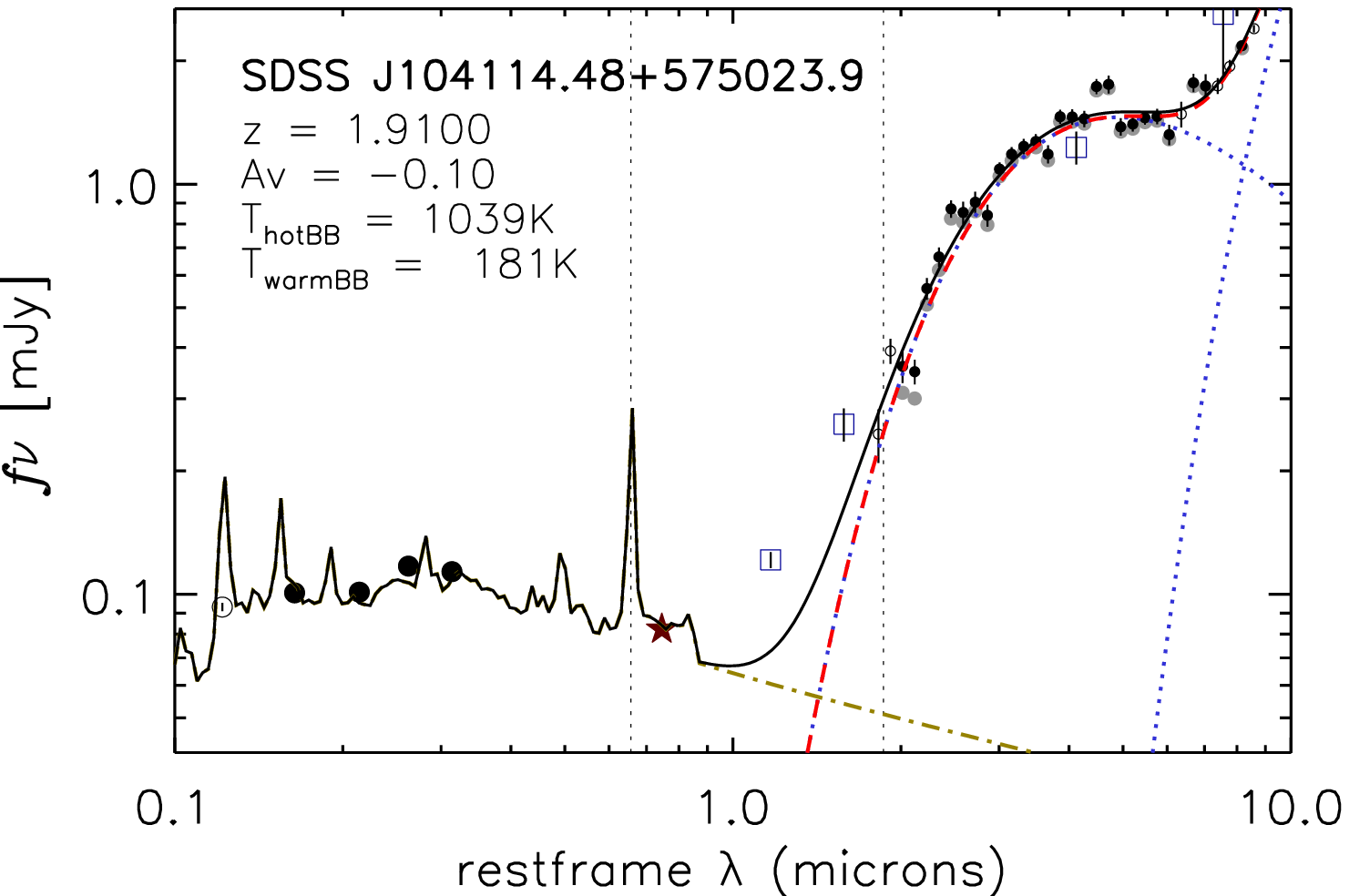}\vspace{0.4cm}
\includegraphics[width=8.4cm]{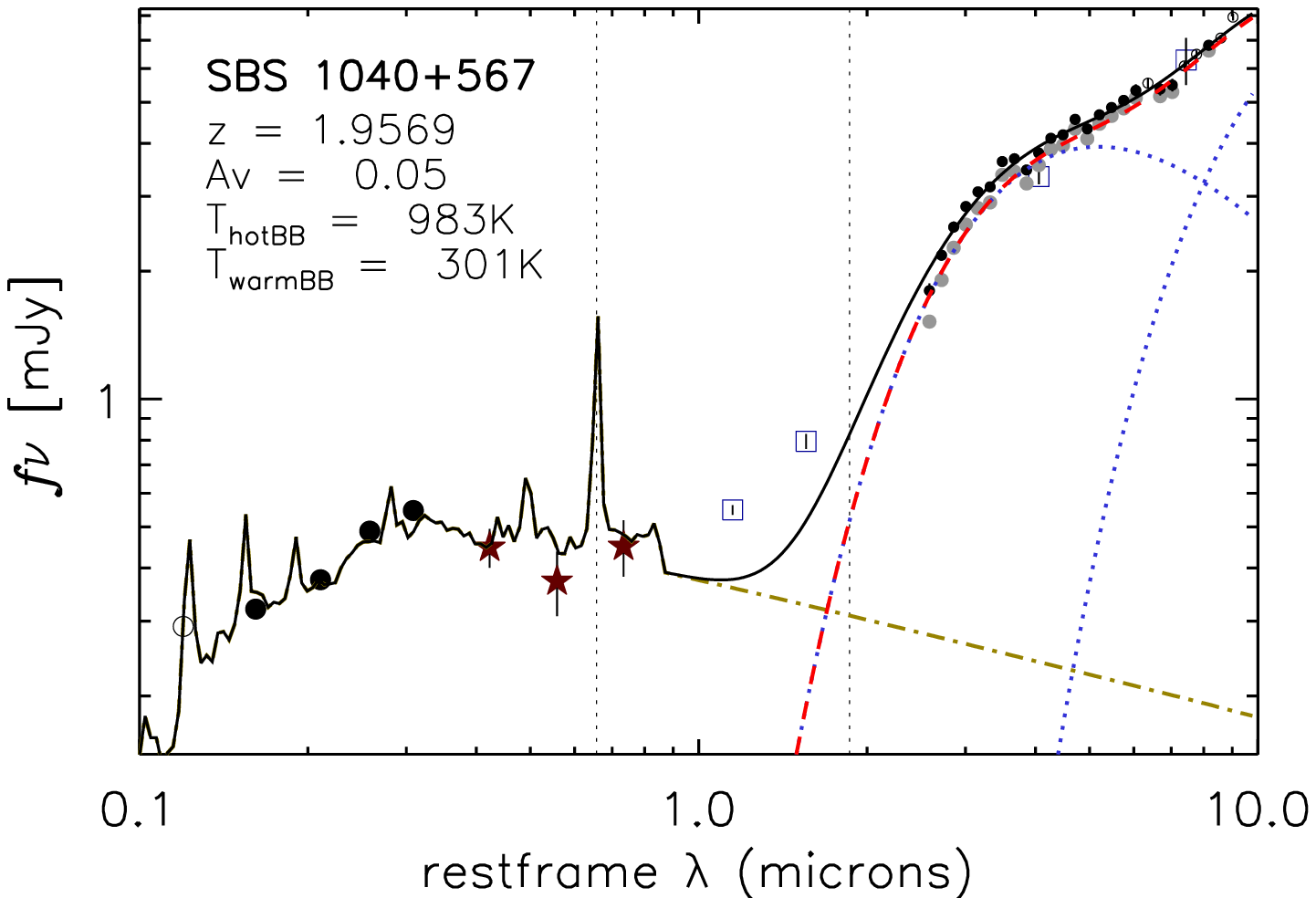}
\includegraphics[width=8.4cm]{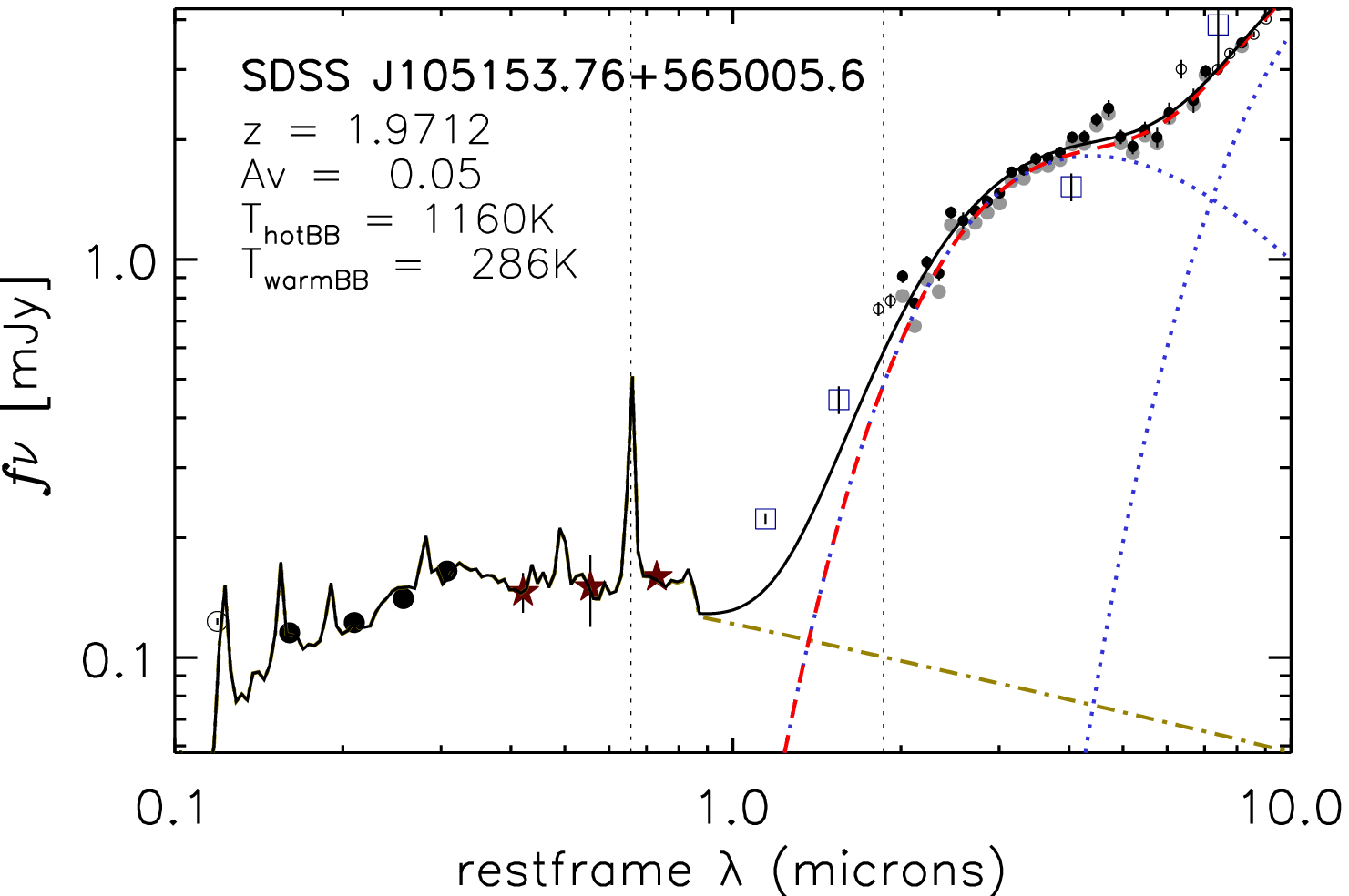}
\caption{continued}
\end{figure*}

\addtocounter{figure}{-1}
\begin{figure*}
\includegraphics[width=8.4cm]{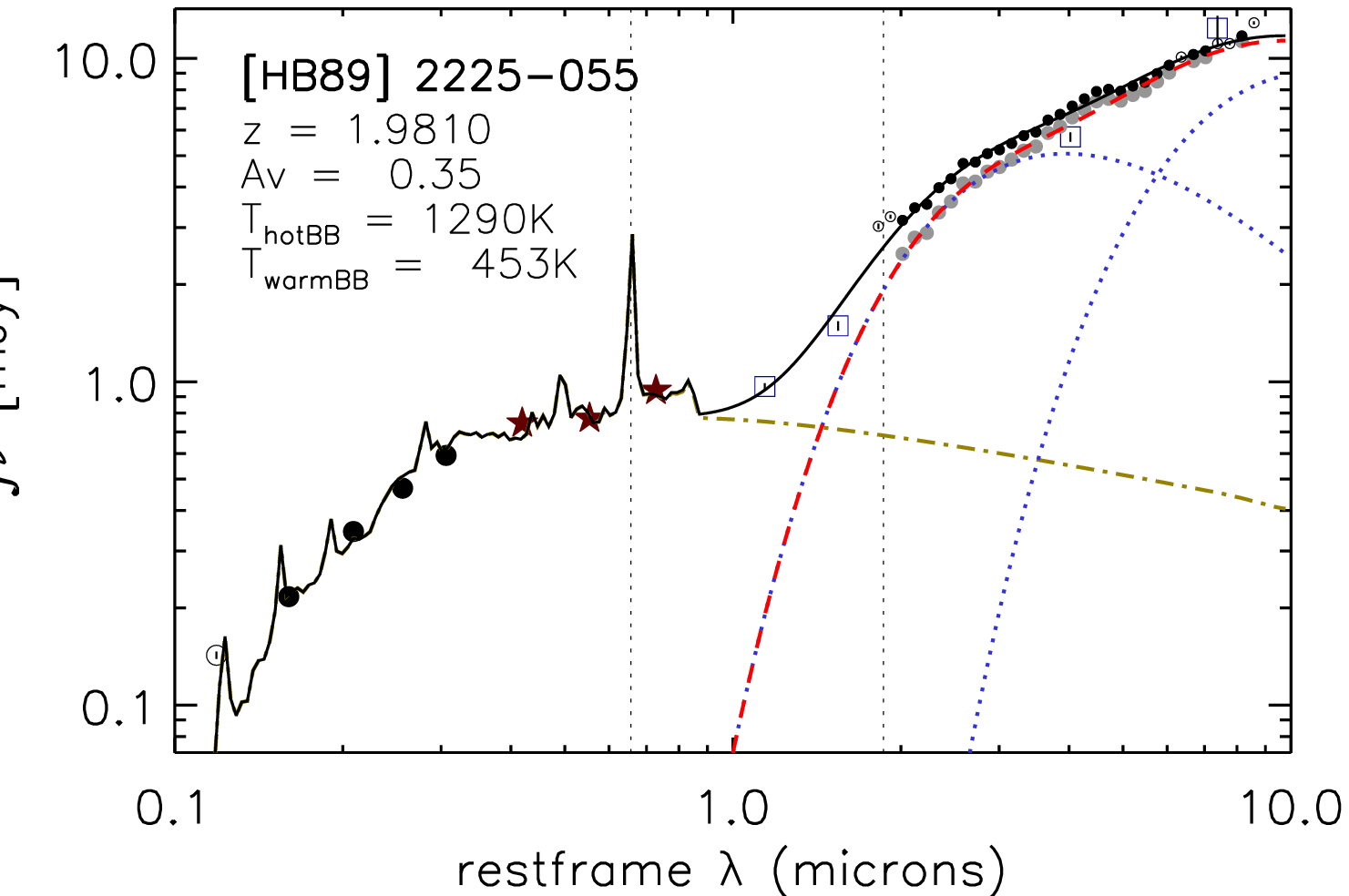}
\includegraphics[width=8.4cm]{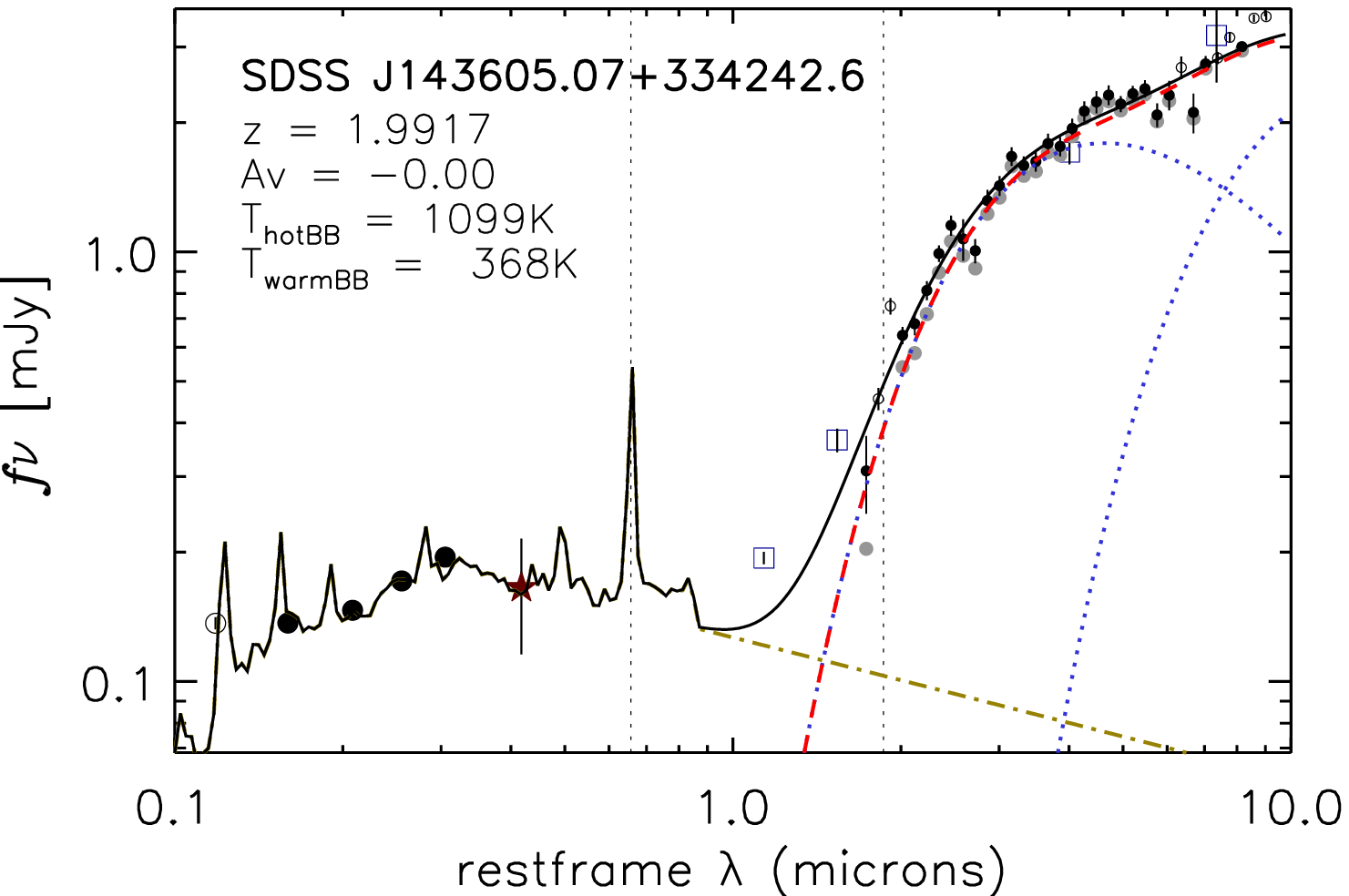}\vspace{0.4cm}
\includegraphics[width=8.4cm]{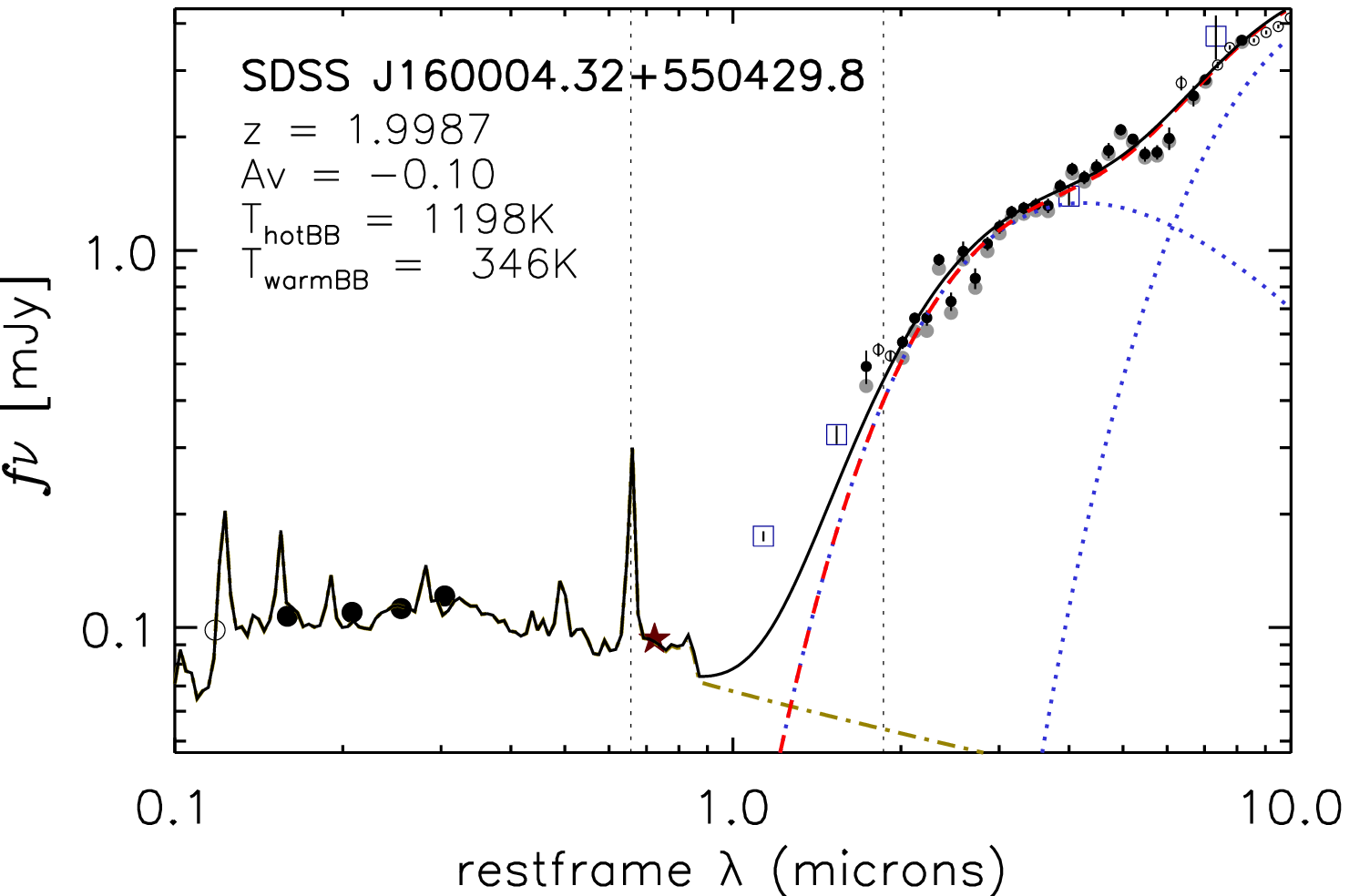}
\includegraphics[width=8.4cm]{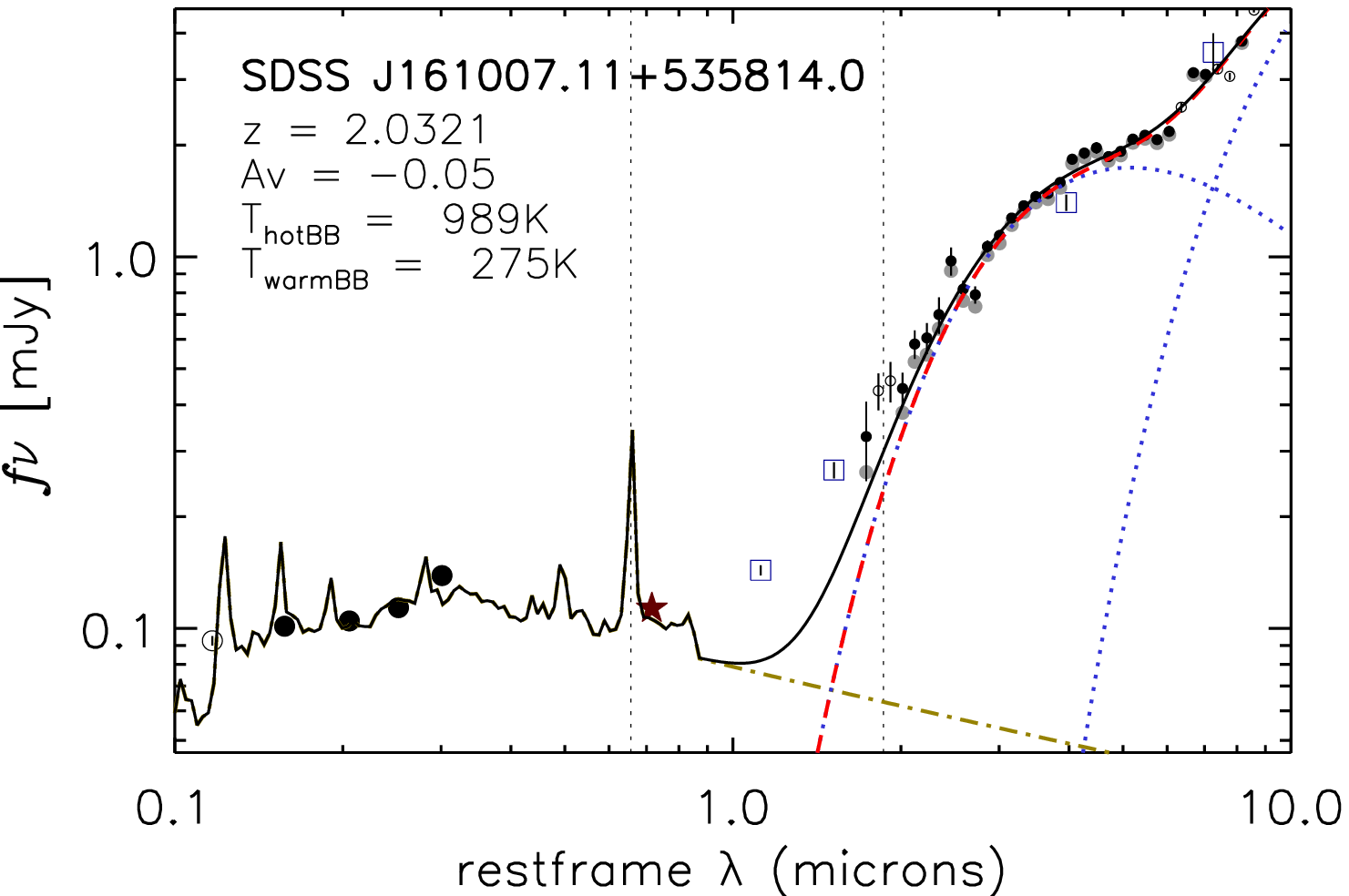}\vspace{0.4cm}
\includegraphics[width=8.4cm]{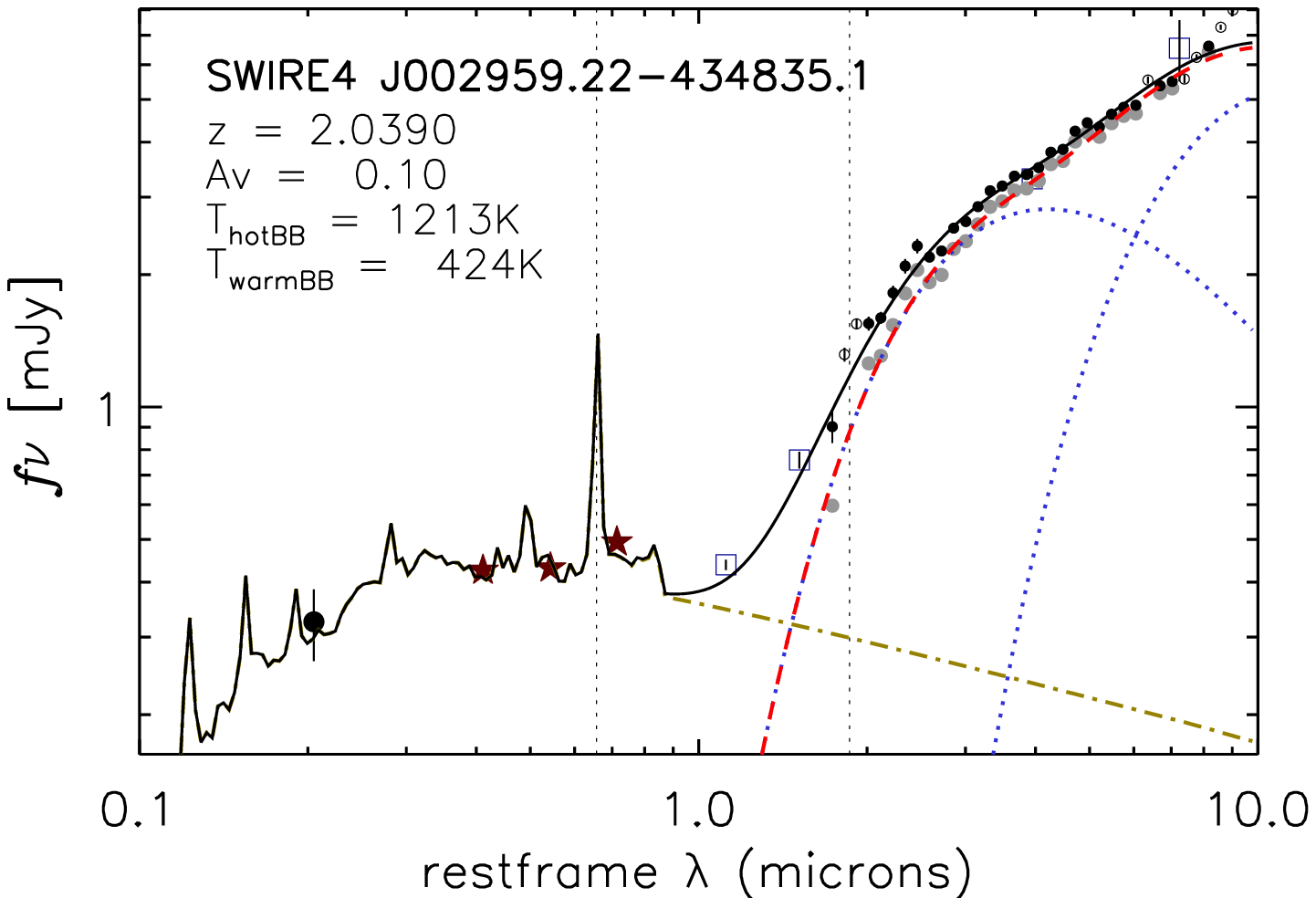}
\includegraphics[width=8.4cm]{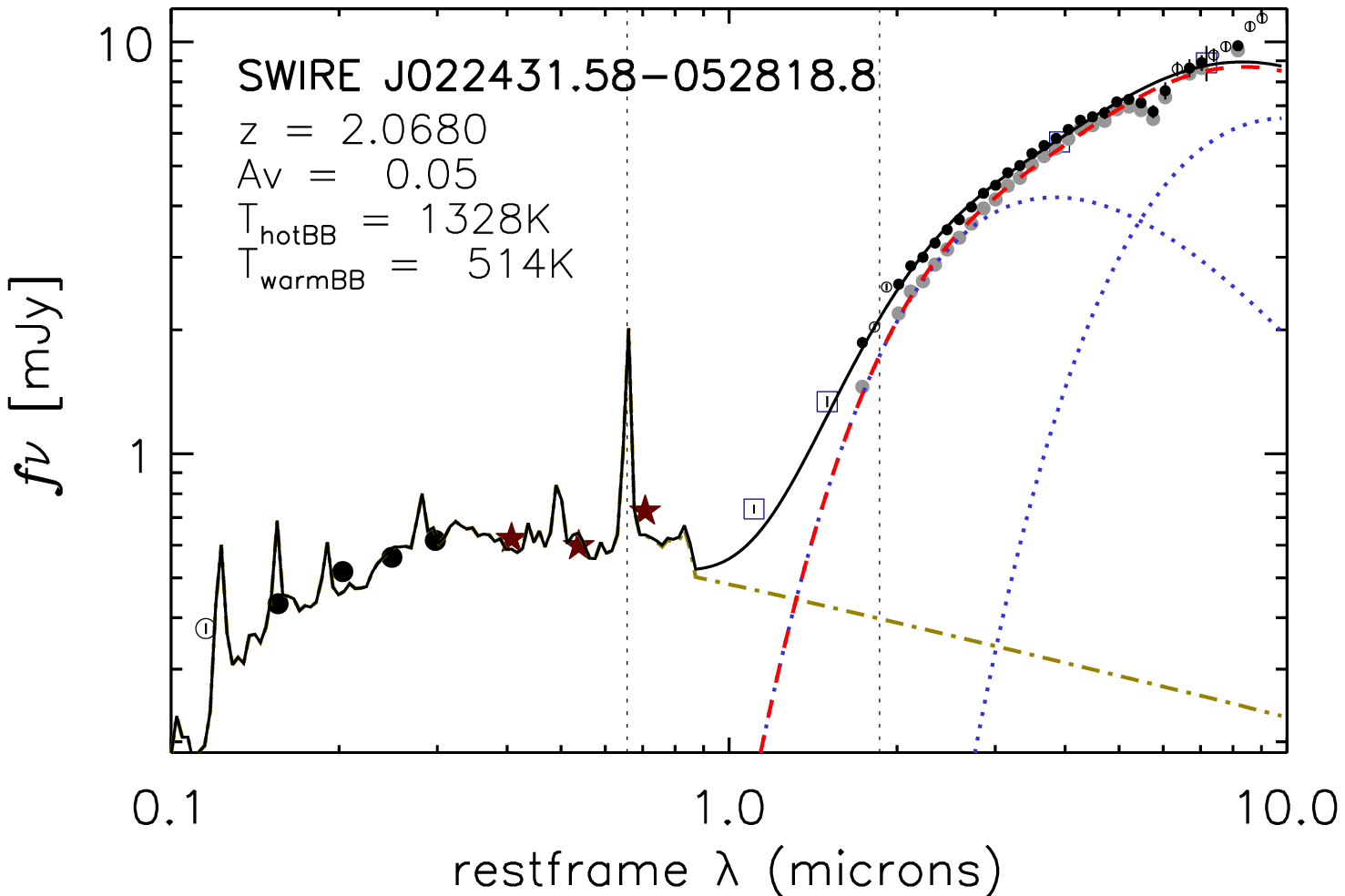}\vspace{0.4cm}
\includegraphics[width=8.4cm]{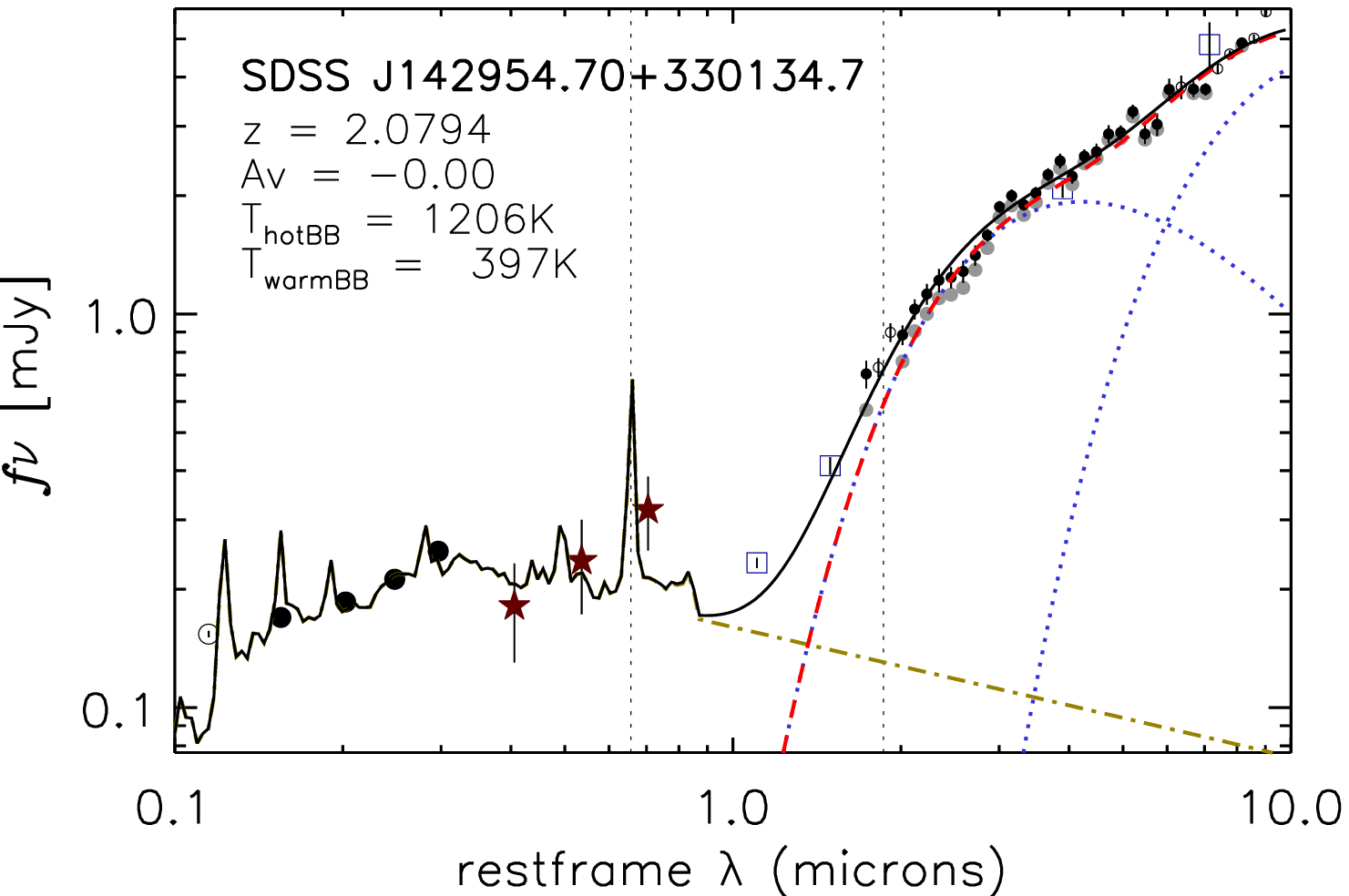}
\includegraphics[width=8.4cm]{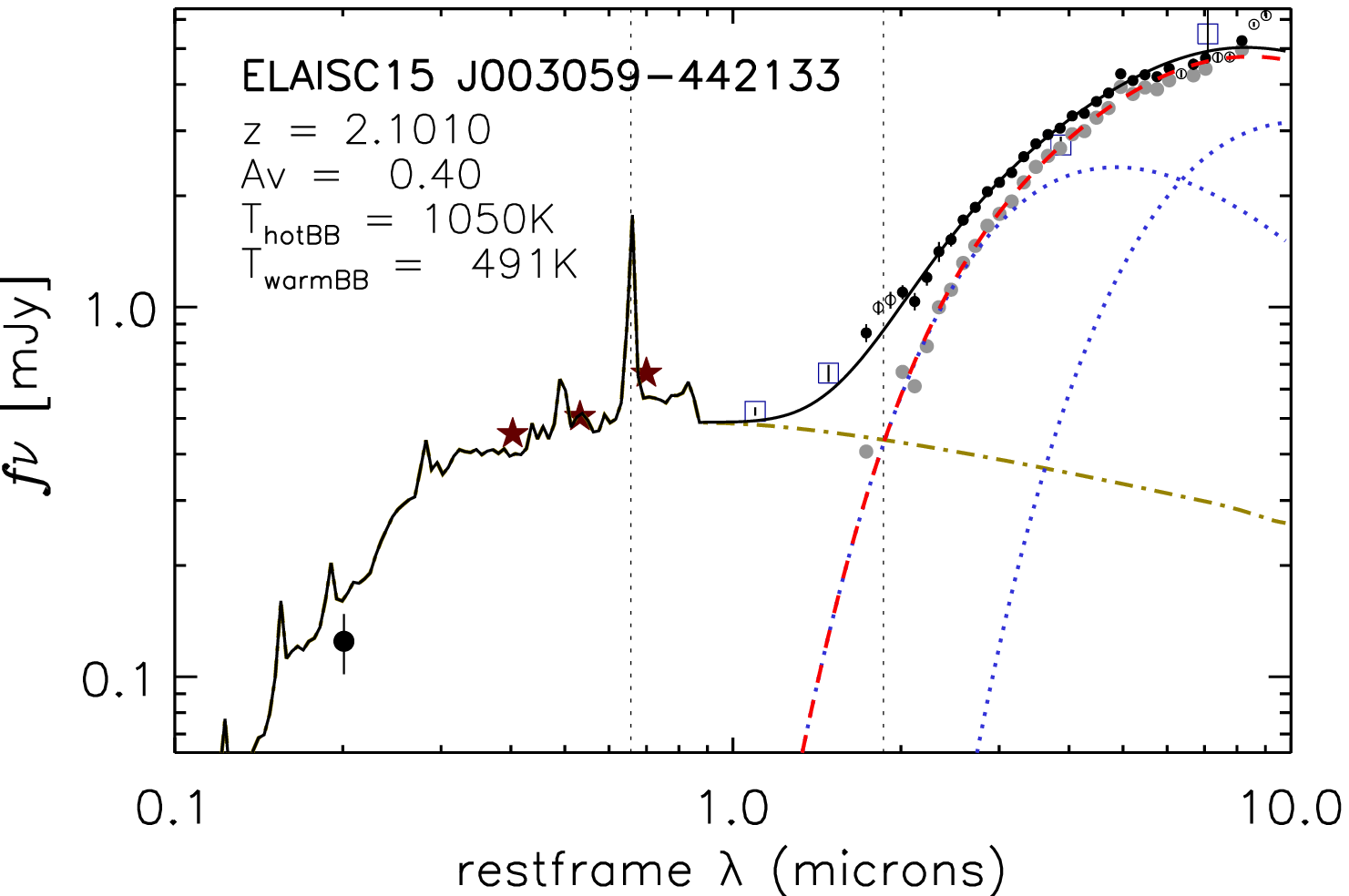}
\caption{continued}
\end{figure*}

\addtocounter{figure}{-1}
\begin{figure*}
\includegraphics[width=8.4cm]{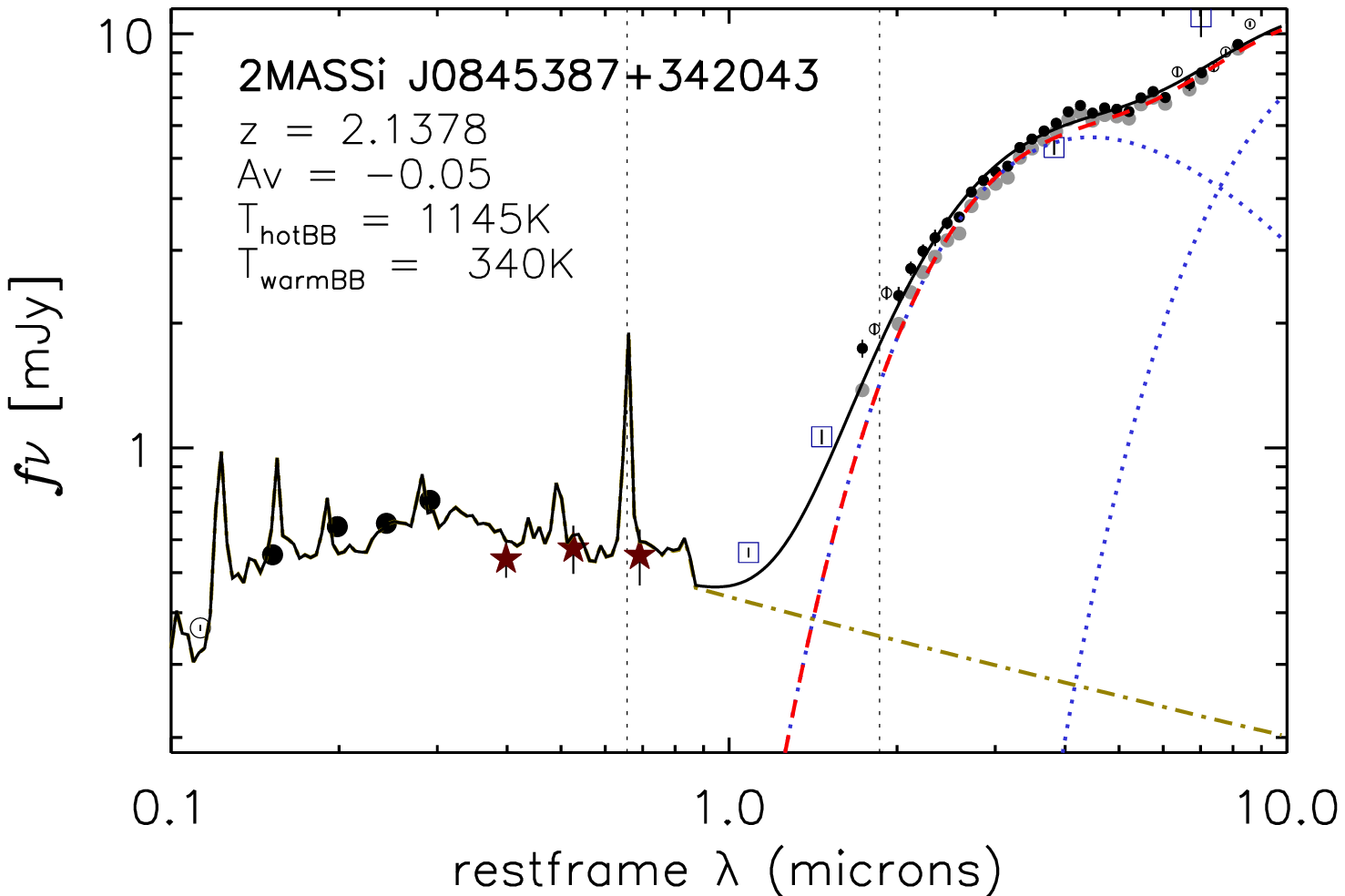}
\includegraphics[width=8.4cm]{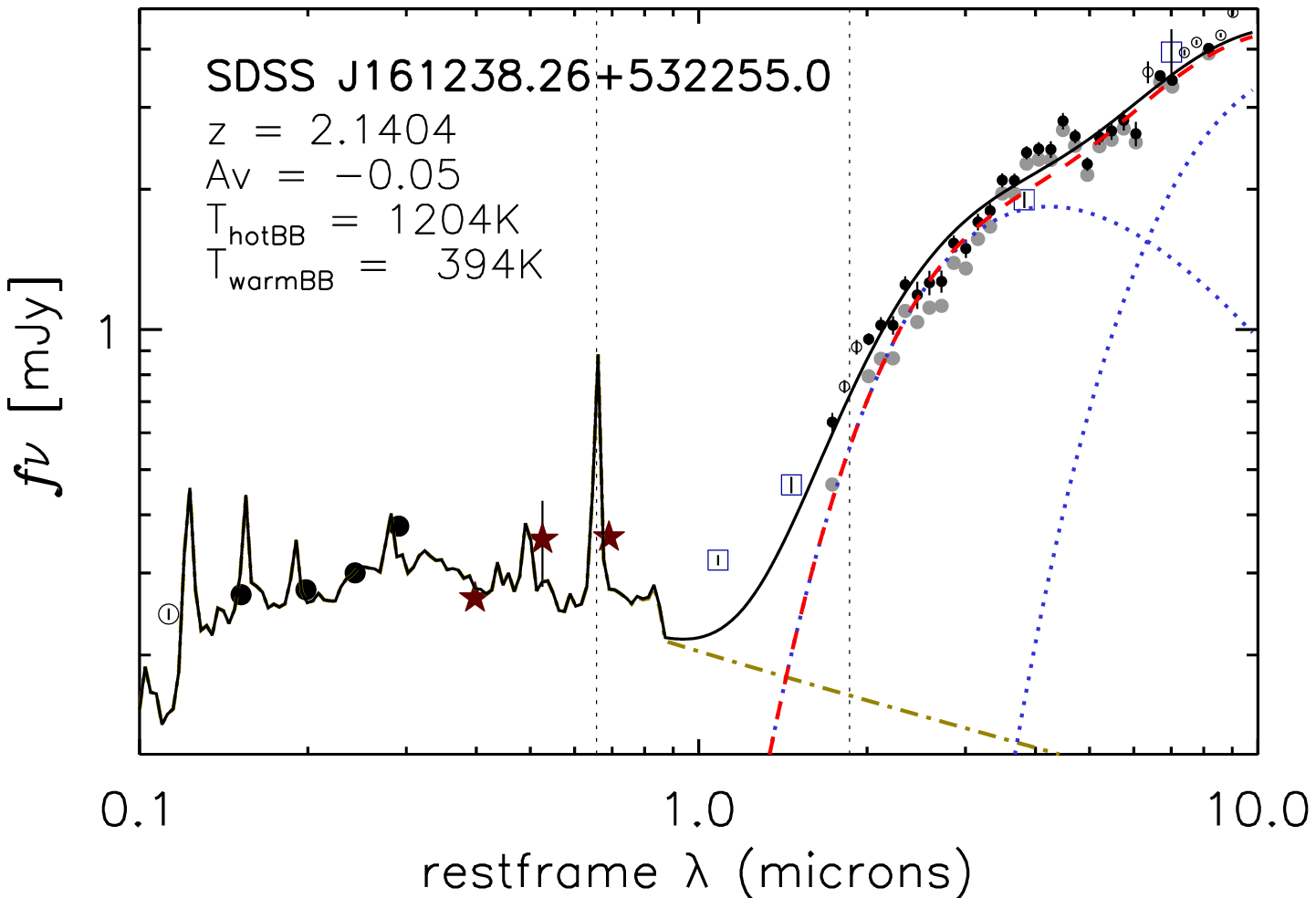}\vspace{0.4cm}
\includegraphics[width=8.4cm]{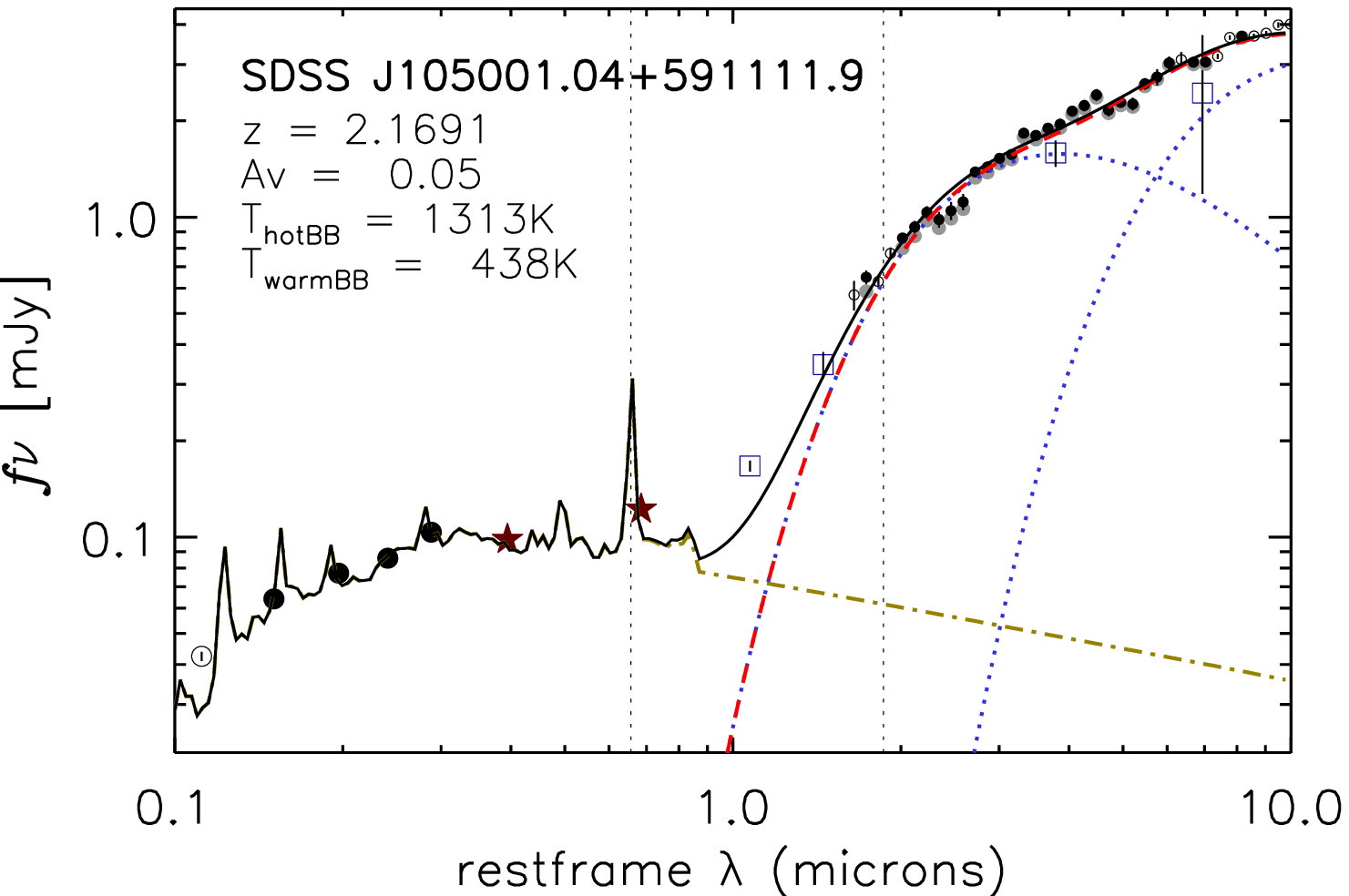}
\includegraphics[width=8.4cm]{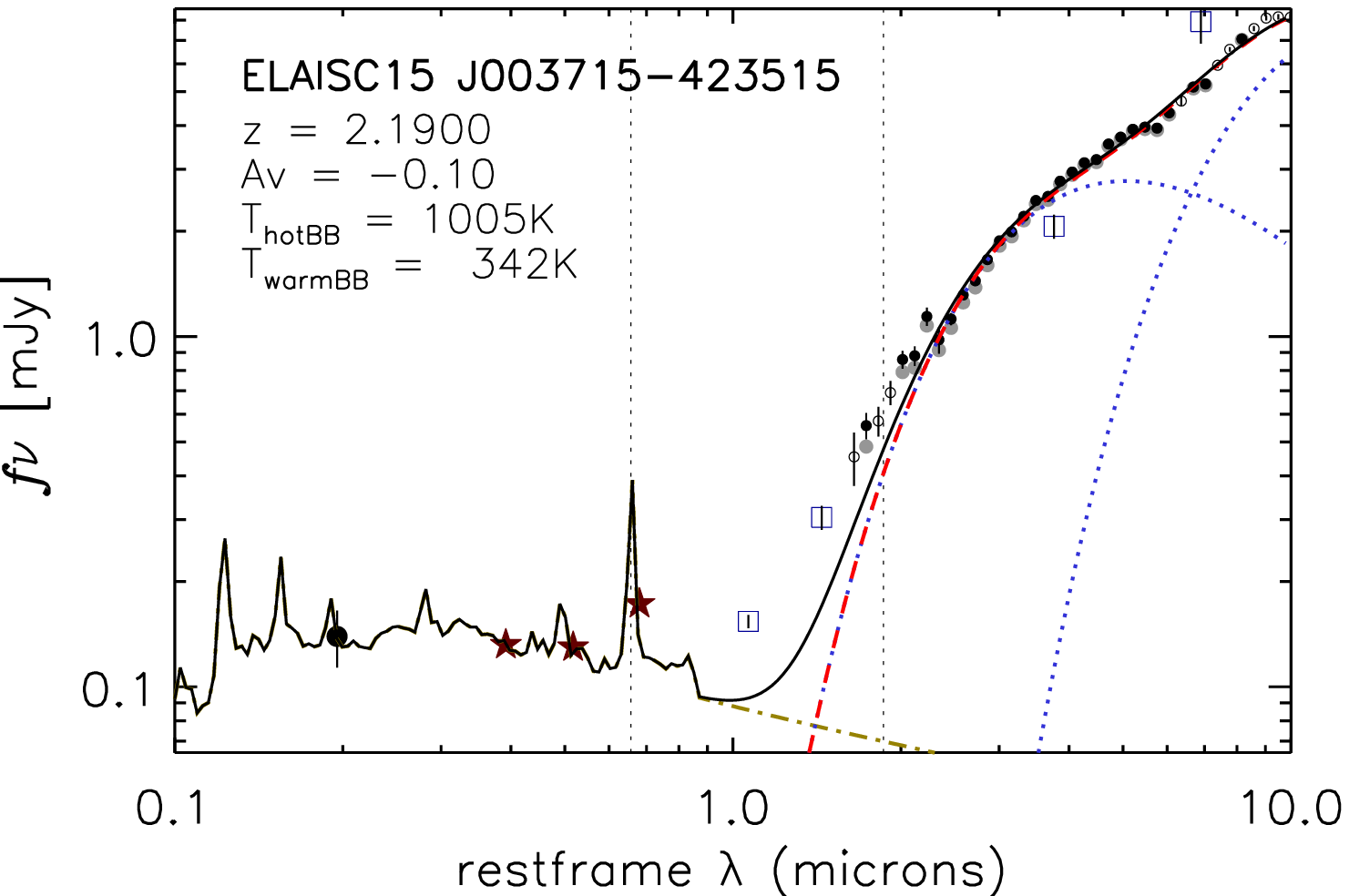}\vspace{0.4cm}
\includegraphics[width=8.4cm]{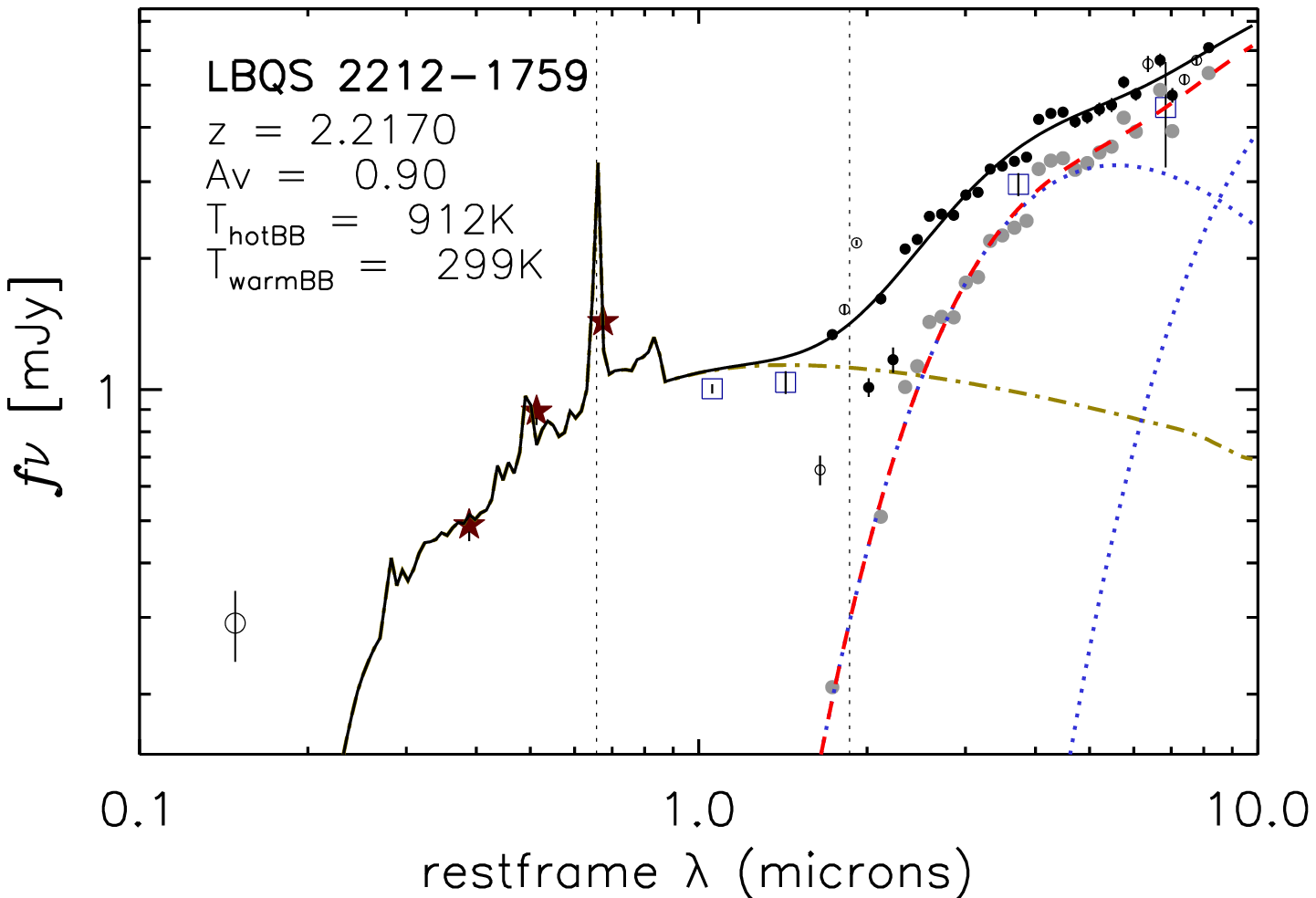}
\includegraphics[width=8.4cm]{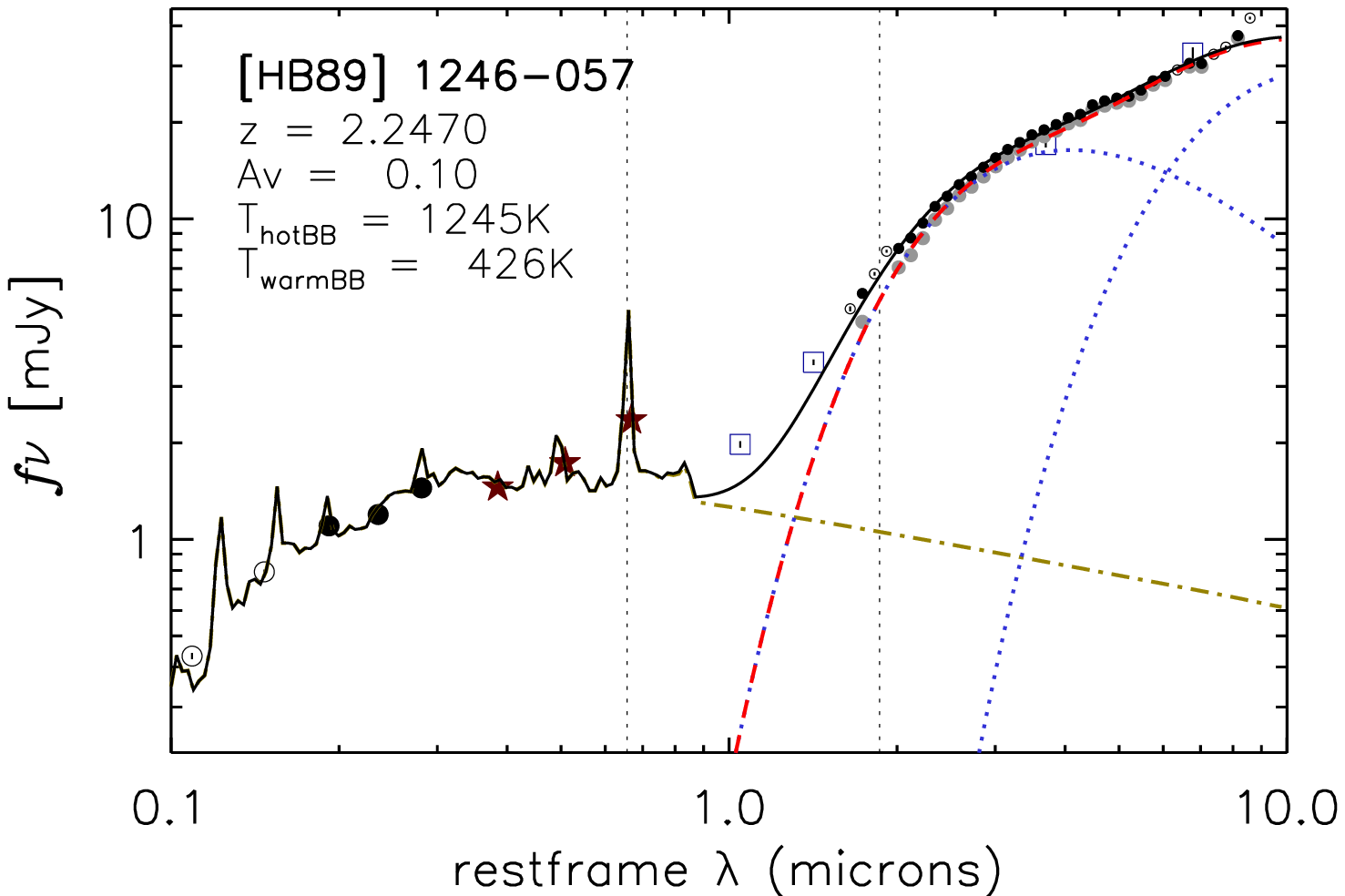}\vspace{0.4cm}
\includegraphics[width=8.4cm]{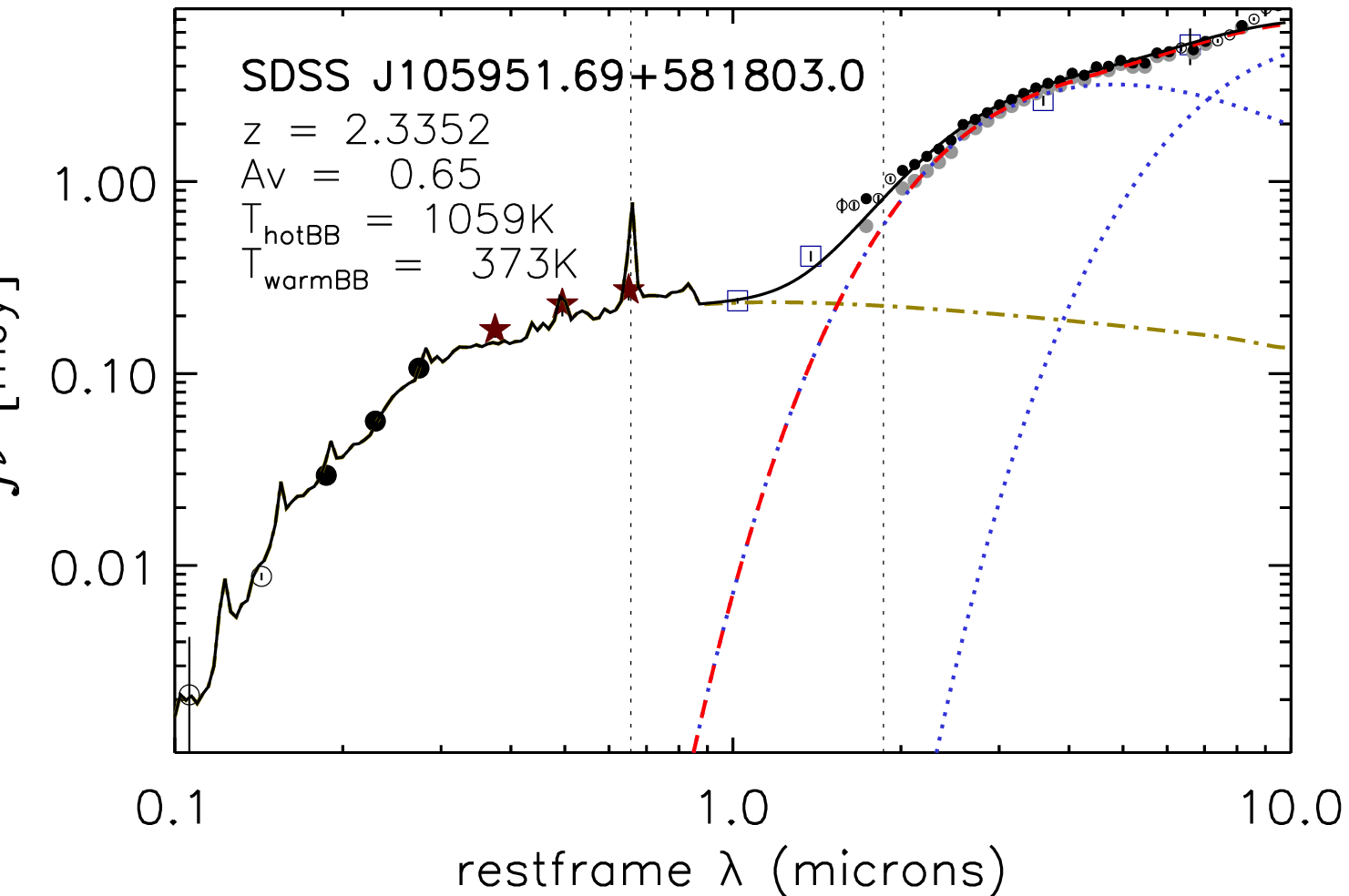}
\includegraphics[width=8.4cm]{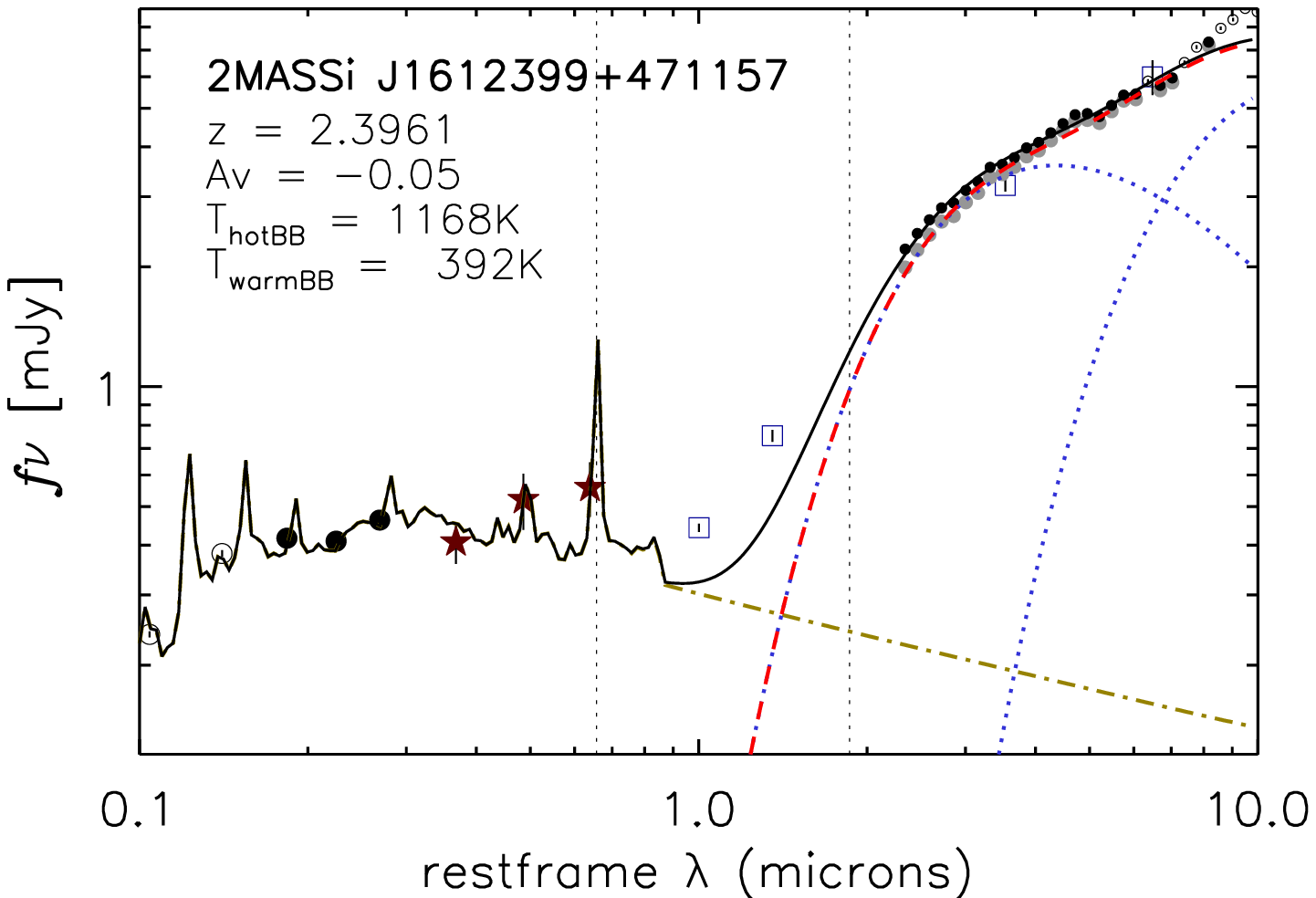}
\caption{continued}
\end{figure*}

\addtocounter{figure}{-1}
\begin{figure*}
\includegraphics[width=8.4cm]{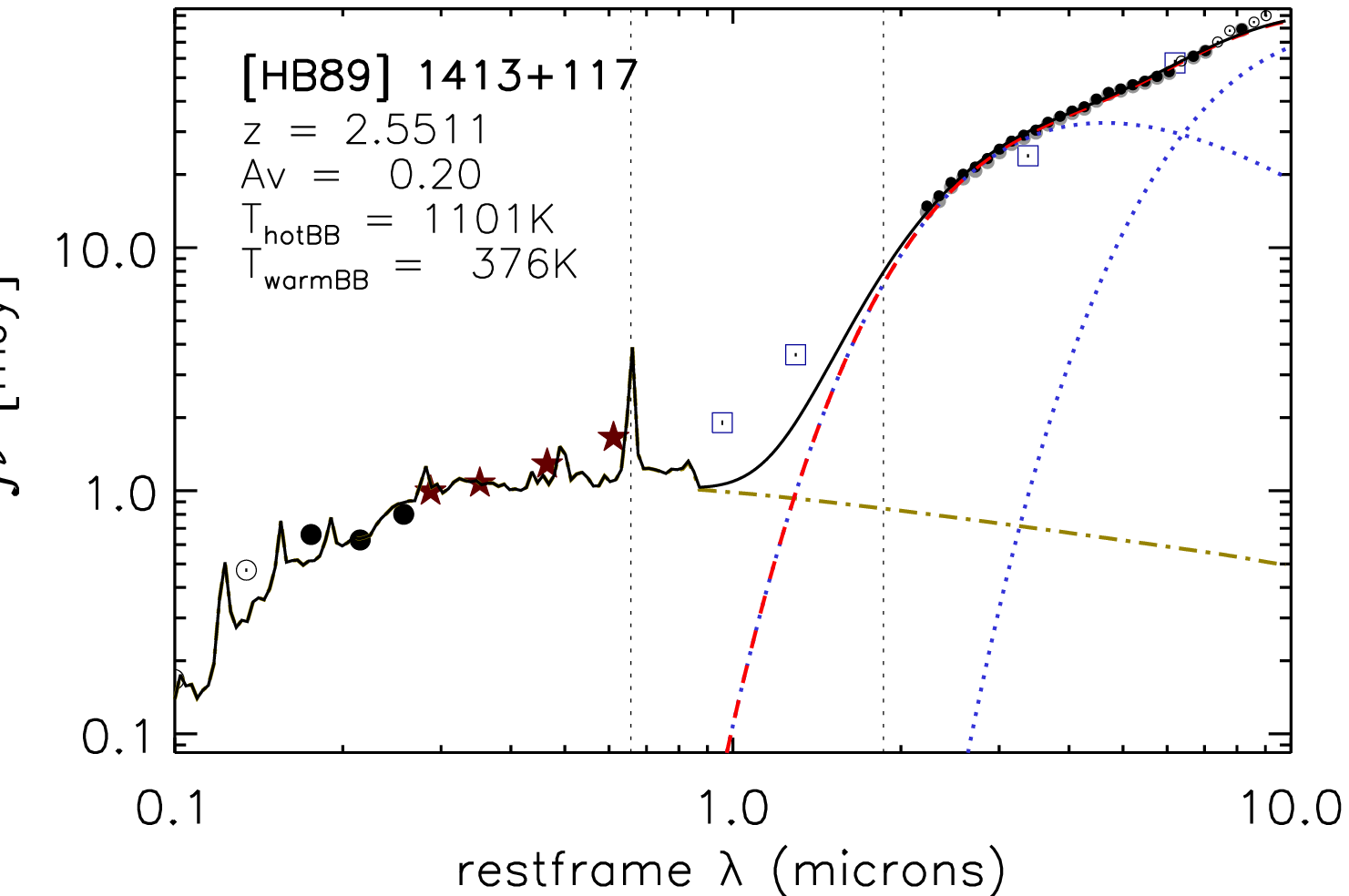}
\includegraphics[width=8.4cm]{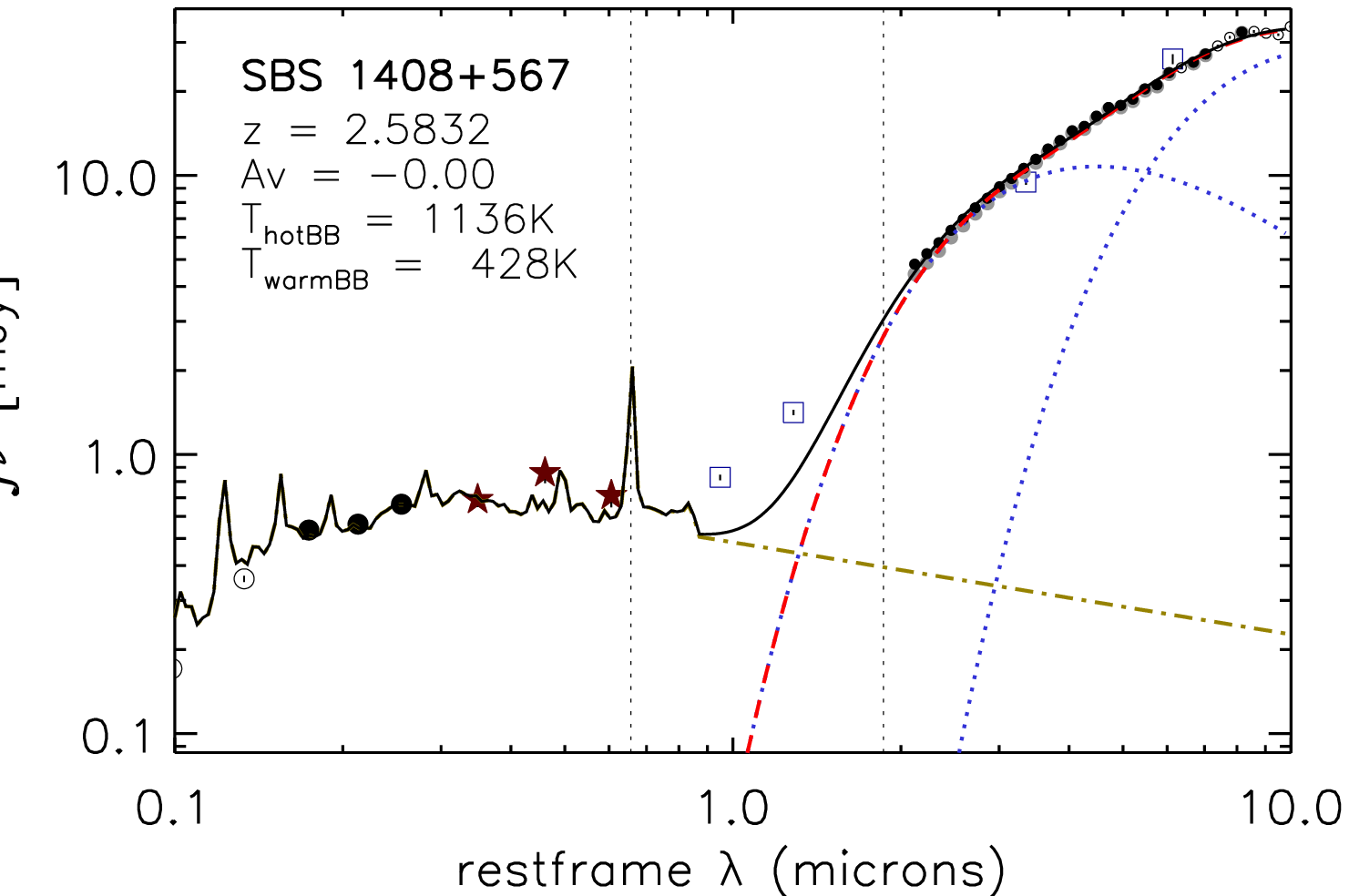}\vspace{0.4cm}
\includegraphics[width=8.4cm]{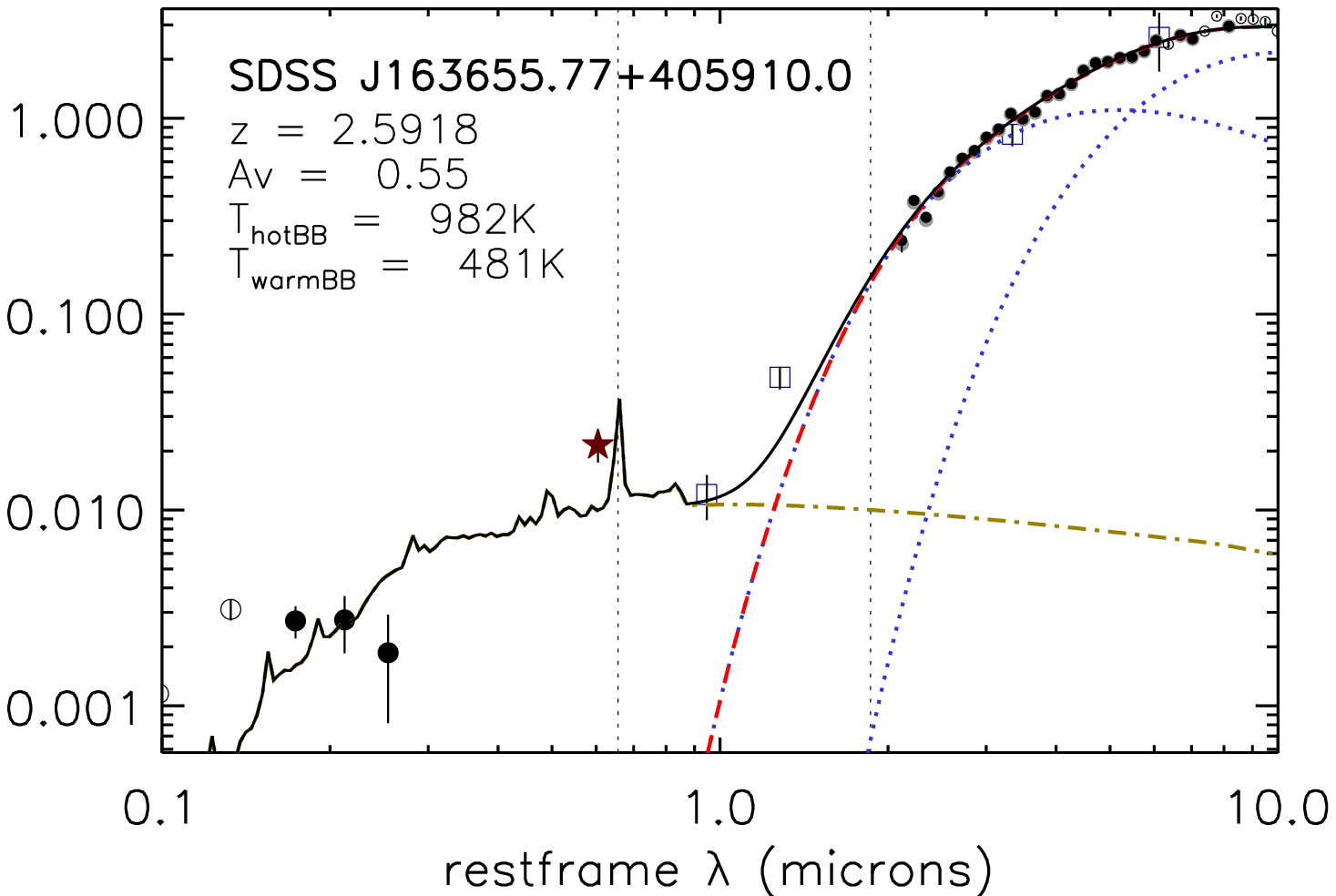}
\includegraphics[width=8.4cm]{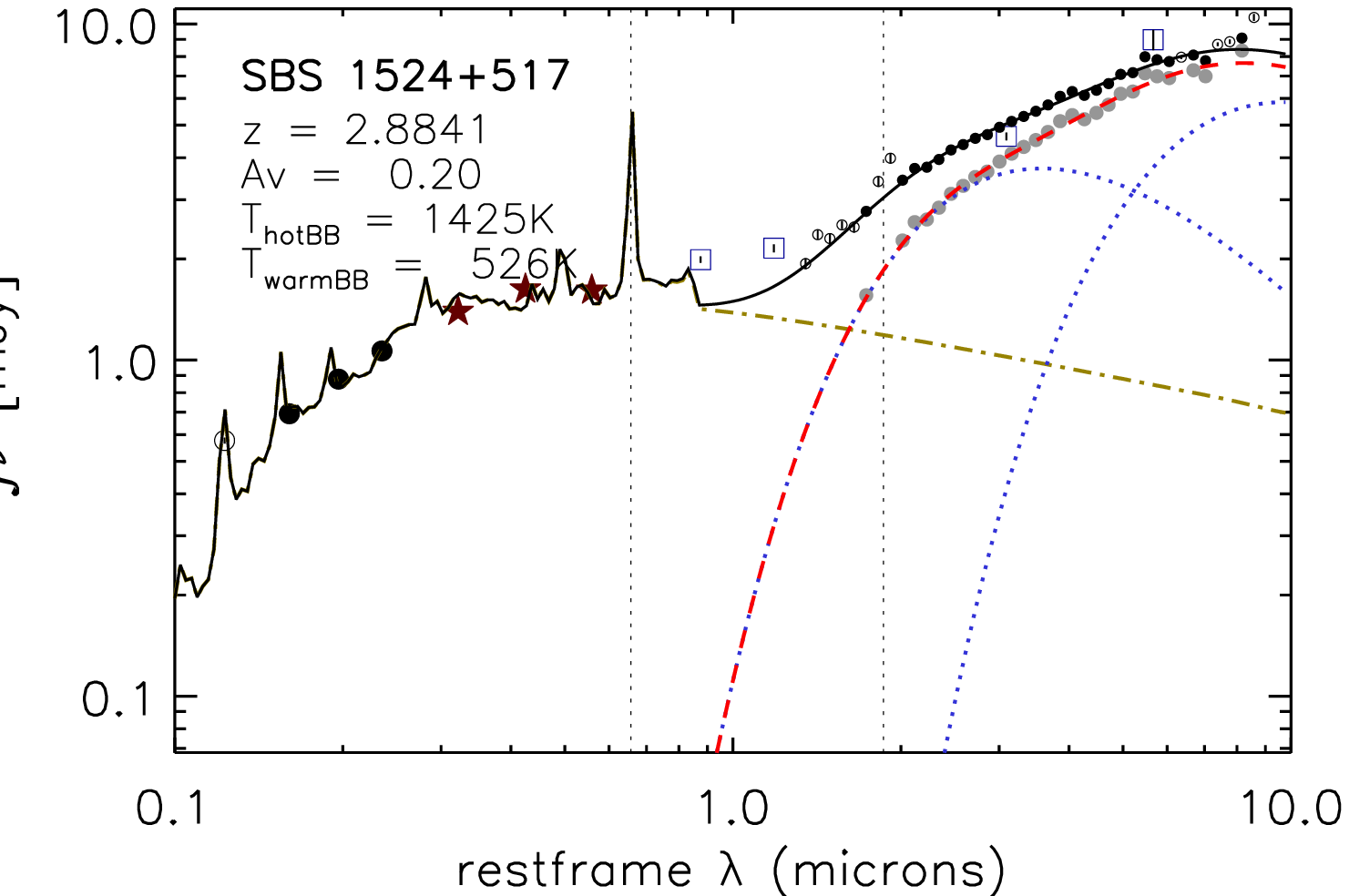}\vspace{0.4cm}
\includegraphics[width=8.4cm]{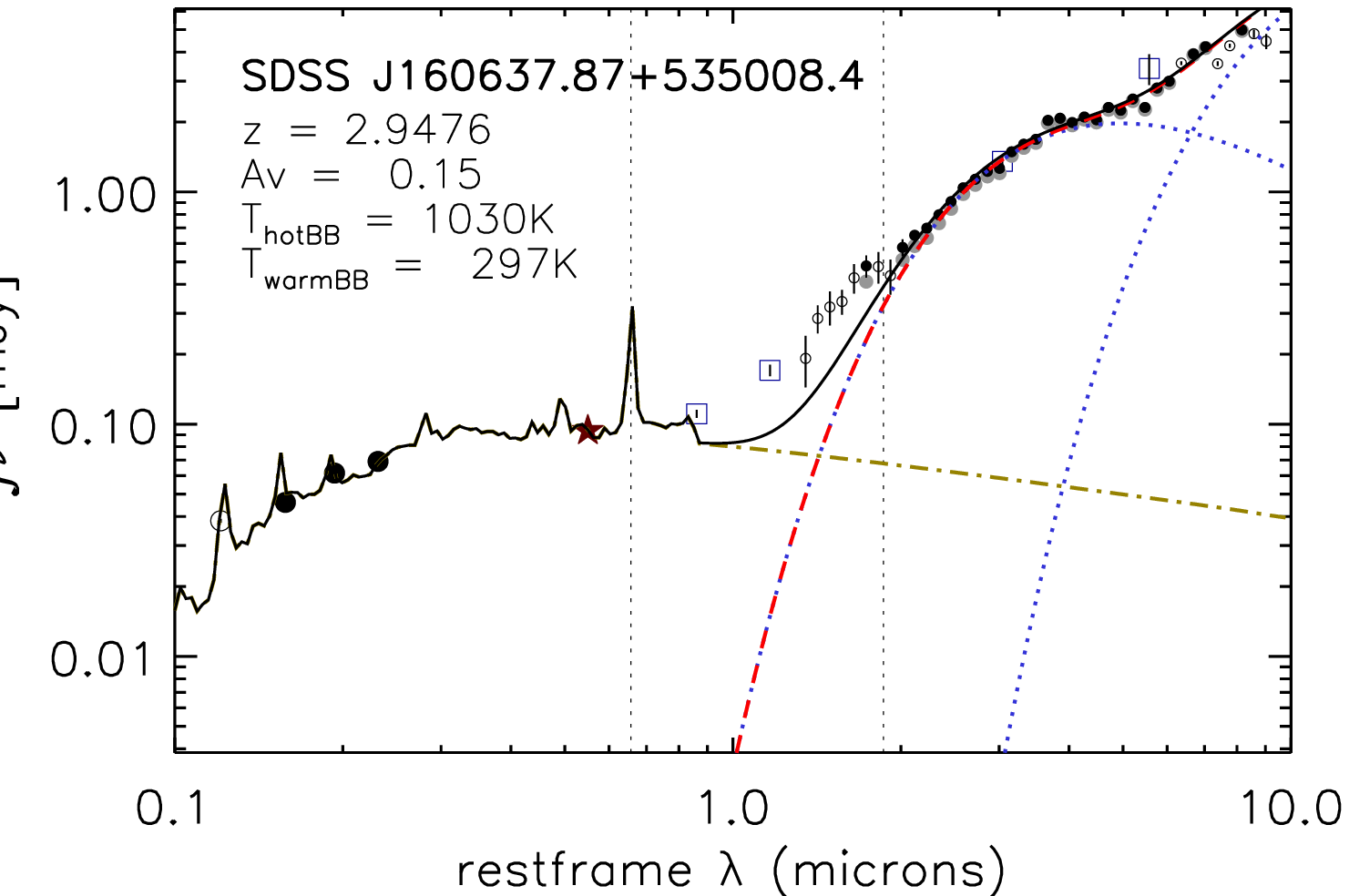}
\includegraphics[width=8.4cm]{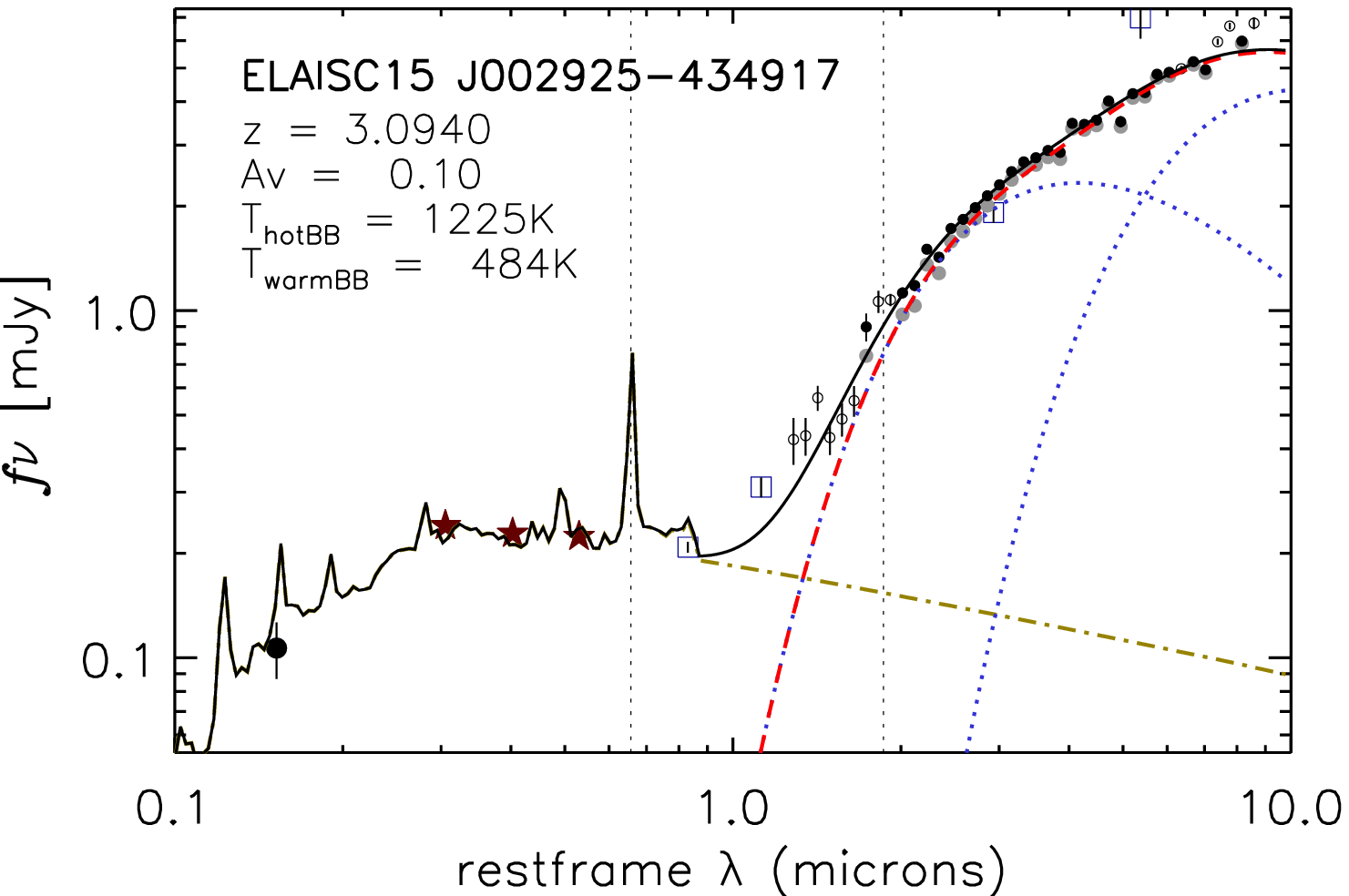}\vspace{0.4cm}
\includegraphics[width=8.4cm]{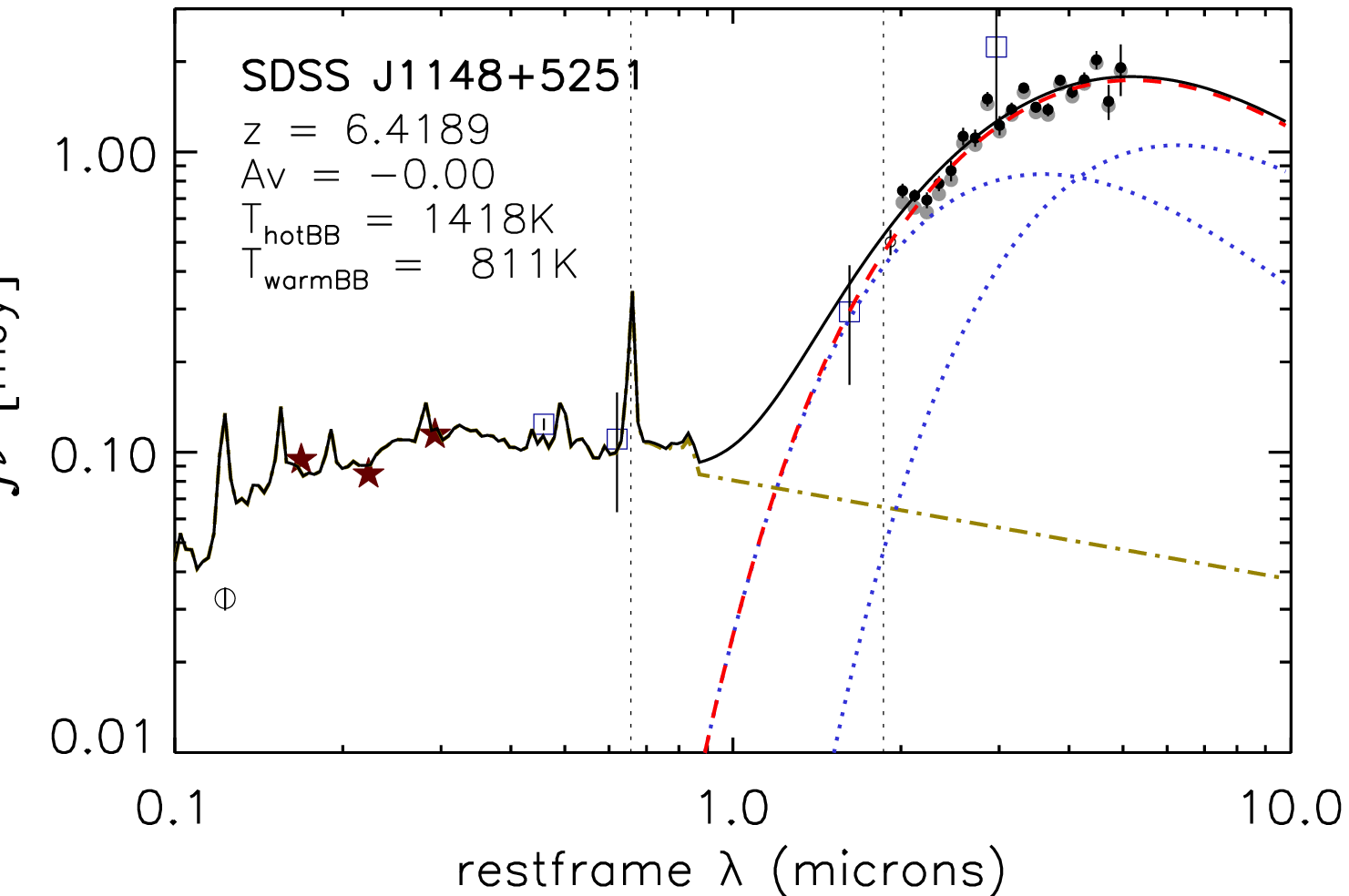}
\caption{continued}
\end{figure*}


\begin{thebibliography}{999}
\bibitem[Alonso-Herrero et al.(2011)]{Alonso-Herrero11}Alonso-Herrero A., et al., 2011, ApJ, 736, 82
\bibitem[Antonucci(1993)]{Antonucci93}Antonucci, R. 1993, ARA\&A, 31, 473
\bibitem[Barger et al.(2015)]{Barger15}Barger A. J., Cowie L. L., Owen, F. N., Chen C.-C., Hasinger G., Hsu L.-Y., Li, Y., 2015, ApJ, 801, 87
\bibitem[Barvainis(1987)]{Barvainis87}Barvainis, R. 1987, ApJ 320, 537
\bibitem[Bongiorno et al.(2012)]{Bongiorno12}Bongiorno, A., et al., 2012, MNRAS, 427, 3103
\bibitem[Burtscher et al.(2015)]{Burtscher15}Burtscher L., et al. 2015, A\&A, 578, 47
\bibitem[Clavel et al.(2006)]{Clavel06}Clavel J., Schartel N., Tomas L., 2006, A\&A, 446, 439
\bibitem[Cutri et al.(2012)]{Cutri12}Cutri R. M., et al. 2012, Explanatory Supplement to the WISE All-sky Data Release Products, http://adsabs.harvard.edu/abs/2012wise.rept....1C
\bibitem[Deo et al.(2011)]{Deo11}Deo, R. P., Richards, G. T., Nikutta, R., Elitzur, M., Gallagher, S. C., Ivezi\'c, Z, Hines, D., 2011, ApJ, 729, 108
\bibitem[Elvis et al.(1994)]{Elvis94}Elvis, M., Wilkes, B. J., McDowell, J. C. et al. 1994, ApJS, 95, 1
\bibitem[Feltre et al.(2012)]{Feltre12}Feltre, A., Hatziminaoglou, E., Fritz, J., Franceschini, A. 2012, MNRAS, 426, 120
\bibitem[Fritz et al.(2006)]{Fritz06}Fritz J., Franceschini A., Hatziminaoglou E. 2006, MNRAS, 366, 767
\bibitem[Gallagher et al.(2007)]{Gallagher07}Gallagher S. C., Richards G. T., Lacy M., Hines D. C., Elitzur, M., Storrie-Lombardi, L. J., 2007, ApJ, 661, 30
\bibitem[Glikman et al.(2006)]{Glikman06}Glikman E., Helfand D. J., White R. L., 2006, ApJ, 640, 579
\bibitem[Gordon et al.(2003)]{Gordon03}Gordon K. D., Clayton G. C., Misselt K. A., Landolt A. U. Wolff M. J., 2003, ApJ, 594, 279
\bibitem[Granato \& Danese(1994)]{Granato94}Granato, G. L. and Danese, L., 1994, MNRAS, 268, 235
\bibitem[Haas et al.(2003)]{Haas03}Haas, M., Klaas, U., M\"uller, S. A. H., Bertoldi, F., Camenzind, M., et al. 2003, A\&A, 402, 87
\bibitem[Hao et al.(2010)]{Hao10}Hao, H., et al., 2010, ApJL, 724, L59
\bibitem[Hatziminaoglou et al.(2005)]{Hatziminaoglou05}Hatziminaoglou, E., P\'erez-Fournon, I., Polletta, M., Afonso-Luis, A., Hern\'an-Caballero, A., Montenegro-Montes, F. M. et al. 2005, AJ, 129, 1198
\bibitem[Hern\'an-Caballero \& Hatziminaoglou(2011)]{Hernan-Caballero11}Hern\'an-Caballero A. \& Hatziminaoglou E., 2011, MNRAS, 414, 500
\bibitem[Hern\'an-Caballero et al.(2009)]{Hernan-Caballero09}Hern\'an-Caballero A. et al., 2009, MNRAS, 395, 1695
\bibitem[Hern\'an-Caballero et al.(2015)]{Hernan-Caballero15}Hern\'an-Caballero A. et al., 2015, ApJ, 803, 109
\bibitem[Hern\'an-Caballero et al.(2016)]{Hernan-Caballero16}Hern\'an-Caballero, A., Spoon, H. W. W., Lebouteiller, V., Barry D. J., 2016, MNRAS, 455, 1796
\bibitem[Hern\'an-Caballero(2012)]{Hernan-Caballero12}Hern\'an-Caballero, A., 2012, MNRAS, 427, 816
\bibitem[Hill et al.(2014)]{Hill14}Hill, A. R., Gallagher, S. C., Deo, R. P., Peeters, E., Richards, G. T., 2014, MNRAS, 438, 2317
\bibitem[H\"onig et al.(2006)]{Hoenig06}H\"onig S. F., Beckert T., Ohnaka K., Weigelt G., 2006, A\&A, 452, 459
\bibitem[Hook et al.(1994)]{Hook94}Hook I. M., McMahon R. G., Boyle B. J., Irwin M. J., 1994, MNRAS, 268, 305
\bibitem[Hubeny et al.(2001)]{Hubeny01}Hubeny I., Blaes O., Krolik J.~H., Agol E., 2001, ApJ, 559, 680
\bibitem[Kim et al.(2015)]{Kim15}Kim, D., et al., 2015, ApJSS, 216, 17
\bibitem[Kishimoto et al.(2008)]{Kishimoto08}Kishimoto, M., Antonucci, R., Blaes, O., et al. 2008, Nature, 454, 492
\bibitem[Koz\l{}owski(2016)]{Kozlowski16}Koz\l{}owski S., 2016, ApJ in press (astro-ph/1604.05858)
\bibitem[Krawczyk et al.(2015)]{Krawczyk15}Krawczyk, C. M., Richards, G. T., Gallagher, S. C., Leighly, K. M., Hewett, P. C., Ross, N. P., Hall, P. B., 2015, AJ, 149, 203
\bibitem[Landt et al.(2008)]{Landt08}Landt H., Padovani P., Giommi P., Perri M., Cheung C. C., 2008, ApJ, 676, 87
\bibitem[Lawrence et al.(2007)]{Lawrence07}Lawrence, A., Warren, S. J., Almaini, O. et al. 2007, MNRAS, 379, 15599
\bibitem[Lebouteiller et al.(2011)]{Lebouteiller11}Lebouteiller V., Barry D.J., Spoon H.W.W., Bernard-Salas J., Sloan G.C., Houck J.R., \& Weedman D., 2011, ApJS, 196, 8
\bibitem[Lebouteiller et al.(2015)]{Lebouteiller15}Lebouteiller V., Barry, D. J., Goes, C., Sloan G. C., Spoon, H. W. W., Weedman, D. W., Bernard-Salas, J., Houck, J. R., 2015, ApJSS, 218, 21
\bibitem[Lutz et al.(2008)]{Lutz08}Lutz D. et al., 2008, ApJ, 684, 853L
\bibitem[Ma \& Wang(2013)]{Ma13}Ma X.-C., Wang T.-G., 2013, MNRAS, 430, 3445
\bibitem[Maiolino et al.(2007)]{Maiolino07}Maiolino R., Shemmer O., Imanishi M., Netzer H., Oliva E., Lutz D., Sturm E., 2007, A\&A, 468, 979
\bibitem[Markwardt(2009)]{Markwardt09}Markwardt, C.~B.\ 2009, Astronomical Data Analysis Software and Systems XVIII, 411, 251
\bibitem[Mateos et al.(2015)]{Mateos15}Mateos, S. et al., 2015, MNRAS, 449, 1422
\bibitem[Mateos et al.(2016)]{Mateos16}Mateos, S. et al., 2016, ApJ, 819, 166
\bibitem[McMahon et al.(2013)]{McMahon13}McMahon R. G., Banerji M., Gonzalez E., Kopsov S. E., B\'ejar V. J., Lodieu N., Rebolo R., the VHS collaboration 2013, The Messenger, 154, 35
\bibitem[Merloni(2015)]{Merloni15}Merloni A., 2015, in Haardt F., ed., Astrophysical Black Holes. Springer,
International Publishing AG, Cham. preprint (arXiv:1505.04940)
\bibitem[Meusinger \& Weiss(2013)]{Meusinger13}Meusinger H., Weiss V., 2013, A\&A, 560, 140
\bibitem[Mor et al.(2009)]{Mor09}Mor R., Netzer H., Elitzur M., 2009, ApJ, 705, 298
\bibitem[Nenkova et al.(2002)]{Nenkova02}Nenkova M., Ivezic Z., Elitzur M., 2002, ApJ, 570, 9
\bibitem[Nenkova et al.(2008)]{Nenkova08}Nenkova M., Sirocky M.M., Ivezic Z., Elitzur M., 208, ApJ, 685, 147
\bibitem[Netzer et al.(2007)]{Netzer07}Netzer H. et al., 2007, ApJ, 666, 806
\bibitem[Netzer(2015)]{Netzer15}Netzer H., 2015, ARA\&A, 53, 365
\bibitem[Neugebauer et al.(1979)]{Neugebauer79}Neugebauer G. et al., 1979, ApJ, 230, 79
\bibitem[P\^aris et al.(2012)]{Paris12}P\^aris I. et al., 2012, A\&A, 548, 66
\bibitem[Rees et al.(1969)]{Rees69}Rees M. J. et al., 1969, Nature, 223, 788
\bibitem[Richards et al.(2001)]{Richards01}Richards G. T. et al., 2001, AJ, 121, 2308 
\bibitem[Richards et al.(2006)]{Richards06}Richards, G. T., et al., 2006, ApJS, 166, 470
\bibitem[Rowan-Robinson(2000)]{Rowan-Robinson00}Rowan-Robinson M., 2000, MNRAS, 316, 885
\bibitem[Salpeter(1955)]{Salpeter55}Salpeter E. E., 1955, ApJ, 121, 161
\bibitem[Schlegel et al.(1998)]{Schlegel98}Schlegel, D. J., Finkbeiner, D. P., and Davis, M. 1998, ApJ, 500, 525
\bibitem[Schmidt \& Hines(1999)]{Schmidt99}Schmidt G. D., Hines D. C., 1999, ApJ, 512, 125
\bibitem[Schneider et al.(2010)]{Schneider10}Schneider D. P., et al., 2010, AJ, 139, 2360
\bibitem[Shakura \& Sunyaev(1973)]{Shakura73}Shakura N. I., Sunyaev R. A., 1973, A\&A, 24, 337
\bibitem[Shen(2016)]{Shen16}Shen Y., 2016, ApJ, 817, 55
\bibitem[Siebenmorgen et al.(2015)]{Siebenmorgen15}Siebenmorgen R., Heymann F., Efstathiou A., 2015, A\&A, 583, 120
\bibitem[Skrutskie et al.(2006)]{Skrutskie06}Skrutskie, M. F., Cutri, R. M., Stiening, R. et al. 2006, AJ, 131, 1163
\bibitem[Soifer et al.(2004)]{Soifer04}Soifer B. T. et al., 2004, ApJSS, 154, 151
\bibitem[Stalevski et al.(2012)]{Stalevski12}Stalevski M., Fritz J., Baes M., Nakos T., Popovi\'c L. C., 2012, MNRAS, 420, 2756
\bibitem[Stalevski et al.(2016)]{Stalevski16}Stalevski M., Ricci C., Ueda Y., Lira P., Fritz J., Baes M., 2016, MNRAS in press (astro-ph/1602.06954)
\bibitem[Suganuma et al.(2006)]{Suganuma06}Suganuma M., et al., 2006, ApJ, 639, 46
\bibitem[Trump et al.(2006)]{Trump06}Trump J. R. et al., 2006, ApJSS, 165, 1
\bibitem[Vanden Berk et al.(2001)]{VandenBerk01}Vanden Berk, D. E., et al. 2001, AJ, 122, 549  
\bibitem[Veron \& Veron(2006)]{Veron06}V\'eron-Cetty M.P., V\'eron P., 2006, A\&A, 455, 773
\bibitem[Willott et al.(2003)]{Willott03}Willott C. J., et al. 2003, MNRAS, 339, 397
\bibitem[Wright et al.(2010)]{Wright10}Wright E. L., et al., 2010, AJ, 140, 1868
\end{thebibliography}
\end{document}